\documentclass[11pt]{amsart}
\usepackage{entete}
\newcommand{\CH}{\Sigma}
\newcommand{\Queue}{\operatorname{Q}}
\newcommand{\lift}[2]{#1(#2)}
\newcommand{\Liftop}[2]{\lambda^{#1}_{#2}}

\newcommand{\vl}[1]{\operatorname{v}_*(#1)}

\newcommand{\central}{c}
\newcommand{\previous}{\ell}
\newcommand{\next}{\ell'}

\newcommand{\pointset}{\mathbb{A}}
\newcommand{\lineset}{\mathbb{L}}

\newcommand{\Map}[3]{#1 : #2 \rightarrow #3}
\newcommand{\MapLight}[3]{#2 \rightarrow #3}

\newcommand{\universal}[1]{\widehat{#1}}

\newcommand{\CoSur}[1]{\mathbb{B}_{#1}}  %% the covering surface

\newcommand{\groundbodies}{\overline{{\mathcal D}}}  %% Set of disks in the ground topological plane
\newcommand{\disks}{{\mathcal D}}     %% Set of disks in the covering surface
\newcommand{\straddle}{\widehat{{\mathcal D}}}     %% the o_i(k) + the central body
\newcommand{\size}{n}      %%% number of disks
\newcommand{\nbsheets}{k}  %%% number of sheets of the ambient surface
\newcommand{\nbbbls}{h}    %%% number of sheets of the ambient surface

\newcommand{\settopface}{{\bf \hat{1}}}
\newcommand{\setbotface}{{\bf \hat{0}}}
\newcommand{\botface}{\hat{0}}

\newcommand{\p}{p}

\newcommand{\pbackop}{\p_{\wordback}}

\newcommand{\pforwop}{\p_{\wordforw}}

\newcommand{\pforwstarop}{\p_{\wordforw_*}}

\newcommand{\Bback}[1]{B_{\wordback}(#1)}
\newcommand{\Bforw}[1]{B_{\wordforw}(#1)}

\newcommand{\Lea}{{\mathscr{A}}}
\newcommand{\Lealea}{\Lea^{\operatorname{lea}}}

\newcommand{\Ent}{{\mathscr{B}}}
\newcommand{\Entent}{\Ent^{\operatorname{ent}}}

\newcommand{\Enttwo}{\Ent^2}
\newcommand{\Entthr}{\Ent^3}
\newcommand{\Entfou}{\Ent^4}

\newcommand{\opLeaveHG}{\Lea_0}
\newcommand{\LeaveHG}[1]{\opLeaveHG(#1)}
\newcommand{\LeaHGone}[1]{\opLeaveHG^1(#1)}
\newcommand{\LeaHGtwo}[1]{\opLeaveHG^2(#1)}
\newcommand{\LeaHGthr}[1]{\opLeaveHG^3(#1)}
\newcommand{\LeaHGfou}[1]{\opLeaveHG^4(#1)}
\newcommand{\LeaHGlea}{\opLeaveHG^{\operatorname{lea}}}
\newcommand{\LeaHGtt}{\opLeaveHG^{23}}

\newcommand{\opEnterHG}{\Ent_0}
\newcommand{\EnterHG}[1]{\opEnterHG(#1)}

\newcommand{\EntHGent}{\opEnterHG^{\operatorname{ent}}}

\newtheorem{theorem}{Theorem}[section]

\newtheorem{exa}{Example}[section]
\newenvironment{example}{\begin{exa}\em}{\end{exa}}

\newcommand{\hourglassRone}[1]{\operatorname{H}^1_{\operatorname{rc}}(#1)}
\newcommand{\hourglassRtwo}[1]{\operatorname{H}^2_{\operatorname{rc}}(#1)}
\newcommand{\hourglassRthr}[1]{\operatorname{H}^3_{\operatorname{rc}}(#1)}
\newcommand{\hourglassRplus}[1]{\operatorname{H}^+_{\operatorname{rc}}(#1)}
\newcommand{\hourglassRminus}[1]{\operatorname{H}^-_{\operatorname{rc}}(#1)}
\newcommand{\hourglassR}[1]{\operatorname{H}_{\operatorname{rc}}(#1)}

\newcommand{\ophourglassR}{\operatorname{H}_{\operatorname{rc}}}
\newcommand{\hourglassLone}[1]{\operatorname{H}^1_{\operatorname{lc}}(#1)}
\newcommand{\hourglassLtwo}[1]{\operatorname{H}^2_{\operatorname{lc}}(#1)}
\newcommand{\hourglassLthr}[1]{\operatorname{H}^3_{\operatorname{lc}}(#1)}
\newcommand{\hourglassL}[1]{\operatorname{H}_{\operatorname{lc}}(#1)}
\newcommand{\hourglassLplus}[1]{\operatorname{H}^{+}_{\operatorname{lc}}(#1)}
\newcommand{\hourglassLminus}[1]{\operatorname{H}^{-}_{\operatorname{lc}}(#1)}
\newcommand{\ophourglassL}{\operatorname{H}_{\operatorname{lc}}}

\newcommand{\SqBit}{\nabla}

\newcommand{\HRsource}{\operatorname{R}_{\wordsour}}
\newcommand{\HLsource}{\operatorname{L}_{\wordsour}}
\newcommand{\HRsink}{\operatorname{R}_{\wordsink}}
\newcommand{\HLsink}{\operatorname{L}_{\wordsink}}
\newcommand{\Hglass}{\operatorname{{\cal H}}}

\newcommand{\absoluteignore}[1]{}

\newcommand{\dgur}{{\cal K}_{\urop}}
\newcommand{\dgdl}{{\cal K}_{\dlop}}

\newcommand{\reverse}{\iota}

\newcommand{\touch}[1]{\operatorname{b}(#1)}
\newcommand{\touchop}{\operatorname{b}}

\newcommand{\bitangente}[1]{\operatorname{b}(#1)}

\newcommand{\vcor}[1]{\mathbb{V}_{#1}}
\newcommand{\vc}[1]{\mathbb{V}_{#1}}
\newcommand{\vctwo}[1]{\mathbb{V}^2_{#1}}
\newcommand{\vcone}[1]{\mathbb{V}^1_{#1}}
\newcommand{\vczer}[1]{\mathbb{V}^0_{#1}}

\newcommand{\wc}[1]{\mathbb{W}_{#1}}

\newcommand{\fs}[1]{\mathbb{F}_{#1}}%% Free space

\newcommand{\sigmaback}{\sigma_{\Sback}}
\newcommand{\sigmaforw}{\sigma_{\Sforw}}
\newcommand{\sigmaleft}{\sigma_{\Sleft}}

\newcommand{\sigmarigh}{\sigma_{\Srigh}}

\newcommand{\xxx}{\alpha}

\newcommand{\alphabet}{\Lambda}

\newcommand{\Ssour}{\epsilon}
\newcommand{\Ssink}{\operatorname{s}}
\newcommand{\Srigh}{\operatorname{r}}
\newcommand{\Sleft}{\operatorname{l}}
\newcommand{\Sforw}{\operatorname{f}}
\newcommand{\Sback}{\operatorname{b}}

\newcommand{\wordsour}{\operatorname{sour}}
\newcommand{\wordsink}{\operatorname{sink}}
\newcommand{\wordback}{\operatorname{back}}
\newcommand{\wordforw}{\operatorname{forw}}

\newcommand{\bitmphi}{\overline{\varphi}}
\newcommand{\mphi}{\varphi}

\newcommand{\phisour}{\mphi_{\Ssour}}
\newcommand{\phisink}{\mphi_{\Ssink}}
\newcommand{\bitphiback}{\bitmphi_{\Sback}}
\newcommand{\phiback}{\mphi_{\Sback}}
\newcommand{\phibackstar}{\mphi_{\Sback_*}}
\newcommand{\bitphiforw}{\bitmphi_{\Sforw}}
\newcommand{\phiforw}{\mphi_{\Sforw}}
\newcommand{\phileft}{\mphi_{\Sleft}}
\newcommand{\bitphirigh}{\bitmphi_{\Srigh}}
\newcommand{\phirigh}{\mphi_{\Srigh}}

\newcommand{\vertex}{v}
\newcommand{\vertexbis}{u}

\newcommand{\G}{\operatorname{G}}%%%

\newcommand{\ur}[1]{\operatorname{ur}(#1)}
\newcommand{\ul}[1]{\operatorname{ul}(#1)}
\newcommand{\dr}[1]{\operatorname{dr}(#1)}
\newcommand{\dl}[1]{\operatorname{dl}(#1)}

\newcommand{\urop}{\operatorname{ur}}

\newcommand{\dlop}{\operatorname{dl}}

\newcommand{\sour}[1]{\operatorname{sour}(#1)}
\newcommand{\sinkop}{\operatorname{sink}}
\newcommand{\sink}[1]{\operatorname{sink}(#1)}
\newcommand{\sinkpar}[1]{\operatorname{sink}_{#1}}

\newcommand{\R}{\operatorname{R}}

\newcommand{\Rtwo}{\R^2}
\newcommand{\Rthr}{\R^3}

\renewcommand{\L}{\operatorname{L}}

\newcommand{\one}{1}
\newcommand{\two}{2}
\newcommand{\thr}{3}
\newcommand{\thrset}{\{1,2,3\}}
\newcommand{\twoset}{\{1,2\}}

\newcommand{\reversetwoset}{\{2,3\}}

\newcommand{\rc}{\operatorname{rc}}

\newcommand{\rcNew}[1]{\operatorname{rc}(#1)}
\newcommand{\rconeNew}[1]{\rc_{1}(#1)}
\newcommand{\rctwoNew}[1]{\rc_{2}(#1)}
\newcommand{\rcthrNew}[1]{\rc_{3}(#1)}

\newcommand{\rconeop}{\rc_{1}}
\newcommand{\rctwoop}{\rc_{2}}
\newcommand{\rcthrop}{\rc_{3}}

\newcommand{\lc}{\operatorname{lc}}

\newcommand{\lcNew}[1]{\operatorname{lc}(#1)}
\newcommand{\lconeNew}[1]{\lc_{1}(#1)}
\newcommand{\lctwoNew}[1]{\lc_{2}(#1)}

\newcommand{\lconeop}{\lc_{1}}
\newcommand{\lctwoop}{\lc_{2}}
\newcommand{\lcthrop}{\lc_{3}}

\newcommand{\order}{\ll}
\newcommand{\orderstar}{\ll_{*}}
\newcommand{\downset}{{\cal J}}

\newcommand{\facesG}{\faces^{\circ}}
\newcommand{\edges}{{\cal E}}

\newcommand{\ufaces}{{\cal U}}
\newcommand{\faces}{{\cal F}}
\newcommand{\facedetype}[2]{\faces_{#1#2}}
\newcommand{\facedetypeG}[2]{\faces^{\circ}_{#1#2}}
\newcommand{\fdt}[2]{\facedetype#1#2}
\newcommand{\fdtG}[2]{\facedetypeG#1#2}

\newcommand{\fdtthrthr}{\fdt{3}{3}}

\newcommand{\fdtthrone}{\fdt{3}{1}}
\newcommand{\fdtthrtwo}{\fdt{3}{2}}

\newcommand{\fdtonethr}{\fdt{1}{3}}
\newcommand{\fdttwothr}{\fdt{2}{3}} 
\newcommand{\fdttwotwo}{\fdt{2}{2}}
\newcommand{\fdtonetwo}{\fdt{1}{2}}
\newcommand{\fdttwoone}{\fdt{2}{1}} 
\newcommand{\fdtoneone}{\fdt{1}{1}}

\newcommand{\phiforwstar}{{\mphi_{\Sforw}}_*}

\newcommand{\toponecell}[1]{\operatorname{top}(#1)}

\newcommand{\urf}[1]{\operatorname{urf}(#1)}
\newcommand{\ulf}[1]{\operatorname{ulf}(#1)}
\newcommand{\drf}[1]{\operatorname{drf}(#1)}
\newcommand{\dlf}[1]{\operatorname{dlf}(#1)}

\newcommand{\urfop}{\operatorname{urf}}
\newcommand{\ulfop}{\operatorname{ulf}}
\newcommand{\drfop}{\operatorname{drf}}
\newcommand{\dlfop}{\operatorname{dlf}}

\newcounter{compteuritem}
{\end{list}}
{\end{list}}
{\end{list}}
\newcommand{\EPTG}{{\cal T}}
\newcommand{\nameM}{M}
\newcommand{\bg}[2]{b_{#1#2}}
\newcommand{\dline}[1]{\tau(#1)}
\newcommand{\onevc}{1}
\newcommand{\twovc}{2}
\newcommand{\thrvc}{3}
\newcommand{\fouvc}{4}
\newcommand{\onevcp}{1'}
\newcommand{\twovcp}{2'}
\newcommand{\thrvcp}{3'}
\newcommand{\fouvcp}{4'}
\newcommand{\unun}{i_1}
\newcommand{\unde}{i_2}
\newcommand{\unth}{i_3}
\newcommand{\unfo}{i_4}
\newcommand{\unfi}{i'_1}
\newcommand{\unsi}{i'_2}
\newcommand{\unse}{i'_3}
\newcommand{\unhe}{i'_4}
\newcommand{\deun}{j_1}
\newcommand{\dede}{j_2}
\newcommand{\deth}{j_3}
\newcommand{\defo}{j_4}
\newcommand{\defi}{j'_1}
\newcommand{\desi}{j'_2}
\newcommand{\dese}{j'_3}
\newcommand{\dehe}{j'_4}
\newcommand{\vertices}{{\cal V}}
\newcommand{\CS}[1]{\Gamma(#1)}

\newcommand{\cal}{\mathcal}

\newcommand{\Hpqs}{${\mathcal H}$-pseudoquadrangles}
\newcommand{\Hpq}{${\mathcal H}$-pseudoquadrangle}
\newcommand{\ABpts}{${\mathcal A \mathcal B}$-pseudotriangles}
\newcommand{\ABzerpts}{${\mathcal A\mathcal B_0}$-pseudotriangles}
\newcommand{\udf}{\text{-}}

\newcommand{\liftoperator}{\mu}
\newcommand{\etasection}{\xi}
\newcommand{\ve}[1]{v_{#1}}
\newcommand{\sm}{\footnotesize}
\newcommand{\paths}{{\cal P}}

\newcommand{\Iwc}{\sofl{\afs}^*}
\newcommand{\Ivc}{\sofl{\afs}}
\newcommand{\Ilineset}{\sofl{\pointset}}
\newcommand{\uniIlineset}{\unisofl{\pointset}}
\newcommand{\Vposet}{{\cal O}}

\newcommand{\MAC}{J}
\newcommand{\MACBIS}{J'}
\newcommand{\shiftop}{\nu}
\newcommand{\vfilter}[1]{#1^{+}}
\newcommand{\systemlines}[1]{{\cal L}(#1)}
\newcommand{\afs}{\mathbb{X}}

\newcommand{\sofl}[1]{{\cal L}(#1)}
\newcommand{\unisofl}[1]{{\cal L}^{\operatorname{u}}(#1)}
\newcommand{\soflbis}[1]{#1^*}
\newcommand{\spaceoflinesof}[1]{{\cal L}(#1)}
\newcommand{\spaceoforientedlinesof}[1]{{\cal L}^{\operatorname{or}}(#1)}

\newcommand{\ccw}[3]{u_{}^{#1}{(#2)}}
\newcommand{\ccwop}[1]{u}

\newcommand{\qq}{q}
\newcommand{\CCWT}[2]{U_{#1}^{#2}}
\newcommand{\TRIA}[2]{T_{#1}^{#2}}
\newcommand{\MISS}[2]{R_{#1}^{#2}}

\newcommand{\burop}[2]{\operatorname{ur}_{#2}}
\newcommand{\bulop}[2]{\operatorname{ul}_{#2}}
\newcommand{\bdrop}[2]{\operatorname{dr}_{#2}}
\newcommand{\bdlop}[2]{\operatorname{dl}_{#2}}

\newcommand{\bur}[2]{\operatorname{ur}_{#2}(#1)}

\newcommand{\bdl}[2]{\operatorname{dl}_{#2}(#1)}

\newcommand{\btoponecellop}[2]{\operatorname{top}_{#2}}
\newcommand{\bbotonecellop}[2]{\operatorname{bot}_{#2}}

\newcommand{\btoponecell}[2]{\operatorname{top}_{#2}(#1)}
\newcommand{\bbotonecell}[2]{\operatorname{bot}_{#2}(#1)}

\title[\today]{
Computing pseudotriangulations \\ via  branched coverings
}
\author{
Luc Habert and 
Michel Pocchiola
}
\thanks{MP was partially supported by the TEOMATRO grant ANR-10-BLAN 0207.}
\address{Luc Habert}
\email{Luc.Habert@normalesup.org}
\address{Michel Pocchiola\\
Universit{\'e} Pierre \&  Marie Curie\\
Institut de Math{\'e}matiques de Jussieu, UMR 7586\\
4 place Jussieu\\
75252 Paris Cedex 05\\
France}
\email{pocchiola@math.jussieu.fr}

\date{\today}

\begin{document}
%%%%%%%%%%%%%%%%%%%%%%%%%%%%%%%%%%%%%%%%%%%%%%%%%%%%%%%%%%%%%%%%%%%%%%%%%%%%%%%%
%%%%%%%%%%%%%%%%%%%%%%%%%%%%%%%%%%%%%%%%%%%%%%%%%%%%%%%%%%%%%%%%%%%%%%%%%%%%%%%%
%%%%%%%%%%%%%%%%%%%%%%%%%%%%%%%%%%%%%%%%%%%%%%%%%%%%%%%%%%%%%%%%%%%%%%%%%%%%%%%%
%%%%%%%%%%%%%%%%%%%%%%%%%%%%%%%%%%%%%%%%%%%%%%%%%%%%%%%%%%%%%%%%%%%%%%%%%%%%%%%%
%%%%%%%%%%%%%%%%%%%%%%%%%%%%%%%%%%%%%%%%%%%%%%%%%%%%%%%%%%%%%%%%%%%%%%%%%%%%%%%%
\begin{abstract}
We describe an efficient algorithm to compute a pseudotriangulation of a finite planar family of pairwise disjoint convex bodies presented by its chirotope.
The design of the algorithm relies on a deepening of the theory of visibility complexes and on the extension of that theory to the setting 
of branched coverings. 
The problem of computing a pseudotriangulation that contains a given set of bitangent line segments is also examined. 

\medskip
\noindent
{\bf Keywords.} Convexity, convex hulls, pseudotriangulations, constrained pseudotriangulations, partial linear spaces, 
visibility complexes, topological planes, branched coverings, geometric predicates, chirotopes,
fundamental and practical algorithms. 
\end{abstract}
%%%%%%%%%%%%%%%%%%%%%%%%%%%%%%%%%%%%%%%%%%%%%%%%%%%%%%%%%%%%%%%%%%%%%%%%%%%%%%%%
%%%%%%%%%%%%%%%%%%%%%%%%%%%%%%%%%%%%%%%%%%%%%%%%%%%%%%%%%%%%%%%%%%%%%%%%%%%%%%%%
%%%%%%%%%%%%%%%%%%%%%%%%%%%%%%%%%%%%%%%%%%%%%%%%%%%%%%%%%%%%%%%%%%%%%%%%%%%%%%%%
%%%%%%%%%%%%%%%%%%%%%%%%%%%%%%%%%%%%%%%%%%%%%%%%%%%%%%%%%%%%%%%%%%%%%%%%%%%%%%%%
%%%%%%%%%%%%%%%%%%%%%%%%%%%%%%%%%%%%%%%%%%%%%%%%%%%%%%%%%%%%%%%%%%%%%%%%%%%%%%%%

\maketitle

\clearpage
\tableofcontents
%%%%%%%%%%%%%%%%%%%%%%%%%%%%%%%%%%%%%%%%%%%%%%%%%%%%%%%%%%%%%%%%%%%%%%%%%%%%%%%
%%%%%%%%%%%%%%%%%%%%%%%%%%%%%%%%%%%%%%%%%%%%%%%%%%%%%%%%%%%%%%%%%%%%%%%%%%%%%%%
%%%%%%%%%%%%%%%%%%%%%%%%%%%%%%%%%%%%%%%%%%%%%%%%%%%%%%%%%%%%%%%%%%%%%%%%%%%%%%%
%%%%%%%%%%%%%%%%%%%%%%%%%%%%%%%%%%%%%%%%%%%%%%%%%%%%%%%%%%%%%%%%%%%%%%%%%%%%%%%
%%%%%%%%%%%%%%%%%%%%%%%%%%%%%%%%%%%%%%%%%%%%%%%%%%%%%%%%%%%%%%%%%%%%%%%%%%%%%%%

\clearpage

%%%%%%%%%%%%%%%%%%%%%%%%%%%%%%%%%%%%%%%%%%%%%%%%%%%%%%%%%%%%%%%%%%%%%%%%%%%%%%%
%%%%%%%%%%%%%%%%%%%%%%%%%%%%%%%%%%%%%%%%%%%%%%%%%%%%%%%%%%%%%%%%%%%%%%%%%%%%%%%
%%%%%%%%%%%%%%%%%%%%%%%%%%%%%%%%%%%%%%%%%%%%%%%%%%%%%%%%%%%%%%%%%%%%%%%%%%%%%%%
%%%%%%%%%%%%%%%%%%%%%%%%%%%%%%%%%%%%%%%%%%%%%%%%%%%%%%%%%%%%%%%%%%%%%%%%%%%%%%%
%%%%%%%%%%%%%%%%%%%%%%%%%%%%%%%%%%%%%%%%%%%%%%%%%%%%%%%%%%%%%%%%%%%%%%%%%%%%%%%

\section{Introduction}
%%%%%%%%%%%%%%%%%%%%%%%%%%%%%%%%%%%%%%%%%%%%%%%%%%%%%%%%%%%%%%%%%%%%%%%%%%%%%%%
%%%%%%%%%%%%%%%%%%%%%%%%%%%%%%%%%%%%%%%%%%%%%%%%%%%%%%%%%%%%%%%%%%%%%%%%%%%%%%%
%%%%%%%%%%%%%%%%%%%%%%%%%%%%%%%%%%%%%%%%%%%%%%%%%%%%%%%%%%%%%%%%%%%%%%%%%%%%%%%
%%%%%%%%%%%%%%%%%%%%%%%%%%%%%%%%%%%%%%%%%%%%%%%%%%%%%%%%%%%%%%%%%%%%%%%%%%%%%%%
\subsection{Main result of the paper} 
Throughout the paper we address the problem of computing efficiently a  
pseudo\-trian\-gulation of a finite planar
 family of pairwise disjoint convex bodies presented by its {\it chirotope}~:  Here 
the term chirotope refers to a natural extension to finite planar families of pairwise disjoint convex bodies 
of the classical notion of chirotope (or order type) of a finite planar family of points~\cite{blswz-om-99,b-com-06}; 
and the  term planar refers to any oriented topological plane on $\mathbb{R}^2$,
 e.g., Euclidean plane, hyperbolic plane, Moulton planes, arc planes, etc.; cf. Appendix~\ref{TopPlanes}.

%%%%%%%%%%%%%%%%%%%%%%%%%%%%%%%%%%%%%%%%%%%%%%%%%%%%%%%%%%%%%%%%%%%%%%%%%%%%%%%
%%%%%%%%%%%%%%%%%%%%%%%%%%%%%%%%%%%%%%%%%%%%%%%%%%%%%%%%%%%%%%%%%%%%%%%%%%%%%%%
%%%%%%%%%%%%%%%%%%%%%%%%%%%%%%%%%%%%%%%%%%%%%%%%%%%%%%%%%%%%%%%%%%%%%%%%%%%%%%%
%%%%%%%%%%%%%%%%%%%%%%%%%%%%%%%%%%%%%%%%%%%%%%%%%%%%%%%%%%%%%%%%%%%%%%%%%%%%%%%
\subsubsection{Chirotopes.}
\begin{figure}[!htb]
\centering
%\small
\psfrag{A}{$A$}
\psfrag{B}{$A'$}
\psfrag{C}{$B$}
\psfrag{D}{$B'$}
\psfrag{E}{$B''$}
\psfrag{poun}{\footnotesize $10000$}
\psfrag{pode}{\footnotesize $01000$}
\psfrag{potr}{\footnotesize $00100$}
\psfrag{poqu}{\footnotesize $00010$}
\psfrag{poci}{\footnotesize $00001$}
\psfrag{1}{$1$} \psfrag{un}{1} \psfrag{2}{$2$} \psfrag{3}{$3$} \psfrag{deux}{$2$} \psfrag{trois}{$3$} \psfrag{infty}{$\infty$}
\includegraphics[width = 0.75\linewidth]{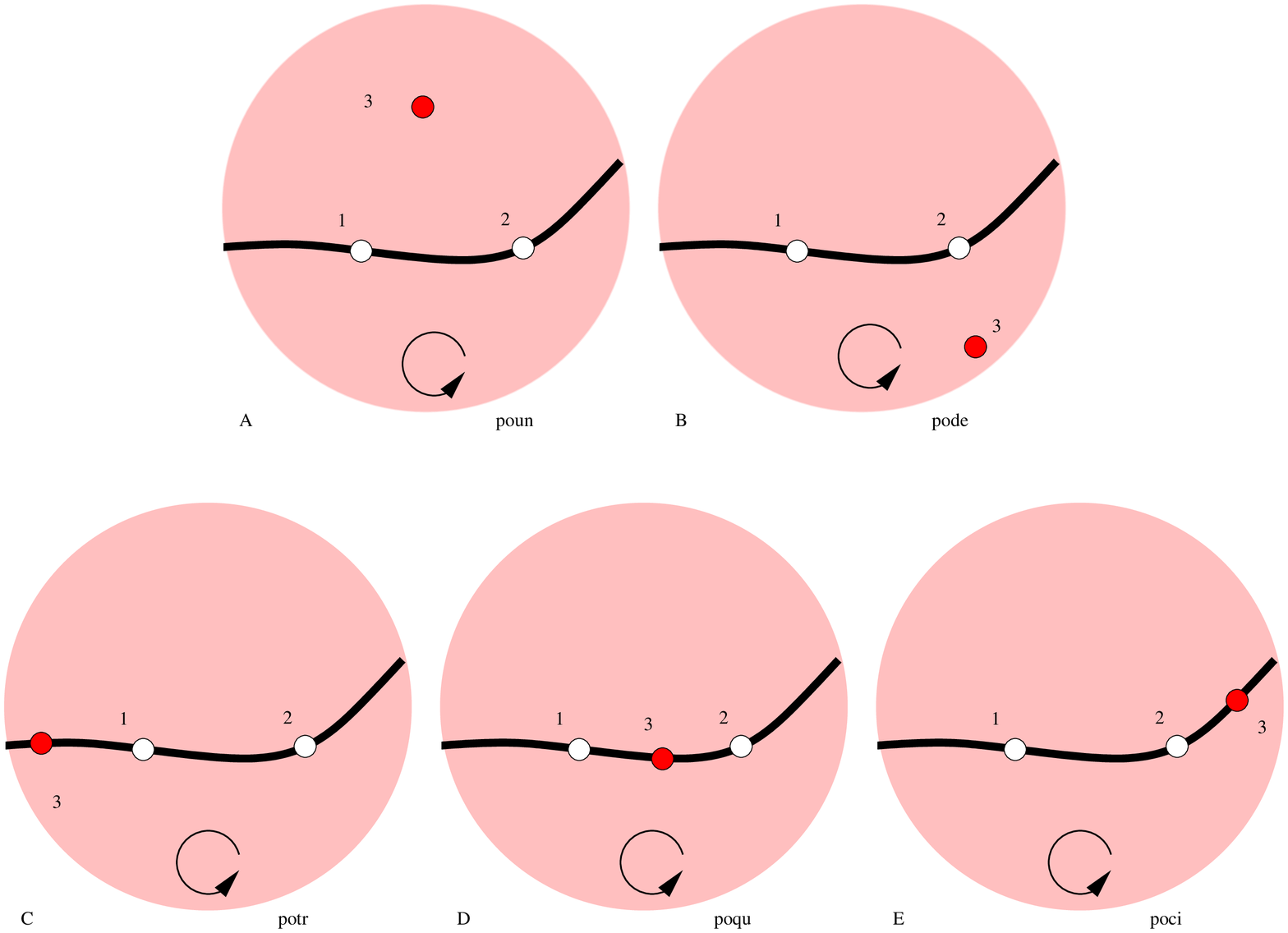}
\caption{\protect \small 
%% Representatives of the five chirotopes of planar families of three points, and
%% representatives of the
%% corresponding five isotopy classes of
%% arrangements of three pseudolines defined on a given set of indices.
\label{thrchipoints}}
\end{figure}

\begin{figure}[!htb]
\centering
\footnotesize
\tiny
\psfrag{oneb}{$1$} \psfrag{twob}{$2$} \psfrag{thrb}{$3$}
\psfrag{one}{right-left} \psfrag{two}{right-right} \psfrag{thr}{left-left}
\psfrag{fou}{left-right} \psfrag{lhp}{left side} \psfrag{rhp}{right side} 
\psfrag{initialsegment}{initial} \psfrag{mseg}{median}
\psfrag{fseg}{final} \psfrag{firstbody}{first body} \psfrag{secondbody}{second body} \psfrag{firstbody}{first}
\psfrag{secondbody}{second} \psfrag{body}{body}
\includegraphics[width=0.98575\linewidth]{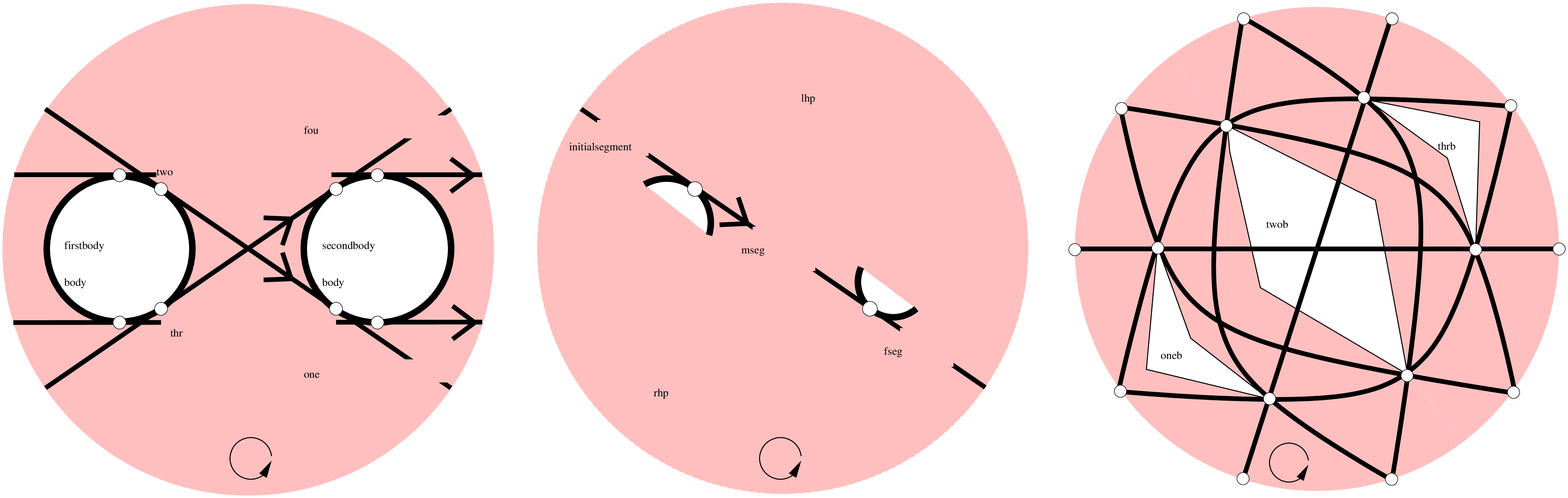}
\caption{\protect \footnotesize \label{Chirotope}}
\end{figure}

Recall that the chirotope of a finite planar family of points is (or can be defined as)  the map that assigns to each ordered triple of distinct indices of the family of points
the {\it position vector} of
the corresponding ordered triple of points, that is, the boolean vector of truth-values 
of the five relations ``the third point of the triple belongs to the open left side
(open right side, initial part, median part, final part) of the directed line joining
the first point of the triple to the second point of the triple."
Figure~\ref{thrchipoints} shows five families of three points realizing the five possible chirotopes on the indexing set $\{1,2,3\}.$ 
%%%%%%%%%%%%%%%%%%%%%%%%%%%%%%%%%%%%%%%%%%%%%%%%%%%%%%%%%%%%%%%%%%%%%%%%%%%%%%
%%%%%%%%%%%%%%%%%%%%%%%%%%%%%%%%%%%%%%%%%%%%%%%%%%%%%%%%%%%%%%%%%%%%%%%%%%%%%%%
%%%%%%%%%%%%%%%%%%%%%%%%%%%%%%%%%%%%%%%%%%%%%%%%%%%%%%%%%%%%%%%%%%%%%%%%%%%%%%%
%%%%%%%%%%%%%%%%%%%%%%%%%%%%%%%%%%%%%%%%%%%%%%%%%%%%%%%%%%%%%%%%%%%%%%%%%%%%%%%
%%%%%%%%%%%%%%%%%%%%%%%%%%%%%%%%%%%%%%%%%%%%%%%%%%%%%%%%%%%%%%%%%%%%%%%%%%%%%%%
%%%%%%%%%%%%%%%%%%%%%%%%%%%%%%%%%%%%%%%%%%%%%%%%%%%%%%%%%%%%%%%%%%%%%%%%%%%%%%%
%%%%%%%%%%%%%%%%%%%%%%%%%%%%%%%%%%%%%%%%%%%%%%%%%%%%%%%%%%%%%%%%%%%%%%%%%%%%%%%
%%%%%%%%%%%%%%%%%%%%%%%%%%%%%%%%%%%%%%%%%%%%%%%%%%%%%%%%%%%%%%%%%%%%%%%%%%%%%%%
In this figure the plane is represented by the interior of a circular diagram, marked with a little oriented circle to indicate its orientation, 
and each diagram is labeled at its left bottom corner with a symbol to name it and at its right bottom corner with the position vector of the ordered triple of points corresponding to the ordered triple of indices $1,2,3$.
%%%%%%%%%%%%%%%%%%%%%%%%%%%%%%%%%%%%%%%%%%%%%%%%%%%%%%%%%%%%%%%%%%%%%%%%%%%%%%%
%%%%%%%%%%%%%%%%%%%%%%%%%%%%%%%%%%%%%%%%%%%%%%%%%%%%%%%%%%%%%%%%%%%%%%%%%%%%%%%
%%%%%%%%%%%%%%%%%%%%%%%%%%%%%%%%%%%%%%%%%%%%%%%%%%%%%%%%%%%%%%%%%%%%%%%%%%%%%%%
%%%%%%%%%%%%%%%%%%%%%%%%%%%%%%%%%%%%%%%%%%%%%%%%%%%%%%%%%%%%%%%%%%%%%%%%%%%%%%%
The notion of chirotope of a planar family of pairwise disjoint convex bodies is defined similarly: as for families of points  we use the notion of 
position vector as a coding of the relative positions of the convex bodies with respect to a line.
To set out the definition we use the following standard terminology: a directed bitangent joining an ordered pair of disjoint convex bodies is,
as illustrated in the left part of Figure~\ref{Chirotope},
classified {\it left-left, right-right, left-right} or {\it right-left} depending on which sides (left or right side) of the bitangent are the convex bodies; 
walking along a directed bitangent we traverse successively, as illustrated in the middle part 
of Figure~\ref{Chirotope}, its {\it initial, median} and {\it final} parts; the median part of a bitangent is called  a {\it  bitangent line segment} thereafter. 
%%%%%%%%%%%%%%%%%%%%%%%%%%%%%%%%%%%%%%%%%%%%%%%%%%%%%%%%%%%%%%%%%%%%%%%%%%%%%%%
%%%%%%%%%%%%%%%%%%%%%%%%%%%%%%%%%%%%%%%%%%%%%%%%%%%%%%%%%%%%%%%%%%%%%%%%%%%%%%%
%%%%%%%%%%%%%%%%%%%%%%%%%%%%%%%%%%%%%%%%%%%%%%%%%%%%%%%%%%%%%%%%%%%%%%%%%%%%%%%
%%%%%%%%%%%%%%%%%%%%%%%%%%%%%%%%%%%%%%%%%%%%%%%%%%%%%%%%%%%%%%%%%%%%%%%%%%%%%%%
%%%%%%%%%%%%%%%%%%%%%%%%%%%%%%%%%%%%%%%%%%%%%%%%%%%%%%%%%%%%%%%%%%%%%%%%%%%%%%%
%%%%%%%%%%%%%%%%%%%%%%%%%%%%%%%%%%%%%%%%%%%%%%%%%%%%%%%%%%%%%%%%%%%%%%%%%%%%%%%
%%%%%%%%%%%%%%%%%%%%%%%%%%%%%%%%%%%%%%%%%%%%%%%%%%%%%%%%%%%%%%%%%%%%%%%%%%%%%%%
%%%%%%%%%%%%%%%%%%%%%%%%%%%%%%%%%%%%%%%%%%%%%%%%%%%%%%%%%%%%%%%%%%%%%%%%%%%%%%%
Using this terminology  we are able to  define the  {\it chirotope} of a finite planar family of pairwise disjoint convex bodies as 
the map 
that assigns to each ordered triple of distinct indices of the family of bodies the position vector of the corresponding ordered triple 
of bodies, that is, the boolean vector of truth-values of the twenty relations 
``the third body of the triple intersects  the open left side (open right side, initial part, median part, final part)
 of the left-left (left-right, right-left, right-right) directed bitangent 
joining the first body of the triple to the second body of the triple.''    
For example, consider the family of three convex bodies on the indexing set $\{1,2,3\}$ 
depicted together with its $3\times 4$  bitangents in the right part of Figure~\ref{Chirotope} ($6$ of the $12$ bitangents are tritangents).
%%%%%%%%%%%%%%%%%%%%%%%%%%%%%%%%%%%%%%%%%%%%%%%%%%%%%%%%%%%%%%%%%%%%%%%%%%%%%%%
%%%%%%%%%%%%%%%%%%%%%%%%%%%%%%%%%%%%%%%%%%%%%%%%%%%%%%%%%%%%%%%%%%%%%%%%%%%%%%%
%%%%%%%%%%%%%%%%%%%%%%%%%%%%%%%%%%%%%%%%%%%%%%%%%%%%%%%%%%%%%%%%%%%%%%%%%%%%%%%
%%%%%%%%%%%%%%%%%%%%%%%%%%%%%%%%%%%%%%%%%%%%%%%%%%%%%%%%%%%%%%%%%%%%%%%%%%%%%%%
Then its  chirotope is the map $\chi$ defined by  
$$
\begin{array}{lcccc}
 & {\text{right-left}} &\text{right-right} &{\text{left-left}}   &\text{left-right} \\
\chi(1,3,2) = \chi(3,1,2) = & 
11010 &
01010 &
10010 &
11010\phantom{.} \\
\chi(1,2,3) = \chi(3,2,1) = & 
10000 &
01001 &
10001 &
01000\phantom{.} \\
\chi(2,3,1) = \chi(2,1,3) =  &
01000 &
01100 &
10100 &
10000.
\end{array}
$$
The number of chirotopes of planar families of $3$ pairwise disjoint convex bodies on a given indexing set of size $3$ is $531$ and  among these $531$ chirotopes $118$ are 
{\it simple} chirotopes, that is, chirotopes of families of convex bodies 
with no tritangent; as for the chirotope of a planar family of points a key feature of the chirotope of a planar family of pairwise disjoint convex bodies is that it encodes 
its {\it dual arrangement}, i.e., 
the arrangement, in the space of lines of the plane, of the curves of tangents to the bodies; cf. Appendix~\ref{TopPlanes}.
Throughout the paper we will assume that the boundaries of the bodies are free of line segments and that there is exactly one tangent through each boundary point; these 
assumptions facilitate the geometric definition of pseudotriangulations without ruling out any chirotope of families of pairwise disjoint convex bodies.
%%%%%%%%%%%%%%%%%%%%%%%%%%%%%%%%%%%%%%%%%%%%%%%%%%%%%%%%%%%%%%%%%%%%%%%%%%%%%%%
%%%%%%%%%%%%%%%%%%%%%%%%%%%%%%%%%%%%%%%%%%%%%%%%%%%%%%%%%%%%%%%%%%%%%%%%%%%%%%%
%%%%%%%%%%%%%%%%%%%%%%%%%%%%%%%%%%%%%%%%%%%%%%%%%%%%%%%%%%%%%%%%%%%%%%%%%%%%%%%
%%%%%%%%%%%%%%%%%%%%%%%%%%%%%%%%%%%%%%%%%%%%%%%%%%%%%%%%%%%%%%%%%%%%%%%%%%%%%%%

%%%%%%%%%%%%%%%%%%%%%%%%%%%%%%%%%%%%%%%%%%%%%%%%%%%%%%%%%%%%%%%%%%%%%%%%%%%%%%%
%%%%%%%%%%%%%%%%%%%%%%%%%%%%%%%%%%%%%%%%%%%%%%%%%%%%%%%%%%%%%%%%%%%%%%%%%%%%%%%
%%%%%%%%%%%%%%%%%%%%%%%%%%%%%%%%%%%%%%%%%%%%%%%%%%%%%%%%%%%%%%%%%%%%%%%%%%%%%%%
%%%%%%%%%%%%%%%%%%%%%%%%%%%%%%%%%%%%%%%%%%%%%%%%%%%%%%%%%%%%%%%%%%%%%%%%%%%%%%%

\subsubsection{Pseudotriangulations.}
\begin{figure}[!htb]
\footnotesize
\psfrag{one}{$1$}\psfrag{two}{$2$}\psfrag{thr}{$3$}\psfrag{fou}{4}\psfrag{fiv}{$5$}\psfrag{six}{$6$}\psfrag{sev}{$7$}\psfrag{8}{8}%%%
\psfrag{central}{a boundary body} \psfrag{sheet2}{sheet 1} \psfrag{sheet3}{sheet 2} \psfrag{sheet4}{sheet 3} \psfrag{tau}{$\tau$}
\psfrag{mfls}{a $2$-cell of the cross-section and its sink bitangent} \psfrag{bitangente}{its sink}
\begin{center}
\includegraphics[width=0.98575\linewidth]{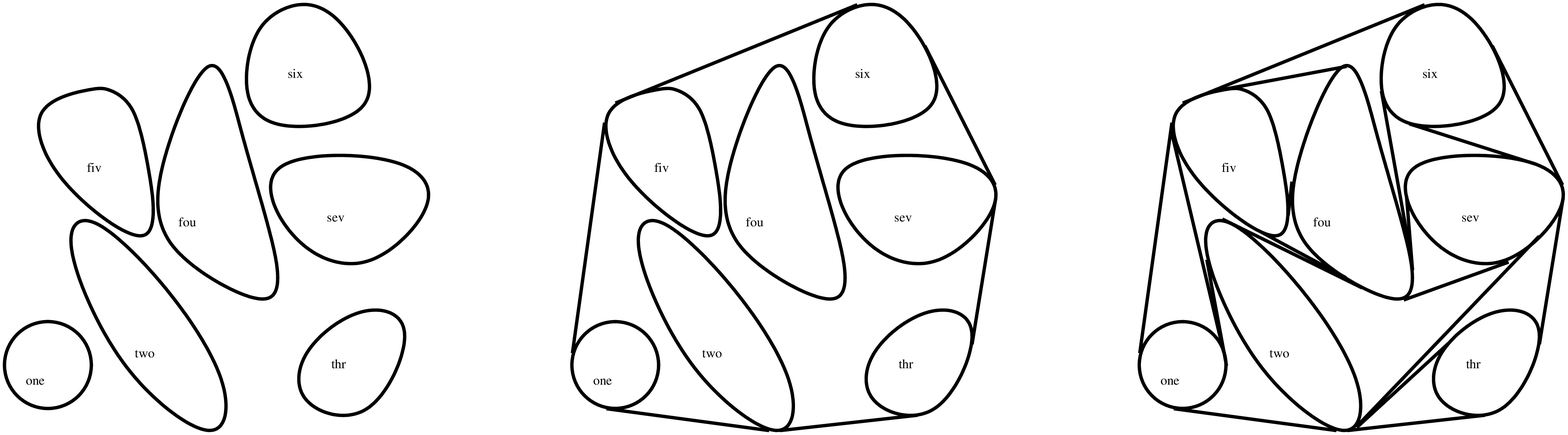}%%%
\end{center}
\caption{\protect \footnotesize 
%%A planar family of seven pairwise disjoint convex bodies
%%and one constraint, the bitangent line segments in the boundary of its convex hull,  
%%and one of its pseudotriangulations which includes the constraint.i
\label{theproblem}}
\end{figure}

Let $o_1,o_2,\ldots,o_n$ be a finite planar family of $n$ pairwise disjoint convex bodies; 
a {\it boundary bitangent line segment} is a bitangent line segment of the $o_i$ contained in the boundary of their convex hull; all other bitangent line segments are said to be {\it interior bitangent line segments}; the number of boundary 
bitangent line segments is denoted $h$;
{\it free space} is the complement in the plane of the interiors of the $o_i$; a 
{\it pseudotriangulation} is a maximal (for the inclusion relation) family of pairwise interior non-crossing free bitangent line segments.
A pseudotriangulation  contains the $h$ boundary bitangent line segments plus $3n-3-h$ interior bitangent line segments (thus $3n-3$ altogether) and 
induces a decomposition of the free part of the convex hull of the $o_i$ into $2n-2$ pseudotriangles~\cite{G-pv-vc-96}.
Figure~\ref{theproblem} shows a family of $7$ pairwise disjoint convex bodies of the real affine plane, 
its ($6$ in number) boundary bitangent line segments, and one 
%%%%%%%%%%%%%%%%%%%%%%%%%%%%%%%%%%%%%%%%%%%%%%%%%%%%%%%%%%%%%%%%%%%%%%%%%%%%%%%
%%%%%%%%%%%%%%%%%%%%%%%%%%%%%%%%%%%%%%%%%%%%%%%%%%%%%%%%%%%%%%%%%%%%%%%%%%%%%%%
%%%%%%%%%%%%%%%%%%%%%%%%%%%%%%%%%%%%%%%%%%%%%%%%%%%%%%%%%%%%%%%%%%%%%%%%%%%%%%%
%%%%%%%%%%%%%%%%%%%%%%%%%%%%%%%%%%%%%%%%%%%%%%%%%%%%%%%%%%%%%%%%%%%%%%%%%%%%%%%
%%%%%%%%%%%%%%%%%%%%%%%%%%%%%%%%%%%%%%%%%%%%%%%%%%%%%%%%%%%%%%%%%%%%%%%%%%%%%%%
%%%%%%%%%%%%%%%%%%%%%%%%%%%%%%%%%%%%%%%%%%%%%%%%%%%%%%%%%%%%%%%%%%%%%%%%%%%%%%%
%%%%%%%%%%%%%%%%%%%%%%%%%%%%%%%%%%%%%%%%%%%%%%%%%%%%%%%%%%%%%%%%%%%%%%%%%%%%%%%
%%%%%%%%%%%%%%%%%%%%%%%%%%%%%%%%%%%%%%%%%%%%%%%%%%%%%%%%%%%%%%%%%%%%%%%%%%%%%%% 
%%%%%%%%%%%%%%%%%%%%%%%%%%%%%%%%%%%%%%%%%%%%%%%%%%%%%%%%%%%%%%%%%%%%%%%%%%%%%%%
%%%%%%%%%%%%%%%%%%%%%%%%%%%%%%%%%%%%%%%%%%%%%%%%%%%%%%%%%%%%%%%%%%%%%%%%%%%%%%%
 of its pseudotriangulation. 
The set of pseudotriangulations of a family of convex bodies depends only on its chirotope; cf.  Appendix~\ref{Perturbations}. 
Therefore it is sensible to ask if a pseudotriangulation  of a family of convex bodies presented  by its
chirotope is efficiently computable and, more generally, it is sensible to ask if a pseudotriangulation that contains a given set of pairwise interior non-crossing distinguished free bitangent line segments is efficiently computable. 
The main result of the paper is a  positive answer to the first question  and, at the same price, a positive answer to a restricted version of the second question.

%%%%%%%%%%%%%%%%%%%%%%%%%%%%%%%%%%%%%%%%%%%%%%%%%%%%%%%%%%%%%%%%%%%%%%%%%%%%%%%
%%%%%%%%%%%%%%%%%%%%%%%%%%%%%%%%%%%%%%%%%%%%%%%%%%%%%%%%%%%%%%%%%%%%%%%%%%%%%%%
%%%%%%%%%%%%%%%%%%%%%%%%%%%%%%%%%%%%%%%%%%%%%%%%%%%%%%%%%%%%%%%%%%%%%%%%%%%%%%%
%%%%%%%%%%%%%%%%%%%%%%%%%%%%%%%%%%%%%%%%%%%%%%%%%%%%%%%%%%%%%%%%%%%%%%%%%%%%%%%
%%%%%%%%%%%%%%%%%%%%%%%%%%%%%%%%%%%%%%%%%%%%%%%%%%%%%%%%%%%%%%%%%%%%%%%%%%%%%%%
\begin{theorem}\label{mainresult}
A pseudotriangulation (and in particular the boundary bitan\-gent line segments) of a finite planar family of $n$ pairwise disjoint convex bodies presented  by its chirotope
is computable in $O(n\log n)$ time and linear space.  
A similar result holds for the problem of computing a pseudotriangulation that contains a given set of pairwise interior non-crossing distinguished free bitangent line segments,
 under the assumption that the number of distinguished bitangent line segments that appear consecutively on the boundary of any pseudotriangle of any pseudotriangulation 
of the family of convex bodies containing the distinguished bitangent line segments is a constant. \qed
\end{theorem}
%%%%%%%%%%%%%%%%%%%%%%%%%%%%%%%%%%%%%%%%%%%%%%%%%%%%%%%%%%%%%%%%%%%%%%%%%%%%%%%
%%%%%%%%%%%%%%%%%%%%%%%%%%%%%%%%%%%%%%%%%%%%%%%%%%%%%%%%%%%%%%%%%%%%%%%%%%%%%%%
%%%%%%%%%%%%%%%%%%%%%%%%%%%%%%%%%%%%%%%%%%%%%%%%%%%%%%%%%%%%%%%%%%%%%%%%%%%%%%%
%%%%%%%%%%%%%%%%%%%%%%%%%%%%%%%%%%%%%%%%%%%%%%%%%%%%%%%%%%%%%%%%%%%%%%%%%%%%%%
\subsubsection{Three independent algorithms.}
Subsequently we use the term {\it family of pairwise disjoint convex bodies with constraints} for a finite planar family of pairwise disjoint convex bodies together with a, 
possibly empty, set of pairwise interior non-crossing  
distinguished free bitangent line segments, the {\it constraints} for short; 
in this context, {\it free space} is the space obtained by cutting the complement in the plane of the interiors of the 
convex bodies along the constraints:  this is the disjoint union of two-dimensional surfaces whose cuffs contain exactly one cusp point per endpoint of constraint (counting multiplicities); 
in particular if the set of constraints is a pseudotriangulation, free space is the disjoint union of the pseudotriangles of the pseudotriangulation plus the complement in the plane of the interior of the 
convex hull of the bodies.
%these cusp points split the cuffs of these surfaces into a finite family of convex curves 
%whose number is twice the number of constraints plus the number of bodies free of constraints; these convex curves are called the {\it views} of the constrained family of bodies, 
%and  the complexity of a view is defined to be one plus its number of constraints; in 
%particular if the set of constraints is empty,  free space is the complement of the interiors of the bodies, 
%the views are simply the boundaries of the bodies and the complexity of a view is a constant. %%% see Figure~\ref{cuffs} for an illustration.) 
%%%%%%%%%%%%%%%%%%%%%%%%%%%%%%%%%%%%%%%%%%%%%%%%%%%%%%%%%%%%%%%%%%%%%%%%%%%%%%%
%%%%%%%%%%%%%%%%%%%%%%%%%%%%%%%%%%%%%%%%%%%%%%%%%%%%%%%%%%%%%%%%%%%%%%%%%%%%%%%
The family will be said {\it  well-constrained} if it satisfies the condition stated in the theorem above, that is, if the number of constraints 
that appear consecutively on the boundary of any pseudotriangle of any completion of the set of constraints into a pseudotriangulation
of the family of convex bodies is a constant. 
%%%%%%%%%%%%%%%%%%%%%%%%%%%%%%%%%%%%%%%%%%%%%%%%%%%%%%%%%%%%%%%%%%%%%%%%%%%%%%%
%%%%%%%%%%%%%%%%%%%%%%%%%%%%%%%%%%%%%%%%%%%%%%%%%%%%%%%%%%%%%%%%%%%%%%%%%%%%%%%
%%%%%%%%%%%%%%%%%%%%%%%%%%%%%%%%%%%%%%%%%%%%%%%%%%%%%%%%%%%%%%%%%%%%%%%%%%%%%%%
%%%%%%%%%%%%%%%%%%%%%%%%%%%%%%%%%%%%%%%%%%%%%%%%%%%%%%%%%%%%%%%%%%%%%%%%%%%%%%%
Our pseudotriangulation algorithm is the composition of three independent algorithms:
\begin{enumerate} 
\item an algorithm to compute the convex hull, i.e., the boundary bitangent line segments,  of a planar family of pairwise disjoint convex bodies;
% as usual ``computing the convex hull'' means ``computing its boundary bitan\-gent line
%segments; 
\item an algorithm to compute a {\it cross-section} of the {\it  visibility complex} of a family of pairwise disjoint convex bodies with constraints;
and 
\item an algorithm to compute the {\it greedy pseudotriangulation} associated to a given cross-section of the visibility complex of a 
family of pairwise disjoint convex bodies with constraints
whose set of constraints contains the boundary bitangent line segments of the family of bodies.
\end{enumerate}
%%%%%%%%%%%%%%%%%%%%%%%%%%%%%%%%%%%%%%%%%%%%%%%%%%%%%%%%%%%%%%%%%%%%%%%%%%%%%%%
%%%%%%%%%%%%%%%%%%%%%%%%%%%%%%%%%%%%%%%%%%%%%%%%%%%%%%%%%%%%%%%%%%%%%%%%%%%%%%%
%%%%%%%%%%%%%%%%%%%%%%%%%%%%%%%%%%%%%%%%%%%%%%%%%%%%%%%%%%%%%%%%%%%%%%%%%%%%%%%
%%%%%%%%%%%%%%%%%%%%%%%%%%%%%%%%%%%%%%%%%%%%%%%%%%%%%%%%%%%%%%%%%%%%%%%%%%%%%%%

Before recalling the definitions of the terms {\it visibility complex}, {\it cross-section}, and {\it greedy pseudotriangulation}, 
 we add to our two (non-restrictive) assumptions concerning the boundaries of the convex bodies---recall that one of these two assumptions says that the boundaries are free of line segments and the other one says 
that there is exactly one tangent through each boundary point---the assumption that the family of convex bodies has no triple tangent. 
This additional assumption is not a restriction on the possible inputs of our algorithm since for any non-simple chirotope there exists a simple chirotope, 
computable in constant time, 
such that the non-simple chirotope and the simple chirotope have the same set of free bitangent line segments and the same set of pseudotriangulations; cf. Appendix~\ref{Perturbations}.  
%%%%%%%%%%%%%%%%%%%%%%%%%%%%%%%%%%%%%%%%%%%%%%%%%%%%%%%%%%%%%%%%%%%%%%%%%%%%%%%
%%%%%%%%%%%%%%%%%%%%%%%%%%%%%%%%%%%%%%%%%%%%%%%%%%%%%%%%%%%%%%%%%%%%%%%%%%%%%%%
%%%%%%%%%%%%%%%%%%%%%%%%%%%%%%%%%%%%%%%%%%%%%%%%%%%%%%%%%%%%%%%%%%%%%%%%%%%%%%%
%%%%%%%%%%%%%%%%%%%%%%%%%%%%%%%%%%%%%%%%%%%%%%%%%%%%%%%%%%%%%%%%%%%%%%%%%%%%%%%
%%%%%%%%%%%%%%%%%%%%%%%%%%%%%%%%%%%%%%%%%%%%%%%%%%%%%%%%%%%%%%%%%%%%%%%%%%%%%%%

\subsubsection{Visibility complexes.}
Let $\afs$ be a connected component or a  union of connected components of the free space of a 
given family of pairwise disjoint convex bodies with constraints living in a topological plane~$\pointset$. 
 We denote by 
$\spaceoflinesof{\mathbb{A}}$ and $\spaceoforientedlinesof{\mathbb{A}}$ the spaces of lines and directed lines of $\pointset$
 and we take for granted that the canonical projection  
$\MapLight{}{\spaceoforientedlinesof{\mathbb{A}}}{\spaceoflinesof{\mathbb{A}}}$ is a two-covering. 
The space $\afs$ inherits from the topological point-line incidence geometry  of~$\pointset$ a natural partial topological point-line incidence geometry  whose
system of lines
%%%, denoted  $\spaceoflinesof{\mathbb{X}}$, 
$\spaceoflinesof{\mathbb{X}}$ 
is defined as the space of pairs $(x,\ell)$ where
$\ell$ ranges over the space of lines 
%% $\spaceoflinesof{\pointset}$ 
of $\pointset$ and $x$ 
the set of connected components of the pre-image of the line $\ell$ under the
canonical projection $\mathbb{X} \rightarrow \mathbb{A}$, 
and whose set of incidences is the set of point-line pairs $(p,(x,l)) \in \mathbb{X} \times \spaceoflinesof{\mathbb{X}}$ with $p\in x$.  Note that  
the second component $\ell$ of a pair $(x,\ell) \in \spaceoflinesof{\mathbb{X}}$ is determined by its first component $x$ unless $x$ is reduced to a point,
which happens precisely when $x$ is a cusp point of the boundary of $\mathbb{X}$.  
%%%{\sc Should I speak of the topology put on the set of lines?} 
Except in the case where $\afs$ is the complement of the interior of the convex hull of the bodies, in which case $\spaceoflinesof{\mathbb{X}}$ is a torus to which is attached, along one of its 
non trivial closed simple curve, a one-punctured disk, the space of lines of  $\mathbb{X}$ 
has a natural structure of (possibly one-punctured) two-dimensional cell complex:  its $1$-skeleton is the set of tangents to the boundary of $\mathbb{X}$---which includes the lines 
through the cusp points of the boundary of $\mathbb{X}$---and its $0$-skeleton is the set of bitangents of $\mathbb{X}$.  
The  {\it visibility complex} of $\mathbb{X}$ is its space of lines $\spaceoflinesof{\mathbb{X}}$ endowed with its natural structure of cell complex; furthermore 
we add to the definition that the one-skeleton of the visibility complex is endowed with the orientation inherited by duality from the orientation of the underlying topological plane; cf. \cite{G-pv-vc-96,G-pv-ptta-96,ap-sstvc-03}.
Similarly we introduce the space $\spaceoforientedlinesof{\mathbb{X}}$ of directed lines of $\mathbb{X}$,  endowed with its natural structure of cell complex together 
with the natural orientation of its one-skeleton inherited by duality from the orientation of the underlying topological plane, and we take for granted that the natural projection $\MapLight{}{\spaceoforientedlinesof{\mathbb{X}}}{\spaceoflinesof{\mathbb{X}}}$ is a two-covering in picture of the two-covering $\MapLight{}{\spaceoforientedlinesof{\mathbb{A}}}{\spaceoflinesof{\mathbb{A}}}$ and that 
the cell structure on $\spaceoforientedlinesof{\mathbb{X}}$ is regular contrary, in general, to that of $\spaceoflinesof{\mathbb{X}}$. 
%The {\it label} of a directed line is the sequence of bodies intersected by the line ordered as they appear along the line 
%and prefixed or postfixed or both prefixed and postfixed by the symbol $\infty$ in case the line is (orientation preserving) homeomorphic 
%to $\mathbb{R}^+,\mathbb{R}^-$ or $\mathbb{R}$ endowed with their natural orientations. 
%%%%%%%%%%%%%%%%%%%%%%%%%%%%%%%%%%%%%%%%%%%%%%%%%%%%%%%%%%%%%%%%%%%%%%%%%%%%%%%
%%%%%%%%%%%%%%%%%%%%%%%%%%%%%%%%%%%%%%%%%%%%%%%%%%%%%%%%%%%%%%%%%%%%%%%%%%%%%%%
%%%%%%%%%%%%%%%%%%%%%%%%%%%%%%%%%%%%%%%%%%%%%%%%%%%%%%%%%%%%%%%%%%%%%%%%%%%%%%%
%%%%%%%%%%%%%%%%%%%%%%%%%%%%%%%%%%%%%%%%%%%%%%%%%%%%%%%%%%%%%%%%%%%%%%%%%%%%%%%
\begin{example}
The visibility complex $\sofl{\afs}$ of the free space $\afs$ of a family of two disjoint convex bodies $o_i,o_j$ is composed~of
 \begin{enumerate}
\item four $0$-cells: the four bitangents $t_1,t_2,t_3,t_4$ of the family of bodies; 
\item eight oriented $1$-cells: the four connected components of $\soflbis{o_i}\setminus \{t_1,t_2,t_3,t_4\}$ 
and the four connected components of $\soflbis{o_j}\setminus \{t_1,t_2,t_3,t_4\}$, where $\soflbis{o_i}$ denotes the set of tangents to $o_i$; and   
\item five $2$-cells : the sets of lines with labels---in the context of a family of convex bodies with empty set of constraints, the {\it label} of a directed line is the sequence of bodies intersected by the line ordered as they appear along the line
and prefixed or postfixed or both prefixed and postfixed by the symbol $\infty$ in case the line is (orientation preserving) homeomorphic
to $\mathbb{R}^+,\mathbb{R}^-$ or $\mathbb{R}$ endowed with their natural orientations---with labels $ij$, $i\infty$, $j\infty$, the set of lines with label $\infty\infty$ that separate the two bodies, and the set of lines with label 
$\infty\infty$ that do not separate the two bodies; 
\end{enumerate}
put together as
indicated in Figure~\ref{VCtwobodies} where we write $i$ for the bitangent $t_i$; 
%%%%%%%%%%%%%%%%%%%%%%%%%%%%%%%%%%%%%%%%%%%%%%%%%%%%%%%%%%%%%%%%%%%%%%%%%%%%%%%
%%%%%%%%%%%%%%%%%%%%%%%%%%%%%%%%%%%%%%%%%%%%%%%%%%%%%%%%%%%%%%%%%%%%%%%%%%%%%%%
%%%%%%%%%%%%%%%%%%%%%%%%%%%%%%%%%%%%%%%%%%%%%%%%%%%%%%%%%%%%%%%%%%%%%%%%%%%%%%%
%%%%%%%%%%%%%%%%%%%%%%%%%%%%%%%%%%%%%%%%%%%%%%%%%%%%%%%%%%%%%%%%%%%%%%%%%%%%%%%
%%%%%%%%%%%%%%%%%%%%%%%%%%%%%%%%%%%%%%%%%%%%%%%%%%%%%%%%%%%%%%%%%%%%%%%%%%%%%%%
%%%%%%%%%%%%%%%%%%%%%%%%%%%%%%%%%%%%%%%%%%%%%%%%%%%%%%%%%%%%%%%%%%%%%%%%%%%%%%%
%%%%%%%%%%%%%%%%%%%%%%%%%%%%%%%%%%%%%%%%%%%%%%%%%%%%%%%%%%%%%%%%%%%%%%%%%%%%%%%
%%%%%%%%%%%%%%%%%%%%%%%%%%%%%%%%%%%%%%%%%%%%%%%%%%%%%%%%%%%%%%%%%%%%%%%%%%%%%%%
%%\begin{figure}[p]
\begin{figure}[!htb]
\begin{center}
\psfrag{one}{$1$}\psfrag{two}{$2$}\psfrag{thr}{$3$}\psfrag{fou}{$4$}
\psfrag{unun}{$i_1$}
\psfrag{deun}{$j_1$}
\psfrag{unde}{$i_2$}
\psfrag{dede}{$j_2$}
\psfrag{unth}{$i_3$}
\psfrag{deth}{$j_3$}
\psfrag{unfo}{$i_4$}
\psfrag{defo}{$j_4$}
\psfrag{un}{$i$}
\psfrag{de}{$j$}
\psfrag{firstbody}{$i$}
\psfrag{secondbody}{$j$}
\psfrag{lunde}{$ij$}
\psfrag{inin}{$\infty\infty$}
\psfrag{unin}{$i\infty$}
\psfrag{dein}{$j\infty$}
\psfrag{infty}{$\infty$}
\psfrag{body}{}
\includegraphics[width = 0.958575\linewidth]{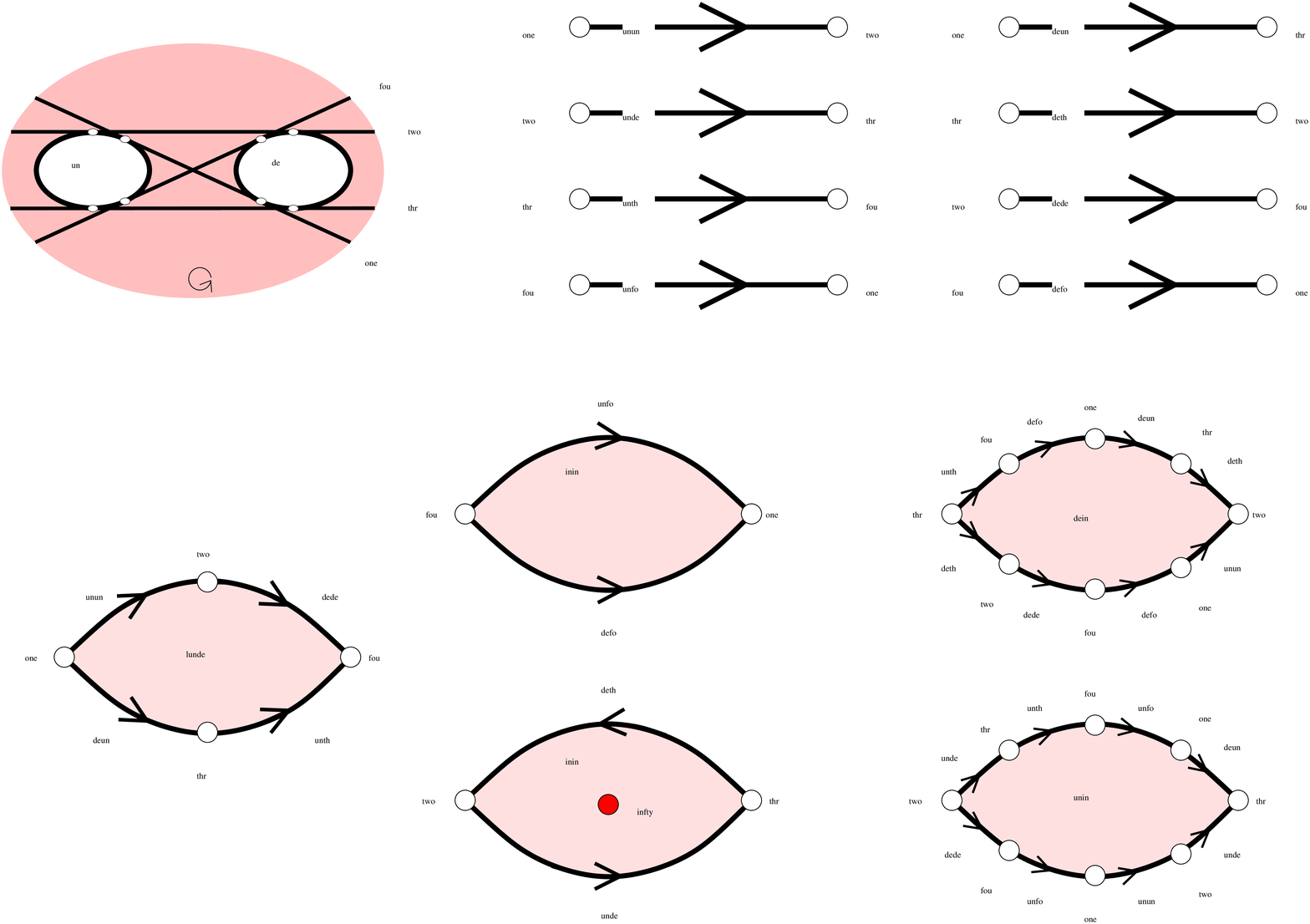}%xfig
\end{center}
\caption{\protect \footnotesize 
%%The visibility complex of a family of two disjoint convex bodies is composed of four $0$-cells, eight $1$-cells,  and five $2$-cells put together as
%%indicated in the figure.
\label{VCtwobodies}}
\end{figure}
%%%%%%%%%%%%%%%%%%%%%%%%%%%%%%%%%%%%%%%%%%%%%%%%%%%%%%%%%%%%%%%%%%%%%%%%%%%%%%%
%%%%%%%%%%%%%%%%%%%%%%%%%%%%%%%%%%%%%%%%%%%%%%%%%%%%%%%%%%%%%%%%%%%%%%%%%%%%%%%
%%%%%%%%%%%%%%%%%%%%%%%%%%%%%%%%%%%%%%%%%%%%%%%%%%%%%%%%%%%%%%%%%%%%%%%%%%%%%%%
%%%%%%%%%%%%%%%%%%%%%%%%%%%%%%%%%%%%%%%%%%%%%%%%%%%%%%%%%%%%%%%%%%%%%%%%%%%%%%%
%%%%%%%%%%%%%%%%%%%%%%%%%%%%%%%%%%%%%%%%%%%%%%%%%%%%%%%%%%%%%%%%%%%%%%%%%%%%%%%
%%%%%%%%%%%%%%%%%%%%%%%%%%%%%%%%%%%%%%%%%%%%%%%%%%%%%%%%%%%%%%%%%%%%%%%%%%%%%%%
%%%%%%%%%%%%%%%%%%%%%%%%%%%%%%%%%%%%%%%%%%%%%%%%%%%%%%%%%%%%%%%%%%%%%%%%%%%%%%%
%%%%%%%%%%%%%%%%%%%%%%%%%%%%%%%%%%%%%%%%%%%%%%%%%%%%%%%%%%%%%%%%%%%%%%%%%%%%%%%
%%similarly the visibility complex of a `generic' pseudotriangle is composed of three $2$-cells, six $1$-cells, and three $0$-cells put together as depicted in Figure~\ref{soflpt}.
this complex is not regular~: the boundaries of the $2$-cells 
with label $i\infty$ and $j\infty$ are complete graphs on four elements; and this complex has one end, indicated by a marked point $\infty$ (red in pdf color) in the figure.  
\end{example}
\begin{example}
The visibility complex $\sofl{\afs}$  of a generic pseudotriangle $\afs$ with cusp points $a, b, c$ consists of 
\begin{enumerate} 
\item three $0$-cells : the tangents $t_a, t_b$  and $t_c$ 
at the cusp points $a, b$, and $c$;
\item six oriented $1$-cells : the $\soflbis{x} = \sofl{x}\setminus \{t_x\}$, $x\in \{a,b,c\}$, where $\sofl{x}$ denotes the set of lines through the point $x$, and 
the three connected components $\alpha,\beta$ and $\gamma$ of the curve $\sofl{\partial\mathbb{X}}$ of tangent lines to the
pseudotriangle minus $t_a, t_b$ , and $t_c$; and 
\item three $2$-cells : the interiors of the $\sofl{x,x'}$ where $\sofl{x,x'}$ denotes the set of lines 
joining the sides opposite to the pair of cusp points $x$ and~$x'$; 
\end{enumerate} 
%These three $2$-cells, six $1$-cells, and three $0$-cells are put together as indicated in  Figure~~\ref{soflpt}; 
put together as indicated in  Figure~\ref{soflpt}; 
again observe that this complex is non regular (its one-skeleton is already non regular); this complex has no end.
%%%%%%%%%%%%%%%%%%%%%%%%%%%%%%%%%%%%%%%%%%%%%%%%%%%%%%%%%%%%%%%%%%%%%%%%%%%%%%%
%%%%%%%%%%%%%%%%%%%%%%%%%%%%%%%%%%%%%%%%%%%%%%%%%%%%%%%%%%%%%%%%%%%%%%%%%%%%%%%
%%%%%%%%%%%%%%%%%%%%%%%%%%%%%%%%%%%%%%%%%%%%%%%%%%%%%%%%%%%%%%%%%%%%%%%%%%%%%%%
%%%%%%%%%%%%%%%%%%%%%%%%%%%%%%%%%%%%%%%%%%%%%%%%%%%%%%%%%%%%%%%%%%%%%%%%%%%%%%%
%%%%%%%%%%%%%%%%%%%%%%%%%%%%%%%%%%%%%%%%%%%%%%%%%%%%%%%%%%%%%%%%%%%%%%%%%%%%%%%
%%\begin{figure}[p]
\begin{figure}[!htb]
\begin{center}
\psfrag{PT}{a generic pseudotriangle $\mathbb{X}$}
\psfrag{one}{the one-skeleton of $\sofl{\mathbb{X}}$}
\psfrag{twoskeleton}{the two-skeleton of $\sofl{\mathbb{X}}$}
\psfrag{vc}{the visibility complex $\sofl{\mathbb{X}}$ of $\mathbb{X}$}
\psfrag{lla}{$\sofl{b,c}$}
\psfrag{llb}{$\sofl{c,a}$}
\psfrag{llc}{$\sofl{a,b}$}
\psfrag{ptstar}{$\partial {\mathbb{X}}^*$}
\psfrag{ptstar}{$\sofl{\partial\mathbb{X}}$}
\psfrag{lab}{$\sofl{a,b}$}
\psfrag{t1}{$t_a$}
\psfrag{t2}{$t_b$}
\psfrag{t3}{$t_c$}
\psfrag{a}{$a$}
\psfrag{b}{$b$}
\psfrag{c}{$c$}
\psfrag{ts}{$\sofl{\partial \mathbb{X}}$}
\psfrag{as}{$\sofl{a}$}
\psfrag{bs}{$\sofl{b}$}
\psfrag{cs}{$\sofl{c}$}
\psfrag{astar}{$\soflbis{a}$}
\psfrag{bstar}{$\soflbis{b}$}
\psfrag{cstar}{$\soflbis{c}$}
\psfrag{1}{1}
\psfrag{2}{2}
\psfrag{3}{3}
\psfrag{4}{4}
\psfrag{alpha}{$\alpha$}\psfrag{beta}{$\beta$}\psfrag{gamma}{$\gamma$}
\includegraphics[width = 0.958575\linewidth]{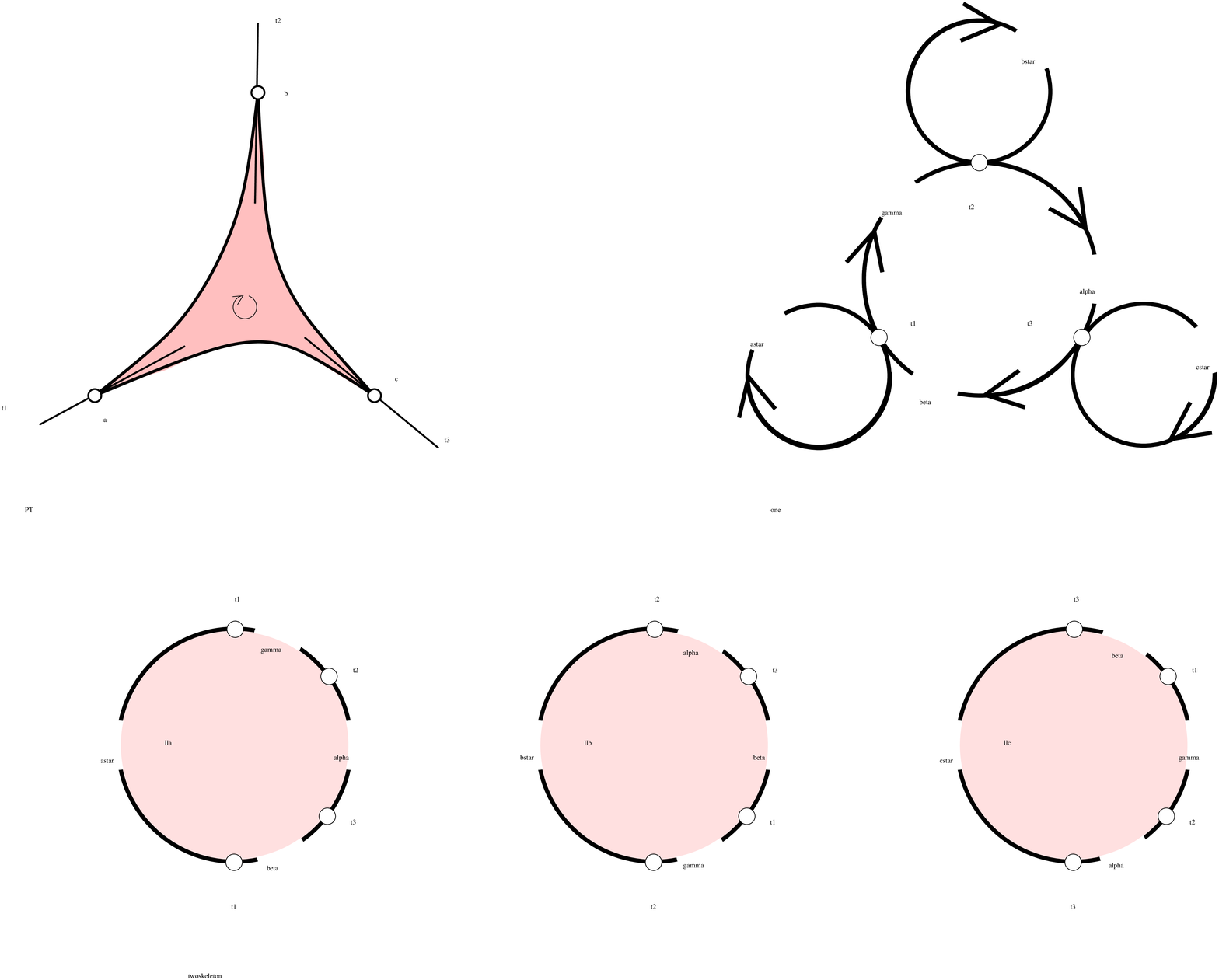}%xfig
\caption{\protect \small 
%% The visibility complex $\sofl{\afs}$  of a generic pseudotriangle $\afs$ with cusp points $a, b, c$ consists of three $0$-cells (the tangents $t_a, t_b$  and $t_c$ 
%% at the cusp points $a, b$, and $c$),  six $1$-cells (the $\soflbis{x} = \sofl{x}\setminus \{t_x\}$, $x\in \{a,b,c\}$, where $\sofl{x}$ denotes the set of lines through the point $x$, 
%% and the three connected components $\alpha,\beta$ and $\gamma$ of the curve $\sofl{\partial\mathbb{X}}$ of tangent lines to the
%% pseudotriangle minus $t_a, t_b$ , and $t_c$), and three $2$-cells (the interiors of the $\sofl{x,x'}$ where $\sofl{x,x'}$ denotes the set of lines 
%% joining the sides opposite to the pair of cusp points $x$ and $x'$).  
\label{soflpt}}
\end{center}
\end{figure}
%%%%%%%%%%%%%%%%%%%%%%%%%%%%%%%%%%%%%%%%%%%%%%%%%%%%%%%%%%%%%%%%%%%%%%%%%%%%%%%
%%%%%%%%%%%%%%%%%%%%%%%%%%%%%%%%%%%%%%%%%%%%%%%%%%%%%%%%%%%%%%%%%%%%%%%%%%%%%%%
%%%%%%%%%%%%%%%%%%%%%%%%%%%%%%%%%%%%%%%%%%%%%%%%%%%%%%%%%%%%%%%%%%%%%%%%%%%%%%%
%%%%%%%%%%%%%%%%%%%%%%%%%%%%%%%%%%%%%%%%%%%%%%%%%%%%%%%%%%%%%%%%%%%%%%%%%%%%%%%
\end{example}

%\clearpage
%We now recall the definition of the cross-sections (of a visibility complex) and that of their associated greedy
%pseudotriangulations. 
%The orientation of the plane $\pointset$ induces, via the duality map, the choice of a generator of the
%fundamental group of  its space of lines $\systemlines{\pointset}$, and therefore an orientation of the $1$-skeleton of
%$\systemlines{\mathbb{X}}$. 
\subsubsection{Cross-sections.}
%A remarkable property of the orientation of the $1$-skeleton of the visibility complex is that the  boundary  
The boundary  of any bounded $2$-cell of $\spaceoforientedlinesof{\mathbb{X}}$---bounded in the
sense that the cell contains no end of $\systemlines{\mathbb{X}}$---has a unique
vertex of outdegree two and a unique vertex of indegree two; therefore one can
speak of the source and sink vertices of a $1$-  or  bounded $2$-cell of $\spaceoforientedlinesof{\mathbb{X}}$,
 and one can speak of the left and right boundary chains of a $2$-cell.
Let $\Iwc{}\rightarrow \Ivc{}$ be the inverse image of a universal cover $\uniIlineset$ of $\Ilineset$  
under the natural projection $\MapLight{\linecoo}{\Ivc}{\Ilineset}$---that is $\Iwc{}$ is the set of pairs $(v,l) \in \Ivc\times \uniIlineset$ such that 
the image of $v$ under $\MapLight{\linecoo}{\Ivc}{\Ilineset}$ coincides with the image of $l$ 
under $\MapLight{}{\uniIlineset}{\Ilineset}$, and $\MapLight{}{\Iwc{}}{\Ivc{}}$ is the first projection,
cf.~\cite[pages 113-114]{g-eta-71}---let $\Vposet(\mathbb{X})$ be the set of 
 cells of 
$\Iwc{}$ endowed with the partial order generated by the relations 
\begin{equation}
\sour{\sigma} \prec \sigma \prec \sink{\sigma}
\end{equation}
 where $\sigma$ ranges over the
set  of $1$- and bounded $2$-cells of $\Vposet(\mathbb{X})$ and 
where $\sour{\sigma}$ and $\sink{\sigma}$ denote the source and the sink of
the cell $\sigma$, let $\shiftop$  be the generator of the automorphism group of the covering $\Iwc{} \rightarrow \Ivc{}$ defined by the condition that 
$\sigma \prec \shiftop(\sigma)$: the shift operator for short, and finally let $\MAC$ be a maximal antichain of $\Vposet(\mathbb{X})$.
%Since the lines of $\Iwc{}$ are oriented one can define canonically the left and right boundaries of a  bounded $2$-cell and one can define the left
%or (exclusive) right boundary of an unbounded $2$-cell.
The {\it cross-section}, denoted  $\CS{\MAC}$, of the visibility complex of $\mathbb{X}$ at the maximal antichain $\MAC$ is the directed multigraph whose 
set of arcs is the set of $2$-cells of $\MAC$ and whose
set of nodes is the set of $0$- and $1$-cells of $\MAC$,  the source node of an arc
being defined as the unique node included in its right boundary (if any) and
its sink node being defined as the unique node included in its left boundary (if
any).
%%%%%%%%%%%%%%%%%%%%%%%%%%%%%%%%%%%%%%%%%%%%%%%%%%%%%%%%%%%%%%%%%%%%%%%%%%%%%%%
%%%%%%%%%%%%%%%%%%%%%%%%%%%%%%%%%%%%%%%%%%%%%%%%%%%%%%%%%%%%%%%%%%%%%%%%%%%%%%%
%%%%%%%%%%%%%%%%%%%%%%%%%%%%%%%%%%%%%%%%%%%%%%%%%%%%%%%%%%%%%%%%%%%%%%%%%%%%%%%
%%%%%%%%%%%%%%%%%%%%%%%%%%%%%%%%%%%%%%%%%%%%%%%%%%%%%%%%%%%%%%%%%%%%%%%%%%%%%%%
\begin{example} \label{filterantichain}
Let $I$ be a proper filter of the subposet of $0$-cells of $\Vposet(\afs)$. Then the set 
 of $1$- and $2$-cells of $\Vposet(\afs)$  whose sinks belong to $I$ but not their sources is a maximal antichain; the corresponding cross-section 
is called the canonical cross-section associated with the filter $I$.
\end{example}
%%%%%%%%%%%%%%%%%%%%%%%%%%%%%%%%%%%%%%%%%%%%%%%%%%%%%%%%%%%%%%%%%%%%%%%%%%%%%%%
%%%%%%%%%%%%%%%%%%%%%%%%%%%%%%%%%%%%%%%%%%%%%%%%%%%%%%%%%%%%%%%%%%%%%%%%%%%%%%%
%%%%%%%%%%%%%%%%%%%%%%%%%%%%%%%%%%%%%%%%%%%%%%%%%%%%%%%%%%%%%%%%%%%%%%%%%%%%%%%
%%%%%%%%%%%%%%%%%%%%%%%%%%%%%%%%%%%%%%%%%%%%%%%%%%%%%%%%%%%%%%%%%%%%%%%%%%%%%%%
\begin{example} \label{defccsangle}
Figure~\ref{FinalSlopeCS} depicts  a family of $7$ convex bodies of the real affine plane with one constraint 
(the bodies are numbered from $1$ to $7$ and the constraint is the undirected version of the right-right bitangent line segment joining the third body of the family to the fourth body) 
and (an upward drawing of) the canonical cross-section of its visibility complex associated 
with the filter of the subposet of vertices 
of $\Vposet(\afs)$ with angle $\geq 0$. 
%and the canonical cross-section $\Gamma(0)$ of its visibility complex associated with the null angle . 
%%%%%%%%%%%%%%%%%%%%%%%%%%%%%%%%%%%%%%%%%%%%%%%%%%%%%%%%%%%%%%%%%%%%%%%%%%%%%%%
%%%%%%%%%%%%%%%%%%%%%%%%%%%%%%%%%%%%%%%%%%%%%%%%%%%%%%%%%%%%%%%%%%%%%%%%%%%%%%%
%%%%%%%%%%%%%%%%%%%%%%%%%%%%%%%%%%%%%%%%%%%%%%%%%%%%%%%%%%%%%%%%%%%%%%%%%%%%%%%
%%%%%%%%%%%%%%%%%%%%%%%%%%%%%%%%%%%%%%%%%%%%%%%%%%%%%%%%%%%%%%%%%%%%%%%%%%%%%%%
\begin{figure}[!htb]
%%%\footnotesize
\centering
\psfrag{one}{\footnotesize $1$}\psfrag{two}{\footnotesize $2$}\psfrag{thr}{\footnotesize $3$}\psfrag{fou}{\footnotesize $4$}\psfrag{fiv}{\footnotesize $5$}\psfrag{six}{\footnotesize $6$}\psfrag{sev}{\footnotesize $7$}\psfrag{8}{8}%%%
\psfrag{Gu}{$\G(u)$}
\psfrag{CSu}{$\Gamma(u)$}
\psfrag{ez}{$e_0$}
\psfrag{eo}{$e_1$}
\psfrag{aone}{$1$}
\psfrag{atwo}{$2$}
\psfrag{athr}{$3$}
\psfrag{afou}{$4$}
\psfrag{afiv}{$5$}
\psfrag{safiv}{\footnotesize $5$}
\psfrag{asix}{$6$}
\psfrag{asev}{$7$}
\psfrag{ahei}{$8$}
\psfrag{anin}{$9$}
\psfrag{aten}{$10$}
\psfrag{aele}{$11$}
\psfrag{atwe}{$12$}
\psfrag{athrt}{$13$}
\psfrag{afout}{$14$}
\psfrag{safout}{\footnotesize $14$}
\psfrag{afivt}{$15$}
\psfrag{asixt}{$16$}
\psfrag{asevt}{$17$}
\psfrag{aheit}{$18$}
\psfrag{anint}{$19$}
\psfrag{atwt}{$20$}
\psfrag{atwtone}{$21$}
\psfrag{satwtone}{\footnotesize $21$}
\includegraphics[width=0.9958575\linewidth]{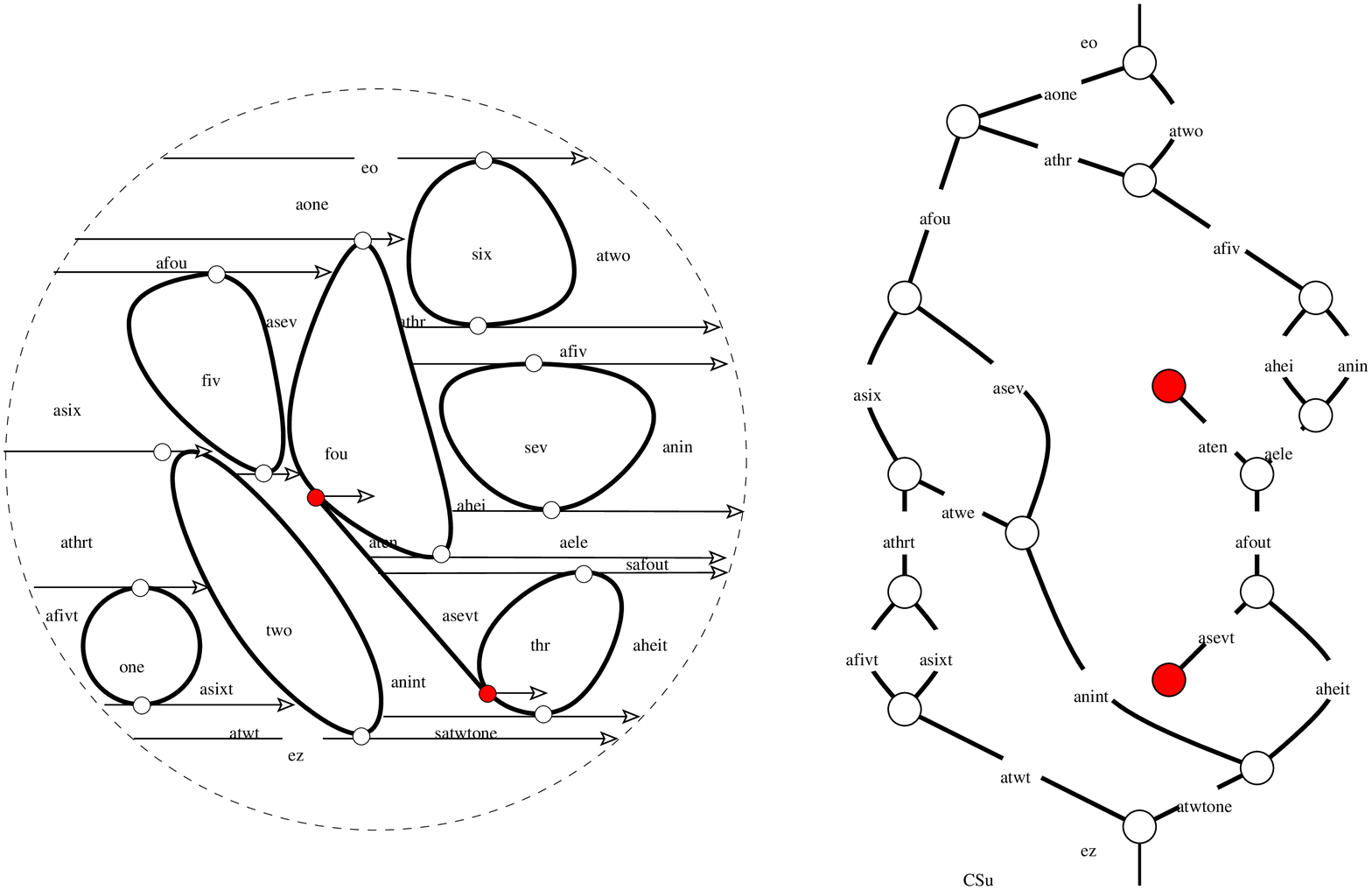}
\caption{\protect \small \label{FinalSlopeCS}}
\end{figure}
%%%%%%%%%%%%%%%%%%%%%%%%%%%%%%%%%%%%%%%%%%%%%%%%%%%%%%%%%%%%%%%%%%%%%%%%%%%%%%%
%%%%%%%%%%%%%%%%%%%%%%%%%%%%%%%%%%%%%%%%%%%%%%%%%%%%%%%%%%%%%%%%%%%%%%%%%%%%%%%
%%%%%%%%%%%%%%%%%%%%%%%%%%%%%%%%%%%%%%%%%%%%%%%%%%%%%%%%%%%%%%%%%%%%%%%%%%%%%%%
%%%%%%%%%%%%%%%%%%%%%%%%%%%%%%%%%%%%%%%%%%%%%%%%%%%%%%%%%%%%%%%%%%%%%%%%%%%%%%%
The family of convex bodies is augmented for each $1$-cell $e$ of the cross-section with the horizontal line $t(e) \in e$.  The horizontal lines 
$t(e)$ induce a trapezoidal decomposition of free space whose trapezoids ($23$ in number) are in one-to-one correspondence with the arcs of the cross-section :  $21$ of these $23$ 
trapezoids are labeled in the Figure 
and these labels are reported on the corresponding arcs of the cross-section.
\end{example}
%%%%%%%%%%%%%%%%%%%%%%%%%%%%%%%%%%%%%%%%%%%%%%%%%%%%%%%%%%%%%%%%%%%%%%%%%%%%%%%
%%%%%%%%%%%%%%%%%%%%%%%%%%%%%%%%%%%%%%%%%%%%%%%%%%%%%%%%%%%%%%%%%%%%%%%%%%%%%%%
%%%%%%%%%%%%%%%%%%%%%%%%%%%%%%%%%%%%%%%%%%%%%%%%%%%%%%%%%%%%%%%%%%%%%%%%%%%%%%%
%%%%%%%%%%%%%%%%%%%%%%%%%%%%%%%%%%%%%%%%%%%%%%%%%%%%%%%%%%%%%%%%%%%%%%%%%%%%%%%
%%%%%%%%%%%%%%%%%%%%%%%%%%%%%%%%%%%%%%%%%%%%%%%%%%%%%%%%%%%%%%%%%%%%%%%%%%%%%%%
%%%%%%%%%%%%%%%%%%%%%%%%%%%%%%%%%%%%%%%%%%%%%%%%%%%%%%%%%%%%%%%%%%%%%%%%%%%%%%%
%%%%%%%%%%%%%%%%%%%%%%%%%%%%%%%%%%%%%%%%%%%%%%%%%%%%%%%%%%%%%%%%%%%%%%%%%%%%%%%
%%%%%%%%%%%%%%%%%%%%%%%%%%%%%%%%%%%%%%%%%%%%%%%%%%%%%%%%%%%%%%%%%%%%%%%%%%%%%%%
\begin{example} \label{defccs}
Figure~\ref{FinalCCSFNnew} depicts a family of $7$ convex bodies of the real affine plane 
with one constraint (the bodies are numbered from $1$ to $7$  and the constraint is the undirected version of the right-left bitangent line segment joining the second body of the family to the fourth body) 
and the canonical cross-section associated with the filter of $0$-cells of $\Vposet(\afs)$ generated by the lift in $\Iwc{}$ of 
the principal filter of any left-left lift in $\uniIlineset$ of a left-left boundary bitangent (the one joining the first body to the second body).   
%%%%%%%%%%%%%%%%%%%%%%%%%%%%%%%%%%%%%%%%%%%%%%%%%%%%%%%%%%%%%%%%%%%%%%%%%%%%%%%
%%%%%%%%%%%%%%%%%%%%%%%%%%%%%%%%%%%%%%%%%%%%%%%%%%%%%%%%%%%%%%%%%%%%%%%%%%%%%%%
%%%%%%%%%%%%%%%%%%%%%%%%%%%%%%%%%%%%%%%%%%%%%%%%%%%%%%%%%%%%%%%%%%%%%%%%%%%%%%%
%%%%%%%%%%%%%%%%%%%%%%%%%%%%%%%%%%%%%%%%%%%%%%%%%%%%%%%%%%%%%%%%%%%%%%%%%%%%%%%
\begin{figure}[!htb]
%%%\footnotesize
\centering
\psfrag{one}{\footnotesize $1$}\psfrag{two}{\footnotesize $2$}\psfrag{thr}{\footnotesize $3$}\psfrag{fou}{\footnotesize $4$}\psfrag{fiv}{\footnotesize $5$}\psfrag{six}{\footnotesize $6$}\psfrag{sev}{\footnotesize $7$}\psfrag{8}{8}%%%
\psfrag{a}{$u$}\psfrag{b}{$1$}\psfrag{c}{$2$}\psfrag{d}{$3$}\psfrag{e}{$4$}\psfrag{f}{$v$}
\psfrag{u}{$e_0$}
\psfrag{v}{$e_1$}
\psfrag{w}{$e$}
\psfrag{tw}{$t(e)$}
\psfrag{x}{$5$}
\psfrag{g}{$5$}
\psfrag{ssigp}{$6$}
\psfrag{sigp}{$6$}
\psfrag{sig}{$7$}
\psfrag{mfls}{$7$}
\psfrag{Gu}{$\G(u)$}
\psfrag{CSu}{$\Gamma(u)$}
\psfrag{ahei}{$8$}
\psfrag{anin}{$9$}
\includegraphics[width=0.958575\linewidth]{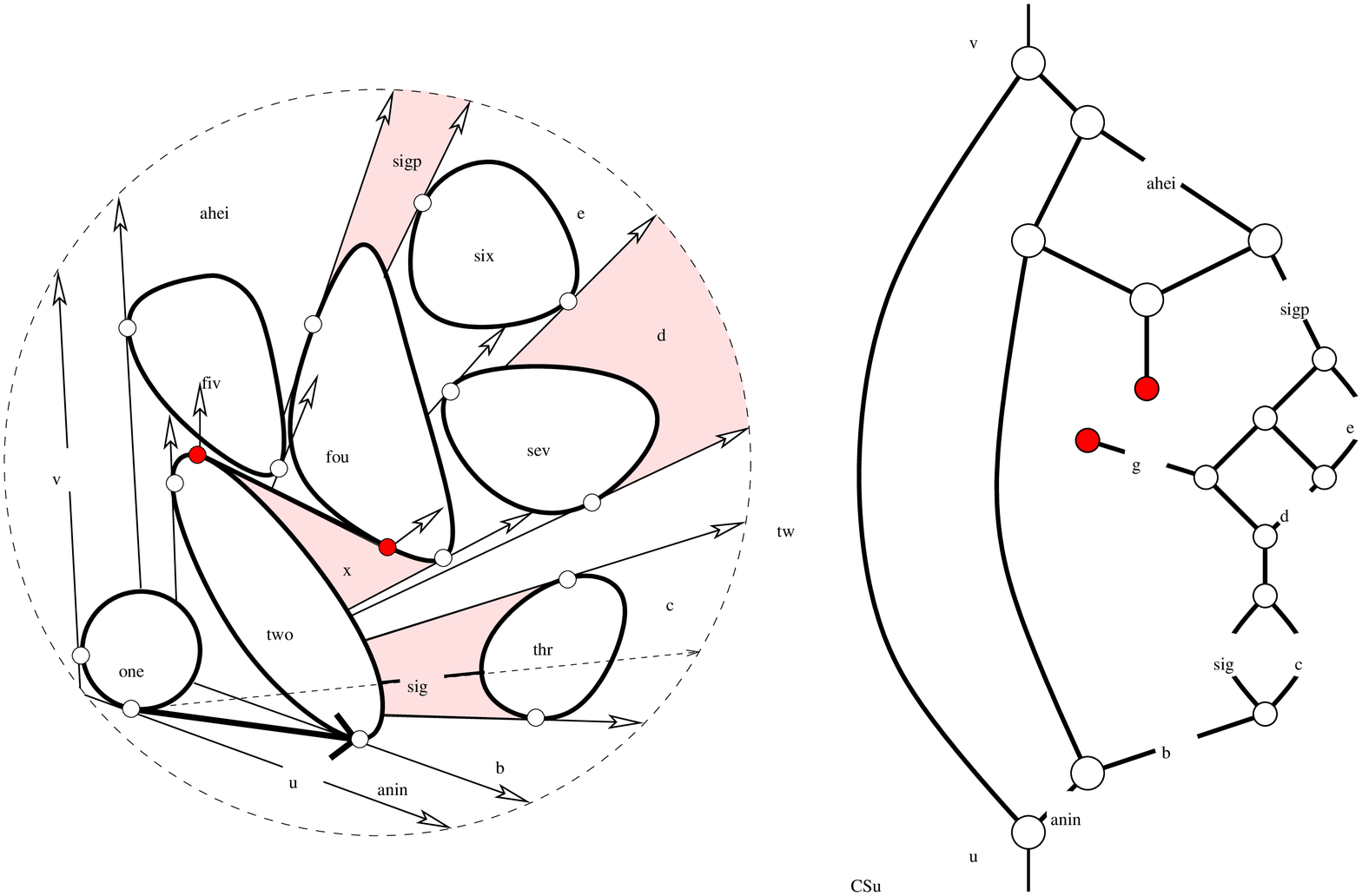}
\caption{\protect \small \label{FinalCCSFNnew}}
\end{figure}
The family of convex bodies is augmented for each $1$-cell $e$ of the cross-section with a line $t(e) \in e$.  
The $t(e)$ induce a trapezoidal
decomposition of free space whose trapezoids ($23$ in number) are in one-to-one correspondence with the arcs of the cross-section :  $9$ of these $23$ 
trapezoids are labeled in the Figure 
and these labels are reported on the corresponding arcs of the cross-section.
\end{example}
%%%%%%%%%%%%%%%%%%%%%%%%%%%%%%%%%%%%%%%%%%%%%%%%%%%%%%%%%%%%%%%%%%%%%%%%%%%%%%%
%%%%%%%%%%%%%%%%%%%%%%%%%%%%%%%%%%%%%%%%%%%%%%%%%%%%%%%%%%%%%%%%%%%%%%%%%%%%%%%
%%%%%%%%%%%%%%%%%%%%%%%%%%%%%%%%%%%%%%%%%%%%%%%%%%%%%%%%%%%%%%%%%%%%%%%%%%%%%%%
%%%%%%%%%%%%%%%%%%%%%%%%%%%%%%%%%%%%%%%%%%%%%%%%%%%%%%%%%%%%%%%%%%%%%%%%%%%%%%%

%\clearpage
\subsubsection{Greedy pseudotriangulations.}
One of the key results of the theory of visibility complexes is that the set of sink bitangent line segments{}\footnote{Since there are no tritangent
the map that assigns to a free bitangent line segment its supporting line realizes a one-to-one and onto correspondence between the set of free bitangent line segments
 and the set of vertices of the visibility complex; thus one can speak of the  sink bitangent line segment of a $0$-, $1$-, or $2$-cell.}
of the cells of a cross-section of a visibility complex is a  pseudotriangulation; cf.~\cite[Theorem 6, Claim 1]{ap-sstvc-03}.
This pseudotriangulation is called {\it greedy} because it can also be defined as the set of bitangent line segments 
of the sequence $v_1,v_2,\ldots, v_h$ of vertices of $\Vposet(\mathbb{X})$ 
defined inductively by $v_i$ is a $\prec$-minimal element in the poset of vertices of the filter generated by the cross-section 
crossing none of the elements of the set $\{v_1,v_2,\ldots,v_{i-1}\}.$

%%%%%%%%%%%%%%%%%%%%%%%%%%%%%%%%%%%%%%%%%%%%%%%%%%%%%%%%%%%%%%%%%%%%%%%%%%%%%%%
%%%%%%%%%%%%%%%%%%%%%%%%%%%%%%%%%%%%%%%%%%%%%%%%%%%%%%%%%%%%%%%%%%%%%%%%%%%%%%%
%%%%%%%%%%%%%%%%%%%%%%%%%%%%%%%%%%%%%%%%%%%%%%%%%%%%%%%%%%%%%%%%%%%%%%%%%%%%%%%
%%%%%%%%%%%%%%%%%%%%%%%%%%%%%%%%%%%%%%%%%%%%%%%%%%%%%%%%%%%%%%%%%%%%%%%%%%%%%%%
\begin{example} 
Figure~\ref{CrossSectionImprovedFirst} depicts the greedy pseudotriangulations associated with the two cross-sections introduced in Examples~\ref{defccsangle} and~\ref{defccs}.  
%%%%%%%%%%%%%%%%%%%%%%%%%%%%%%%%%%%%%%%%%%%%%%%%%%%%%%%%%%%%%%%%%%%%%%%%%%%%%%%
%%%%%%%%%%%%%%%%%%%%%%%%%%%%%%%%%%%%%%%%%%%%%%%%%%%%%%%%%%%%%%%%%%%%%%%%%%%%%%%
%%%%%%%%%%%%%%%%%%%%%%%%%%%%%%%%%%%%%%%%%%%%%%%%%%%%%%%%%%%%%%%%%%%%%%%%%%%%%%%
%%%%%%%%%%%%%%%%%%%%%%%%%%%%%%%%%%%%%%%%%%%%%%%%%%%%%%%%%%%%%%%%%%%%%%%%%%%%%%%
\begin{figure}[!htb]
%%%\footnotesize
\centering
\psfrag{one}{\footnotesize $1$}\psfrag{two}{\footnotesize $2$}\psfrag{thr}{\footnotesize $3$}\psfrag{fou}{\footnotesize $4$}\psfrag{fiv}{\footnotesize $5$}\psfrag{six}{\footnotesize $6$}\psfrag{sev}{\footnotesize $7$}\psfrag{8}{8}
%%%
\psfrag{a}{$u$}\psfrag{b}{$b$}\psfrag{c}{$c$}\psfrag{d}{$d$}\psfrag{e}{$e$}\psfrag{f}{$v$}
\psfrag{u}{$u$}
\psfrag{v}{$v$}
\psfrag{w}{$w$}
\psfrag{tw}{$w$}
\psfrag{x}{$g$}
\psfrag{g}{$g$}
\psfrag{ssigp}{$\sigma'$}
\psfrag{sigp}{$\sigma'$}
\psfrag{sig}{$\sigma$}
\psfrag{mfls}{$\sigma$}
\psfrag{Gu}{Example~\ref{defccsangle}}
\psfrag{Gutwo}{Example~\ref{defccs}}
\psfrag{aone}{$1$}
\psfrag{atwo}{$2$}
\psfrag{athr}{$3$}
\psfrag{afou}{$4$}
\psfrag{afiv}{$5$}
\psfrag{safiv}{\footnotesize $5$}
\psfrag{asix}{$6$}
\psfrag{asev}{$7$}
\psfrag{ahei}{$8$}
\psfrag{anin}{$9$}
\psfrag{aten}{$10$}
\psfrag{aele}{$11$}
\psfrag{atwe}{$12$}
\psfrag{athrt}{$13$}
\psfrag{afout}{$14$}
\psfrag{safout}{\footnotesize $14$}
\psfrag{afivt}{$15$}
\psfrag{asixt}{$16$}
\psfrag{asevt}{$17$}
\psfrag{aheit}{$18$}
\psfrag{anint}{$19$}
\psfrag{atwt}{$20$}
\psfrag{atwtone}{$21$}
\psfrag{satwtone}{\footnotesize $21$}
\includegraphics[width=0.575\linewidth]{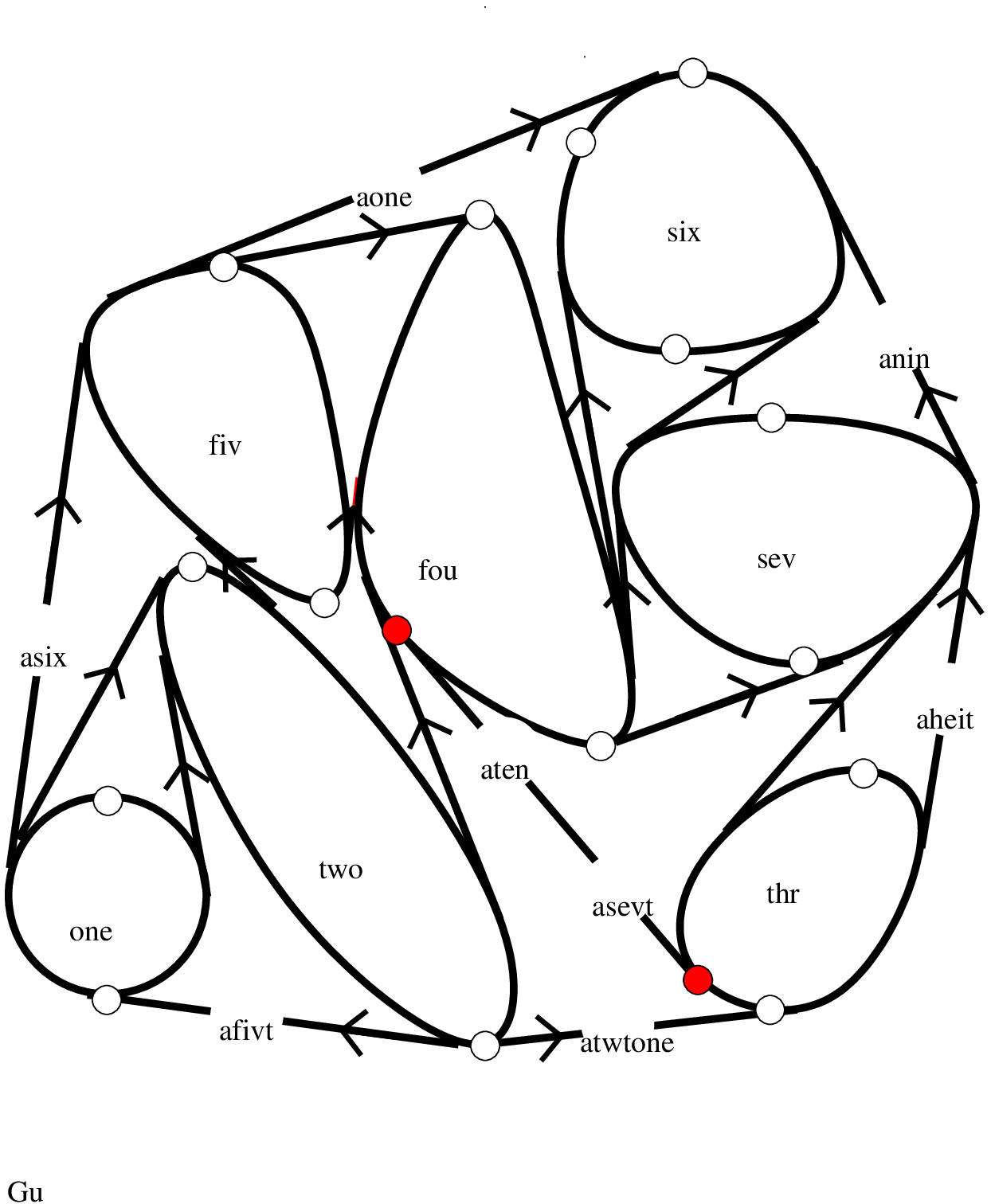}
\includegraphics[width=0.575\linewidth]{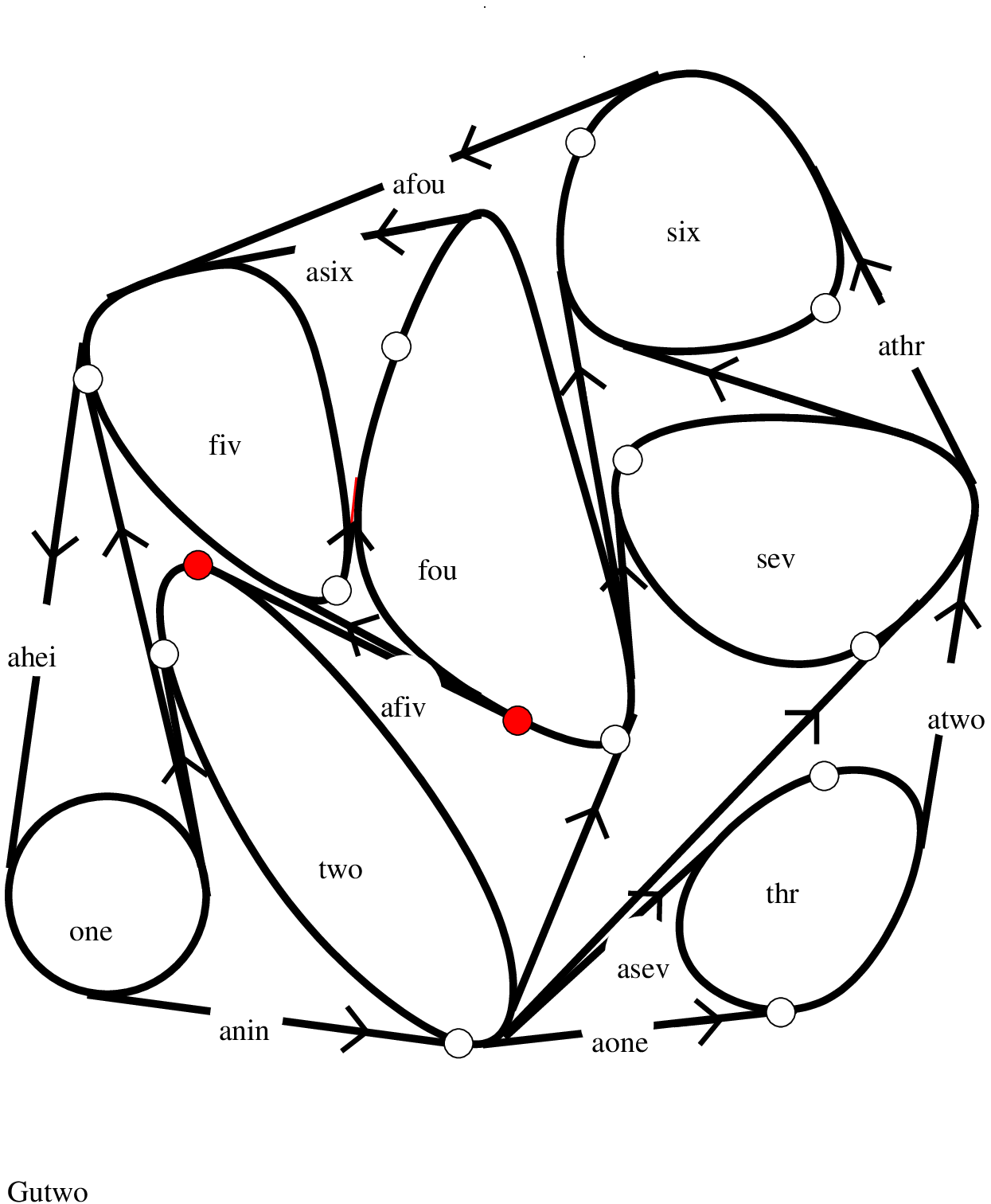}
\caption{\protect \small \label{CrossSectionImprovedFirst}}
\end{figure}
Some of the labels of the arcs of the cross-sections are reported on the corresponding bitangent line segments of the associated greedy pseudotriangulations. 
\end{example}
%%%%%%%%%%%%%%%%%%%%%%%%%%%%%%%%%%%%%%%%%%%%%%%%%%%%%%%%%%%%%%%%%%%%%%%%%%%%%%%
%%%%%%%%%%%%%%%%%%%%%%%%%%%%%%%%%%%%%%%%%%%%%%%%%%%%%%%%%%%%%%%%%%%%%%%%%%%%%%%
%%%%%%%%%%%%%%%%%%%%%%%%%%%%%%%%%%%%%%%%%%%%%%%%%%%%%%%%%%%%%%%%%%%%%%%%%%%%%%%
%%%%%%%%%%%%%%%%%%%%%%%%%%%%%%%%%%%%%%%%%%%%%%%%%%%%%%%%%%%%%%%%%%%%%%%%%%%%%%%

\subsubsection{Declination of the main result.}
%With this terminology in mind one can reformulate more precisely our main result as follows. 
Theorem~\ref{mainresult} can then be declined as follows. 
 
%%%%%%%%%%%%%%%%%%%%%%%%%%%%%%%%%%%%%%%%%%%%%%%%%%%%%%%%%%%%%%%%%%%%%%%%%%%%%%%
%%%%%%%%%%%%%%%%%%%%%%%%%%%%%%%%%%%%%%%%%%%%%%%%%%%%%%%%%%%%%%%%%%%%%%%%%%%%%%%
%%%%%%%%%%%%%%%%%%%%%%%%%%%%%%%%%%%%%%%%%%%%%%%%%%%%%%%%%%%%%%%%%%%%%%%%%%%%%%%
%%%%%%%%%%%%%%%%%%%%%%%%%%%%%%%%%%%%%%%%%%%%%%%%%%%%%%%%%%%%%%%%%%%%%%%%%%%%%%%
%%%%%%%%%%%%%%%%%%%%%%%%%%%%%%%%%%%%%%%%%%%%%%%%%%%%%%%%%%%%%%%%%%%%%%%%%%%%%%%

%%%%%%%%%%%%%%%%%%%%%%%%%%%%%%%%%%%%%%%%%%%%%%%%%%%%%%%%%%%%%%%%%%%%%%%%%%%%%%%
%%%%%%%%%%%%%%%%%%%%%%%%%%%%%%%%%%%%%%%%%%%%%%%%%%%%%%%%%%%%%%%%%%%%%%%%%%%%%%%
%%%%%%%%%%%%%%%%%%%%%%%%%%%%%%%%%%%%%%%%%%%%%%%%%%%%%%%%%%%%%%%%%%%%%%%%%%%%%%%
%%%%%%%%%%%%%%%%%%%%%%%%%%%%%%%%%%%%%%%%%%%%%%%%%%%%%%%%%%%%%%%%%%%%%%%%%%%%%%%
\begin{theorem}\label{veryfirstmainresult}
The convex hull of a planar family of $n$ pairwise disjoint convex bodies 
presented  by its chirotope is computable (under the guise of the circular sequence of boundary bitangent line segments of the family of bodies) in $O(n\log n)$ time and linear space. \qed
\end{theorem}
%%%%%%%%%%%%%%%%%%%%%%%%%%%%%%%%%%%%%%%%%%%%%%%%%%%%%%%%%%%%%%%%%%%%%%%%%%%%%%%
%%%%%%%%%%%%%%%%%%%%%%%%%%%%%%%%%%%%%%%%%%%%%%%%%%%%%%%%%%%%%%%%%%%%%%%%%%%%%%%
%%%%%%%%%%%%%%%%%%%%%%%%%%%%%%%%%%%%%%%%%%%%%%%%%%%%%%%%%%%%%%%%%%%%%%%%%%%%%%%
%%%%%%%%%%%%%%%%%%%%%%%%%%%%%%%%%%%%%%%%%%%%%%%%%%%%%%%%%%%%%%%%%%%%%%%%%%%%%%%

%%%%%%%%%%%%%%%%%%%%%%%%%%%%%%%%%%%%%%%%%%%%%%%%%%%%%%%%%%%%%%%%%%%%%%%%%%%%%%%
%%%%%%%%%%%%%%%%%%%%%%%%%%%%%%%%%%%%%%%%%%%%%%%%%%%%%%%%%%%%%%%%%%%%%%%%%%%%%%%
%%%%%%%%%%%%%%%%%%%%%%%%%%%%%%%%%%%%%%%%%%%%%%%%%%%%%%%%%%%%%%%%%%%%%%%%%%%%%%%
%%%%%%%%%%%%%%%%%%%%%%%%%%%%%%%%%%%%%%%%%%%%%%%%%%%%%%%%%%%%%%%%%%%%%%%%%%%%%%%
\begin{theorem}\label{firstmainresult}
The canonical cross-section associated with a given boundary bitangent line segment (as defined in Example~\ref{defccs}) of the visibility complex 
 of a family of $n$ pairwise disjoint convex bodies with constraints
presented by its chirotope is computable in $O(n\log n)$ time and linear space. \qed
\end{theorem}
%%%%%%%%%%%%%%%%%%%%%%%%%%%%%%%%%%%%%%%%%%%%%%%%%%%%%%%%%%%%%%%%%%%%%%%%%%%%%%%
%%%%%%%%%%%%%%%%%%%%%%%%%%%%%%%%%%%%%%%%%%%%%%%%%%%%%%%%%%%%%%%%%%%%%%%%%%%%%%%
%%%%%%%%%%%%%%%%%%%%%%%%%%%%%%%%%%%%%%%%%%%%%%%%%%%%%%%%%%%%%%%%%%%%%%%%%%%%%%%
%%%%%%%%%%%%%%%%%%%%%%%%%%%%%%%%%%%%%%%%%%%%%%%%%%%%%%%%%%%%%%%%%%%%%%%%%%%%%%%

%%%%%%%%%%%%%%%%%%%%%%%%%%%%%%%%%%%%%%%%%%%%%%%%%%%%%%%%%%%%%%%%%%%%%%%%%%%%%%%
%%%%%%%%%%%%%%%%%%%%%%%%%%%%%%%%%%%%%%%%%%%%%%%%%%%%%%%%%%%%%%%%%%%%%%%%%%%%%%%
%%%%%%%%%%%%%%%%%%%%%%%%%%%%%%%%%%%%%%%%%%%%%%%%%%%%%%%%%%%%%%%%%%%%%%%%%%%%%%%
%%%%%%%%%%%%%%%%%%%%%%%%%%%%%%%%%%%%%%%%%%%%%%%%%%%%%%%%%%%%%%%%%%%%%%%%%%%%%%%
\begin{theorem}\label{secondmainresult}
The greedy pseudotriangulation associated with  a given cross-section of the visibility complex of a family of $n$ pairwise disjoint convex bodies with constraints 
 presented by its chirotope is computable in linear time under the assumptions 
that the family is well-constrained and that the set of constraints contains the boundary bitangent line segments of the family of bodies. 
\qed
\end{theorem}
%%%%%%%%%%%%%%%%%%%%%%%%%%%%%%%%%%%%%%%%%%%%%%%%%%%%%%%%%%%%%%%%%%%%%%%%%%%%%%%
%%%%%%%%%%%%%%%%%%%%%%%%%%%%%%%%%%%%%%%%%%%%%%%%%%%%%%%%%%%%%%%%%%%%%%%%%%%%%%%
%%%%%%%%%%%%%%%%%%%%%%%%%%%%%%%%%%%%%%%%%%%%%%%%%%%%%%%%%%%%%%%%%%%%%%%%%%%%%%%
%%%%%%%%%%%%%%%%%%%%%%%%%%%%%%%%%%%%%%%%%%%%%%%%%%%%%%%%%%%%%%%%%%%%%%%%%%%%%%%

%%%%%%%%%%%%%%%%%%%%%%%%%%%%%%%%%%%%%%%%%%%%%%%%%%%%%%%%%%%%%%%%%%%%%%%%%%%%%%%
%%%%%%%%%%%%%%%%%%%%%%%%%%%%%%%%%%%%%%%%%%%%%%%%%%%%%%%%%%%%%%%%%%%%%%%%%%%%%%%
%%%%%%%%%%%%%%%%%%%%%%%%%%%%%%%%%%%%%%%%%%%%%%%%%%%%%%%%%%%%%%%%%%%%%%%%%%%%%%%
%%%%%%%%%%%%%%%%%%%%%%%%%%%%%%%%%%%%%%%%%%%%%%%%%%%%%%%%%%%%%%%%%%%%%%%%%%%%%%%
%%%%%%%%%%%%%%%%%%%%%%%%%%%%%%%%%%%%%%%%%%%%%%%%%%%%%%%%%%%%%%%%%%%%%%%%%%%%%%%
Of course it is also sensitive to ask if the (cell structure of the) visibility complex of (the free space of) a family  of convex bodies with  constraints 
presented by its chirotope is efficiently computable. 
Under the assumption that the family is well-constrained, a positive answer to that question is given by Angelier and Pocchiola~\cite[Theorem 1]{ap-sstvc-03} 
modulo the efficient computation of a cross-section and 
the efficient computation of its associated greedy pseudotriangulation. (The notion of chirotope used in~\cite{ap-sstvc-03} 
is finer than the notion of chirotope that we are using here---however the algorithmic technique developed in~\cite[page 117]{ap-sstvc-03}, 
called the $\chi_1$-{\sc Walk} procedure, can be adapted to the present situation; details on this point will be reported in a different paper.) 
Therefore combining our Theorems~\ref{veryfirstmainresult}, \ref{firstmainresult}, and~\ref{secondmainresult}  with  Theorem~1 of Angelier and Pocchiola~\cite{ap-sstvc-03} 
we get the following theorem. 

%%%%%%%%%%%%%%%%%%%%%%%%%%%%%%%%%%%%%%%%%%%%%%%%%%%%%%%%%%%%%%%%%%%%%%%%%%%%%%%
%%%%%%%%%%%%%%%%%%%%%%%%%%%%%%%%%%%%%%%%%%%%%%%%%%%%%%%%%%%%%%%%%%%%%%%%%%%%%%%
%%%%%%%%%%%%%%%%%%%%%%%%%%%%%%%%%%%%%%%%%%%%%%%%%%%%%%%%%%%%%%%%%%%%%%%%%%%%%%%
%%%%%%%%%%%%%%%%%%%%%%%%%%%%%%%%%%%%%%%%%%%%%%%%%%%%%%%%%%%%%%%%%%%%%%%%%%%%%%%
%%%%%%%%%%%%%%%%%%%%%%%%%%%%%%%%%%%%%%%%%%%%%%%%%%%%%%%%%%%%%%%%%%%%%%%%%%%%%%%
\begin{theorem}\label{bigresult} The visibility complex of a planar family of $n$ pairwise disjoint convex bodies 
presented by its chirotope is computable in $O(k + n \log n)$ time and linear working space where $k$ is the size of the visibility complex.
A similar result holds for the visibility complex of a family of  pairwise disjoint convex bodies with constraints under the assumption that the family is well-constrained. \qed
\end{theorem}
%%%%%%%%%%%%%%%%%%%%%%%%%%%%%%%%%%%%%%%%%%%%%%%%%%%%%%%%%%%%%%%%%%%%%%%%%%%%%%%
%%%%%%%%%%%%%%%%%%%%%%%%%%%%%%%%%%%%%%%%%%%%%%%%%%%%%%%%%%%%%%%%%%%%%%%%%%%%%%%
%%%%%%%%%%%%%%%%%%%%%%%%%%%%%%%%%%%%%%%%%%%%%%%%%%%%%%%%%%%%%%%%%%%%%%%%%%%%%%%
%%%%%%%%%%%%%%%%%%%%%%%%%%%%%%%%%%%%%%%%%%%%%%%%%%%%%%%%%%%%%%%%%%%%%%%%%%%%%%%
%%%%%%%%%%%%%%%%%%%%%%%%%%%%%%%%%%%%%%%%%%%%%%%%%%%%%%%%%%%%%%%%%%%%%%%%%%%%%%%
In particular the well-constrained chapter of the above result can be used to show that the visibility graph of a finite planar family of pairwise 
interior non-crossing line segments presented by the chirotope of the endpoints of the line segments is efficiently  computable; cf.  Appendix~\ref{CompVisGraphs}.

%%%%%%%%%%%%%%%%%%%%%%%%%%%%%%%%%%%%%%%%%%%%%%%%%%%%%%%%%%%%%%%%%%%%%%%%%%%%%%%
%%%%%%%%%%%%%%%%%%%%%%%%%%%%%%%%%%%%%%%%%%%%%%%%%%%%%%%%%%%%%%%%%%%%%%%%%%%%%%%
%%%%%%%%%%%%%%%%%%%%%%%%%%%%%%%%%%%%%%%%%%%%%%%%%%%%%%%%%%%%%%%%%%%%%%%%%%%%%%%
%%%%%%%%%%%%%%%%%%%%%%%%%%%%%%%%%%%%%%%%%%%%%%%%%%%%%%%%%%%%%%%%%%%%%%%%%%%%%%%
%%%%%%%%%%%%%%%%%%%%%%%%%%%%%%%%%%%%%%%%%%%%%%%%%%%%%%%%%%%%%%%%%%%%%%%%%%%%%%%
\subsection{Previous work}
The convex hull and pseudotriangulation problems 
have been addressed in the past only for families of pairwise disjoint convex bodies of an affine topological plane---strictly speaking the problems 
have only been studied in the real affine plane, however it is simple
exercise to adapt the arguments to affine topological planes---the
following solutions have been reported:
the set of  boundary bitangent line segments can be computed as the set of breakpoints of the
upper envelope of the support functions of the bodies  using a
divide-and-conquer algorithm, cf.~\cite[chap. 6]{sa-dsstg-95} and~\cite{r-chada-92}, and a pseudotriangulation can be computed using a
straight sweep {\`a} la Bentley-Ottmann 
from the positive
horizontal direction to the negative horizontal direction
of a dynamically changing visibility complex, cf.~\cite{G-pv-tsvcp-96}.  
Both algorithms run in $O(n \log n)$ time using  not only the chirotope of the family of convex bodies 
but also the direction or slope order on the set of bitangents of the family augmented 
with a point outside the convex hull of the bodies, an information which is meaningless in a topological plane 
which is not affine; the situation is even worse for the constrained pseudotriangulation problem since the algorithm uses 
also the chirotope of the family of bodies and constraints, that is, also the relative positions of the endpoints of the constraints with respect to the 
bitangents. (To fix the ideas we mention that given  four  pairwise  disjoint ellipses in the real affine plane 
 evaluating the position of an endpoint of a bitangent line segment joining the
first two ellipses  with respect to a bitangent joining the last two ellipses 
is out of the reach of the current practical techniques in formal calculus: Gröbner bases and so one~\cite{t-lgc-06}.)  
More sophisticated techniques---using even more involved
predicates like slicing the bodies---have been developed to design output
sensitive convex hull algorithm, cf.~\cite{ny-oscha-98}.  The related but different problem
of computing the convex hull of a simple curved polygon is addressed in~\cite{bk-choba-91}.

We mention that  our pseudotriangulation algorithm accepts a larger set of inputs, uses simpler data-structures 
and simpler geometric predicates, has fewer degenerate cases to handle, 
 and is faster by a $\log n$ 
factor in its main phase (which consists of deriving a pseudotriangulation 
from a cross-section of the visibility complex of the family of convex bodies with constraints) than the 
one developed in~\cite{G-pv-tsvcp-96}  and currently implemented in the visibility complex 
package of the CGAL library~\cite{G-ecg:ap-civc-03}.

For families  of points the situation is different:  
Graham's scan~\cite{g-eadch-72} and the Knuth's two incremental algorithms~\cite[pages 45--61]{k-ah-92}
compute in $O(n \log n)$ time 
 the convex hull of a family of points using only its chirotope;
on the other hand neither the Chan's output sensitive
convex hull algorithm~\cite{c-oosch-96} nor the one of 
Kirkpatrick and Seidel~\cite{ks-upcha-86}
are only based  on the chirotope since a preliminary step of 
both algorithms is to compute in linear time an extreme 
point of the family (the one with minimum horizontal coordinate),
a problem known to be open for families of points only given by their 
chirotopes~\cite[page 98]{k-ah-92}.  
Similarly a greedy pseudotriangulation
of a finite planar family of points can be computed in $O(n \log n)$ time 
using only the chirotope of the family of points as we explain in Appendix~\ref{ptpoints}.

%%%%%%%%%%%%%%%%%%%%%%%%%%%%%%%%%%%%%%%%%%%%%%%%%%%%%%%%%%%%%%%%%%%%%%%%%%%%%%%
%%%%%%%%%%%%%%%%%%%%%%%%%%%%%%%%%%%%%%%%%%%%%%%%%%%%%%%%%%%%%%%%%%%%%%%%%%%%%%%
%%%%%%%%%%%%%%%%%%%%%%%%%%%%%%%%%%%%%%%%%%%%%%%%%%%%%%%%%%%%%%%%%%%%%%%%%%%%%%%
%%%%%%%%%%%%%%%%%%%%%%%%%%%%%%%%%%%%%%%%%%%%%%%%%%%%%%%%%%%%%%%%%%%%%%%%%%%%%%%
%%%%%%%%%%%%%%%%%%%%%%%%%%%%%%%%%%%%%%%%%%%%%%%%%%%%%%%%%%%%%%%%%%%%%%%%%%%%%%%
\subsection{Outline of our pseudotriangulation algorithm}
The design and correction of our pseudotriangulation algorithm relies 
on an extension of the theory of visibility complexes of families of pairwise disjoint convex bodies of the real affine plane 
to families of pairwise disjoint convex bodies of topological planes and of their branched
coverings.  In particular our Theorem~\ref{secondmainresult} is not only valid 
for families of pairwise disjoint convex bodies with constraints of topological planes 
but also for families of pairwise disjoint convex bodies with constraints 
of branched covering of topological planes 
(under the mild assumption that the convex bodies cover the branch points of
the covering space). A similar observation can be made regarding  
Theorem~1 of Angelier and Pocchiola~\cite{ap-sstvc-03}. 
While the use of universal coverings, or portions of universal coverings, in the design of geometric algorithms 
had already appeared in the early days of the computational geometry literature, e.g.,~\cite{gjkmrs-cholr-4scg-88,hs-cmlpg-94}, 
it seems to be the first time that branched coverings 
are used in the design of a geometric algorithm. 
(Branched coverings are used in~\cite{s-cdtvd-88} to define the dual Voronoi
diagram of a constrained Delaunay triangulation in the plane, but apparently without
algorithmic consequences---see also the discussion in~\cite[page 30]{e-gtmg-01}.) 
%%We are convinced that  branched coverings will find other applications  in the design of geometric algorithms.
We refer to \cite[page 145]{m-bcat-91}, \cite{j-cts-79},\cite[page 18]{lz-gsa-04} and the 
references cited therein for background material on branched coverings.

Our algorithm proceeds in three steps:
we first compute the  convex hull of the family of convex bodies,
then the cross-section of the visibility complex of the family of convex bodies with constraints assigned to a distinguished boundary bitangent line segment, and 
finally the greedy pseudotriangulation associated with that cross-section, that is,
the set of sinks of its  $2$-cells, cf.~\cite[Theorem 12]{G-pv-tsvcp-96}
and more generally~\cite[Theorem 6, Claim 1]{ap-sstvc-03} in the case where we look for a constrained pseudotriangulation.

%%%%%%%%%%%%%%%%%%%%%%%%%%%%%%%%%%%%%%%%%%%%%%%%%%%%%%%%%%%%%%%%%%%%%%%%%%%%%%%
%%%%%%%%%%%%%%%%%%%%%%%%%%%%%%%%%%%%%%%%%%%%%%%%%%%%%%%%%%%%%%%%%%%%%%%%%%%%%%%
%%%%%%%%%%%%%%%%%%%%%%%%%%%%%%%%%%%%%%%%%%%%%%%%%%%%%%%%%%%%%%%%%%%%%%%%%%%%%%%
%%%%%%%%%%%%%%%%%%%%%%%%%%%%%%%%%%%%%%%%%%%%%%%%%%%%%%%%%%%%%%%%%%%%%%%%%%%%%%%
%%%%%%%%%%%%%%%%%%%%%%%%%%%%%%%%%%%%%%%%%%%%%%%%%%%%%%%%%%%%%%%%%%%%%%%%%%%%%%%
\subsubsection{Convex hull algorithm.} 
\begin{figure}[!htb]
\footnotesize
\psfrag{one}{$1$}\psfrag{two}{$2$}\psfrag{thr}{$3$}\psfrag{fou}{$4$}\psfrag{fiv}{$5$}\psfrag{six}{$6$}\psfrag{sev}{$7$}\psfrag{8}{8}%%%
\psfrag{plane}{the plane}  \psfrag{sheet0}{sheet 0} \psfrag{sheet1}{sheet 1} \psfrag{sheet2}{sheet 2} \psfrag{sheet3}{sheet 3}
\begin{center}
\includegraphics[width=0.958575\linewidth]{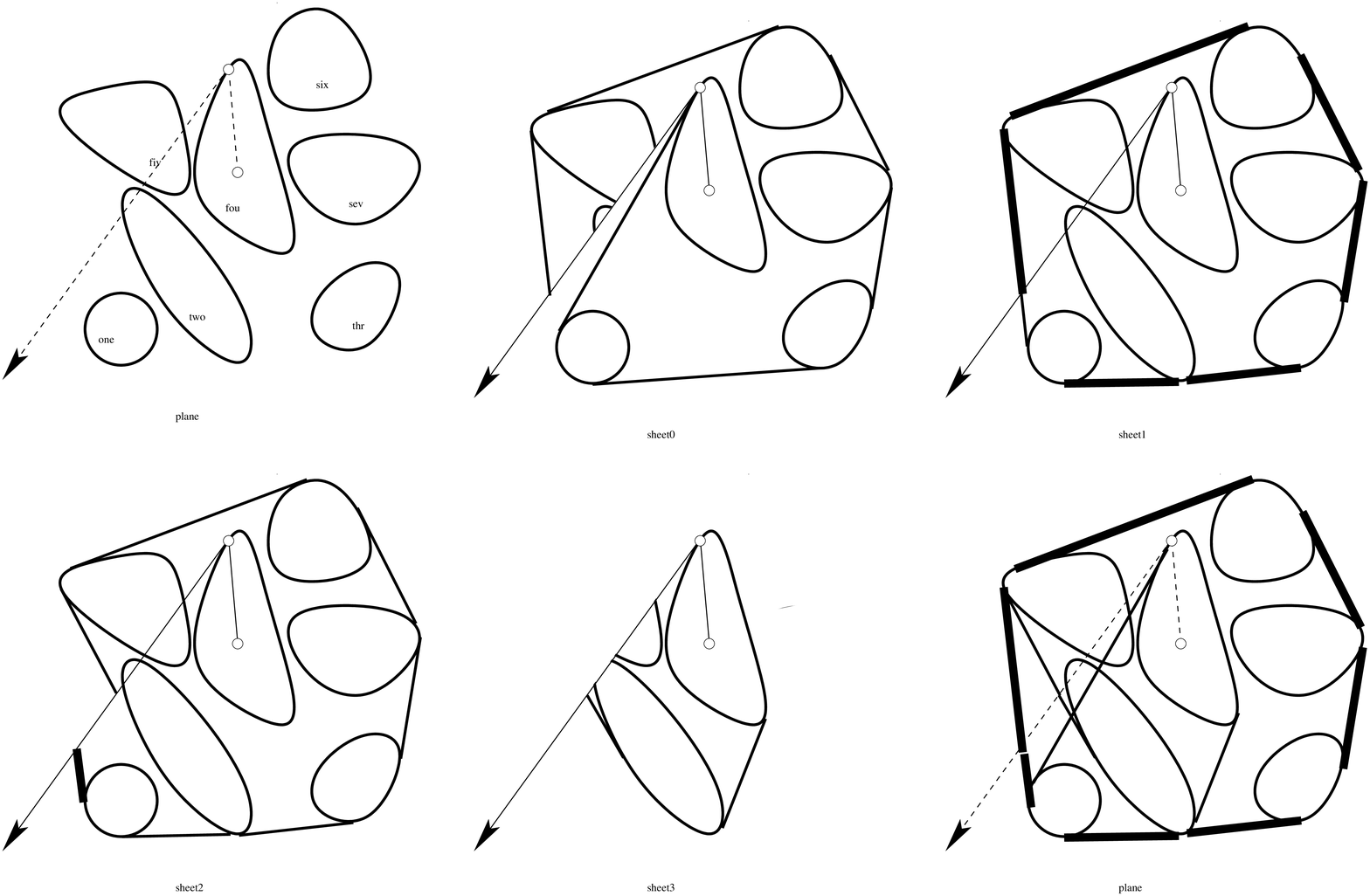}
\end{center}
\caption{\protect \small \label{convexhull}}
\end{figure}

Our convex-hull algorithm  is a sweep  of a connected $4$-sheeted branched covering
of the underlying plane ramified over any interior point of an arbitrarily distinguished 
body: 
we sweep the $4$-sheeted covering surface 
with a half-line whose supporting line is a left tangent at the origin of the half-line
to the lift of the distinguished body. Any body, except the distinguished one, 
has  four lifts in the $4$-sheeted covering surface; we only keep 
the lifts 
either lying in one of the first three sheets, 
either straddling 
the first two sheets or the second and third sheets or the last two sheets, as illustrated in Figure~\ref{convexhull} where the bodies numbered $1,3,6$ and $7$ 
%%%%%%%%%%%%%%%%%%%%%%%%%%%%%%%%%%%%%%%%%%%%%%%%%%%%%%%%%%%%%%%%%%%%%%%%%%%%%%%
%%%%%%%%%%%%%%%%%%%%%%%%%%%%%%%%%%%%%%%%%%%%%%%%%%%%%%%%%%%%%%%%%%%%%%%%%%%%%%%
%%%%%%%%%%%%%%%%%%%%%%%%%%%%%%%%%%%%%%%%%%%%%%%%%%%%%%%%%%%%%%%%%%%%%%%%%%%%%%%
%%%%%%%%%%%%%%%%%%%%%%%%%%%%%%%%%%%%%%%%%%%%%%%%%%%%%%%%%%%%%%%%%%%%%%%%%%%%%%%
%%%%%%%%%%%%%%%%%%%%%%%%%%%%%%%%%%%%%%%%%%%%%%%%%%%%%%%%%%%%%%%%%%%%%%%%%%%%%%%
%%%%%%%%%%%%%%%%%%%%%%%%%%%%%%%%%%%%%%%%%%%%%%%%%%%%%%%%%%%%%%%%%%%%%%%%%%%%%%%
%%%%%%%%%%%%%%%%%%%%%%%%%%%%%%%%%%%%%%%%%%%%%%%%%%%%%%%%%%%%%%%%%%%%%%%%%%%%%%%
%%%%%%%%%%%%%%%%%%%%%%%%%%%%%%%%%%%%%%%%%%%%%%%%%%%%%%%%%%%%%%%%%%%%%%%%%%%%%%%
%%%%%%%%%%%%%%%%%%%%%%%%%%%%%%%%%%%%%%%%%%%%%%%%%%%%%%%%%%%%%%%%%%%%%%%%%%%%%%%
are lifted only in the first three sheets and where the bodies numbered $2$ and $5$ are lifted astride the first two sheets, the second and third sheets, the last two sheets but not astride 
the last and first sheets.  
%%%%%%%%%%%%%%%%%%%%%%%%%%%%%%%%%%%%%%%%%%%%%%%%%%%%%%%%%%%%%%%%%%%%%%%%%%%%%%%
The sweep starts at a boundary tangent and induces a total order on the lifted bodies with the
property that a body contributes to one or zero connected piece to the boundary of
its convex hull  with its predecessors in the total order. 
During the sweep we maintain the convex hull of 
the lifts of the bodies that have been entirely swept or partially swept by the sweeping half-line; 
the convex hull of the family of bodies is then extracted from the convex
hull of the lifts, as illustrated in Figure~\ref{convexhull} where one can read the convex hull of the family of bodies  as the 
boundary bitangents of the lifts drawn with a bold line. 

%%%%%%%%%%%%%%%%%%%%%%%%%%%%%%%%%%%%%%%%%%%%%%%%%%%%%%%%%%%%%%%%%%%%%%%%%%%%%%%
%%%%%%%%%%%%%%%%%%%%%%%%%%%%%%%%%%%%%%%%%%%%%%%%%%%%%%%%%%%%%%%%%%%%%%%%%%%%%%%
%%%%%%%%%%%%%%%%%%%%%%%%%%%%%%%%%%%%%%%%%%%%%%%%%%%%%%%%%%%%%%%%%%%%%%%%%%%%%%%
%%%%%%%%%%%%%%%%%%%%%%%%%%%%%%%%%%%%%%%%%%%%%%%%%%%%%%%%%%%%%%%%%%%%%%%%%%%%%%%
%%%%%%%%%%%%%%%%%%%%%%%%%%%%%%%%%%%%%%%%%%%%%%%%%%%%%%%%%%%%%%%%%%%%%%%%%%%%%%%
\subsubsection{Cross-section algorithm.} 
Our cross-section algorithm is again a sweep but now a simple sweep of the convex hull of the bodies by a half-line 
whose supporting line is a 
left tangent at the origin of the half-line to one of the bodies appearing on the boundary of the convex hull---a boundary body, for short. 
The sweep starts at one of the boundary bitangent line segments leaving the distinguished boundary body.
During the sweep we construct the canonical cross-section of the visibility complex of the family of convex bodies with constraints assigned to the distinguished boundary bitangent line segment,
cf. Example~\ref{defccs}.
%we identify the $2$-cells whose sinks are hull-bitangent line segments or constraints 
%and we compute their sink
%bitangent line segments 
%(this identification and the computation that follows are not strictly necessary since they can be also done during the third step of the algorithm using as input any cross-section), 
The method  presents some interesting and novel features due to the fact that the relative
positions of the constraints with respect to the bitangents are not
completely determined by the chirotope of the convex bodies. It is also interesting
to mention that this second step is implementable in $O(n \log n)$ without
restriction on the possible sets of constraints.

%%%%%%%%%%%%%%%%%%%%%%%%%%%%%%%%%%%%%%%%%%%%%%%%%%%%%%%%%%%%%%%%%%%%%%%%%%%%%%%
%%%%%%%%%%%%%%%%%%%%%%%%%%%%%%%%%%%%%%%%%%%%%%%%%%%%%%%%%%%%%%%%%%%%%%%%%%%%%%%
%%%%%%%%%%%%%%%%%%%%%%%%%%%%%%%%%%%%%%%%%%%%%%%%%%%%%%%%%%%%%%%%%%%%%%%%%%%%%%%
%%%%%%%%%%%%%%%%%%%%%%%%%%%%%%%%%%%%%%%%%%%%%%%%%%%%%%%%%%%%%%%%%%%%%%%%%%%%%%%
%%%%%%%%%%%%%%%%%%%%%%%%%%%%%%%%%%%%%%%%%%%%%%%%%%%%%%%%%%%%%%%%%%%%%%%%%%%%%%%
\subsubsection{Greedy pseudotriangulation algorithm.} 
Our third and last algorithm---which consists of deriving the greedy pseudotriangulation associated to a given cross-section
of the visibility complex of a family of pairwise disjoint convex bodies with constraints whose set of constraints contains 
the boundary bitangent line segments of the bodies---is the most elaborate and fully benefits from the idea of using branched coverings.
A preliminary version of this third algorithm---of which the idea of using branched coverings was unfortunately missing---was discussed several years ago by the second author of the paper with 
his PhD student Pierre Angelier, see~\cite[pages 83--92]{G-a-agv-02} and compare with~\cite[Appendix A]{G-pv-vc-96}.

%%%%%%%%%%%%%%%%%%%%%%%%%%%%%%%%%%%%%%%%%%%%%%%%%%%%%%%%%%%%%%%%%%%%%%%%%%%%%%%
%%%%%%%%%%%%%%%%%%%%%%%%%%%%%%%%%%%%%%%%%%%%%%%%%%%%%%%%%%%%%%%%%%%%%%%%%%%%%%%
%%%%%%%%%%%%%%%%%%%%%%%%%%%%%%%%%%%%%%%%%%%%%%%%%%%%%%%%%%%%%%%%%%%%%%%%%%%%%%%
%%%%%%%%%%%%%%%%%%%%%%%%%%%%%%%%%%%%%%%%%%%%%%%%%%%%%%%%%%%%%%%%%%%%%%%%%%%%%%%
%%%%%%%%%%%%%%%%%%%%%%%%%%%%%%%%%%%%%%%%%%%%%%%%%%%%%%%%%%%%%%%%%%%%%%%%%%%%%%%
We define a partial order $<$   
on the set of $2$-cells $\sigma$ of the input cross-section whose
sink bitangent line segment $t(\sigma)$ is not a  constraint (and thus not a boundary bitangent line segment), and  for each $\sigma$ we define a pair 
of adjacent pseudotriangles, called the \ABpts\ of $\sigma$, made  with the $t(\sigma')$, $\sigma' < \sigma$, and
with auxiliary bitangent line segments $s_{j}(\sigma)$, $1\leq j\leq \sigma^*$,  such that 
a representation of the \ABpts\ of $\sigma$ by a linked structure ${\cal R}_\sigma$---that is, collections of nodes interconnected by pointers; cf.~\cite[page 8]{t-dsna-83}---is computable in
constant amortized time  and  such that $t(\sigma)$ is computable as the bitangent line segment joining the \ABpts\ of $\sigma$ 
in constant amortized time starting from the knowledge of the linked 
structure ${\cal R}_{\sigma}$.
%%%%%%%%%%%%%%%%%%%%%%%%%%%%%%%%%%%%%%%%%%%%%%%%%%%%%%%%%%%%%%%%%%%%%%%%%%%%%%%
%%%%%%%%%%%%%%%%%%%%%%%%%%%%%%%%%%%%%%%%%%%%%%%%%%%%%%%%%%%%%%%%%%%%%%%%%%%%%%%
%%%%%%%%%%%%%%%%%%%%%%%%%%%%%%%%%%%%%%%%%%%%%%%%%%%%%%%%%%%%%%%%%%%%%%%%%%%%%%%
%%%%%%%%%%%%%%%%%%%%%%%%%%%%%%%%%%%%%%%%%%%%%%%%%%%%%%%%%%%%%%%%%%%%%%%%%%%%%%%
%%%%%%%%%%%%%%%%%%%%%%%%%%%%%%%%%%%%%%%%%%%%%%%%%%%%%%%%%%%%%%%%%%%%%%%%%%%%%%%

\begin{figure}[htb!]
%\begin{figure}[p]
\centering
\psfrag{A}{$\mathbb{A}$}
\psfrag{B}{$\mathbb{B}$}
\psfrag{Ag}{$\mathbb{A}_{\gamma'}$}
\psfrag{Bg}{$\mathbb{B}_{\gamma}$}
\psfrag{g}{$\gamma$}
\psfrag{gp}{$\gamma'$}
\psfrag{gg}{$\gamma_{\text{left}}$}
\psfrag{gd}{$\gamma_{\text{righ}}$}
\psfrag{gp}{$\gamma'$}
\psfrag{gpg}{$\gamma'_{\text{left}}$}
\psfrag{gpd}{$\gamma'_{\text{righ}}$}
\includegraphics[width=0.95875\linewidth]{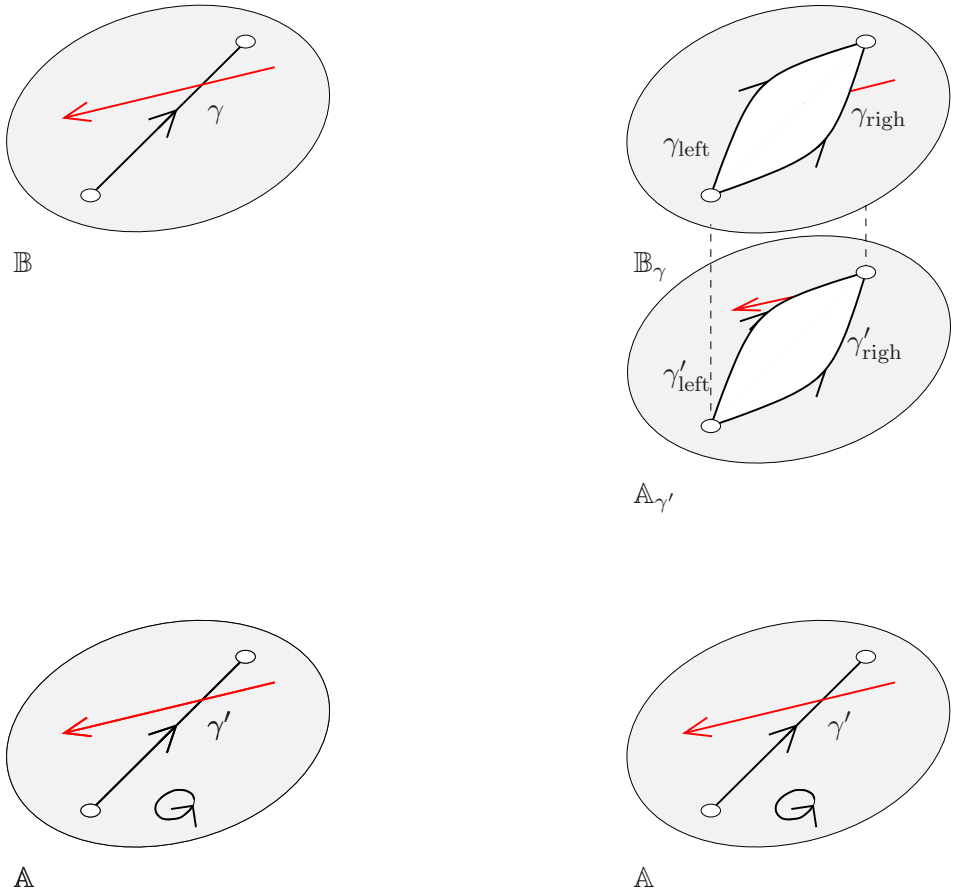}
\caption{\protect \small Given a branched covering $\CoSur{}$ of a topological plane $\pointset$ 
and a simple oriented curve $\gamma$ in $\CoSur{}$ homeomorphic to 
its projection $\gamma'$ in $\pointset$ under the covering map
$\CoSur{}\rightarrow \pointset$ 
we construct a new branched covering 
$\Liftop{}{\gamma}{(\CoSur{})}$ of $\pointset$ as follows.
Cut $\mathbb{C} = \CoSur{} \sqcup \pointset$ along $\gamma$ and $\gamma'$, call $\mathbb{C}_{\gamma}$ the resulting surface
and $\Map{q}{\mathbb{C}_{\gamma}}{\mathbb{C}}$ the induced projection. Then 
we define $\Liftop{}{\gamma}{(\CoSur{})}$ as the quotient space of $\mathbb{C}_{\gamma}$ 
by identification of the left lift of $\gamma$ under $q$ with the right 
lift of $\gamma'$ under $q$ as well as the right lift of $\gamma$ under $q$ with the left lift of $\gamma'$ under $q.$ 
The operator $\liftoperator$ is defined as the composition of several $\Liftop{}{\gamma}{}$ operators. 
\label{ConnectedSum}}
\end{figure}

%%%%%%%%%%%%%%%%%%%%%%%%%%%%%%%%%%%%%%%%%%%%%%%%%%%%%%%%%%%%%%%%%%%%%%%%%%%%%%%
%%%%%%%%%%%%%%%%%%%%%%%%%%%%%%%%%%%%%%%%%%%%%%%%%%%%%%%%%%%%%%%%%%%%%%%%%%%%%%%
%%%%%%%%%%%%%%%%%%%%%%%%%%%%%%%%%%%%%%%%%%%%%%%%%%%%%%%%%%%%%%%%%%%%%%%%%%%%%%%
%%%%%%%%%%%%%%%%%%%%%%%%%%%%%%%%%%%%%%%%%%%%%%%%%%%%%%%%%%%%%%%%%%%%%%%%%%%%%%%
%%%%%%%%%%%%%%%%%%%%%%%%%%%%%%%%%%%%%%%%%%%%%%%%%%%%%%%%%%%%%%%%%%%%%%%%%%%%%%%
A key feature of our method
is that the \ABpts\ are defined as projections in the plane 
of pseudotriangles of pseudotriangulations of sets of 
lifts of bodies in certain branched coverings of the plane. 
(Some of the $s_{j}(\sigma)$, $1\leq j\leq \sigma^*$, are computed by a recursive
application of the procedure to compute the $t(\sigma)$.)
More precisely,
given a finite family of pairwise disjoint convex bodies with constraints (including the boundary bitangent line segments of the convex bodies) of a branched covering $\CoSur{}$ of a 
topological plane $\pointset$,
we  associate to each bounded $2$-cell $\sigma$ 
of its  visibility complex
whose source bitangent line segment is not a constraint a pseudoquadrangle containing $\bigcup \sigma$, called the \Hpq\ of $\sigma$ and denoted  
$\Hglass(\sigma)$, whose diagonals are the source and the sink bitangent line segments of~$\sigma$; pseudoquadrangle 
from which we derive, once a cross-section $\Gamma$ containing $\sigma$ is chosen,  
a pair 
of pseudotriangles 
adjacent along the source bitangent line segment of $\sigma$, called the  \ABzerpts\ of $\sigma$, with the property that 
the bitangent line segment joining the \ABzerpts\ of $\sigma$  is the sink bitangent line segment of $\sigma$; 
the definition of the  \ABzerpts\  
depends on the {\it type} of $\sigma$ in $\Gamma$ which is a pair $ij$, $i,j \in \{1,2,3\}$, that encodes the position of the source and sink nodes of $\sigma$  
in the decomposition of the left and right boundaries of $\sigma$ into convex chains ($3$ in number at most).
Then we assign to the $2$-cell $\sigma$, element of the cross-section $\Gamma$,  a $2$-cell  $\liftoperator(\sigma)$, element of a 
certain cross-section $\liftoperator(\Gamma)$  
of the visibility complex of a certain family of convex bodies and constraints of a certain branched covering  
$\liftoperator(\CoSur{})$ of the topological plane $\pointset$---obtained as connected sum of $\CoSur{}$ and
copies of the plane $\pointset$ as indicated in  Figure~\ref{ConnectedSum}---so that, among other things, 
$\sigma$ and $\liftoperator(\sigma)$ have the same sink. 
The \ABpts\  of $\sigma$ are then defined as the \ABzerpts\ of $\liftoperator(\sigma)$.
The correction of the method relies on several 
new properties of cross-sections of visibility complexes. 
%%%%%%%%%%%%%%%%%%%%%%%%%%%%%%%%%%%%%%%%%%%%%%%%%%%%%%%%%%%%%%%%%%%%%%%%%%%%%%%
%%%%%%%%%%%%%%%%%%%%%%%%%%%%%%%%%%%%%%%%%%%%%%%%%%%%%%%%%%%%%%%%%%%%%%%%%%%%%%%
%%%%%%%%%%%%%%%%%%%%%%%%%%%%%%%%%%%%%%%%%%%%%%%%%%%%%%%%%%%%%%%%%%%%%%%%%%%%%%%
%%%%%%%%%%%%%%%%%%%%%%%%%%%%%%%%%%%%%%%%%%%%%%%%%%%%%%%%%%%%%%%%%%%%%%%%%%%%%%%
%%%%%%%%%%%%%%%%%%%%%%%%%%%%%%%%%%%%%%%%%%%%%%%%%%%%%%%%%%%%%%%%%%%%%%%%%%%%%%%

%%%%%%%%%%%%%%%%%%%%%%%%%%%%%%%%%%%%%%%%%%%%%%%%%%%%%%%%%%%%%%%%%%%%%%%%%%%%%%%
%%%%%%%%%%%%%%%%%%%%%%%%%%%%%%%%%%%%%%%%%%%%%%%%%%%%%%%%%%%%%%%%%%%%%%%%%%%%%%%
%%%%%%%%%%%%%%%%%%%%%%%%%%%%%%%%%%%%%%%%%%%%%%%%%%%%%%%%%%%%%%%%%%%%%%%%%%%%%%%
%%%%%%%%%%%%%%%%%%%%%%%%%%%%%%%%%%%%%%%%%%%%%%%%%%%%%%%%%%%%%%%%%%%%%%%%%%%%%%%
%%%%%%%%%%%%%%%%%%%%%%%%%%%%%%%%%%%%%%%%%%%%%%%%%%%%%%%%%%%%%%%%%%%%%%%%%%%%%%%
\subsection{Organization of the paper}
In the next section we extend the
theory of pseudotriangulations and visibility complexes to the setting of  
branched coverings of topological planes (no proofs will be given since one can adapt easily to that setting the proofs given in~\cite{G-pv-vc-96,G-pv-tsvcp-96,ap-sstvc-03}), we establish 
several new properties of cross-sections of visibility complexes, and we
introduce the  main ingredients of our pseudotriangulation algorithm
mentioned in the previous sections.
%% , that is, the pseudoquadrangle $\Hglass(\sigma)$, the pseudotriangles $\opLeaveHG(\sigma)$ and $\opEnterHG(\sigma)$, and the 
%% lift operator $\liftoperator$.  
In the third section we describe our pseudotriangulation algorithm, we analyze
its complexity, and we 
conclude in the fourth and last section. 
%%To keep as simple as possible (essentially in the necessary notations) 
%%we only expose our generalization of the theory of visibility complexes to the case where the set of constraints is empty, 
%%the general case can be treated very similarly 
%%using the definition of visibility complexes of families of convex
%%bodies and constraints given in~\cite{ap-sstvc-03}.  

\absoluteignore{
Autres references: Pseudotriangulations are useful 
data structures for visibility graph computations~\cite{G-pv-vc-96,G-pv-tsvcp-96}, 
ray-shooting~\cite{cegghss-rspug-94,gt-drssp-97},
collision detection~\cite{abghz-dfstk-01,ks-kmcsh-02},
and planning expansive motions of polygonal chains~\cite{rss-ecpmp-03}.
}%%%%%%%%%%%%%%%%%%%

\clearpage
%%%%%%%%%%%%%%%%%%%%%%%%%%%%%%%%%%%%%%%%%%%%%%%%%%%%%%%%%%%%%%%%%%%%%%%%%%%%%%%%%%%%%%%%%%%%%%%%%%%%%%%%%%%
%%%%%%%%%%%%%%%%%%%%%%%%%%%%%%%%%%%%%%%%%%%%%%%%%%%%%%%%%%%%%%%%%%%%%%%%%%%%%%%%%%%%%%%%%%%%%%%%%%%%%%%%%%%
%%%%%%%%%%%%%%%%%%%%%%%%%%%%%%%%%%%%%%%%%%%%%%%%%%%%%%%%%%%%%%%%%%%%%%%%%%%%%%%%%%%%%%%%%%%%%%%%%%%%%%%%%%%
%%%%%%%%%%%%%%%%%%%%%%%%%%%%%%%%%%%%%%%%%%%%%%%%%%%%%%%%%%%%%%%%%%%%%%%%%%%%%%%%%%%%%%%%%%%%%%%%%%%%%%%%%%%
%%%%%%%%%%%%%%%%%%%%%%%%%%%%%%%%%%%%%%%%%%%%%%%%%%%%%%%%%%%%%%%%%%%%%%%%%%%%%%%%%%%%%%%%%%%%%%%%%%%%%%%%%%%
\section{Visibility in branched coverings}
In this section we extend the theory of pseudotriangulations and visibility complexes to the setting of 
%%%finite families of pairwise disjoint convex bodies of branched coverings of topological planes, we establish 
branched coverings of topological planes; we establish 
several new properties of cross-sections of visibility complexes; and we
introduce the key ingredients of our algorithm mentioned in the introduction~: \Hpqs, \ABzerpts, and \ABpts.
For the sake of simplicity and clarity we only went over the case where the set of constraints is empty, 
the general case can be treated very similarly 
using the definition of visibility complexes of families of pairwise disjoint convex bodies with constraints 
given in the introduction.
%% (we refer to~\cite{ap-sstvc-03} for a description of the local topology of visibility complexes in the presence of constraints).  
%Let $\CoSur{}$ be a  finite connected branched covering space of an oriented topological plane~$\pointset$ equipped with the partial topological point-line incidence
%structure, with singularities at the branch points,
%inherited from the point-line incidence structure of $\pointset$.
%%Any subset $\mathbb{X}$ of $\CoSur{}$, and $\CoSur{}$ in particular, is endowed with the 
%%point-line incidence 
%%structure
%%inherited from the point-line incidence structure of $\pointset$, that is, 
%%the lines of $\mathbb{X}$ are  the traces on $\mathbb{X}$ of the subsets of $\CoSur{}$ 
%%homeomorphic to the lines of $\pointset$ via the covering map $\CoSur{}
%%\rightarrow \pointset$, and two points of $\mathbb{X}$ are termed mutually
%%visible if they belong to a same line of $\mathbb{X}$.

%%%%%%%%%%%%%%%%%%%%%%%%%%%%%%%%%%%%%%%%%%%%%%%%%%%%%%%%%%%%%%%%%%%%%%%%%%%%%%%%%%%%%%%%%%%%%%%%%%%%%%%%%%%
%%%%%%%%%%%%%%%%%%%%%%%%%%%%%%%%%%%%%%%%%%%%%%%%%%%%%%%%%%%%%%%%%%%%%%%%%%%%%%%%%%%%%%%%%%%%%%%%%%%%%%%%%%%
%%%%%%%%%%%%%%%%%%%%%%%%%%%%%%%%%%%%%%%%%%%%%%%%%%%%%%%%%%%%%%%%%%%%%%%%%%%%%%%%%%%%%%%%%%%%%%%%%%%%%%%%%%%
%%%%%%%%%%%%%%%%%%%%%%%%%%%%%%%%%%%%%%%%%%%%%%%%%%%%%%%%%%%%%%%%%%%%%%%%%%%%%%%%%%%%%%%%%%%%%%%%%%%%%%%%%%%
%%%%%%%%%%%%%%%%%%%%%%%%%%%%%%%%%%%%%%%%%%%%%%%%%%%%%%%%%%%%%%%%%%%%%%%%%%%%%%%%%%%%%%%%%%%%%%%%%%%%%%%%%%%
Let $\disks$ be a finite family of pairwise disjoint convex bodies
of  a finite connected branched covering space $\CoSur{}$ of an oriented topological plane~$\pointset$ equipped with the partial topological point-line incidence
structure, with singularities at the branch points,
inherited from the point-line incidence structure of $\pointset$.
We assume that the boundaries of the convex bodies are free of line segments, 
that there is exactly one tangent line through each boundary point, 
that the bodies surround the branch points of the covering space, and we use the following associated terminology and notations:
{\it free space} is the complement of the interiors of the bodies;
a {\it bitangent line segment} is a closed line segment of free space tangent to two bodies at its endpoints; 
a {\it boundary bitangent line segment} is a bitangent line segment contained in the boundary of the convex hull of the bodies;
all other bitangent line segments are said to be {\it interior bitangent line segments};
a {\it primitive arc} is a connected component of the boundary of the bodies minus the
bitangent line segments; $\nbbbls_{\disks}$ is the number of boundary bitangent line segments;
$\size_{\disks}$  is the sum of the orders of the branch points plus the number of bodies
surrounding no branch points; $\nbsheets_{\disks}$ is 
the number of sheets of the branched covering space.
\subsection{Pseudotriangulations}
A {\it pseudotriangulation} is a maximal, for the inclusion relation,
collection of pairwise interior non-crossing  bitangent line segments.
As in the case where the covering map $\MapLight{}{\CoSur{}}{\pointset}$ is the
identity map of the  real affine plane,  a pseudotriangulation induces a subdivision of free space
whose bounded regions are pseudotriangles, that is,
subsets of free space homeomorphic via the covering map to
pseudotriangles of the topological plane.

%%%%%%%%%%%%%%%%%%%%%%%%%%%%%%%%%%%%%%%%%%%%%%%%%%%%%%%%%%%%%%%%%%%%%%%%%%%%%%%%%%%%%%%%%%%%%%%%%%%%%%%%%%%
%%%%%%%%%%%%%%%%%%%%%%%%%%%%%%%%%%%%%%%%%%%%%%%%%%%%%%%%%%%%%%%%%%%%%%%%%%%%%%%%%%%%%%%%%%%%%%%%%%%%%%%%%%%
%%%%%%%%%%%%%%%%%%%%%%%%%%%%%%%%%%%%%%%%%%%%%%%%%%%%%%%%%%%%%%%%%%%%%%%%%%%%%%%%%%%%%%%%%%%%%%%%%%%%%%%%%%%
%%%%%%%%%%%%%%%%%%%%%%%%%%%%%%%%%%%%%%%%%%%%%%%%%%%%%%%%%%%%%%%%%%%%%%%%%%%%%%%%%%%%%%%%%%%%%%%%%%%%%%%%%%%
%%%%%%%%%%%%%%%%%%%%%%%%%%%%%%%%%%%%%%%%%%%%%%%%%%%%%%%%%%%%%%%%%%%%%%%%%%%%%%%%%%%%%%%%%%%%%%%%%%%%%%%%%%%
\begin{theorem} Let ${\cal T}$ be a pseudotriangulation of $\disks$.
Then the bounded faces of the subdivision of free space induced by ${\cal T}$ are pseudotriangles, their number is $2\size_{\disks}-2\nbsheets_{\disks}$ and
the size of ${\cal T}$ is $3\size_\disks-3\nbsheets_\disks$.
Furthermore any interior bitangent line segment of ${\cal T}$ can be flipped, that is, replaced by an interior bitangent line segment to obtain a
new pseudotriangulation.
 \qed
\end{theorem}
\begin{proof} One can repeat the proof given for the real affine plane in~\cite{G-pv-vc-96}
since the lines of a topological plane---and consequently the lines of free space---are geodesics for an ad hoc metric on
the topological plane; cf.~\cite[Theorem 11.2, page 56]{b-gg-55}.
\end{proof}

Two pseudotriangulations are said to be adjacent (or related by a flip) if they differ by a single (necessarily interior) bitangent line segment. 
The adjacency graph on the set of pseudotriangulations is a connected regular graph of degree $3\size_\disks-3\nbsheets_\disks - \nbbbls_{\disks}$.
More generally the collection, ordered by inclusion, of subsets of pairwise interior non-crossing free interior bitangent line segments 
is a  strongly flag-connected pure simplicial complex of dimension $3\size_\disks-3\nbsheets_\disks - \nbbbls_{\disks}$ which satisfies the diamond property.  
This simplicial complex will be called thereafter the complex of pseudotriangulations of the family of convex bodies.

\begin{example} Figure~\ref{twobodytwosheets} depicts a family of two convex bodies of a $2$-sheeted branched covering of $\pointset$ 
with two branch points (the two sheets are obtained by cutting the covering space along the two line segments joining the two branch points).
%%%%%%%%%%%%%%%%%%%%%%%%%%%%%%%%%%%%%%%%%%%%%%%%%%%%%%%%%%%%%%%%%%%%%%%%%%%%%%%
%%%%%%%%%%%%%%%%%%%%%%%%%%%%%%%%%%%%%%%%%%%%%%%%%%%%%%%%%%%%%%%%%%%%%%%%%%%%%%%
%%%%%%%%%%%%%%%%%%%%%%%%%%%%%%%%%%%%%%%%%%%%%%%%%%%%%%%%%%%%%%%%%%%%%%%%%%%%%%%
%%%%%%%%%%%%%%%%%%%%%%%%%%%%%%%%%%%%%%%%%%%%%%%%%%%%%%%%%%%%%%%%%%%%%%%%%%%%%%%
\begin{figure}[!htb]
\begin{center}
\psfrag{one}{1} \psfrag{two}{2} \psfrag{thr}{3} \psfrag{fou}{4}
\psfrag{six}{6} \psfrag{sev}{7} \psfrag{hei}{8} \psfrag{fiv}{5}
\psfrag{unun}{$i_1$} \psfrag{unde}{$i_2$} \psfrag{unth}{$i_3$} \psfrag{unfo}{$i_4$}
\psfrag{unfi}{$i_5$} \psfrag{unsi}{$i_6$} \psfrag{unse}{$i_7$} \psfrag{unhe}{$i_8$}
\psfrag{deun}{$j_1$} \psfrag{dede}{$j_2$} \psfrag{deth}{$j_3$} \psfrag{defo}{$j_4$}
\psfrag{defi}{$j_5$} \psfrag{desi}{$j_6$} \psfrag{dese}{$j_7$} \psfrag{dehe}{$j_8$}
\psfrag{un}{$i$} \psfrag{de}{$j$} \psfrag{firstbody}{$i$} \psfrag{secondbody}{$j$}
\psfrag{lunde}{$ij$}
\psfrag{inin}{$\infty\infty$}
\psfrag{unin}{$i\infty$}
\psfrag{dein}{$j\infty$}
\psfrag{infty}{$\infty$}
\psfrag{body}{}
\includegraphics[width = 0.375575\linewidth]{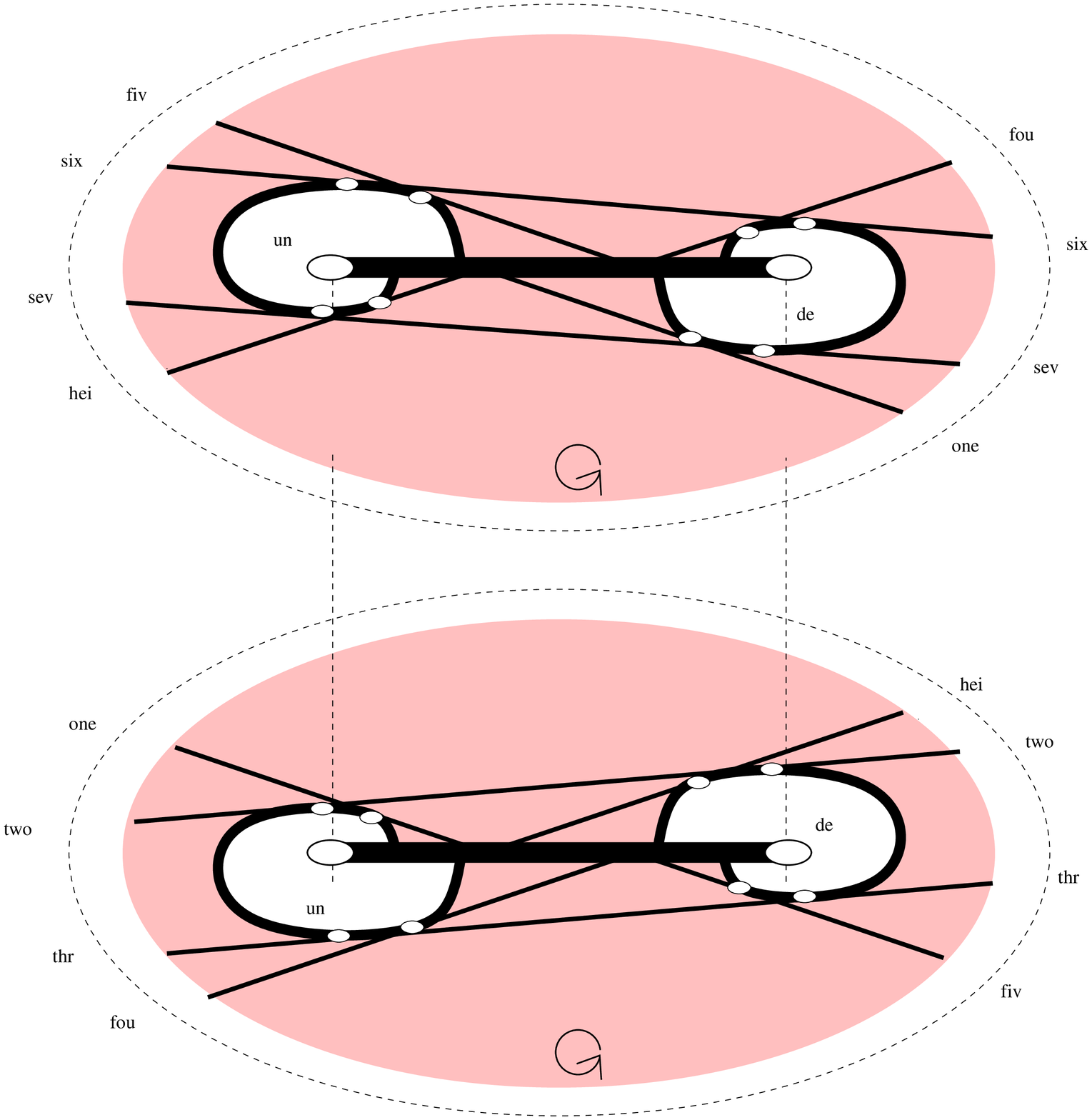}%xfigfinal
\end{center}
\caption{\protect \footnotesize
%%The visibility complex of a family of two disjoint convex bodies is composed of four 0-cells, eight 1-cells,  and five 2-cells put together as
%%indicated in the figure.
\label{twobodytwosheets}}
\end{figure}
%%%%%%%%%%%%%%%%%%%%%%%%%%%%%%%%%%%%%%%%%%%%%%%%%%%%%%%%%%%%%%%%%%%%%%%%%%%%%%%
%%%%%%%%%%%%%%%%%%%%%%%%%%%%%%%%%%%%%%%%%%%%%%%%%%%%%%%%%%%%%%%%%%%%%%%%%%%%%%%
%%%%%%%%%%%%%%%%%%%%%%%%%%%%%%%%%%%%%%%%%%%%%%%%%%%%%%%%%%%%%%%%%%%%%%%%%%%%%%%
Its complex of pseudotriangulations is the cocube of dimension $2$. 
More generally the complex of pseudotriangulations of a family of two convex bodies of a $n$-sheeted covering surface of the plane with two branch points 
is the cocube of dimension $n$. 
\end{example}

\subsection{Visibility complexes}
We now assume that there is no tritangent.
Free space in denoted $\fs{}$.
The space $\fs{}$ inherits from the point-line incidence structure of~$\pointset$ a natural partial point-line incidence structure  whose
system of lines
is defined as the space of
connected components of the pre-images of the lines of $\pointset$ under the
canonical projection $\mathbb{X} \rightarrow \mathbb{A}$,
and whose set of incidences is the set of point-line pairs $(p,x)$
%%% \in \mathbb{X} \times \spaceoflinesof{\mathbb{X}}$ with $p\in x$.
with $p\in x$.
%%%%%%%%%%%%%%%%%%%%%%%%%%%%%%%%%%%%%%%%%%%%%%%%%%%%%%%%%%%%%%%%%%%%%%%%%%%%%%%%%%%%%%%%%%%%%%%%%%%%%%%%%%%
%%%%%%%%%%%%%%%%%%%%%%%%%%%%%%%%%%%%%%%%%%%%%%%%%%%%%%%%%%%%%%%%%%%%%%%%%%%%%%%%%%%%%%%%%%%%%%%%%%%%%%%%%%%
%%%%%%%%%%%%%%%%%%%%%%%%%%%%%%%%%%%%%%%%%%%%%%%%%%%%%%%%%%%%%%%%%%%%%%%%%%%%%%%%%%%%%%%%%%%%%%%%%%%%%%%%%%%
%%%%%%%%%%%%%%%%%%%%%%%%%%%%%%%%%%%%%%%%%%%%%%%%%%%%%%%%%%%%%%%%%%%%%%%%%%%%%%%%%%%%%%%%%%%%%%%%%%%%%%%%%%%
%%%%%%%%%%%%%%%%%%%%%%%%%%%%%%%%%%%%%%%%%%%%%%%%%%%%%%%%%%%%%%%%%%%%%%%%%%%%%%%%%%%%%%%%%%%%%%%%%%%%%%%%%%%
The {\it label} of a directed line of $\fs{}$ is the sequence of bodies
%%(and more generally with the sequences of views in case we work in the setting of constrained family of convex bodies) 
intersected by the line ordered as they appear along the line and prefixed or postfixed or both prefixed and postfixed
with the symbol $\infty$ in case the line is
(orientation preserving) homeomorphic to the curves $\mathbb{R}^+$, $\mathbb{R}^-$ or $\mathbb{R}$ endowed with their natural orientations.
%The {\it backward/forward view} of a line %%% is denoted $\Fv{r}$ ($\Bv{r}$)  and 
%is the first/last  element of its label. 
A directed line of $\fs{}$ touching tangentially a body $o$ is called a {\it left} or {\it right tangent} to $o$
depending on whether  $o$ lies, locally around the touching point, on the left side or on the right side of the line.
A directed line of $\fs{}$ joining tangentially a body $o$ to a body $o'$ is said to {\it leave} $o$ and to {\it reach} (or {\it enter})  
$o'$ and is called a {\it left-left}, {\it left-right}, {\it right-left}, or {\it right-right bitangent}
depending on whether the line is a left tangent to both $o$ and $o'$, a left tangent to $o$ and a right tangent to $o'$, a right tangent to  $o$ and  a left tangent to $o'$, 
or a right tangent to both $o$ and $o'$.
%The left cycle of a body is its set of left tangents; similarly the right cycle of a
%body is its set of right tangents; 
%a cycle is a simple closed curve  
%to which we assign the  orientation inherited from the counterclockwise
%orientation of the boundary of its underlying body.  
The sets of left and right tangents to a body are simple closed curves
to which we assign the  orientation inherited by duality from the orientation of the ground topological plane $\pointset$.  
%%These curves will be called cycles thereafter.
%%% ({\it est-ce vraiment necessaire d'introduire la terminologie des cycles?}).

%Let $\vc{}$ be the space of lines of $\fs{}$, $\vczer{}$ the space of bitangents, $\vcone{}$ the space of tangents,
%and let $\widetilde{\vc{}}$ be the Freudenthal compactification of $\vc{}$---that is, the
%space obtained by adding to $\vc{}$ its ends. The space $\vc{}$
%satisfies the following properties:

\subsubsection{Cell structure}
Let $\vc{} = \vctwo{}$ be the space of directed lines of $\fs{}$, $\vcone{}$ its space of left and right tangents, and $\vczer{}$ its space of left-left, left-right, right-left and right-right bitangents. 
The operator that reverses the direction of a directed line  is denoted $\reverse$ and we take for granted that the natural projection 
${\vc{}} \rightarrow {\vc{}/\reverse}$ is a $2$-covering. The increasing sequence 
\begin{equation} 
\vczer{} \subset \vcone{} \subset \vctwo{} = \vc{}
\end{equation}
is, modulo the adjunction of a point at infinity in each connected component of $\vc{}\setminus \vcone{}$ whose topological closure is noncompact, 
the sequence of $0$-, $1$-, and $2$-skeletons of a natural structure of finite $2$-dimensional regular cell complex on $\vc{}$: since the curves of tangents to the bodies 
are oriented curves one can speak of the source and sink vertices or $0$-cells of a $1$-cell; as usual a chain of $\vcone{}$ is a sequence
%The source and sink vertices or $0$-cells of a $1$-cell are defined with respect of the orientation of the cycles; as usual a chain of $\vcone{}$ is a sequence
of $0$- and $1$-cells such that the predecessor (if any) and the successor (if any)
of a $1$-cell are its source and its sink, respectively; as usual the points added at infinity are called the ends of $\vc{}$; and a $2$-cell is said bounded if it contains no end. This complex 
satisfies the following properties:
\begin{enumerate}
%\item The natural projection ${\vc{}} \rightarrow {\vc{}/\reverse}$ is a 2-covering;
%By a slight abuse of terminology the closure of a connected component of
%$\vc{}\setminus\vcone{}$ is still called a $2$-cell.  A $2$-cell of
%$\vc{}$ is  either a closed 2-ball or a closed 2-ball minus an interior point (which in that case correspond to an end of $\vc{}$). 
%In the first case the $2$-cell will be termed bounded and in the second case the $2$-cell  will be termed  unbounded. 

\item A $0$-cell is the source and the sink of two $1$-cells;
%we denote by $\ccLop$ the operator that assigns to a $0$-cell  
%the sink of the $1$-cell supported by the cycle that the $0$-cell leaves; 
%similarly  we denote by $\ccRop$ the operator that assigns to a $0$-cell  
%the sink of the $1$-cell supported by the cycle that the $0$-cell enters.{}\footnote{This terminology is inherited from the terminology on 
%the bitangent line segments : a bitangent joining a first body to a second body is said to
%leave the first body and to enter the second one.}

\item The boundary of a bounded $2$-cell
is composed of two
chains that share the same source/sink, called the source/sink of the $2$-cell.
Conversely any vertex is the source/sink of a bounded $2$-cell.
 By convention the right/left boundary chain of a bounded $2$-cell $\sigma$ with source $v$, denoted $\rcNew{\sigma}$/$\lcNew{\sigma}$, 
is the boundary chain of $\sigma$  whose first
$1$-cell is supported by the curve of tangents to the body reached/left by $v$  and supporting~$v$;

\item The boundary of an unbounded $2$-cell is composed of a single chain;
An unbounded $2$-cell is said to be {\it left }  or {\it right
unbounded} depending on whether its boundary is composed of right or left tangents,
respectively.
The number of left  unbounded $2$-cells and the number of right unbounded $2$-cells are both equal to
the number of connected components of the complement of the convex hull
of the family of convex bodies;

%%%%%%%%%%%%%%%%%%%%%%%%%%%%%%%%%%%%%%%%%%%%%%%%%%%%%%%%%%%%%%%%%%%%%%%%%%%%%%%%%%%%%%%%%%%%%%%%%%%%%%%%%%%
%%%%%%%%%%%%%%%%%%%%%%%%%%%%%%%%%%%%%%%%%%%%%%%%%%%%%%%%%%%%%%%%%%%%%%%%%%%%%%%%%%%%%%%%%%%%%%%%%%%%%%%%%%%
%%%%%%%%%%%%%%%%%%%%%%%%%%%%%%%%%%%%%%%%%%%%%%%%%%%%%%%%%%%%%%%%%%%%%%%%%%%%%%%%%%%%%%%%%%%%%%%%%%%%%%%%%%%
%%%%%%%%%%%%%%%%%%%%%%%%%%%%%%%%%%%%%%%%%%%%%%%%%%%%%%%%%%%%%%%%%%%%%%%%%%%%%%%%%%%%%%%%%%%%%%%%%%%%%%%%%%%
%%%%%%%%%%%%%%%%%%%%%%%%%%%%%%%%%%%%%%%%%%%%%%%%%%%%%%%%%%%%%%%%%%%%%%%%%%%%%%%%%%%%%%%%%%%%%%%%%%%%%%%%%%%
\item a $1$-cell is incident to three $2$-cells with labels the subsequences
of length two of its label;
The three $2$-cells incident to the $1$-cell $\sigma$ of $\vcor{}$ with label $ijk$ are denoted
$\sigma_{\alpha}$,  $\alpha \in \{\Srigh,\Sleft, \Sback,\Sforw\}$,
according to the following rule:
the $2$-cells with label $ij$ and $jk$ are denoted $\sigma_{\Sback}$ and
$\sigma_{\Sforw}$, respectively; and  the remaining $2$-cell, whose  label is  $ik$, is denoted $\sigma_{\Srigh}$  or
$\sigma_{\Sleft}$ depending on whether the lines of $\sigma$ are left or right tangents, as illustrated in Figure~\ref{vicinityonecell}.
%%%%%%%%%%%%%%%%%%%%%%%%%%%%%%%%%%%%%%%%%%%%%%%%%%%%%%%%%%%%%%%%%%%%%%%%%%%%%%%%%%%%%%%%%%%%%%%%%%%%%%%%%%%
%%%%%%%%%%%%%%%%%%%%%%%%%%%%%%%%%%%%%%%%%%%%%%%%%%%%%%%%%%%%%%%%%%%%%%%%%%%%%%%%%%%%%%%%%%%%%%%%%%%%%%%%%%%
%%%%%%%%%%%%%%%%%%%%%%%%%%%%%%%%%%%%%%%%%%%%%%%%%%%%%%%%%%%%%%%%%%%%%%%%%%%%%%%%%%%%%%%%%%%%%%%%%%%%%%%%%%%
%%%%%%%%%%%%%%%%%%%%%%%%%%%%%%%%%%%%%%%%%%%%%%%%%%%%%%%%%%%%%%%%%%%%%%%%%%%%%%%%%%%%%%%%%%%%%%%%%%%%%%%%%%%
%%%%%%%%%%%%%%%%%%%%%%%%%%%%%%%%%%%%%%%%%%%%%%%%%%%%%%%%%%%%%%%%%%%%%%%%%%%%%%%%%%%%%%%%%%%%%%%%%%%%%%%%%%%
\begin{figure}[!htb]
\begin{center}
\psfrag{sig}{\sm $\sigma$}
\psfrag{i}{\sm$i$}
\psfrag{j}{\sm $j$}
\psfrag{k}{\sm $k$}
\psfrag{l}{\sm $l$}
\psfrag{left}{\sm $\sigma_{\Sleft}$} \psfrag{back}{\sm $\sigma_{\Sback}$} \psfrag{righ}{\sm $\sigma_{\Srigh}$} \psfrag{forw}{\sm $\sigma_{\Sforw}$}
\psfrag{sour}{\sm $\sigma_{\Ssour}$} \psfrag{sink}{\sm $\sigma_{\Ssink}$} \psfrag{4}{\sm $\sigma_{\Sforw}$}
\includegraphics[width = 0.8575\linewidth]{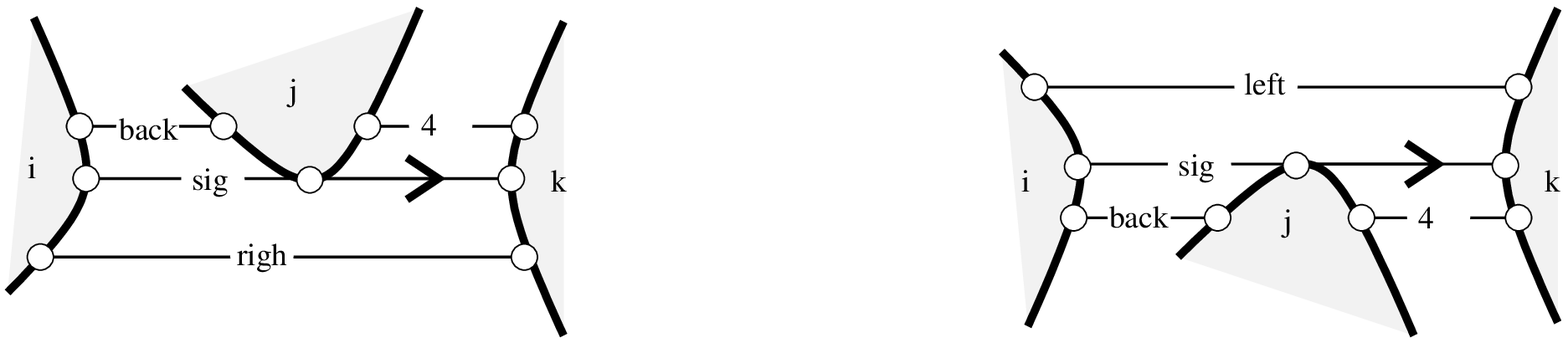}%xfig
\caption{\protect \small
%A left $1$-cell $\sigma$ with label $ijk$ is incident to three $2$-cells $\sigmaback,\sigmaforw$
%and $\sigmarigh$ with labels $ij,jk$ and $ik$, respectively.
%Similar notations are used for right $1$-cells.
\label{vicinityonecell}}
\end{center}
\end{figure}

%%%%%%%%%%%%%%%%%%%%%%%%%%%%%%%%%%%%%%%%%%%%%%%%%%%%%%%%%%%%%%%%%%%%%%%%%%%%%%%%%%%%%%%%%%%%%%%%%%%%%%%%%%
%%%%%%%%%%%%%%%%%%%%%%%%%%%%%%%%%%%%%%%%%%%%%%%%%%%%%%%%%%%%%%%%%%%%%%%%%%%%%%%%%%%%%%%%%%%%%%%%%%%%%%%%%%%
%%%%%%%%%%%%%%%%%%%%%%%%%%%%%%%%%%%%%%%%%%%%%%%%%%%%%%%%%%%%%%%%%%%%%%%%%%%%%%%%%%%%%%%%%%%%%%%%%%%%%%%%%%%
%%%%%%%%%%%%%%%%%%%%%%%%%%%%%%%%%%%%%%%%%%%%%%%%%%%%%%%%%%%%%%%%%%%%%%%%%%%%%%%%%%%%%%%%%%%%%%%%%%%%%%%%%%%
%%%%%%%%%%%%%%%%%%%%%%%%%%%%%%%%%%%%%%%%%%%%%%%%%%%%%%%%%%%%%%%%%%%%%%%%%%%%%%%%%%%%%%%%%%%%%%%%%%%%%%%%%%%
\item
a $0$-cell is incident to four $1$-cells with labels
the subsequences of length~3 of its label.
\item
a $0$-cell is incident to  six $2$-cells with labels
the subsequences of length~2 of its label.
The six $2$-cells incident
to the  $0$-cell $\sigma$ of $\vcor{}$ with label $ijkl$
are denoted $\sigma_{\alpha}$, $\alpha \in \alphabet = \{\Ssour,\Ssink,\Srigh,\Sleft, \Sback,\Sforw\}$
according to the following rules:
\begin{enumerate}
\item $\sigma_{\Sback}$ is the $2$-cell with label $ij$;
\item $\sigma_{\Sforw}$ is the $2$-cell with label $kl$;
%\item the two $2$-cells with labels $ij$ and $kl$ are denoted $\sigma_{\Sback}$ and $\sigma_{\Sforw}$, respectively;
\item $\sigma_{\Ssour}$ is the $2$-cell with label
$jk,jl,il$, or $ik$ depending on whether
$\sigma$ is a left-right, left-left, right-left or right-right bitangent;
\item $\sigma_{\Ssink}$ is the $2$-cell with label
$il,ik,jk$, or $jl$ depending on whether
$\sigma$ is a left-right, left-left, right-left or right-right bitangent;
\item $\sigma_{\Srigh}$ is the $2$-cell with label
$ik,il,jl$, or $jk$ depending on whether
$\sigma$ is a left-right, left-left, right-left or right-right bitangent;
\item $\sigma_{\Sleft}$ is the $2$-cell with label
$jl,jk,ik$, or $il$ depending on whether
$\sigma$ is a left-right, left-left, right-left or right-right bitangent,
as illustrated in Figure~\ref{vicinity}.
%%%%%%%%%%%%%%%%%%%%%%%%%%%%%%%%%%%%%%%%%%%%%%%%%%%%%%%%%%%%%%%%%%%%%%%%%%%%%%%%%%%%%%%%%%%%%%%%%%%%%%%%%%%
%%%%%%%%%%%%%%%%%%%%%%%%%%%%%%%%%%%%%%%%%%%%%%%%%%%%%%%%%%%%%%%%%%%%%%%%%%%%%%%%%%%%%%%%%%%%%%%%%%%%%%%%%%%
%%%%%%%%%%%%%%%%%%%%%%%%%%%%%%%%%%%%%%%%%%%%%%%%%%%%%%%%%%%%%%%%%%%%%%%%%%%%%%%%%%%%%%%%%%%%%%%%%%%%%%%%%%%
%%%%%%%%%%%%%%%%%%%%%%%%%%%%%%%%%%%%%%%%%%%%%%%%%%%%%%%%%%%%%%%%%%%%%%%%%%%%%%%%%%%%%%%%%%%%%%%%%%%%%%%%%%%
%%%%%%%%%%%%%%%%%%%%%%%%%%%%%%%%%%%%%%%%%%%%%%%%%%%%%%%%%%%%%%%%%%%%%%%%%%%%%%%%%%%%%%%%%%%%%%%%%%%%%%%%%%%
\begin{figure}[!htb]
\begin{center}
\psfrag{sig}{\sm $\sigma$}
\psfrag{i}{\sm$i$}
\psfrag{j}{\sm $j$}
\psfrag{k}{\sm $k$}
\psfrag{l}{\sm $l$}
\psfrag{left}{\sm $\sigma_{\Sleft}$} \psfrag{back}{\sm $\sigma_{\Sback}$} \psfrag{righ}{\sm $\sigma_{\Srigh}$} \psfrag{forw}{\sm $\sigma_{\Sforw}$}
\psfrag{sour}{\sm $\sigma_{\Ssour}$} \psfrag{sink}{\sm $\sigma_{\Ssink}$} \psfrag{4}{\sm $\sigma_{\Sforw}$}
\includegraphics[width = 0.8575\linewidth]{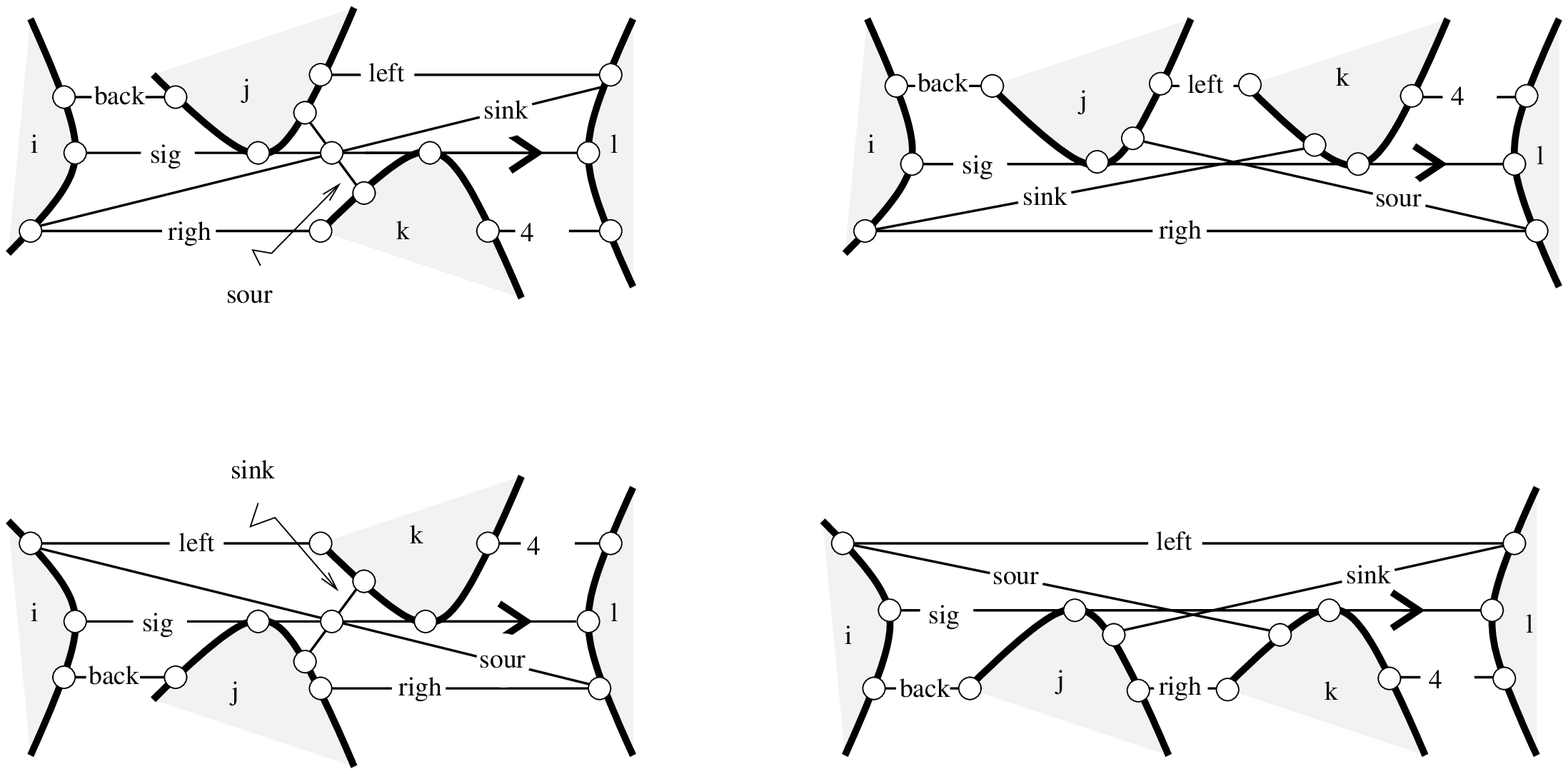}%xfig
\caption{\protect \small
%A left-right vertex $\sigma$ with label $ijkl$ is incident to six $2$-cells
%$\sigmaback, \sigmaforw, \sigmaleft,\sigmarigh$,
%$\sigmasour$ and $\sigmasink$
%with labels $ij, kl,ik,jl,jk,il$, respectively.
%Similar notations are used for left-left, right-left and right-right
%vertices.
\label{vicinity}}
\end{center}
\end{figure}
\end{enumerate}
\end{enumerate}
%%%%%%%%%%%%%%%%%%%%%%%%%%%%%%%%%%%%%%%%%%%%%%%%%%%%%%%%%%%%%%%%%%%%%%%%%%%%%%%%%%%%%%%%%%%%%%%%%%%%%%%%%%%
%%%%%%%%%%%%%%%%%%%%%%%%%%%%%%%%%%%%%%%%%%%%%%%%%%%%%%%%%%%%%%%%%%%%%%%%%%%%%%%%%%%%%%%%%%%%%%%%%%%%%%%%%%%
%%%%%%%%%%%%%%%%%%%%%%%%%%%%%%%%%%%%%%%%%%%%%%%%%%%%%%%%%%%%%%%%%%%%%%%%%%%%%%%%%%%%%%%%%%%%%%%%%%%%%%%%%%%
%%%%%%%%%%%%%%%%%%%%%%%%%%%%%%%%%%%%%%%%%%%%%%%%%%%%%%%%%%%%%%%%%%%%%%%%%%%%%%%%%%%%%%%%%%%%%%%%%%%%%%%%%%%
%%%%%%%%%%%%%%%%%%%%%%%%%%%%%%%%%%%%%%%%%%%%%%%%%%%%%%%%%%%%%%%%%%%%%%%%%%%%%%%%%%%%%%%%%%%%%%%%%%%%%%%%%%%
%%%%%%%%%%%%%%%%%%%%%%%%%%%%%%%%%%%%%%%%%%%%%%%%%%%%%%%%%%%%%%%%%%%%%%%%%%%%%%%%%%%%%%%%%%%%%%%%%%%%%%%%%%%
%%%%%%%%%%%%%%%%%%%%%%%%%%%%%%%%%%%%%%%%%%%%%%%%%%%%%%%%%%%%%%%%%%%%%%%%%%%%%%%%%%%%%%%%%%%%%%%%%%%%%%%%%%%
%%%%%%%%%%%%%%%%%%%%%%%%%%%%%%%%%%%%%%%%%%%%%%%%%%%%%%%%%%%%%%%%%%%%%%%%%%%%%%%%%%%%%%%%%%%%%%%%%%%%%%%%%%%
%%%%%%%%%%%%%%%%%%%%%%%%%%%%%%%%%%%%%%%%%%%%%%%%%%%%%%%%%%%%%%%%%%%%%%%%%%%%%%%%%%%%%%%%%%%%%%%%%%%%%%%%%%%
%%%%%%%%%%%%%%%%%%%%%%%%%%%%%%%%%%%%%%%%%%%%%%%%%%%%%%%%%%%%%%%%%%%%%%%%%%%%%%%%%%%%%%%%%%%%%%%%%%%%%%%%%%%
By definition the visibility complex of the family of pairwise disjoint convex bodies $\disks$ is the regular 
cell-complex $\vczer{} \subset \vcone{}\subset \vctwo{} = \vc{} $ endowed with the orientation of its one-skeleton $\vcone{}$
inherited from the orientation of the ground topological plane $\pointset$.
%\clearpage

\begin{example} \label{vctwobodies}
The visibility complex a family of two disjoint convex bodies $o_i,o_j$ of $\pointset$
 is composed of eight $0$-cells, sixteen (oriented) $1$-cells and ten $2$-cells
(the sets of lines with labels $ij$, $ji$, $i\infty$, $\infty i$, $j\infty$, $\infty j$, and  $\infty\infty$ four times)
put together as
%%%%%%%%%%%%%%%%%%%%%%%%%%%%%%%%%%%%%%%%%%%%%%%%%%%%%%%%%%%%%%%%%%%%%%%%%%%%%%%
%%%%%%%%%%%%%%%%%%%%%%%%%%%%%%%%%%%%%%%%%%%%%%%%%%%%%%%%%%%%%%%%%%%%%%%%%%%%%%%
%%%%%%%%%%%%%%%%%%%%%%%%%%%%%%%%%%%%%%%%%%%%%%%%%%%%%%%%%%%%%%%%%%%%%%%%%%%%%%%
%%%%%%%%%%%%%%%%%%%%%%%%%%%%%%%%%%%%%%%%%%%%%%%%%%%%%%%%%%%%%%%%%%%%%%%%%%%%%%%
\begin{figure}[!htb]
\begin{center}
\psfrag{one}{$\onevc$}  \psfrag{two}{$\twovc$}  \psfrag{thr}{$\thrvc$}  \psfrag{fou}{$\fouvc$}
\psfrag{onep}{$\onevcp$}\psfrag{twop}{$\twovcp$}\psfrag{thrp}{$\thrvcp$}\psfrag{foup}{$\fouvcp$}
\psfrag{six}{6}       \psfrag{sev}{7}       \psfrag{hei}{8}       \psfrag{fiv}{5}
\psfrag{unun}{$\unun$}\psfrag{unde}{$\unde$}\psfrag{unth}{$\unth$}\psfrag{unfo}{$\unfo$}
\psfrag{unfi}{$\unfi$}\psfrag{unsi}{$\unsi$}\psfrag{unse}{$\unse$}\psfrag{unhe}{$\unhe$}
\psfrag{deun}{$\deun$}\psfrag{dede}{$\dede$}\psfrag{deth}{$\deth$}\psfrag{defo}{$\defo$}
\psfrag{defi}{$\defi$}\psfrag{desi}{$\desi$}\psfrag{dese}{$\dese$}\psfrag{dehe}{$\dehe$}
\psfrag{un}{$i$}      \psfrag{de}{$j$}      \psfrag{firstbody}{$i$}  \psfrag{secondbody}{$j$}
\psfrag{lunde}{$ij$}
\psfrag{ldeun}{$ji$}
\psfrag{inin}{$\infty\infty$}
\psfrag{unin}{$i\infty$}
\psfrag{inun}{$\infty i$}
\psfrag{inde}{$\infty j$}
\psfrag{dein}{$j\infty$}
\psfrag{infty}{$\infty$}
\psfrag{body}{}
\includegraphics[width = 0.8575\linewidth]{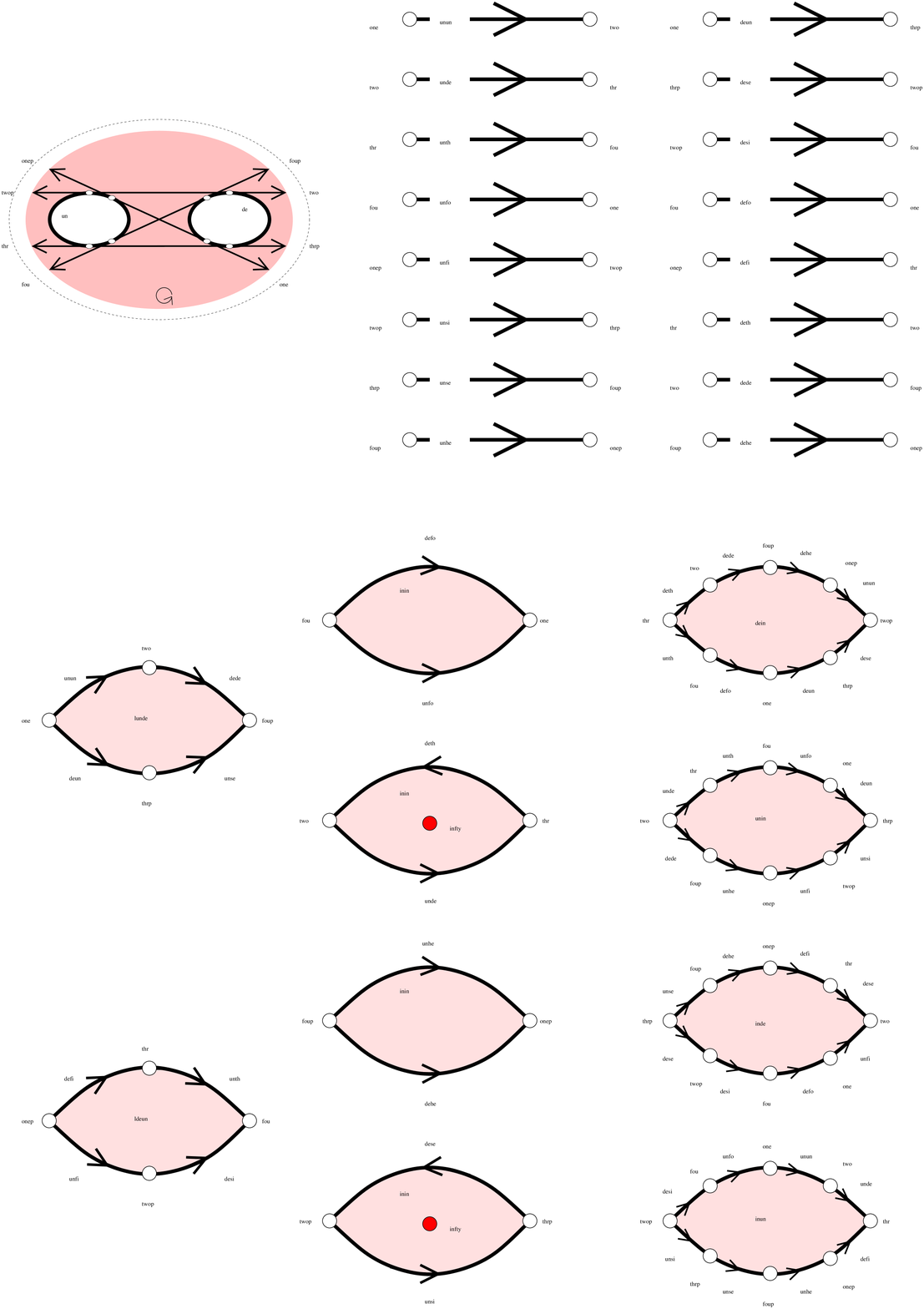}%xfig
\end{center}
\caption{\protect \footnotesize \label{DirectedVCtwobodies}}
\end{figure}
%%%%%%%%%%%%%%%%%%%%%%%%%%%%%%%%%%%%%%%%%%%%%%%%%%%%%%%%%%%%%%%%%%%%%%%%%%%%%%%
%%%%%%%%%%%%%%%%%%%%%%%%%%%%%%%%%%%%%%%%%%%%%%%%%%%%%%%%%%%%%%%%%%%%%%%%%%%%%%%
%%%%%%%%%%%%%%%%%%%%%%%%%%%%%%%%%%%%%%%%%%%%%%%%%%%%%%%%%%%%%%%%%%%%%%%%%%%%%%%
indicated in  Figure~\ref{DirectedVCtwobodies} where, by convention, the left boundary chain of a bounded $2$-cell is  above its right boundary chain 
(thus, one can read on the Figure that the right boundary chain of the $2$-cell with label $ij$ is $\onevc\deun\thrvcp\unse\fouvcp$ and that its left boundary chain is 
$\onevc\unun\twovc\dede\fouvcp$); 
in the introduction section we observed that the quotient of this complex under $\reverse$ is not regular.
\end{example}
%\clearpage
\begin{example}
Figure~\ref{VCTwoBodiesTwoSheets} depicts the cell decomposition of the quotient under $\reverse$ of the visibility complex of a family of two convex bodies of a 2-sheeted branched covering of $\pointset$ with two branch points (the two sheets are obtained by cutting the covering space along the two line segments joining the two branch points).

%%%%%%%%%%%%%%%%%%%%%%%%%%%%%%%%%%%%%%%%%%%%%%%%%%%%%%%%%%%%%%%%%%%%%%%%%%%%%%%
%%%%%%%%%%%%%%%%%%%%%%%%%%%%%%%%%%%%%%%%%%%%%%%%%%%%%%%%%%%%%%%%%%%%%%%%%%%%%%%
%%%%%%%%%%%%%%%%%%%%%%%%%%%%%%%%%%%%%%%%%%%%%%%%%%%%%%%%%%%%%%%%%%%%%%%%%%%%%%%
%%%%%%%%%%%%%%%%%%%%%%%%%%%%%%%%%%%%%%%%%%%%%%%%%%%%%%%%%%%%%%%%%%%%%%%%%%%%%%%
\begin{figure}[!htb]
\begin{center}
\psfrag{one}{1} \psfrag{two}{2} \psfrag{thr}{3} \psfrag{fou}{4}
\psfrag{six}{6} \psfrag{sev}{7} \psfrag{hei}{8} \psfrag{fiv}{5}
\psfrag{unun}{$i_1$} \psfrag{unde}{$i_2$} \psfrag{unth}{$i_3$} \psfrag{unfo}{$i_4$}
\psfrag{unfi}{$i_5$} \psfrag{unsi}{$i_6$} \psfrag{unse}{$i_7$} \psfrag{unhe}{$i_8$}
\psfrag{deun}{$j_1$} \psfrag{dede}{$j_2$} \psfrag{deth}{$j_3$} \psfrag{defo}{$j_4$}
\psfrag{defi}{$j_5$} \psfrag{desi}{$j_6$} \psfrag{dese}{$j_7$} \psfrag{dehe}{$j_8$}
\psfrag{un}{$i$} \psfrag{de}{$j$} \psfrag{firstbody}{$i$} \psfrag{secondbody}{$j$}
\psfrag{lunde}{$ij$}
\psfrag{inin}{$\infty\infty$}
\psfrag{unin}{$i\infty$}
\psfrag{dein}{$j\infty$}
\psfrag{infty}{$\infty$}
\psfrag{body}{}
\includegraphics[width = 0.8575\linewidth]{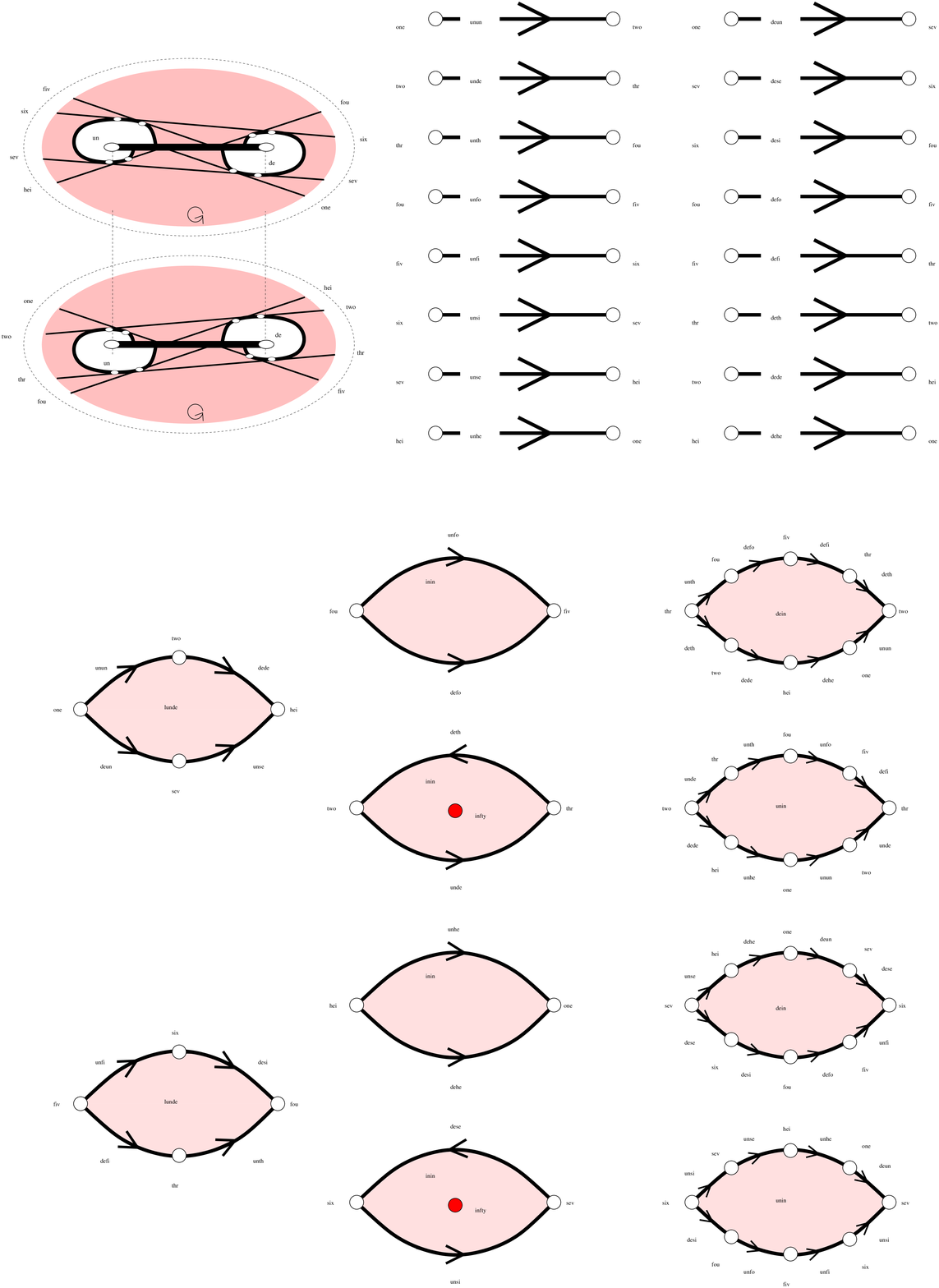}%xfig
\end{center}
\caption{\protect \footnotesize
\label{VCTwoBodiesTwoSheets}}
\end{figure}
%%%%%%%%%%%%%%%%%%%%%%%%%%%%%%%%%%%%%%%%%%%%%%%%%%%%%%%%%%%%%%%%%%%%%%%%%%%%%%%
%%%%%%%%%%%%%%%%%%%%%%%%%%%%%%%%%%%%%%%%%%%%%%%%%%%%%%%%%%%%%%%%%%%%%%%%%%%%%%%
%%%%%%%%%%%%%%%%%%%%%%%%%%%%%%%%%%%%%%%%%%%%%%%%%%%%%%%%%%%%%%%%%%%%%%%%%%%%%%%

\end{example}

\clearpage
\subsubsection{Horizon operators}
%%%%%%%%%%%%%%%%%%%%%%%%%%%%%%%%%%%%%%%%%%%%%%%%%%%%%%%%%%%%%%%%%%%%%%%%%%%%%%%%%%%%%%%%%%%%%%%%%%%%%%%%%%%
%%%%%%%%%%%%%%%%%%%%%%%%%%%%%%%%%%%%%%%%%%%%%%%%%%%%%%%%%%%%%%%%%%%%%%%%%%%%%%%%%%%%%%%%%%%%%%%%%%%%%%%%%%%
%%%%%%%%%%%%%%%%%%%%%%%%%%%%%%%%%%%%%%%%%%%%%%%%%%%%%%%%%%%%%%%%%%%%%%%%%%%%%%%%%%%%%%%%%%%%%%%%%%%%%%%%%%%
%%%%%%%%%%%%%%%%%%%%%%%%%%%%%%%%%%%%%%%%%%%%%%%%%%%%%%%%%%%%%%%%%%%%%%%%%%%%%%%%%%%%%%%%%%%%%%%%%%%%%%%%%%%
%%%%%%%%%%%%%%%%%%%%%%%%%%%%%%%%%%%%%%%%%%%%%%%%%%%%%%%%%%%%%%%%%%%%%%%%%%%%%%%%%%%%%%%%%%%%%%%%%%%%%%%%%%%
We now describe, in preparation for the section on \ABpts, the boundary chains of the $1$- and $2$-cells in terms of the 
operators, denoted $\mphi_{\xxx}$, $\xxx\in \{\Ssour,\Ssink,\Srigh,\Sleft, \Sback,\Sforw\}$, 
that assign to a $0$- or $1$-cell $\sigma$ the sink vertices of its incident $2$-cells $\sigma_{\xxx}$ (if defined); 
%The sink and source of the cell $\sigma_{\xxx}$, $\xxx\in \{\Ssour,\Ssink,\Srigh,\Sleft, \Sback,\Sforw\}$, are  denoted $\mphi_{\xxx}(\sigma)$ and $\mphi_{\xxx_*}(\sigma)$,
in particular,
 $\phisink(\vertex)$ is the sink of the $2$-cell with source~$v$, $\phisour(v)$ is the identity operator, and, for a left $1$-cell~$e$, supported by the curve of left tangents to the body~$o$, the bitangent 
$\phiforw(e)$ is the first bitangent leaving~$o$ encountered when we traverse 
the curve of left tangents to~$o$ starting from~$e$.
%Note that $\phisink \circ \mphi_{\xxx_*} = \mphi_{\xxx}$, and
Note 
that  $\phileft$ and $\phiforw$ are the conjugates of $\phirigh$ and $ \phiback$ under the reorientation operator $\reverse$, that is,
$\reverse \circ \phileft =   \phirigh \circ\reverse$,
$\reverse\circ\phiforw =  \phiback \circ \reverse$.
We name these operators the {\it horizon operators} in reference to the operators underlying the definition
of the  horizon trees of Edelsbrunner and Guibas~\cite{eg-tsa-86}.
%(See Figure~\ref{facebis-figure} for an illustration.)
%The conjugate
%of the operator $\mphi_{\xxx}$ under the touching operator
%$\touchop$ is denoted $\bitmphi_{\xxx}.$
For example the table of the horizon operators on the set of bitangents of the visibility complex of
two convex bodies of the plane is the following 
$$\begin{array}{l|llllllll}
         & \onevc  & \twovc  & \thrvc  & \fouvc  & \onevcp & \twovcp & \thrvcp & \fouvcp \\
\hline
\reverse & \onevcp & \twovcp & \thrvcp & \fouvcp & \onevc  & \twovc  & \thrvc  & \fouvc \\ 
%\mphi   & \fouvcp& \twovcp& \thrvcp& \onevc  & \fouvc & \twovc & \thrvc & \onevcp\\
\phisink & \fouvcp & \thrvcp & \twovcp & \onevc  & \fouvc  & \thrvc  & \twovc  & \onevcp\\
\phileft & \twovc  & \udf    &   \udf     & \twovcp & \thrvc  & \fouvc  & \fouvcp & \thrvcp\\
\phirigh & \thrvcp & \fouvc  & \onevcp & \thrvc  & \twovcp &  \udf      &  \udf       & \twovc \\
\phiforw & \twovcp & \twovcp & \twovc  & \twovc  & \thrvcp & \thrvcp & \twovcp & \twovcp\\ 
\phiback & \thrvc  & \thrvc  & \twovc  & \twovc  & \twovc  & \twovc  & \twovcp & \twovcp\\ 
\end{array}
$$
where $\udf$ stands for undefined and where we use the notations of Example~\ref{vctwobodies}.
%The descriptions of the boundary  chains of the $1$- and $2$-cells in terms of the horizon operators are reported in the following two theorems. 
%%%Appendix~\ref{tt}.  
The proofs of the two following theorems are easy (using continuity arguments) and are left to the reader.

%%%%%%%%%%%%%%%%%%%%%%%%%%%%%%%%%%%%%%%%%%%%%%%%%%%%%%%%%%%%%%%%%%%%%%%%%%%%%%%%%%%%%%%%%%%%%%%%%%%%%%%%%%%
%%%%%%%%%%%%%%%%%%%%%%%%%%%%%%%%%%%%%%%%%%%%%%%%%%%%%%%%%%%%%%%%%%%%%%%%%%%%%%%%%%%%%%%%%%%%%%%%%%%%%%%%%%%
%%%%%%%%%%%%%%%%%%%%%%%%%%%%%%%%%%%%%%%%%%%%%%%%%%%%%%%%%%%%%%%%%%%%%%%%%%%%%%%%%%%%%%%%%%%%%%%%%%%%%%%%%%%
%%%%%%%%%%%%%%%%%%%%%%%%%%%%%%%%%%%%%%%%%%%%%%%%%%%%%%%%%%%%%%%%%%%%%%%%%%%%%%%%%%%%%%%%%%%%%%%%%%%%%%%%%%%
%%%%%%%%%%%%%%%%%%%%%%%%%%%%%%%%%%%%%%%%%%%%%%%%%%%%%%%%%%%%%%%%%%%%%%%%%%%%%%%%%%%%%%%%%%%%%%%%%%%%%%%%%%%
\begin{theorem}\label{comput-sink-edge}
Let $e$ be a left $1$-cell supported by the curve of left tangents to the body $o$. Then 
$\sink{e} = \phirigh(e)$ if $\sink{e}$ reaches $o$; otherwise $\sink{e} = \phiforw(e).$ 
Furthermore $\phiforw(e)$ is the first $0$-cell leaving $o$ encountered when we traverse 
its curve of left tangents starting from $e$.
A similar result holds for right $1$-cells using conjugation under~$\reverse.$
\qed
\end{theorem}%%%%
%%%%%%%%%%%%%%%%%%%%%%%%%%%%%%%%%%%%%%%%%%%%%%%%%%%%%%%%%%%%%%%%%%%%%%%%%%%%%%%%%%%%%%%%%%%%%%%%%%%%%%%%%%%
%%%%%%%%%%%%%%%%%%%%%%%%%%%%%%%%%%%%%%%%%%%%%%%%%%%%%%%%%%%%%%%%%%%%%%%%%%%%%%%%%%%%%%%%%%%%%%%%%%%%%%%%%%%
%%%%%%%%%%%%%%%%%%%%%%%%%%%%%%%%%%%%%%%%%%%%%%%%%%%%%%%%%%%%%%%%%%%%%%%%%%%%%%%%%%%%%%%%%%%%%%%%%%%%%%%%%%%
%%%%%%%%%%%%%%%%%%%%%%%%%%%%%%%%%%%%%%%%%%%%%%%%%%%%%%%%%%%%%%%%%%%%%%%%%%%%%%%%%%%%%%%%%%%%%%%%%%%%%%%%%%%
%%%%%%%%%%%%%%%%%%%%%%%%%%%%%%%%%%%%%%%%%%%%%%%%%%%%%%%%%%%%%%%%%%%%%%%%%%%%%%%%%%%%%%%%%%%%%%%%%%%%%%%%%%%

%%%%%%%%%%%%%%%%%%%%%%%%%%%%%%%%%%%%%%%%%%%%%%%%%%%%%%%%%%%%%%%%%%%%%%%%%%%%%%%%%%%%%%%%%%%%%%%%%%%%%%%%%%%
%%%%%%%%%%%%%%%%%%%%%%%%%%%%%%%%%%%%%%%%%%%%%%%%%%%%%%%%%%%%%%%%%%%%%%%%%%%%%%%%%%%%%%%%%%%%%%%%%%%%%%%%%%%
%%%%%%%%%%%%%%%%%%%%%%%%%%%%%%%%%%%%%%%%%%%%%%%%%%%%%%%%%%%%%%%%%%%%%%%%%%%%%%%%%%%%%%%%%%%%%%%%%%%%%%%%%%%
%%%%%%%%%%%%%%%%%%%%%%%%%%%%%%%%%%%%%%%%%%%%%%%%%%%%%%%%%%%%%%%%%%%%%%%%%%%%%%%%%%%%%%%%%%%%%%%%%%%%%%%%%%%
%%%%%%%%%%%%%%%%%%%%%%%%%%%%%%%%%%%%%%%%%%%%%%%%%%%%%%%%%%%%%%%%%%%%%%%%%%%%%%%%%%%%%%%%%%%%%%%%%%%%%%%%%%%
\begin{theorem}\label{face-description} 
Let $\sigma$  be a bounded $2$-cell of $\vcor{}$, let $o$ be the body that the source of $\sigma$ reaches, let $o'$ be the body that the sink of $\sigma$ leaves, 
let $c$ be the curve of tangents to $o$ supporting the source of $\sigma$, and let $c'$ be the curve of tangents to $o'$ supporting the sink of $\sigma$.  
Then the right boundary chain of $\sigma$ is the concatenation 
of three (convex) chains 
$\rconeNew{\sigma}, \rctwoNew{\sigma}$ and  $\rcthrNew{\sigma}$ 
whose atoms $a$, except $\sour{\sigma}$ and  $\sink{\sigma}$, 
are characterized 
by $\phiback(a)$,
$\phileft(a)$ and $\phiforw(a) = \sink{\sigma}$, respectively. 
Furthermore
\begin{enumerate}
\item if $c$ is the curve of right tangents to the body $o$ then $\rconeNew{\sigma} = \sour{\sigma}$; 
otherwise 
$$\rconeNew{\sigma}=  \sour{\sigma}e_{10}v_{11}e_{11}\ldots e_{1k_1}, \qquad (k_1 \geq 0), $$
where $v_{11}v_{12}\ldots v_{1k_1}$ is the maximal sequence of consecutive $0$-cells leaving $o$  that 
follow $\sour{\sigma}$ on $c$;
\item if $c'$ is the curve of right tangents to the body $o'$ then $\rcthrNew{\sigma} = \sink{\sigma}$; otherwise 
$$\rcthrNew{\sigma}=  e_{3k_3}\ldots e_{31}v_{31}e_{30} \sink{\sigma}, \qquad (k_3 \geq 0), $$
where $v_{3k_3}\ldots v_{31}$ is the maximal sequence of consecutive $0$-cells reaching $o'$ that
precede $\sink{\sigma}$ on $c'$.

\item if $c=c'$ then  $c$ is the curve of right tangents to $o$ and $\rctwoNew{\sigma}$ is a $1$-cell 
whose source and sink are the  source and the sink of $\sigma$, respectively;
otherwise
$$\rctwoNew{\sigma}= e_{20}v_{21}e_{21}v_{22} \ldots v_{2k_2} e_{2k_2}, \qquad (k_2 \geq 1),$$ 
where 
\begin{enumerate}
\item $e_{20}$ is the empty chain if and only if $c$ is a curve of left tangents;
\item $e_{2k_2}$ is the empty chain if and only if  $c'$ is a curve of left tangents;
\item $v_{21}$ is the first $0$-cell reaching $o$ that follows
$\sour{\sigma}$ on $c$;
\item $v_{2k_2}$ is the first $0$-cell leaving $o'$ that
precedes $\sink{\sigma}$ on $c'$;
\item $v_{2,i+1} = \phiback(v_{2i})$ and the $e_{2i}$ are right $1$-cells. 
\end{enumerate}
\end{enumerate}
A similar result holds for the left boundary chain of $\sigma$  using conjugation under~$\reverse$.\qed 
%$\rcop \circ\ \reverse = \reverse \circ\ \lcop. $ \qed
\end{theorem}

%%%%%%%%%%%%%%%%%%%%%%%%%%%%%%%%%%%%%%%%%%%%%%%%%%%%%%%%%%%%%%%%%%%%%%%%%%%%%%%
%%%%%%%%%%%%%%%%%%%%%%%%%%%%%%%%%%%%%%%%%%%%%%%%%%%%%%%%%%%%%%%%%%%%%%%%%%%%%%%
%%%%%%%%%%%%%%%%%%%%%%%%%%%%%%%%%%%%%%%%%%%%%%%%%%%%%%%%%%%%%%%%%%%%%%%%%%%%%%%
%%%%%%%%%%%%%%%%%%%%%%%%%%%%%%%%%%%%%%%%%%%%%%%%%%%%%%%%%%%%%%%%%%%%%%%%%%%%%%%

\begin{example} The convex decompositions of the left  and right  boundary chains of the bounded $2$-cells of the visibility complex 
of two convex bodies of the plane are given in the following table 
$$\begin{array}{c|ccc||ccc}
\sigma & \rconeop(\sigma) & \rctwoop(\sigma) & \rcthrop(\sigma) & \lconeop(\sigma) & \lctwoop(\sigma) & \lcthrop(\sigma)  \\
\hline
ij & \onevc \deun & \thrvcp  & \unse \fouvcp & \onevc\unun & \twovc  & \dede \fouvcp \\ 
ji & \onevcp\unfi & \twovcp  & \desi \fouvc & \onevcp \defi & \thrvc  & \unth \fouvc \\ 
i\infty & \twovc & \dede \fouvcp  &\unhe \onevcp \unfi \twovcp\unsi \thrvcp & \twovc\unde \thrvc\unth \fouvc \unfo & \onevc\deun  &\thrvcp\\
\infty i& \twovcp\unsi \thrvcp\unse \fouvcp \unhe & \onevcp\defi  &\thrvc & \twovcp & \desi \fouvc  &\unfo \onevc \unun \twovc\unde \thrvc\\
j\infty &\thrvc & \unth \fouvc & \defo \onevc  \deun \thrvcp \dese  \twovcp &\thrvc \deth \twovc \dede \fouvcp  \dehe & \onevcp \unun & \twovcp\\
\infty j &\thrvcp \dese \twovcp \desi \fouvc  \defo & \onevc \unfi & \twovc &\thrvcp & \unse \fouvcp & \dehe \onevcp  \defi \thrvc \deth  \twovc\\
\infty\infty & \fouvcp & \dehe & \onevcp & \fouvcp & \unhe & \onevcp\\
\infty\infty & \fouvc & \unfo & \onevc & \fouvc & \defo & \onevc \\
\end{array}
$$
where we use the notations of Example~\ref{vctwobodies}.
\end{example}
%%%%%%%%%%%%%%%%%%%%%%%%%%%%%%%%%%%%%%%%%%%%%%%%%%%%%%%%%%%%%%%%%%%%%%%%%%%%%%%%%%%%%%%%%%%%%%%%%%%%%%%%%%%
%%%%%%%%%%%%%%%%%%%%%%%%%%%%%%%%%%%%%%%%%%%%%%%%%%%%%%%%%%%%%%%%%%%%%%%%%%%%%%%%%%%%%%%%%%%%%%%%%%%%%%%%%%%
%%%%%%%%%%%%%%%%%%%%%%%%%%%%%%%%%%%%%%%%%%%%%%%%%%%%%%%%%%%%%%%%%%%%%%%%%%%%%%%%%%%%%%%%%%%%%%%%%%%%%%%%%%%
%%%%%%%%%%%%%%%%%%%%%%%%%%%%%%%%%%%%%%%%%%%%%%%%%%%%%%%%%%%%%%%%%%%%%%%%%%%%%%%%%%%%%%%%%%%%%%%%%%%%%%%%%%%

%%%%%%%%%%%%%%%%%%%%%%%%%%%%%%%%%%%%%%%%%%%%%%%%%%%%%%%%%%%%%%%%%%%%%%%%%%%%%%%%%%%%%%%%%%%%%%%%%%%%%%%%%%%
%%%%%%%%%%%%%%%%%%%%%%%%%%%%%%%%%%%%%%%%%%%%%%%%%%%%%%%%%%%%%%%%%%%%%%%%%%%%%%%%%%%%%%%%%%%%%%%%%%%%%%%%%%%
%%%%%%%%%%%%%%%%%%%%%%%%%%%%%%%%%%%%%%%%%%%%%%%%%%%%%%%%%%%%%%%%%%%%%%%%%%%%%%%%%%%%%%%%%%%%%%%%%%%%%%%%%%%
%%%%%%%%%%%%%%%%%%%%%%%%%%%%%%%%%%%%%%%%%%%%%%%%%%%%%%%%%%%%%%%%%%%%%%%%%%%%%%%%%%%%%%%%%%%%%%%%%%%%%%%%%%%
\begin{example} 
Consider the family of $7$ convex bodies $o_1,o_2,\ldots, o_7$ of the real affine plane depicted in Figure~\ref{convdec} and let $\sigma$ 
be the $2$-cell of its visibility complex that contains the directed line  labeled $\sigma$.  Then, 
 using the notations  
$b_{ij}, b_{\overline{i}j}, b_{i\overline{j}}, b_{\overline{ij}}$ for the left-left, right-left, left-right, right-right bitangents joining 
$o_i$ to $o_j$,
%%%%%%%%%%%%%%%%%%%%%%%%%%%%%%%%%%%%%%%%%%%%%%%%%%%%%%%%%%%%%%%%%%%%%%%%%%%%%%%%%%%%%%%%%%%%%%%%%%%%%%%%%%%
%%%%%%%%%%%%%%%%%%%%%%%%%%%%%%%%%%%%%%%%%%%%%%%%%%%%%%%%%%%%%%%%%%%%%%%%%%%%%%%%%%%%%%%%%%%%%%%%%%%%%%%%%%%
%%%%%%%%%%%%%%%%%%%%%%%%%%%%%%%%%%%%%%%%%%%%%%%%%%%%%%%%%%%%%%%%%%%%%%%%%%%%%%%%%%%%%%%%%%%%%%%%%%%%%%%%%%%
%%%%%%%%%%%%%%%%%%%%%%%%%%%%%%%%%%%%%%%%%%%%%%%%%%%%%%%%%%%%%%%%%%%%%%%%%%%%%%%%%%%%%%%%%%%%%%%%%%%%%%%%%%%
\begin{figure}[!htb]
\begin{center}
\small
\psfrag{Lsi}{$\HLsink$} \psfrag{Rsi}{$\HRsink$} \psfrag{Lso}{$\HLsource$} \psfrag{Rso}{$\HRsource$}
\psfrag{deltaone}{$\HLsource(\sigma) = \HLsink(\sigma)$}
\psfrag{deltaonep}{$\HRsource(\sigma) = \HRsink(\sigma)$}
\psfrag{source}{source} \psfrag{sink}{sink}
\psfrag{vertex}{$\hourglassL{\sigma}$}\psfrag{vertexright}{$\hourglassR{\sigma}$}
\psfrag{sigma}{$\sigma$}
\psfrag{Done}{$\Delta_1$}
\psfrag{Gpone}{$\rc$}
\psfrag{Dpone}{$\Delta'_1$}
\psfrag{Gone}{$\lc$}
\psfrag{one}{$1$}
\psfrag{two}{$2$}
\psfrag{thr}{$3$}
\psfrag{fou}{$4$}
\psfrag{fiv}{$5$}
\psfrag{six}{$6$}
\psfrag{sev}{$7$}
%%\psfrag{HRso}{$\HRsource$} \psfrag{HRsi}{$\HRsink$} \psfrag{HLso}{$\HLsource$} \psfrag{HLsi}{$\HLsink$}
\includegraphics[width= 0.98575\linewidth]{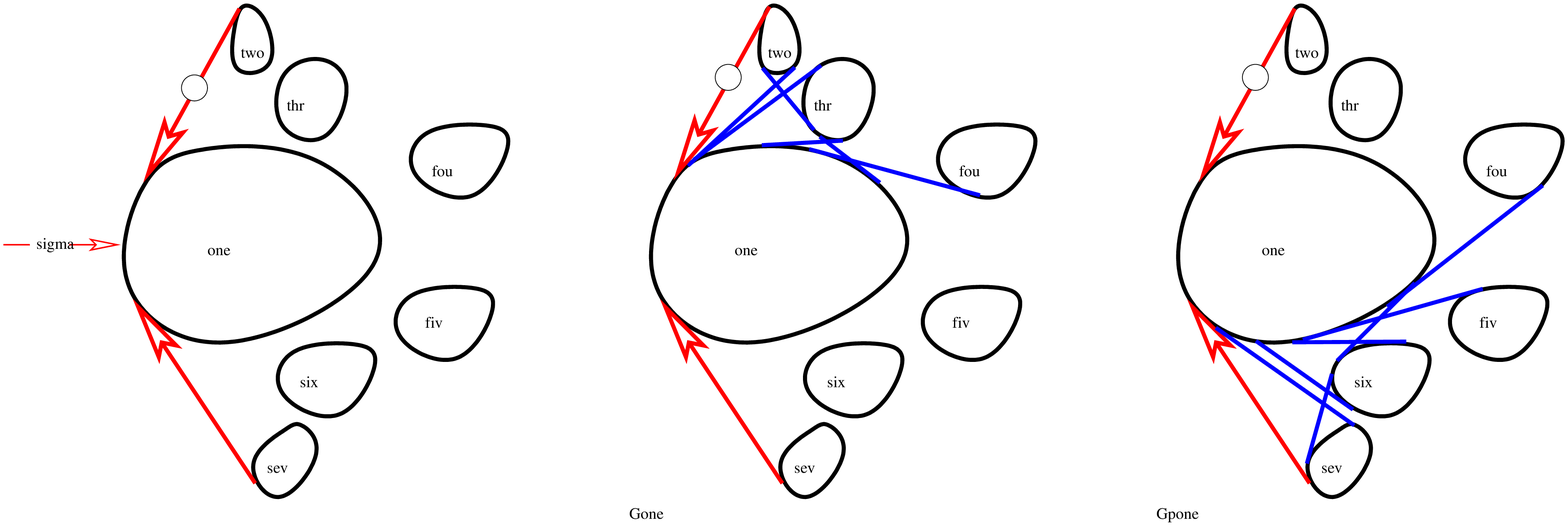}%%figures-xfig
\end{center}
\caption{\protect \small \label{convdec}}
\end{figure}
%%%%%%%%%%%%%%%%%%%%%%%%%%%%%%%%%%%%%%%%%%%%%%%%%%%%%%%%%%%%%%%%%%%%%%%%%%%%%%%%%%%%%%%%%%%%%%%%%%%%%%%%%%%
%%%%%%%%%%%%%%%%%%%%%%%%%%%%%%%%%%%%%%%%%%%%%%%%%%%%%%%%%%%%%%%%%%%%%%%%%%%%%%%%%%%%%%%%%%%%%%%%%%%%%%%%%%%
%%%%%%%%%%%%%%%%%%%%%%%%%%%%%%%%%%%%%%%%%%%%%%%%%%%%%%%%%%%%%%%%%%%%%%%%%%%%%%%%%%%%%%%%%%%%%%%%%%%%%%%%%%%
%%%%%%%%%%%%%%%%%%%%%%%%%%%%%%%%%%%%%%%%%%%%%%%%%%%%%%%%%%%%%%%%%%%%%%%%%%%%%%%%%%%%%%%%%%%%%%%%%%%%%%%%%%%
its source and sink are the bitangents  $\bg{2}{1}$ and $\bg{\overline{7}}{\overline{1}}$ and 
the convex decompositions of its left and right boundary chains are given in the following table 
$$\begin{array}{ccc||ccc}
\rconeop(\sigma) & \rctwoop(\sigma) & \rcthrop(\sigma) & \lconeop(\sigma) & \lctwoop(\sigma) & \lcthrop(\sigma)  \\
\hline
\bg{2}{1} \bg{1}{7} \bg{1}{7} \bg{1}{6} \bg{1}{\overline{6}} \bg{1}{\overline{5}}  \bg{1}{4}&  \bg{\overline{6}}{1} \bg{\overline{7}}{\overline{6}}& \bg{\overline{7}}{\overline{1}} & 
\bg{2}{1} & \bg{2}{3}\bg{3}{\overline{1}}  & \bg{\overline{1}}{4}
\bg{\overline{1}}{3}\bg{\overline{1}}{\overline{3}}\bg{\overline{1}}{2}\bg{\overline{1}}{\overline{2}}\bg{\overline{7}}{\overline{1}}   \\ 
\end{array}
$$
where we only indicate the bitangents of the chains.
\end{example}
%%%%%%%%%%%%%%%%%%%%%%%%%%%%%%%%%%%%%%%%%%%%%%%%%%%%%%%%%%%%%%%%%%%%%%%%%%%%%%%%%%%%%%%%%%%%%%%%%%%%%%%%%%%
%%%%%%%%%%%%%%%%%%%%%%%%%%%%%%%%%%%%%%%%%%%%%%%%%%%%%%%%%%%%%%%%%%%%%%%%%%%%%%%%%%%%%%%%%%%%%%%%%%%%%%%%%%%
%%%%%%%%%%%%%%%%%%%%%%%%%%%%%%%%%%%%%%%%%%%%%%%%%%%%%%%%%%%%%%%%%%%%%%%%%%%%%%%%%%%%%%%%%%%%%%%%%%%%%%%%%%%
%%%%%%%%%%%%%%%%%%%%%%%%%%%%%%%%%%%%%%%%%%%%%%%%%%%%%%%%%%%%%%%%%%%%%%%%%%%%%%%%%%%%%%%%%%%%%%%%%%%%%%%%%%%

%\clearpage
\subsubsection{Greedy pseudotriangulations}
Let $\lineset$ be the set of directed lines of $\pointset$.
Let $\MapLight{}{\wc{}}{\vc{}}$ be the inverse image
of a universal cover $\universal{\lineset}$ of $\lineset$
under the natural projection $\MapLight{\linecoo}{\vc{}}{\lineset}$---that is, $\wc{}$ is the set of pairs $(v,l) \in \vc{}\times \universal{\lineset}$ such that
the image of $v$ under $\MapLight{\linecoo}{\vc{}}{\lineset}$ coincides with the image of $l$
under $\MapLight{}{\universal{\lineset}}{\lineset}$, and $\MapLight{}{\wc{}}{\vc{}}$ is the first projection, cf~\cite[pages 113-114]{g-eta-71}---let $\Vposet$ be the set of
 cells of $\wc{}$ endowed with the partial order generated by the relations
\begin{equation}
\sour{\sigma} \prec \sigma \prec \sink{\sigma}
\end{equation}
 where $\sigma$ ranges over the
set  of $1$- and bounded $2$-cells of $\Vposet$ and where $\sour{\sigma}$ and $\sink{\sigma}$ stand, respectively, for the  source and the sink of~$\sigma$.
The sets of left and right
unbounded $2$-cells of $\wc{}$ are denoted $\setbotface$ and $\settopface$, respectively;  note that the elements of $\setbotface$ and $\settopface$ are isolated elements 
in $\Vposet$ and that the sizes of $\setbotface$ and $\settopface$ are both equal to twice the number of sheets of the branched covering space~$\CoSur{}$.
Finally we denote by  $\shiftop$ the generator of the (infinite cyclic) automorphism group of the covering $\wc{} \rightarrow \vc{}/\reverse$ defined by the condition
that $\sigma \prec \shiftop(\sigma)$, the shift operator for short, and  we keep the same symbol to denote a horizon operator 
and its lift in $\wc{}$; thus 
$\phisink$ is the  map that assigns to a vertex $\vertex$ of $\Vposet$ the sink of the $2$-cell of $\Vposet$ whose source is~$\vertex$. 
Two vertices of $\Vposet$ are said {\it crossing} if their corresponding bitangent line segments are crossing.

%%%%%%%%%%%%%%%%%%%%%%%%%%%%%%%%%%%%%%%%%%%%%%%%%%%%%%%%%%%%%%%%%%%%%%%%%%%%%%%%%%%%%%%%%%%%%%%%%%%%%%%%%%%
%%%%%%%%%%%%%%%%%%%%%%%%%%%%%%%%%%%%%%%%%%%%%%%%%%%%%%%%%%%%%%%%%%%%%%%%%%%%%%%%%%%%%%%%%%%%%%%%%%%%%%%%%%%
%%%%%%%%%%%%%%%%%%%%%%%%%%%%%%%%%%%%%%%%%%%%%%%%%%%%%%%%%%%%%%%%%%%%%%%%%%%%%%%%%%%%%%%%%%%%%%%%%%%%%%%%%%%
%%%%%%%%%%%%%%%%%%%%%%%%%%%%%%%%%%%%%%%%%%%%%%%%%%%%%%%%%%%%%%%%%%%%%%%%%%%%%%%%%%%%%%%%%%%%%%%%%%%%%%%%%%%
%%%%%%%%%%%%%%%%%%%%%%%%%%%%%%%%%%%%%%%%%%%%%%%%%%%%%%%%%%%%%%%%%%%%%%%%%%%%%%%%%%%%%%%%%%%%%%%%%%%%%%%%%%%
\begin{theorem}[{\cite[Theorem 5]{ap-sstvc-03} and \cite[Lemma 8]{G-pv-tsvcp-96}}]\label{mpb}
%\begin{theorem}\label{mpb}
Two crossing vertices  are comparable with respect to the  partial order 
$\prec$ and the map $\mphi : \Vposet^0 \rightarrow \Vposet^0$ that associates with 
$\vertex \in \Vposet^0$ the  minimum element  of the set of $\vertexbis \in \Vposet^0$ 
such that $\vertexbis$ crosses~$\vertex$ and  $\vertex \prec \vertexbis$ is 
well-defined, one-to-one and onto.  
Furthermore 
if $\vertex$ is an interior vertex then  
$\mphi(\vertex) = \phisink(\vertex)$; otherwise 
%$\mphi(\vertex)$ is the sink of the $2$-cell with source $\vertex$; otherwise 
$\mphi(\vertex)= \shiftop(\vertex)$. \qed
%%%(and $\phisink(\vertex) = \reverse \circ \successor(v)$). {\sc what is $ \successor$?}\qed
%$\mphi(\vertex) = \phisink(\vertex)$; otherwise 
%$\mphi(\vertex)$ is the image of $\vertex$ under the shift operator or the sink of the $2$-cell whose source is $\vertex$   
%depending on whether~$\vertex$ is a hull vertex or not. \qed
\end{theorem}
%%%%%%%%%%%%%%%%%%%%%%%%%%%%%%%%%%%%%%%%%%%%%%%%%%%%%%%%%%%%%%%%%%%%%%%%%%%%%%%%%%%%%%%%%%%%%%%%%%%%%%%%%%%
%%%%%%%%%%%%%%%%%%%%%%%%%%%%%%%%%%%%%%%%%%%%%%%%%%%%%%%%%%%%%%%%%%%%%%%%%%%%%%%%%%%%%%%%%%%%%%%%%%%%%%%%%%%
%%%%%%%%%%%%%%%%%%%%%%%%%%%%%%%%%%%%%%%%%%%%%%%%%%%%%%%%%%%%%%%%%%%%%%%%%%%%%%%%%%%%%%%%%%%%%%%%%%%%%%%%%%%
%%%%%%%%%%%%%%%%%%%%%%%%%%%%%%%%%%%%%%%%%%%%%%%%%%%%%%%%%%%%%%%%%%%%%%%%%%%%%%%%%%%%%%%%%%%%%%%%%%%%%%%%%%%
%%%%%%%%%%%%%%%%%%%%%%%%%%%%%%%%%%%%%%%%%%%%%%%%%%%%%%%%%%%%%%%%%%%%%%%%%%%%%%%%%%%%%%%%%%%%%%%%%%%%%%%%%%%
Let $\MAC$ be a maximal antichain of $\Vposet$, let $\MAC^+$ be the filter of cells $z \in \Vposet$ such that $x\preceq z$ for some $x \in \MAC$,
 and let  
\begin{equation}
\G(\MAC)  = \bigcup_{i\geq 1} B_i(\MAC)
\end{equation}
where 
%$B_1(\MAC)$ is the set of minimal elements of $\MAC$, and $B_{i+1}(\MAC)$ 
$B_{i}(\MAC)$ 
is the set of minimal elements of the set of vertices of $\vfilter{\MAC}$ that do not cross any element of $\bigcup_{1}^{i-1} B_j(\MAC)$ where as usual $\bigcup_{1}^{0} B_j(\MAC)=\emptyset$.
We denote by $\touchop$ the operator that assigns to a vertex of $\Vposet$ its corresponding bitangent line segment. 

%%%%%%%%%%%%%%%%%%%%%%%%%%%%%%%%%%%%%%%%%%%%%%%%%%%%%%%%%%%%%%%%%%%%%%%%%%%%%%%%%%%%%%%%%%%%%%%%%%%%%%%%%%%
%%%%%%%%%%%%%%%%%%%%%%%%%%%%%%%%%%%%%%%%%%%%%%%%%%%%%%%%%%%%%%%%%%%%%%%%%%%%%%%%%%%%%%%%%%%%%%%%%%%%%%%%%%%
%%%%%%%%%%%%%%%%%%%%%%%%%%%%%%%%%%%%%%%%%%%%%%%%%%%%%%%%%%%%%%%%%%%%%%%%%%%%%%%%%%%%%%%%%%%%%%%%%%%%%%%%%%%
%%%%%%%%%%%%%%%%%%%%%%%%%%%%%%%%%%%%%%%%%%%%%%%%%%%%%%%%%%%%%%%%%%%%%%%%%%%%%%%%%%%%%%%%%%%%%%%%%%%%%%%%%%%
\begin{theorem}[{\cite[Theorem 5]{ap-sstvc-03} and \cite[Theorem 12]{G-pv-tsvcp-96}}]\label{mpbbis}
%\begin{theorem}\label{mpbbis}
Let $\MAC$ be a maximal antichain of $\Vposet.$
Then $\touchop \circ \G(\MAC)$ is a well-defined  pseudotriangulation and 
%\item $\G(\MAC) = \MAC\setminus \mphi(\MAC) = \sink{\facesG(\MAC)}$ where $\facesG(\MAC)$ is the set of bounded $2$-cells of $\widehat{\MAC}$ minus 
\begin{equation}
\G(\MAC) = \vfilter{\MAC}_0\setminus \mphi(\vfilter{\MAC}_0) = \vertices(\MAC) + \sink{\facesG(\MAC)}
\end{equation}
where $\vfilter{\MAC}_0$ is the set of vertices of $\vfilter{\MAC}$, $\vertices(\MAC)$ is the set of $0$-cells of $\MAC$, and  $\facesG(\MAC)$ is the set of bounded 
$2$-cells of $\MAC$ minus the $\sigma_{\Sforw}$ and  $\sigma'_{\Sback}$  where $\sigma$ ranges over the set of right-right boundary $0$-cells and 
right boundary $1$-cells of $\MAC$ and where $\sigma'$ ranges over the set of left-left boundary $0$-cells  and left boundary $1$-cells of~$\MAC$. \qed \end{theorem}
%%%%%%%%%%%%%%%%%%%%%%%%%%%%%%%%%%%%%%%%%%%%%%%%%%%%%%%%%%%%%%%%%%%%%%%%%%%%%%%%%%%%%%%%%%%%%%%%%%%%%%%%%%%
%%%%%%%%%%%%%%%%%%%%%%%%%%%%%%%%%%%%%%%%%%%%%%%%%%%%%%%%%%%%%%%%%%%%%%%%%%%%%%%%%%%%%%%%%%%%%%%%%%%%%%%%%%%
%%%%%%%%%%%%%%%%%%%%%%%%%%%%%%%%%%%%%%%%%%%%%%%%%%%%%%%%%%%%%%%%%%%%%%%%%%%%%%%%%%%%%%%%%%%%%%%%%%%%%%%%%%%
%%%%%%%%%%%%%%%%%%%%%%%%%%%%%%%%%%%%%%%%%%%%%%%%%%%%%%%%%%%%%%%%%%%%%%%%%%%%%%%%%%%%%%%%%%%%%%%%%%%%%%%%%%%
%%%%%%%%%%%%%%%%%%%%%%%%%%%%%%%%%%%%%%%%%%%%%%%%%%%%%%%%%%%%%%%%%%%%%%%%%%%%%%%%%%%%%%%%%%%%%%%%%%%%%%%%%%%
The pseudotriangulation $\G(\MAC)$ is called the {\it greedy pseudotriangulation  at $\MAC$}.

%%%%%%%%%%%%%%%%%%%%%%%%%%%%%%%%%%%%%%%%%%%%%%%%%%%%%%%%%%%%%%%%%%%%%%%%%%%%%%%%%%%%%%%%%%%%%%%%%%%%%%%%%%%
%%%%%%%%%%%%%%%%%%%%%%%%%%%%%%%%%%%%%%%%%%%%%%%%%%%%%%%%%%%%%%%%%%%%%%%%%%%%%%%%%%%%%%%%%%%%%%%%%%%%%%%%%%%
%%%%%%%%%%%%%%%%%%%%%%%%%%%%%%%%%%%%%%%%%%%%%%%%%%%%%%%%%%%%%%%%%%%%%%%%%%%%%%%%%%%%%%%%%%%%%%%%%%%%%%%%%%%
%%%%%%%%%%%%%%%%%%%%%%%%%%%%%%%%%%%%%%%%%%%%%%%%%%%%%%%%%%%%%%%%%%%%%%%%%%%%%%%%%%%%%%%%%%%%%%%%%%%%%%%%%%%

We describe, again in preparation for the section on \ABpts, the boundary chains of the pseudotriangles 
of the greedy pseudotriangulations in terms of the horizon operators.
%%%%%%%%%%%%%%%%%%%%%%%%%%%%%%%%%%%%%%%%%%%%%%%%%%%%%%%%%%%%%%%%%%%%%%%%%%%%%%%%%%%%%%%%%%%%%%%%%%%%%%%%%%%
%%%%%%%%%%%%%%%%%%%%%%%%%%%%%%%%%%%%%%%%%%%%%%%%%%%%%%%%%%%%%%%%%%%%%%%%%%%%%%%%%%%%%%%%%%%%%%%%%%%%%%%%%%%
%%%%%%%%%%%%%%%%%%%%%%%%%%%%%%%%%%%%%%%%%%%%%%%%%%%%%%%%%%%%%%%%%%%%%%%%%%%%%%%%%%%%%%%%%%%%%%%%%%%%%%%%%%%
%%%%%%%%%%%%%%%%%%%%%%%%%%%%%%%%%%%%%%%%%%%%%%%%%%%%%%%%%%%%%%%%%%%%%%%%%%%%%%%%%%%%%%%%%%%%%%%%%%%%%%%%%%%
%%%%%%%%%%%%%%%%%%%%%%%%%%%%%%%%%%%%%%%%%%%%%%%%%%%%%%%%%%%%%%%%%%%%%%%%%%%%%%%%%%%%%%%%%%%%%%%%%%%%%%%%%%%
Let $\MAC$ be a  maximal antichain of $\Vposet$.  
Let $\vertex$ be a minimal element of the subposet of vertices of $\vfilter{\MAC}$ that is not a left-left boundary bitangent and let $\R(\vertex)$ be the
pseudotriangle 
of the pseudotriangulation $\touchop\circ \G(\MAC)$ lying locally 
on the right side of the bitangent line segment~$\bitangente{\vertex}.$ 
One can easily show that the pseudotriangle $\R(\vertex)$ is independent of the choice of the maximal antichain $\MAC$.
Walking in counterclockwise order along  the boundary of $\R(\vertex)$ starting 
at the tail of $\bitangente{\vertex}$ we traverse successively 4 convex chains $\R^j(\vertex)$ ($j=1,2,3,4$). 
A description of these chains in terms of horizon operators is given in the following theorem
 where  $\bitmphi_{\xxx}$ denotes the conjugate of the operator $\mphi_{\xxx}$ under $\touchop.$  

%%%%%%%%%%%%%%%%%%%%%%%%%%%%%%%%%%%%%%%%%%%%%%%%%%%%%%%%%%%%%%%%%%%%%%%%%%%%%%%%%%%%%%%%%%%%%%%%%%%%%%%%%%%
%%%%%%%%%%%%%%%%%%%%%%%%%%%%%%%%%%%%%%%%%%%%%%%%%%%%%%%%%%%%%%%%%%%%%%%%%%%%%%%%%%%%%%%%%%%%%%%%%%%%%%%%%%%
%%%%%%%%%%%%%%%%%%%%%%%%%%%%%%%%%%%%%%%%%%%%%%%%%%%%%%%%%%%%%%%%%%%%%%%%%%%%%%%%%%%%%%%%%%%%%%%%%%%%%%%%%%%
%%%%%%%%%%%%%%%%%%%%%%%%%%%%%%%%%%%%%%%%%%%%%%%%%%%%%%%%%%%%%%%%%%%%%%%%%%%%%%%%%%%%%%%%%%%%%%%%%%%%%%%%%%%
%%%%%%%%%%%%%%%%%%%%%%%%%%%%%%%%%%%%%%%%%%%%%%%%%%%%%%%%%%%%%%%%%%%%%%%%%%%%%%%%%%%%%%%%%%%%%%%%%%%%%%%%%%%
\begin{theorem}[{\cite[Theorem~10]{ap-sstvc-03}}]
\label{orbit-theorem}
Let $\vertex$ be a  bitangent that is not a left-left boundary bitangent and let 
$$
\R^j(\vertex) = e_{j0}\vertex_{j1}e_{j1},
\ldots, \vertex_{jk_j}e_{jk_j}, \qquad k_j \geq 0$$
where
$\vertex_{jk}$ stands for a bitangent line segment and $e_{jk}$ for an arc. 
Then  
\begin{enumerate}
\item $\vertex_{11} = \bitangente{\vertex}$ and $\vertex_{1,j+1} = \bitphiforw(\vertex_{1j})$;
\item $\vertex_{31} = \bitphiback(\vertex_{11})$ and $\vertex_{3,i+1} = \bitphiforw(\vertex_{3i})$
(assuming that $\vertex_{31}$ is well-defined); 
\item $\vertex_{21} = \bitphirigh(\vertex_{11})$ and $\vertex_{2,i+1} = \bitphiback(\vertex_{2i})$
(assuming that $\vertex_{21}$ is well-defined);
\item $\bitmphi(\vertex)$ leaves an arc of $\Rtwo(\vertex)$ or the first arc of $\Rthr(\vertex)$. \qed
\end{enumerate}
%%and that $e_{20}\vertex_{21}=\lcthr(\sigmarigh(\vertex)).$ 
\end{theorem}

%%%%%%%%%%%%%%%%%%%%%%%%%%%%%%%%%%%%%%%%%%%%%%%%%%%%%%%%%%%%%%%%%%%%%%%%%%%%%%%%%%%%%%%%%%%%%%%%%%%%%%%%%%%
%%%%%%%%%%%%%%%%%%%%%%%%%%%%%%%%%%%%%%%%%%%%%%%%%%%%%%%%%%%%%%%%%%%%%%%%%%%%%%%%%%%%%%%%%%%%%%%%%%%%%%%%%%%
%%%%%%%%%%%%%%%%%%%%%%%%%%%%%%%%%%%%%%%%%%%%%%%%%%%%%%%%%%%%%%%%%%%%%%%%%%%%%%%%%%%%%%%%%%%%%%%%%%%%%%%%%%%
%%%%%%%%%%%%%%%%%%%%%%%%%%%%%%%%%%%%%%%%%%%%%%%%%%%%%%%%%%%%%%%%%%%%%%%%%%%%%%%%%%%%%%%%%%%%%%%%%%%%%%%%%%%
%%%%%%%%%%%%%%%%%%%%%%%%%%%%%%%%%%%%%%%%%%%%%%%%%%%%%%%%%%%%%%%%%%%%%%%%%%%%%%%%%%%%%%%%%%%%%%%%%%%%%%%%%%%

\subsubsection{Cross-sections}
Let $\MAC$ be a maximal antichain of $\Vposet$ and let $\vertices(\MAC)$, $\edges(\MAC)$, $\faces(\MAC)$, and $\ufaces(\MAC)$ be its sets of $0$-, $1$-, bounded and unbounded $2$-cells.
Using the simple fact that a maximal antichain and a maximal chain intersect in a single element  one can easily check that  
\begin{enumerate}
\item $\ufaces(\MAC)$ is the whole set of unbounded $ 2$-cells and its size is $2\nbsheets_{\disks}$.
\item for any bounded $2$-cell $\sigma$ of $\MAC$ there is exactly 
one atom (a $0$- or $1$-cell)  of its right/left boundary chain---denoted $\bbotonecell{\sigma}{\MAC}$/$\btoponecell{\sigma}{\MAC}$ thereafter---that belongs to $J$;
\item for any left/right unbounded $2$-cell $\sigma$ of $\MAC$ there is exactly 
one atom (a $0$- or $1$-cell) of its right/left boundary chain---denoted $\bbotonecell{\sigma}{\MAC}$/$\btoponecell{\sigma}{\MAC}$ thereafter---that belongs to $J$;
\item the $2$-cells of $\MAC$ are exactly the 
$\sigma_{\Srigh},\sigma_{\Sforw}, \sigma_{\Sleft}$, and $\sigma_{\Sback}$,  where $\sigma$ ranges over $ \vertices(\MAC)+ \edges(\MAC)$ and 
where by convention we ignore $\sigma_{\Srigh}$ or (exclusive) $\sigma_{\Sleft}$ if one of them is
not defined, that is, if $\sigma$ is a right  $1$-cell or a left $1$-cell;
\item the size of $\edges(\MAC)$ is  $2 \size_{\disks} - 2 \#\vertices(\MAC)$; 
\item the size of $\faces(\MAC)$ is $3\size_{\disks} - \nbsheets_{\disks} - \# \vertices(\MAC)$.
\end{enumerate}

The {\it cross-section} of the visibility complex of the family of convex bodies $\disks$ at the maximal antichain $\MAC$,  denoted $\CS{\MAC}$,
 is the directed multigraph whose
set of nodes is the set of $0$- and $1$-cells of $\MAC$ and whose
set of arcs is the set of $2$-cells $\sigma$ of $\MAC$ directed from $\bbotonecell{\sigma}{\MAC}$ to $\btoponecell{\sigma}{\MAC}$. 
We use the notation $\fdt{i}{j}(\MAC)$, $i, j \in \{1,2,3\}$, for the set of $\sigma \in \faces(\MAC)$ such that
$\bbotonecell{\sigma}{\MAC} \in \rc_i(\sigma)$, $\btoponecell{\sigma}{\MAC} \in \lc_j(\sigma)$, and both $\bbotonecell{\sigma}{\MAC}$ and $\btoponecell{\sigma}{\MAC}$ 
are $1$-cells; the pair $ij$ is called the {\it type} of the arc $\sigma$; 
%%%%%%%%%%%%%%%%%%%%%%%%%%%%%%%%%%%%%%%%%%%%%%%%%%%%%%%%%%%%%%%%%%%%%%%%%%%%%%%
%%%%%%%%%%%%%%%%%%%%%%%%%%%%%%%%%%%%%%%%%%%%%%%%%%%%%%%%%%%%%%%%%%%%%%%%%%%%%%%
%%%%%%%%%%%%%%%%%%%%%%%%%%%%%%%%%%%%%%%%%%%%%%%%%%%%%%%%%%%%%%%%%%%%%%%%%%%%%%%
\begin{figure}[!htb]
\begin{center}
\footnotesize
\psfrag{sigma}{$\sigma$}
\psfrag{sigmaplus}{$\sigma^+$}
\psfrag{sigmaminus}{$\sigma^-$}
\psfrag{ursigma}{$\ur{\sigma}$}
\psfrag{ulsigma}{$\ul{\sigma}$}
\psfrag{drsigma}{$\dr{\sigma}$}
\psfrag{dlsigma}{$\dl{\sigma}$}
\psfrag{zero}{$\fdtthrthr$}
\psfrag{deux}{$\fdttwotwo$}
\psfrag{four}{$\fdtoneone$}
\psfrag{ureasy}{$\fdtthrtwo$}
\psfrag{urhard}{$\fdtthrone$}
\psfrag{ul}{$\fdtonetwo$}
\psfrag{dr}{$\fdttwoone$}
\psfrag{dleasy}{$\fdttwothr$}
\psfrag{dlhard}{$\fdtonethr$}
\includegraphics[width=0.75\linewidth]{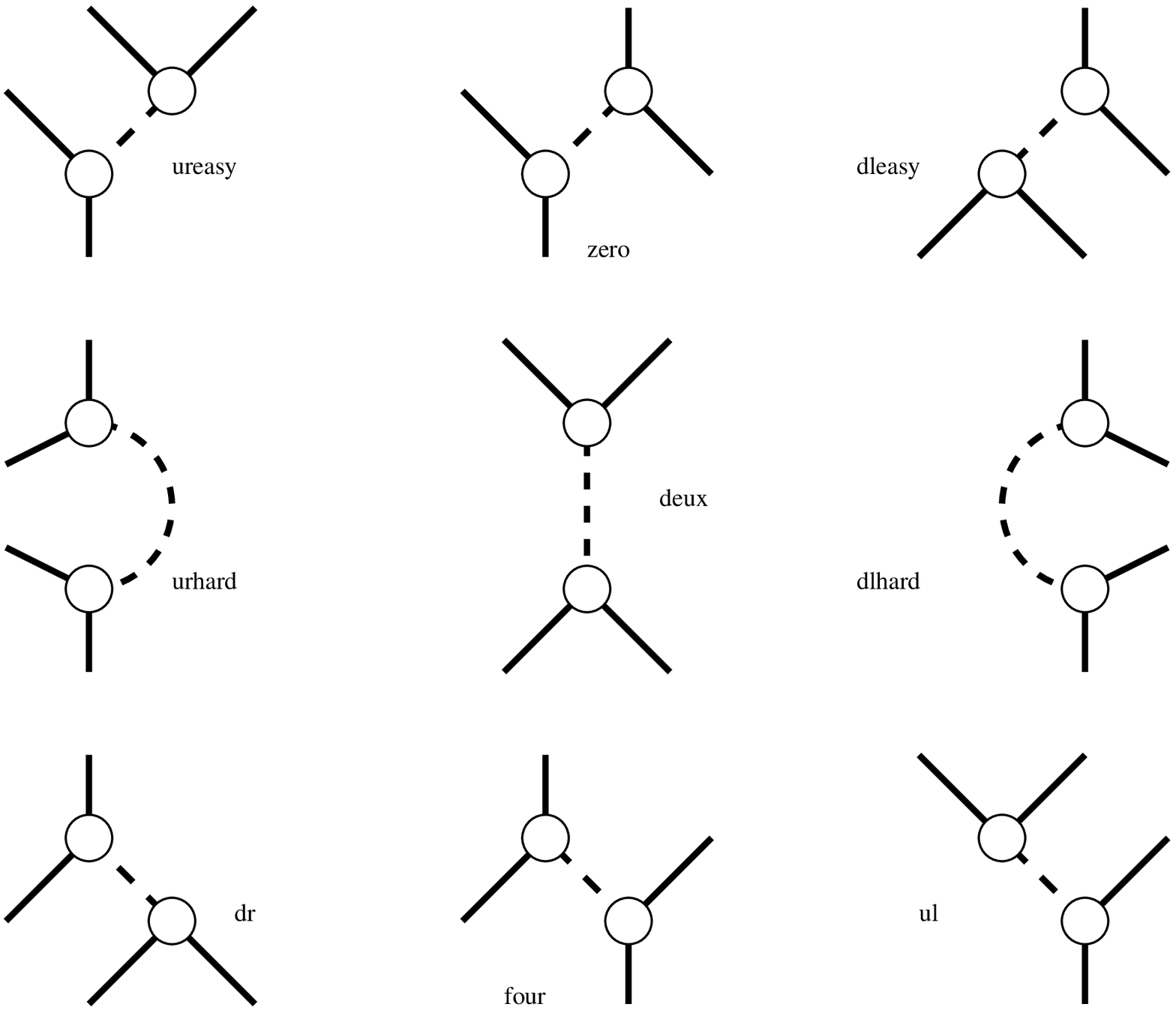}%% ./xfigfinal
\end{center}
\caption{\protect \small \label{incidences}}
\end{figure}
%%%%%%%%%%%%%%%%%%%%%%%%%%%%%%%%%%%%%%%%%%%%%%%%%%%%%%%%%%%%%%%%%%%%%%%%%%%%%%%
%%%%%%%%%%%%%%%%%%%%%%%%%%%%%%%%%%%%%%%%%%%%%%%%%%%%%%%%%%%%%%%%%%%%%%%%%%%%%%%
%%%%%%%%%%%%%%%%%%%%%%%%%%%%%%%%%%%%%%%%%%%%%%%%%%%%%%%%%%%%%%%%%%%%%%%%%%%%%%%
%%%%%%%%%%%%%%%%%%%%%%%%%%%%%%%%%%%%%%%%%%%%%%%%%%%%%%%%%%%%%%%%%%%%%%%%%%%%%%%
the type of an arc captures exactly the upward embedding in the plane of this arc together with its adjacent arcs with the property that the arcs incident to a node $\sigma$ appear in circular order 
$\sigma_{\Srigh},\sigma_{\Sforw}, \sigma_{\Sleft}$, and $\sigma_{\Sback}$ where by convention we ignore $\sigma_{\Srigh}$ or (exclusive) $\sigma_{\Sleft}$ if one of them is 
not defined, that is, if $\sigma$ is a right  $1$-cell or a left $1$-cell, as illustrated in Figure~\ref{incidences}.
We make the set of cross-sections into a poset ${\cal A}(\Vposet)$ by defining
$\CS{\MAC} \preceq \CS{\MAC'}$ in  ${\cal A}(\Vposet)$ by
$\MAC^+ \supseteq \MAC'^+$
where $\MAC^+$ is the filter of cells $z \in \Vposet$ such that $x\preceq z$ for some $x \in \MAC$.
The covering relations in ${\cal A}(\Vposet)$ are described in the following theorem, from which it follows by induction, starting from the
obviously acyclic cross-sections of Examples~\ref{defccsangle} and~\ref{defccs}, that cross-sections are acyclic.

\begin{figure}[!htb]
\small
\psfrag{1}{1} \psfrag{2}{2} \psfrag{3}{3} \psfrag{4}{4}
\psfrag{sigma}{$\sigma$} \psfrag{sigmaplus}{$\sigma^+$} \psfrag{sigmaminus}{$\sigma^-$} \psfrag{ursigma}{$\ur{\sigma}$}
\psfrag{ulsigma}{$\ul{\sigma}$} \psfrag{drsigma}{$\dr{\sigma}$} \psfrag{dlsigma}{$\dl{\sigma}$} \psfrag{zero}{$\sigma \in \fdtthrthr(\MAC)$}
\psfrag{deuxbis}{$\sigma' \in \fdttwotwo(\MAC')$} 
\psfrag{deux}{$\sigma \in \fdttwotwo(\MAC)$} 
\psfrag{four}{$\sigma'\in \fdtoneone(\MAC')$} \psfrag{ureasy}{$\sigma \in \fdtthrtwo(\MAC)$} \psfrag{urhard}{$\fdtthrone$}
\psfrag{ul}{$\sigma'\in \fdtonetwo(\MAC')$} \psfrag{dr}{$\sigma'\in \fdttwoone(\MAC')$} \psfrag{dleasy}{$\sigma \in\fdttwothr(\MAC)$} \psfrag{dlhard}{$\fdtonethr$}
\psfrag{rr}{$\vertices_{rr}$}
\psfrag{ll}{$\vertices_{ll}$}
\psfrag{lr}{$\vertices_{lr}$}
\psfrag{rl}{$\vertices_{rl}$}
\psfrag{rr}{$v$ is right-right}
\psfrag{ll}{$v$ is left-left}
\psfrag{lr}{$v$ is left-right}
\psfrag{rl}{$v$ is right-left}
\psfrag{firstcase}{first case}
\psfrag{secondcase}{second case}
\centering
\includegraphics[width=0.8575\linewidth]{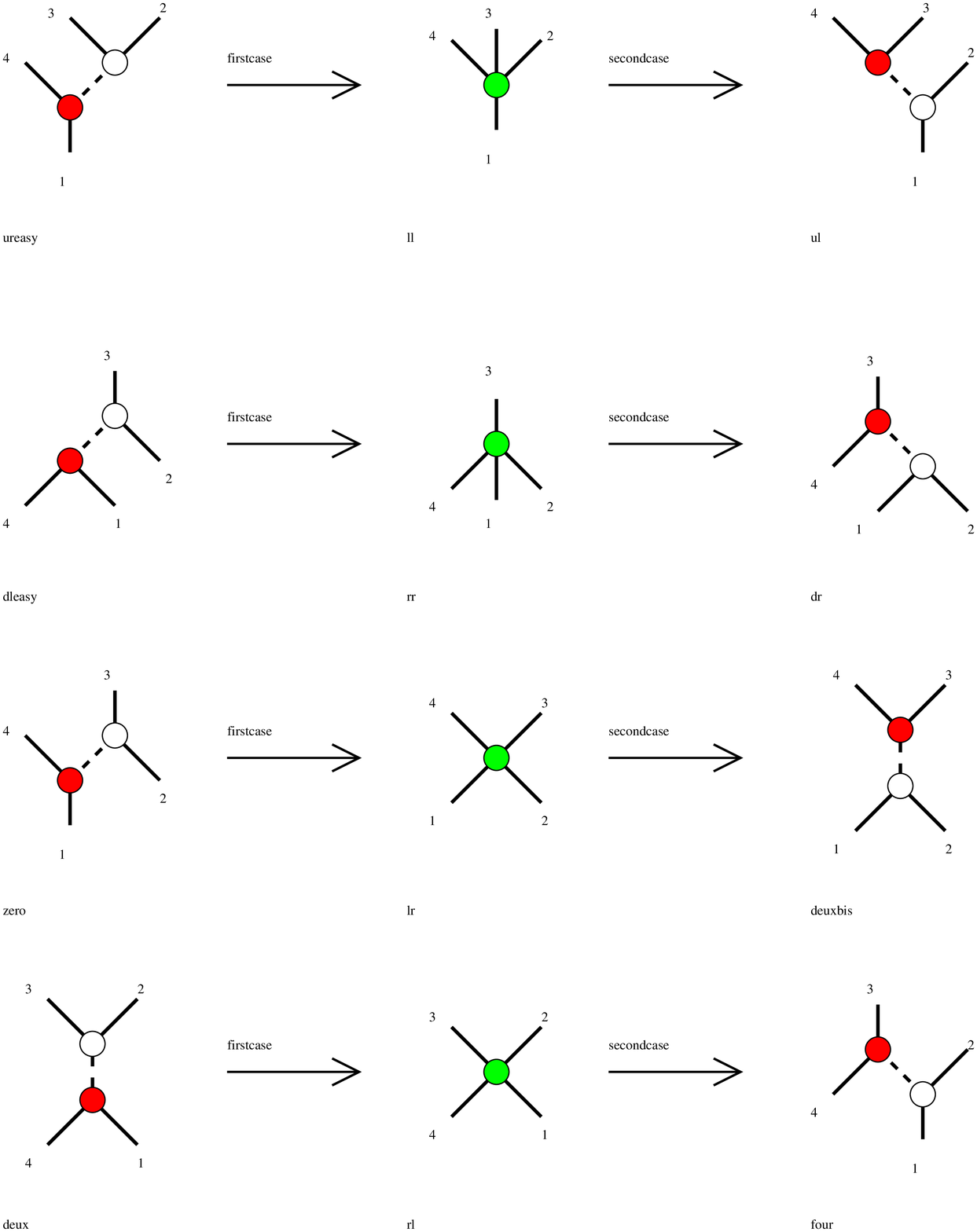}%% xfigfinal
\caption{\protect \small 
\label{localchange}}
\end{figure}

%%%%%%%%%%%%%%%%%%%%%%%%%%%%%%%%%%%%%%%%%%%%%%%%%%%%%%%%%%%%%%%%%%%%%%%%%%%%%%%
%%%%%%%%%%%%%%%%%%%%%%%%%%%%%%%%%%%%%%%%%%%%%%%%%%%%%%%%%%%%%%%%%%%%%%%%%%%%%%%
%%%%%%%%%%%%%%%%%%%%%%%%%%%%%%%%%%%%%%%%%%%%%%%%%%%%%%%%%%%%%%%%%%%%%%%%%%%%%%%
\begin{theorem}
\label{min-criteria}
Let $\MAC$ be a maximal antichain of $\Vposet$, let $v \in \Vposet^0$, $e,e',f,f' \in \Vposet^1$, $\sigma,\sigma' \in \Vposet^2$
with $v = \sink{\sigma}= \sink{e} = \sink{e'} = \sour{\sigma'} = \sour{f} = \sour{f'}$.
Then $v$ is minimal in the subposet of vertices of $\vfilter{\MAC}$ if and only if either (first case) $\sigma, e,e' \in \MAC$ 
or (second case) $v \in \MAC.$ Furthermore  in the first case 
$\MAC' = \MAC - \{\sigma,e,e'\} +v $ is a maximal antichain and $\CS{\MAC'}$ covers $\CS{\MAC}$; 
in the second case
$\MAC' = \MAC - v +  \{f,f',\sigma'\}$ is a maximal antichain and $\CS{\MAC'}$ covers $\CS{\MAC}$; 
and in both case
$\CS{\MAC'}$ is obtained from $\CS{\MAC}$ by local changes as indicated in Figure~\ref{localchange} where the arcs numbered $1,2,3,4$ 
stand for the four $2$-cells $v_{\Srigh},v_{\Sforw}, v_{\Sleft}$, and $v_{\Sback}$ incident to vertex $v$ and where the arcs are oriented upward.\qed
\end{theorem}
%%%%%%%%%%%%%%%%%%%%%%%%%%%%%%%%%%%%%%%%%%%%%%%%%%%%%%%%%%%%%%%%%%%%%%%%%%%%%%%
%%%%%%%%%%%%%%%%%%%%%%%%%%%%%%%%%%%%%%%%%%%%%%%%%%%%%%%%%%%%%%%%%%%%%%%%%%%%%%%
%%%%%%%%%%%%%%%%%%%%%%%%%%%%%%%%%%%%%%%%%%%%%%%%%%%%%%%%%%%%%%%%%%%%%%%%%%%%%%%
%%%%%%%%%%%%%%%%%%%%%%%%%%%%%%%%%%%%%%%%%%%%%%%%%%%%%%%%%%%%%%%%%%%%%%%%%%%%%%%%%%%%%%%%%%%%%%%%%%%%%%%%%%%
%%%%%%%%%%%%%%%%%%%%%%%%%%%%%%%%%%%%%%%%%%%%%%%%%%%%%%%%%%%%%%%%%%%%%%%%%%%%%%%%%%%%%%%%%%%%%%%%%%%%%%%%%%%
%%%%%%%%%%%%%%%%%%%%%%%%%%%%%%%%%%%%%%%%%%%%%%%%%%%%%%%%%%%%%%%%%%%%%%%%%%%%%%%%%%%%%%%%%%%%%%%%%%%%%%%%%%%
%\begin{figure}[!htb]
%%%%%%%%%%%%%%%%%%%%%%%%%%%%%%%%%%%%%%%%%%%%%%%%%%%%%%%%%%%%%%%%%%%%%%%%%%%%%%%
%%%%%%%%%%%%%%%%%%%%%%%%%%%%%%%%%%%%%%%%%%%%%%%%%%%%%%%%%%%%%%%%%%%%%%%%%%%%%%%
%%%%%%%%%%%%%%%%%%%%%%%%%%%%%%%%%%%%%%%%%%%%%%%%%%%%%%%%%%%%%%%%%%%%%%%%%%%%%%%
\begin{theorem}
\label{acyclic}
Let $\MAC$ be a maximal antichain of $\Vposet$.  
Then $\CS{\MAC}$ is acyclic (with set of sources $\setbotface$ and set of sinks $\settopface$) 
and can be embedded in free space 
in such way that the arcs incident to a node $\sigma$ appear in circular order 
$\sigma_{\Srigh},\sigma_{\Sforw}, \sigma_{\Sleft}$, and $\sigma_{\Sback}$ where by convention we ignore $\sigma_{\Srigh}$ or (exclusive) $\sigma_{\Sleft}$ if one of them is 
not defined, that is, if $\sigma$ is a right  $1$-cell or a left $1$-cell.  \qed
\end{theorem}
%%%%%%%%%%%%%%%%%%%%%%%%%%%%%%%%%%%%%%%%%%%%%%%%%%%%%%%%%%%%%%%%%%%%%%%%%%%%%%%
%%%%%%%%%%%%%%%%%%%%%%%%%%%%%%%%%%%%%%%%%%%%%%%%%%%%%%%%%%%%%%%%%%%%%%%%%%%%%%%
%%%%%%%%%%%%%%%%%%%%%%%%%%%%%%%%%%%%%%%%%%%%%%%%%%%%%%%%%%%%%%%%%%%%%%%%%%%%%%%

\begin{sidewaysfigure}[!htb]
%\begin{figure}[!htb]
\centering
\psfrag{Gamma}{$\Gamma$}
\psfrag{iGamma}{$\shiftop(\Gamma)$}
\includegraphics[width = 0.75\linewidth]{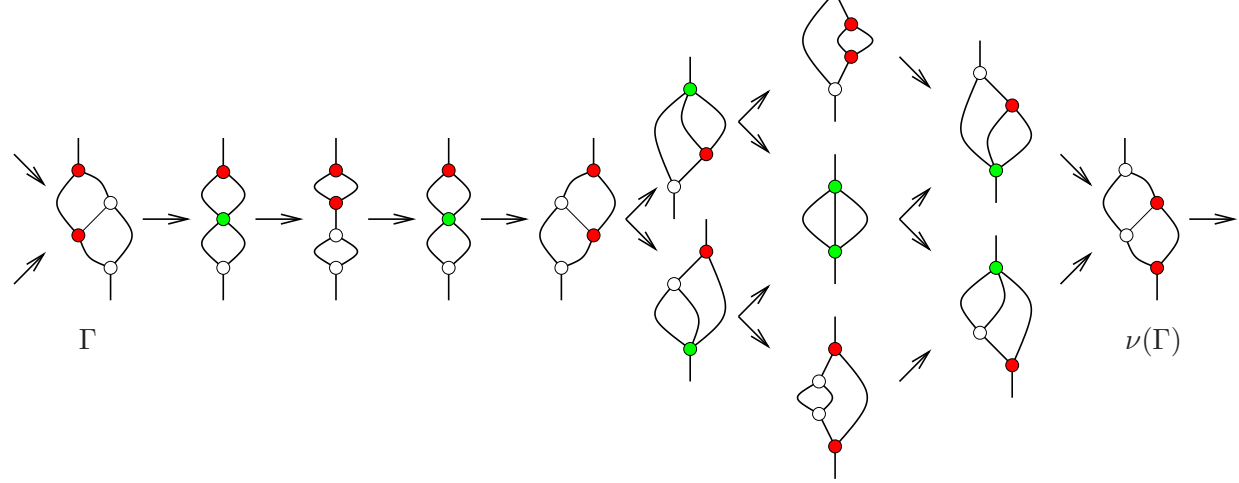}
\caption{\protect \small
The Hasse diagram of the poset of cross-sections of the visibility complex of a family of two bodies: the arcs are oriented upward and the ones incident to a node
$\sigma$ are arranged in the circular order $v_{\Srigh}, v_{\Sforw}, v_{\Sleft}, v_{\Sback}, \ldots$;
the nodes with degree~3 are $1$-cells; the nodes with degree~$4$  are $0$-cells, that is, bitangents;
the arcs incident to a unique node are the unbounded $2$-cells.
\label{vctwoconvexbodies}}
\end{sidewaysfigure}

\begin{example}
The Hasse diagram of the poset of cross-sections of the visibility complex of
a family of two convex bodies is depicted in Figure~\ref{vctwoconvexbodies}: the diagram is of course invariant under the shift operator and its quotient modulo the shift operator
is composed of twelve cross-sections.
\end{example}
%%%%%%%%%%%%%%%%%%%%%%%%%%%%%%%%%%%%%%%%%%%%%%%%%%%%%%%%%%%%%%%%%%%%%%%%%%%%%%%
%%%%%%%%%%%%%%%%%%%%%%%%%%%%%%%%%%%%%%%%%%%%%%%%%%%%%%%%%%%%%%%%%%%%%%%%%%%%%%%
%%%%%%%%%%%%%%%%%%%%%%%%%%%%%%%%%%%%%%%%%%%%%%%%%%%%%%%%%%%%%%%%%%%%%%%%%%%%%%%
%%%%%%%%%%%%%%%%%%%%%%%%%%%%%%%%%%%%%%%%%%%%%%%%%%%%%%%%%%%%%%%%%%%%%%%%%%%%%%%
%\end{figure}
%%%%%%%%%%%%%%%%%%%%%%%%%%%%%%%%%%%%%%%%%%%%%%%%%%%%%%%%%%%%%%%%%%%%%%%%%%%%%%%
%%%%%%%%%%%%%%%%%%%%%%%%%%%%%%%%%%%%%%%%%%%%%%%%%%%%%%%%%%%%%%%%%%%%%%%%%%%%%%%
%%%%%%%%%%%%%%%%%%%%%%%%%%%%%%%%%%%%%%%%%%%%%%%%%%%%%%%%%%%%%%%%%%%%%%%%%%%%%%%
%%%%%%%%%%%%%%%%%%%%%%%%%%%%%%%%%%%%%%%%%%%%%%%%%%%%%%%%%%%%%%%%%%%%%%%%%%%%%%%

\clearpage

%%%%%%%%%%%%%%%%%%%%%%%%%%%%%%%%%%%%%%%%%%%%%%%%%%%%%%%%%%%%%%%%%%%%%%%%%%%%%%%%%%%%%%%%%%%%%%%%%%%%%%%%%%%
%%%%%%%%%%%%%%%%%%%%%%%%%%%%%%%%%%%%%%%%%%%%%%%%%%%%%%%%%%%%%%%%%%%%%%%%%%%%%%%%%%%%%%%%%%%%%%%%%%%%%%%%%%%
%%%%%%%%%%%%%%%%%%%%%%%%%%%%%%%%%%%%%%%%%%%%%%%%%%%%%%%%%%%%%%%%%%%%%%%%%%%%%%%%%%%%%%%%%%%%%%%%%%%%%%%%%%%
%%%%%%%%%%%%%%%%%%%%%%%%%%%%%%%%%%%%%%%%%%%%%%%%%%%%%%%%%%%%%%%%%%%%%%%%%%%%%%%%%%%%%%%%%%%%%%%%%%%%%%%%%%%
\subsection{\ABpts} \label{defabpts}
The definition of the \ABpts\ is given in terms of the horizon operators $\mphi_{\xxx}$ and their {\it dual horizon operators} $\mphi_{\xxx_*}$ which are defined in exactly the same way except 
that we replace in the definition of the horizon operators the sink operator by the source operator, i.e., $\mphi_{\xxx_*}(\sigma)$ is the source of the $2$-cell $\sigma_{\xxx}$; 
note that $\mphi_{\Ssink} \mphi_{\xxx_*} = \mphi_{\xxx}$.
More precisely we are going to use the derived horizon operators $\pforwop$ and $\pbackop$ (and their duals)  which are defined as follows: 
\begin{enumerate} 
\item $\pforwop$ is the operator  that assigns to a bitangent line segment $\vertex$ its image under $\bitphiforw$ if
$\vertex$ is a right-left or left-left bitangent line segment whose image
under $\bitphiforw$ is a  not a boundary bitangent line segment;
the bitangent line segment $\vertex$ otherwise; and 
\item $\pbackop$ is the operator  that assigns to a bitangent line segment $\vertex$ its image under $\bitphiback$ if
$\vertex$ is a right-right or left-right bitangent line segment whose image
under $\bitphiback$ is a  not a boundary  bitangent line segment;
the bitangent line segment $\vertex$ otherwise; note that $\pbackop$ and $\pforwop$  are conjugate under the reorientation operator $\reverse.$
\end{enumerate}
%Thus $\pforwop$ carries a  hull bitangent line segment on itself and  a non hull bitangent line segment on a non hull bitangent line segment.
As a simple consequence of the description of the greedy pseudotriangulations in
terms of the horizon operators given in Theorem~\ref{orbit-theorem} we see that greedy
pseudotriangulations are stable under the operators $\pforwop$ and $\pbackop$.
\subsubsection{\Hpqs} 
For $\sigma$ a bounded $2$-cell we set   
$\hourglassR{\sigma} = \touchop(u)$ if $\rctwoNew{\sigma}$ is reduced 
to a $0$-cell $u$; otherwise we choose a $1$-cell $e$ of $\rctwoNew{\sigma}$, we introduce the 
sequence $u_0 = \touchop\phiback(e), u_{i+1} = \pbackop(u_i)$, the  sequence 
$v_0 = \touchop\phiforwstar(e),v_{i+1} = \pforwstarop(v_i)$,
and we set 
\begin{equation}
\label{equone}
\hourglassR{\sigma} = \hourglassRminus{\sigma,e} \hourglassRplus{\sigma, e} 
\end{equation}
where  
$\hourglassRplus{\sigma, e}$ is, depending on whether $u_0$ is a  boundary bitangent line segment or not, 
the empty sequence or the sequence of $u_i$ truncated just after the first index $i$ such that $u_{i} = u_{i+1}$,
and where, similarly, 
$\hourglassRminus{\sigma, e}$ is, depending on whether $v_0$ is a boundary bitangent line segment or not,   
the empty sequence or the reversal of the sequence of $v_i$ truncated just after the first index $i$ such that $v_i = v_{i+1}$.
The bitangent line segments corresponding to the vertices of $\rctwoNew{\sigma}$ are consecutive elements 
of the sequence $\hourglassR{\sigma}$, from which it follows that 
$\hourglassR{\sigma}$ is independent of the choice of the $1$-cell $e$.
It is convenient to extend the definitions of $\hourglassRminus{\sigma,.}$ and $\hourglassRplus{\sigma,.}$
 to
the whole set of  $1$-cells of $\rcNew{\sigma}$ as follows 
$$\hourglassRminus{\sigma, e} =  \begin{cases} 
\hourglassR{\sigma} & \text{if $e$ is a $1$-cell of $\rcthrNew{\sigma}$}; \\
\epsilon & \text{if $e$ is a $1$-cell of $\rconeNew{\sigma}$};
\end{cases}
$$
%similarly 
$$\hourglassRplus{\sigma, e} =  \begin{cases} 
\hourglassR{\sigma} & \text{if $e$ is a $1$-cell of $\rconeNew{\sigma}$}; \\
\epsilon & \text{if $e$ is a $1$-cell of $\rcthrNew{\sigma}$}; 
\end{cases}
$$
where $\epsilon$ is the empty chain, so that equation (\ref{equone}) holds  
for any $1$-cell $e$ of $\rc(\sigma)$. 
Let $u_0, u_1, \ldots$ be the sequence of bitangent line segments defined by $u_0$ is the source bitangent line segment of $\sigma$ and $u_{i+1} = \pforwstarop(u_i)$; 
similarly, let $v_0,v_1,\ldots$ be the sequence of bitangent line segments defined by $v_0$ is the sink bitangent line segment of $\sigma$ and $v_{i+1} = \pbackop(v_i)$;
clearly by construction one has
\begin{equation}
\hourglassR{\sigma} =  
\hourglassRone{\sigma}   
\hourglassRtwo{\sigma}   
\hourglassRthr{\sigma}  
\end{equation}
where
(1) $\hourglassRone{\sigma}$ is the empty sequence if $u_1$ is a boundary bitangent line segment or if $u_0=u_1$;  
the  reversal of the sequence $u_1, u_2, \ldots$, truncated just after the first index $i$ such that $u_{i} = u_{i+1}$, otherwise; 
(2) $\hourglassRtwo{\sigma}$ is the
sequence of bitangent line segments corresponding to the sequence of vertices of the chain $\rctwoNew{\sigma}$;
and (3) 
$\hourglassRthr{\sigma}$ is the empty sequence if $v_1$ is a boundary  bitangent line segment or if $v_0=v_1$; the sequence 
$v_1, v_2, \ldots$, truncated just after  the first index such that $v_{i} = v_{i+1}$,
 otherwise.  
We consider $\ophourglassR$ as an operator on the set of bounded $2$-cells and we define $\ophourglassL$ to be its conjugate under $\reverse.$

\begin{example}
The table of the operators $\ophourglassR^i$ on the $2$-cells of the visibility complex 
of two convex bodies of the plane is the following 
$$\begin{array}{c|ccc||ccc}
\sigma & \hourglassRone{\sigma} & \hourglassRtwo{\sigma} & \hourglassRthr{\sigma} & \hourglassLone{\sigma} & \hourglassLtwo{\sigma} & \hourglassLthr{\sigma}  \\
\hline
ij & \epsilon & \thrvcp  & \epsilon  & \epsilon & \twovc  & \epsilon \\ 
ji & \epsilon & \twovcp  & \epsilon & \epsilon & \thrvc  & \epsilon \\ 
i\infty & \epsilon & \fouvcp  &\epsilon & \epsilon & \onevc  & \epsilon \\
\infty i& \epsilon & \onevcp  &\epsilon & \epsilon  &  \fouvc  &\epsilon \\
j\infty &\epsilon & \fouvc & \epsilon &\epsilon  & \onevcp  & \epsilon \\
\infty j &\epsilon & \onevc & \epsilon &\epsilon  & \fouvcp & \epsilon \\
\infty\infty & \epsilon & \epsilon& \epsilon & \epsilon & \epsilon & \epsilon\\
\infty\infty & \epsilon & \epsilon& \epsilon & \epsilon & \epsilon & \epsilon \\
\end{array}
$$
where we use the notations of Figure~\ref{DirectedVCtwobodies} and where $\epsilon$ stands for the empty sequence. 
\end{example}
\begin{example}\label{hourglassoneexample}
Consider the family of $7$ convex bodies $o_1,o_2, o_3,\ldots, o_7$ of the real affine plane depicted in the 
left part of Figure~\ref{hourglassone} and let $\sigma$ be the $2$-cell of its visibility complex 
that contains the directed line segment labeled  $\sigma$. 
%%%%%%%%%%%%%%%%%%%%%%%%%%%%%%%%%%%%%%%%%%%%%%%%%%%%%%%%%%%%%%%%%%%%%%%%%%%%%%%
%%%%%%%%%%%%%%%%%%%%%%%%%%%%%%%%%%%%%%%%%%%%%%%%%%%%%%%%%%%%%%%%%%%%%%%%%%%%%%%
%%%%%%%%%%%%%%%%%%%%%%%%%%%%%%%%%%%%%%%%%%%%%%%%%%%%%%%%%%%%%%%%%%%%%%%%%%%%%%%
%%%%%%%%%%%%%%%%%%%%%%%%%%%%%%%%%%%%%%%%%%%%%%%%%%%%%%%%%%%%%%%%%%%%%%%%%%%%%%%
\begin{figure}[!htb]
\begin{center}
\small
\psfrag{sg}{$\sigma$}
\psfrag{un}{$1$}
\psfrag{de}{$2$}
\psfrag{tr}{$3$}
\psfrag{qu}{$4$}
\psfrag{ci}{$5$}
\psfrag{si}{$6$}
\psfrag{se}{$7$}
\psfrag{ze}{$6$}
\psfrag{ma}{$7$}
\psfrag{Lsi}{$\HLsink$} \psfrag{Rsi}{$\HRsink$} \psfrag{Lso}{$\HLsource$} \psfrag{Rso}{$\HRsource$}
\psfrag{deltaone}{$\HLsource(\sigma) = \HLsink(\sigma)$}
\psfrag{deltaonep}{$\HRsource(\sigma) = \HRsink(\sigma)$}
\psfrag{source}{source} \psfrag{sink}{sink}
\psfrag{vertex}{$\hourglassL{\sigma}$}\psfrag{vertexright}{$\hourglassR{\sigma}$}
\psfrag{sigma}{$\sigma$}
%%\psfrag{HRso}{$\HRsource$} \psfrag{HRsi}{$\HRsink$} \psfrag{HLso}{$\HLsource$} \psfrag{HLsi}{$\HLsink$}
\includegraphics[width= 0.8575\linewidth]{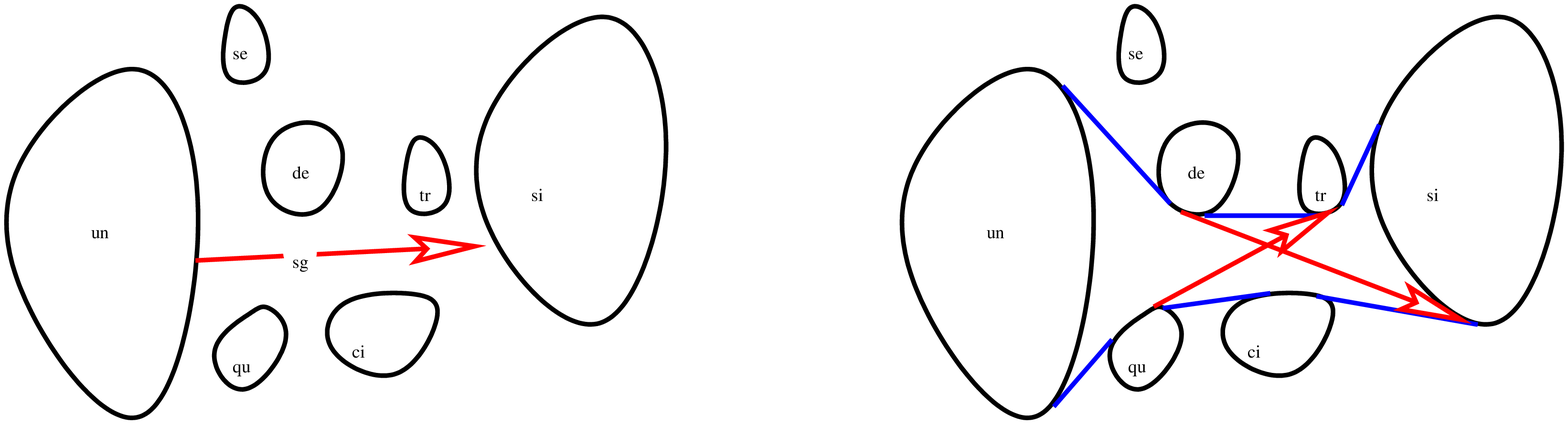}%%figures-xfig
\end{center}
\caption{ \label{hourglassone}}
\end{figure}
%%%%%%%%%%%%%%%%%%%%%%%%%%%%%%%%%%%%%%%%%%%%%%%%%%%%%%%%%%%%%%%%%%%%%%%%%%%%%%%%%%%%%%%%%%%%%%%%%%%%%%%%%%%
%%%%%%%%%%%%%%%%%%%%%%%%%%%%%%%%%%%%%%%%%%%%%%%%%%%%%%%%%%%%%%%%%%%%%%%%%%%%%%%%%%%%%%%%%%%%%%%%%%%%%%%%%%%
%%%%%%%%%%%%%%%%%%%%%%%%%%%%%%%%%%%%%%%%%%%%%%%%%%%%%%%%%%%%%%%%%%%%%%%%%%%%%%%%%%%%%%%%%%%%%%%%%%%%%%%%%%%
%%%%%%%%%%%%%%%%%%%%%%%%%%%%%%%%%%%%%%%%%%%%%%%%%%%%%%%%%%%%%%%%%%%%%%%%%%%%%%%%%%%%%%%%%%%%%%%%%%%%%%%%%%%
Then, 
using the notations  
$b_{ij}, b_{\overline{i}j}, b_{i\overline{j}}, b_{\overline{ij}}$ for the left-left, right-left, left-right, right-right bitangent line segments joining 
$o_i$ to $o_j$,
 its source and sink  bitangent line segments are the line segments $b_{26}$ and $b_{\overline{4}3}$ 
and the $\ophourglassR^i(\sigma)$ and $\ophourglassL^i(\sigma)$ are given in the following table 
$$\begin{array}{ccc||ccc}
\hourglassRone{\sigma} & \hourglassRtwo{\sigma} & \hourglassRthr{\sigma} & \hourglassLone{\sigma} & \hourglassLtwo{\sigma} & \hourglassLthr{\sigma}  \\
\hline 
\epsilon & b_{\overline{5}6}b_{\overline{45}} & b_{1\overline{4}} & b_{\overline{1}2} & b_{23} & b_{3\overline{4}} \\ 
\end{array}
$$
Observe, as illustrated in the right part of the figure, that the bitangent line segments of $\hourglassR{\sigma}$ and $\hourglassL{\sigma}$ are the bitangent line segments of 
a pseudo-quadrangle with diagonals the source and sink bitangent line segments of $\sigma$. 
\end{example}
%%%%%%%%%%%%%%%%%%%%%%%%%%%%%%%%%%%%%%%%%%%%%%%%%%%%%%%%%%%%%%%%%%%%%%%%%%%%%%%%%%%%%%%%%%%%%%%%%%%%%%%%%%%
%%%%%%%%%%%%%%%%%%%%%%%%%%%%%%%%%%%%%%%%%%%%%%%%%%%%%%%%%%%%%%%%%%%%%%%%%%%%%%%%%%%%%%%%%%%%%%%%%%%%%%%%%%%
%%%%%%%%%%%%%%%%%%%%%%%%%%%%%%%%%%%%%%%%%%%%%%%%%%%%%%%%%%%%%%%%%%%%%%%%%%%%%%%%%%%%%%%%%%%%%%%%%%%%%%%%%%%
%%%%%%%%%%%%%%%%%%%%%%%%%%%%%%%%%%%%%%%%%%%%%%%%%%%%%%%%%%%%%%%%%%%%%%%%%%%%%%%%%%%%%%%%%%%%%%%%%%%%%%%%%%%
\begin{example} \label{hourglasstwoexample}
Consider the family of $7$ convex bodies $o_1,o_2,\ldots, o_7$ of the real affine plane depicted in Figure~\ref{hourglasstwo} and let $\sigma$ 
be the $2$-cell of its visibility complex that contains the directed line segment labeled $\sigma$. 
Then 
%using the notations  
%$b_{ij}, b_{\overline{i}j}, b_{i\overline{j}}, b_{\overline{ij}}$ for the left-left, right-left, left-right, right-right bitangent line segments joining 
%$o_i$ to $o_j$,
%%%%%%%%%%%%%%%%%%%%%%%%%%%%%%%%%%%%%%%%%%%%%%%%%%%%%%%%%%%%%%%%%%%%%%%%%%%%%%%%%%%%%%%%%%%%%%%%%%%%%%%%%%%
%%%%%%%%%%%%%%%%%%%%%%%%%%%%%%%%%%%%%%%%%%%%%%%%%%%%%%%%%%%%%%%%%%%%%%%%%%%%%%%%%%%%%%%%%%%%%%%%%%%%%%%%%%%
%%%%%%%%%%%%%%%%%%%%%%%%%%%%%%%%%%%%%%%%%%%%%%%%%%%%%%%%%%%%%%%%%%%%%%%%%%%%%%%%%%%%%%%%%%%%%%%%%%%%%%%%%%%
%%%%%%%%%%%%%%%%%%%%%%%%%%%%%%%%%%%%%%%%%%%%%%%%%%%%%%%%%%%%%%%%%%%%%%%%%%%%%%%%%%%%%%%%%%%%%%%%%%%%%%%%%%%
\begin{figure}[!htb]
\begin{center}
\small
\psfrag{Lsi}{$\HLsink$} \psfrag{Rsi}{$\HRsink$} \psfrag{Lso}{$\HLsource$} \psfrag{Rso}{$\HRsource$}
\psfrag{deltaone}{$\HLsource(\sigma) = \HLsink(\sigma)$}
\psfrag{deltaonep}{$\HRsource(\sigma) = \HRsink(\sigma)$}
\psfrag{source}{source} \psfrag{sink}{sink}
\psfrag{vertex}{$\hourglassL{\sigma}$}\psfrag{vertexright}{$\hourglassR{\sigma}$}
\psfrag{sigma}{$\sigma$}
\psfrag{Done}{$\Delta_1$}
\psfrag{Gpone}{$\ophourglassR$}
\psfrag{Dpone}{$\Delta'_1$}
\psfrag{Gone}{$\ophourglassL$}
\psfrag{one}{$1$}
\psfrag{two}{$2$}
\psfrag{thr}{$3$}
\psfrag{fou}{$4$}
\psfrag{fiv}{$5$}
\psfrag{six}{$6$}
\psfrag{sev}{$7$}
%%\psfrag{HRso}{$\HRsource$} \psfrag{HRsi}{$\HRsink$} \psfrag{HLso}{$\HLsource$} \psfrag{HLsi}{$\HLsink$}
\includegraphics[width= 0.98575\linewidth]{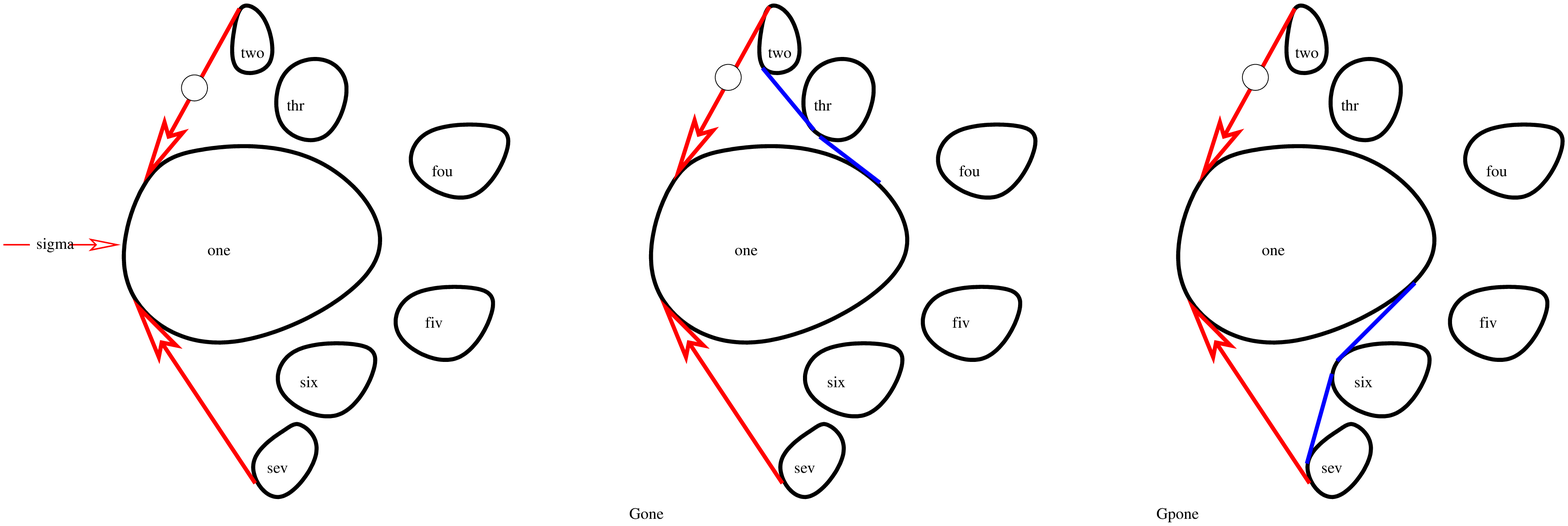}%%figures-xfig
\end{center}
\caption{\protect \small \label{hourglasstwo}}
\end{figure}
%%%%%%%%%%%%%%%%%%%%%%%%%%%%%%%%%%%%%%%%%%%%%%%%%%%%%%%%%%%%%%%%%%%%%%%%%%%%%%%%%%%%%%%%%%%%%%%%%%%%%%%%%%%
%%%%%%%%%%%%%%%%%%%%%%%%%%%%%%%%%%%%%%%%%%%%%%%%%%%%%%%%%%%%%%%%%%%%%%%%%%%%%%%%%%%%%%%%%%%%%%%%%%%%%%%%%%%
%%%%%%%%%%%%%%%%%%%%%%%%%%%%%%%%%%%%%%%%%%%%%%%%%%%%%%%%%%%%%%%%%%%%%%%%%%%%%%%%%%%%%%%%%%%%%%%%%%%%%%%%%%%
%%%%%%%%%%%%%%%%%%%%%%%%%%%%%%%%%%%%%%%%%%%%%%%%%%%%%%%%%%%%%%%%%%%%%%%%%%%%%%%%%%%%%%%%%%%%%%%%%%%%%%%%%%%
its source and sink bitangent line segments are the boundary bitangent line segments  $\bg{2}{1}$ and $\bg{\overline{7}}{\overline{1}}$ 
%and the convex decompositions of its left and right boundary chains are given in the following table 
and the $\ophourglassR^i(\sigma)$ and $\ophourglassL^i(\sigma)$ are given in the following table
$$\begin{array}{ccc||ccc}
\hourglassRone{\sigma} & \hourglassRtwo{\sigma} & \hourglassRthr{\sigma} & \hourglassLone{\sigma} & \hourglassLtwo{\sigma} & \hourglassLthr{\sigma}  \\
\hline
\epsilon &  \bg{\overline{6}}{1} \bg{\overline{7}}{\overline{6}}& \epsilon & 
\epsilon & \bg{2}{3}\bg{3}{\overline{1}}  & \epsilon.
\\
\end{array}
$$
Observe that the bitangent line segments of $\hourglassR{\sigma}/\hourglassL{\sigma}$ augmented with the sink/source of $\sigma$ bound a pseudotriangle. 
\end{example}
%%%%%%%%%%%%%%%%%%%%%%%%%%%%%%%%%%%%%%%%%%%%%%%%%%%%%%%%%%%%%%%%%%%%%%%%%%%%%%%%%%%%%%%%%%%%%%%%%%%%%%%%%%%
%%%%%%%%%%%%%%%%%%%%%%%%%%%%%%%%%%%%%%%%%%%%%%%%%%%%%%%%%%%%%%%%%%%%%%%%%%%%%%%%%%%%%%%%%%%%%%%%%%%%%%%%%%%
%%%%%%%%%%%%%%%%%%%%%%%%%%%%%%%%%%%%%%%%%%%%%%%%%%%%%%%%%%%%%%%%%%%%%%%%%%%%%%%%%%%%%%%%%%%%%%%%%%%%%%%%%%%
%%%%%%%%%%%%%%%%%%%%%%%%%%%%%%%%%%%%%%%%%%%%%%%%%%%%%%%%%%%%%%%%%%%%%%%%%%%%%%%%%%%%%%%%%%%%%%%%%%%%%%%%%%%

%\clearpage
%To prove this in full generality we 
%%%%%%%%%%%%%%%%%%%%%%%%%%%%%%%%%%%%%%%%%%%%%%%%%%%%%%%%%%%%%%%%%%%%%%%%%%%%%%%
%%%%%%%%%%%%%%%%%%%%%%%%%%%%%%%%%%%%%%%%%%%%%%%%%%%%%%%%%%%%%%%%%%%%%%%%%%%%%%%
%%%%%%%%%%%%%%%%%%%%%%%%%%%%%%%%%%%%%%%%%%%%%%%%%%%%%%%%%%%%%%%%%%%%%%%%%%%%%%%
%%%%%%%%%%%%%%%%%%%%%%%%%%%%%%%%%%%%%%%%%%%%%%%%%%%%%%%%%%%%%%%%%%%%%%%%%%%%%%%
We now reinterpret the sequence $\hourglassR{\sigma}$ and $\hourglassL{\sigma}$ through the greedy pseudotriangulations.
 Let $\sigma$ be a bounded $2$-cell, let $e$ be the last $1$-cell of $\lconeNew{\sigma}$ if any, 
the first $1$-cell of $\lctwoNew{\sigma}$ otherwise, 
let $a$ be the initial point of $\touchop\circ\ \sink{e}$ or $\touchop\circ\ \sour{e}$ depending on whether $e$ is the last $1$-cell of $\lconeNew{\sigma}$ 
or the first $1$-cell of $\lctwoNew{\sigma}$,
and let  $\MAC$ be the maximal antichain whose $1$-, and bounded $2$-cells are the cells whose sinks are in the principal filter of the sink of $e$ but not their sources.
 We define inductively a finite sequence of 
pseudotriangulations $G_1, G_2,\ldots, G_p$, 
and a finite sequence of pseudotriangles $\Delta_1, \Delta_2,\ldots, \Delta_p$, $a\in \partial \Delta_i$, as follows
\begin{enumerate}
\item $G_1 = \touchop \circ \G(\MAC)$ and $\Delta_1$ is the pseudotriangle of $G_1$ lying locally to the left of the
sink bitangent line segment of $\sigma$ if any; otherwise (that is, if the sink bitangent line segment of $\sigma$ is a
right-right boundary  bitangent line segment as in the example of Figure~\ref{hourglasstwo}) we define $\Delta_1$ to be the pseudotriangle lying
locally to the right of the image under $\reverse$ of the source bitangent line segment of~$\sigma$; 
walking in counterclockwise order along the boundary of a pseudotriangle $\Delta_i$, $a \in \partial \Delta_i$, 
starting from the point $a$  we traverse successively four convex chains $\Delta^1_i, \Delta^2_i, \Delta^3_i$ 
and $\Delta^4_i$, the last one $\Delta^4_i$ being the empty chain in case $a$ is a cusp point of
$\Delta_i$;

\item 
If $\Delta_i^3$ is an arc or a boundary bitangent line segment we set $i=p$ and we are done;
otherwise  we define $\Delta_{i+1}$ to be the pseudotriangle lying locally to the left of the  sink bitangent line segment of
$\sigma$ in the pseudotriangulation $G_{i+1}$ obtained by flipping clockwise in $G_i$ the first bitangent line segment
$t_i$ of $\Delta_i^3$. 
\end{enumerate}
%%%%%%%%%%%%%%%%%%%%%%%%%%%%%%%%%%%%%%%%%%%%%%%%%%%%%%%%%%%%%%%%%%%%%%%%%%%%%%%
%%%%%%%%%%%%%%%%%%%%%%%%%%%%%%%%%%%%%%%%%%%%%%%%%%%%%%%%%%%%%%%%%%%%%%%%%%%%%%%
%%%%%%%%%%%%%%%%%%%%%%%%%%%%%%%%%%%%%%%%%%%%%%%%%%%%%%%%%%%%%%%%%%%%%%%%%%%%%%%
%%%%%%%%%%%%%%%%%%%%%%%%%%%%%%%%%%%%%%%%%%%%%%%%%%%%%%%%%%%%%%%%%%%%%%%%%%%%%%%
Playing the same game with $e'$ the last $1$-cell of $\rconeNew{\sigma}$ if any, the
first $1$-cell of $\rctwoNew{\sigma}$ otherwise, we define similarly a sequence of
pseudotriangulations  $G'_1,\ldots, G'_{p'}$  and a sequence of pseudotriangles $\Delta'_1,\ldots, \Delta'_{p'}.$  
We denote by $\HLsink$ and $\HRsink$ the operators that assign to a bounded $2$-cell $\sigma$ the pseudotriangle $\Delta_p$ and $\Delta'_{p'}$ 
defined above;  note that  $\HRsink$ and $\HLsink$ are conjugate under $\reverse.$ 
Finally playing the same game with the dual order of the order $\prec$ we  introduce similarly the operators $\HLsource$ and $\HRsource$. 
%%%%%%%%%%%%%%%%%%%%%%%%%%%%%%%%%%%%%%%%%%%%%%%%%%%%%%%%%%%%%%%%%%%%%%%%%%%%%%%
%%%%%%%%%%%%%%%%%%%%%%%%%%%%%%%%%%%%%%%%%%%%%%%%%%%%%%%%%%%%%%%%%%%%%%%%%%%%%%%
%%%%%%%%%%%%%%%%%%%%%%%%%%%%%%%%%%%%%%%%%%%%%%%%%%%%%%%%%%%%%%%%%%%%%%%%%%%%%%%
%%%%%%%%%%%%%%%%%%%%%%%%%%%%%%%%%%%%%%%%%%%%%%%%%%%%%%%%%%%%%%%%%%%%%%%%%%%%%%%

%%%%%%%%%%%%%%%%%%%%%%%%%%%%%%%%%%%%%%%%%%%%%%%%%%%%%%%%%%%%%%%%%%%%%%%%%%%%%%%
%%%%%%%%%%%%%%%%%%%%%%%%%%%%%%%%%%%%%%%%%%%%%%%%%%%%%%%%%%%%%%%%%%%%%%%%%%%%%%%
%%%%%%%%%%%%%%%%%%%%%%%%%%%%%%%%%%%%%%%%%%%%%%%%%%%%%%%%%%%%%%%%%%%%%%%%%%%%%%%
%%%%%%%%%%%%%%%%%%%%%%%%%%%%%%%%%%%%%%%%%%%%%%%%%%%%%%%%%%%%%%%%%%%%%%%%%%%%%%%
\begin{example} Figure~\ref{hourglasshullone} depicts the sequences of pseudotriangulations $G_i, G'_i$ together with the pseudotriangles $\Delta_i, \Delta'_i$ 
for the $2$-cell $\sigma$ of the visibility complex of the family of bodies 
introduced in Example~\ref{hourglasstwoexample} :   
A case where the sink of $\sigma$ is a right-right boundary bitangent : Here $e$ is the first $1$-cell of $\lctwoop(\sigma)$ and its sink is $b_{23}$; 
$e'$ is the last $1$-cell of $\rconeop(\sigma)$ and its sink is $b_{\overline{6}1}$, $p=p'=1$. 
%%%%%%%%%%%%%%%%%%%%%%%%%%%%%%%%%%%%%%%%%%%%%%%%%%%%%%%%%%%%%%%%%%%%%%%%%%%%%%%%%%%%%%%%%%%%%%%%%%%%%%%%%%%
%%%%%%%%%%%%%%%%%%%%%%%%%%%%%%%%%%%%%%%%%%%%%%%%%%%%%%%%%%%%%%%%%%%%%%%%%%%%%%%%%%%%%%%%%%%%%%%%%%%%%%%%%%%
%%%%%%%%%%%%%%%%%%%%%%%%%%%%%%%%%%%%%%%%%%%%%%%%%%%%%%%%%%%%%%%%%%%%%%%%%%%%%%%%%%%%%%%%%%%%%%%%%%%%%%%%%%%
%%%%%%%%%%%%%%%%%%%%%%%%%%%%%%%%%%%%%%%%%%%%%%%%%%%%%%%%%%%%%%%%%%%%%%%%%%%%%%%%%%%%%%%%%%%%%%%%%%%%%%%%%%%
\begin{figure}[!htb]
\begin{center}
\small
\psfrag{Lsi}{$\HLsink$} \psfrag{Rsi}{$\HRsink$} \psfrag{Lso}{$\HLsource$} \psfrag{Rso}{$\HRsource$}
\psfrag{deltaone}{$\HLsource(\sigma) = \HLsink(\sigma)$}
\psfrag{deltaonep}{$\HRsource(\sigma) = \HRsink(\sigma)$}
\psfrag{source}{source} \psfrag{sink}{sink}
\psfrag{vertex}{$\hourglassL{\sigma}$}\psfrag{vertexright}{$\hourglassR{\sigma}$}
\psfrag{sigma}{$\sigma$}
\psfrag{Done}{$\Delta_1$}
\psfrag{Gpone}{$G'_1$}
\psfrag{Dpone}{$\Delta'_1$}
\psfrag{Gone}{$G_1$}
\psfrag{one}{$1$}
\psfrag{two}{$2$}
\psfrag{thr}{$3$}
\psfrag{fou}{$4$}
\psfrag{fiv}{$5$}
\psfrag{six}{$6$}
\psfrag{sev}{$7$}
%%\psfrag{HRso}{$\HRsource$} \psfrag{HRsi}{$\HRsink$} \psfrag{HLso}{$\HLsource$} \psfrag{HLsi}{$\HLsink$}
\includegraphics[width= 0.98575\linewidth]{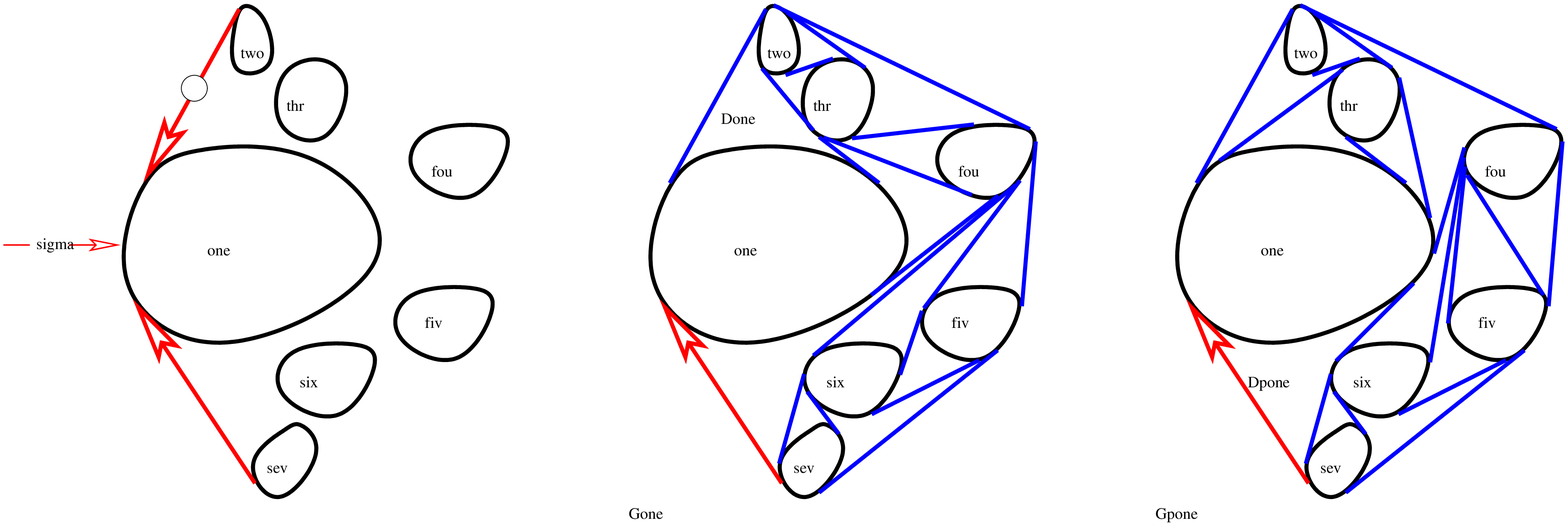}%%figures-xfig
\end{center}
\caption{\protect \small 
%A case where the sink of $\sigma$ is a right-right hull bitangent : Here $e$ is the first $1$-cell of $\lctwoop(\sigma)$ and its sink is $b_{23}$; 
%$e'$ is the last $1$-cell of $\lconeop(\sigma)$ and its sink is $b_{\overline{6}1}$, $p=p'=1$. 
\label{hourglasshullone}}
\end{figure}
%%%%%%%%%%%%%%%%%%%%%%%%%%%%%%%%%%%%%%%%%%%%%%%%%%%%%%%%%%%%%%%%%%%%%%%%%%%%%%%%%%%%%%%%%%%%%%%%%%%%%%%%%%%
%%%%%%%%%%%%%%%%%%%%%%%%%%%%%%%%%%%%%%%%%%%%%%%%%%%%%%%%%%%%%%%%%%%%%%%%%%%%%%%%%%%%%%%%%%%%%%%%%%%%%%%%%%%
%%%%%%%%%%%%%%%%%%%%%%%%%%%%%%%%%%%%%%%%%%%%%%%%%%%%%%%%%%%%%%%%%%%%%%%%%%%%%%%%%%%%%%%%%%%%%%%%%%%%%%%%%%%
%%%%%%%%%%%%%%%%%%%%%%%%%%%%%%%%%%%%%%%%%%%%%%%%%%%%%%%%%%%%%%%%%%%%%%%%%%%%%%%%%%%%%%%%%%%%%%%%%%%%%%%%%%%
\end{example}
%%%%%%%%%%%%%%%%%%%%%%%%%%%%%%%%%%%%%%%%%%%%%%%%%%%%%%%%%%%%%%%%%%%%%%%%%%%%%%%
%%%%%%%%%%%%%%%%%%%%%%%%%%%%%%%%%%%%%%%%%%%%%%%%%%%%%%%%%%%%%%%%%%%%%%%%%%%%%%%
\begin{example} 
Figure~\ref{funnelalgo} depicts the sequences of pseudotriangulations $G_i, G'_i$ together with the pseudotriangles $\Delta_i, \Delta'_i$ 
for the $2$-cell $\sigma$ of the visibility complex of the family of bodies 
introduced in Example~\ref{hourglassoneexample} : here $e$ is the  first $1$-cell of $\lctwoNew{\sigma}$, the sink of $e$ is the bitangent $b_{23}$,  
$p=4$, the sequence 
$t_1,t_2,t_3$ of flipped bitangents is the sequence $ b_{1\overline{2}}, b_{17}, b_{1\overline{7}}$, $e'$ is the last $1$-cell of $\rconeNew{\sigma}$, 
the sink of $e'$ is the bitangent $b_{\overline{5}6}$ and $p'=1$. 

%%%%%%%%%%%%%%%%%%%%%%%%%%%%%%%%%%%%%%%%%%%%%%%%%%%%%%%%%%%%%%%%%%%%%%%%%%%%%%%
%%%%%%%%%%%%%%%%%%%%%%%%%%%%%%%%%%%%%%%%%%%%%%%%%%%%%%%%%%%%%%%%%%%%%%%%%%%%%%%
%%%%%%%%%%%%%%%%%%%%%%%%%%%%%%%%%%%%%%%%%%%%%%%%%%%%%%%%%%%%%%%%%%%%%%%%%%%%%%%
%%%%%%%%%%%%%%%%%%%%%%%%%%%%%%%%%%%%%%%%%%%%%%%%%%%%%%%%%%%%%%%%%%%%%%%%%%%%%%%
\begin{figure}[!htb]
\begin{center}
\small
\psfrag{sg}{$\sigma$}
\psfrag{un}{$1$}
\psfrag{de}{$2$}
\psfrag{tr}{$3$}
\psfrag{qu}{$4$}
\psfrag{ci}{$5$}
\psfrag{si}{$6$}
\psfrag{se}{$7$}
\psfrag{ze}{$6$}
\psfrag{ma}{$7$}
\psfrag{Gpone}{$G'_1$}
\psfrag{Gone}{$G_1$}
\psfrag{Gtwo}{$G_2$}
\psfrag{Gthr}{$G_3$}
\psfrag{Gfou}{$G_4$}
\psfrag{Dpone}{$\Delta'_1$}
\psfrag{Done}{$\Delta_1$}
\psfrag{Dtwo}{$\Delta_2$}
\psfrag{Dthr}{$\Delta_3$}
\psfrag{Dfou}{$\Delta_4$}
\psfrag{Lsi}{$\HLsink$} \psfrag{Rsi}{$\HRsink$} \psfrag{Lso}{\footnotesize $\HLsource$} \psfrag{Rso}{$\HRsource$}
\psfrag{deltaone}{$\HLsource(\sigma) = \HLsink(\sigma)$}
\psfrag{deltaonep}{$\HRsource(\sigma) = \HRsink(\sigma)$}
\psfrag{source}{source} \psfrag{sink}{sink}
\psfrag{vertex}{$\hourglassL{\sigma}$}\psfrag{vertexright}{$\hourglassR{\sigma}$}
\psfrag{sigma}{$\sigma$}
%%\psfrag{HRso}{$\HRsource$} \psfrag{HRsi}{$\HRsink$} \psfrag{HLso}{$\HLsource$} \psfrag{HLsi}{$\HLsink$}
\includegraphics[width= 0.8575\linewidth]{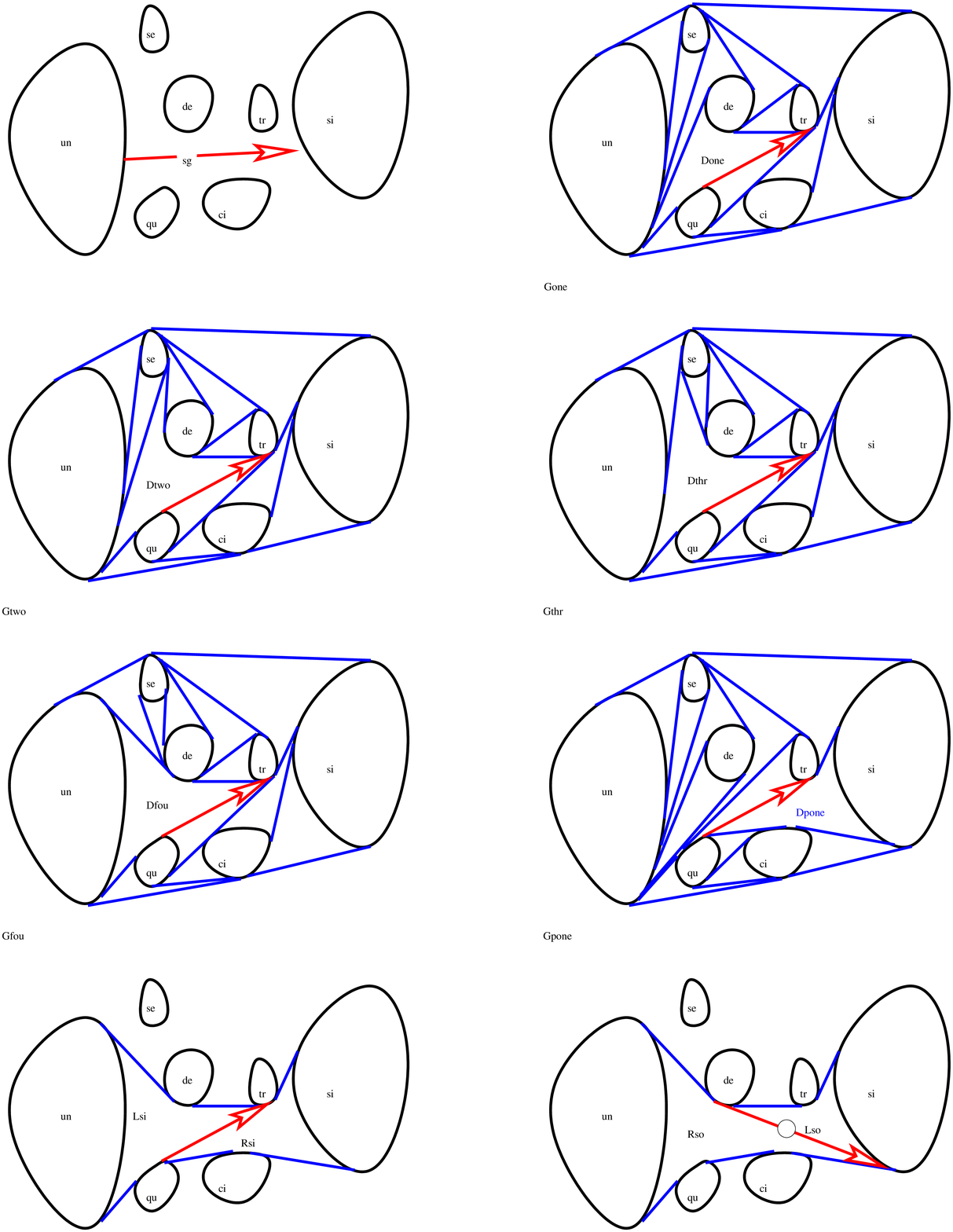}%%figures-xfig
\end{center}
\caption{ \label{funnelalgo}}
\end{figure}
%%%%%%%%%%%%%%%%%%%%%%%%%%%%%%%%%%%%%%%%%%%%%%%%%%%%%%%%%%%%%%%%%%%%%%%%%%%%%%%%%%%%%%%%%%%%%%%%%%%%%%%%%%%
%%%%%%%%%%%%%%%%%%%%%%%%%%%%%%%%%%%%%%%%%%%%%%%%%%%%%%%%%%%%%%%%%%%%%%%%%%%%%%%%%%%%%%%%%%%%%%%%%%%%%%%%%%%
%%%%%%%%%%%%%%%%%%%%%%%%%%%%%%%%%%%%%%%%%%%%%%%%%%%%%%%%%%%%%%%%%%%%%%%%%%%%%%%%%%%%%%%%%%%%%%%%%%%%%%%%%%%
%%%%%%%%%%%%%%%%%%%%%%%%%%%%%%%%%%%%%%%%%%%%%%%%%%%%%%%%%%%%%%%%%%%%%%%%%%%%%%%%%%%%%%%%%%%%%%%%%%%%%%%%%%%
\end{example}

\begin{theorem}
Let $\sigma$ be 
a bounded $2$-cell whose source is an interior vertex.
Then the pseudotriangles $\HLsink(\sigma)$ and $\HRsink(\sigma)$ are adjacent  along $\sink{\sigma}$ and 
their union $\Hglass(\sigma)$ is a free pseudoquadrangle whose
bitangent line segments lying on its boundary are the bitangent line segments of $\hourglassR{\sigma}$ and $\hourglassL{\sigma}$, possibly augmented 
with $1$ or $2$ boundary bitangent line segments in case the first or the last  or both the
first and the last elements of the label of $\sigma$ are the infinity symbol. 
Similarly the pseudotriangles $\HLsource(\sigma)$ and $\HRsource(\sigma)$ are adjacent  along $\sour{\sigma}$ and their union 
is $\Hglass(\sigma)$. \qed
\end{theorem}
\begin{proof}
We denote by $\SqBit_i^j$ the sequence of bitangent line segments of the chain $\Delta_i^j$, $j
\in \{1,2,3,4\}$, $i =1,2,\ldots,p.$
One can easily check that the sequence $\Delta_i$ is well-defined, finite, and
that 
\begin{enumerate}
\item $\Delta_1^1 = \Delta_2^1= \cdots = \Delta_p^1$, $\SqBit_1^1 =\hourglassLtwo{\sigma}$, and  $\SqBit_p^4 = \hourglassLone{\sigma};$
\item $\Delta_i^2$ is a prefix factor of $\Delta_{i+1}^2$, and $\SqBit_p^2 = \sink{\sigma}\hourglassRthr{\sigma}$ 
unless the sink bitangent of $\sigma$ is 
a right-right boundary bitangent, in which case $\Delta_p^2 = \Delta_1^2$ is an arc.
%%\item $\Delta_p^3$ is an arc (supported by the backward view of $\sigma$) or a hull bitangent;
\end{enumerate} 
The theorem follows.\end{proof}

%%%%%%%%%%%%%%%%%%%%%%%%%%%%%%%%%%%%%%%%%%%%%%%%%%%%%%%%%%%%%%%%%%%%%%%%%%%%%%%%%%%%%%%%%%%%%%%%%%%%%%%%%%%
%%%%%%%%%%%%%%%%%%%%%%%%%%%%%%%%%%%%%%%%%%%%%%%%%%%%%%%%%%%%%%%%%%%%%%%%%%%%%%%%%%%%%%%%%%%%%%%%%%%%%%%%%%%
%%%%%%%%%%%%%%%%%%%%%%%%%%%%%%%%%%%%%%%%%%%%%%%%%%%%%%%%%%%%%%%%%%%%%%%%%%%%%%%%%%%%%%%%%%%%%%%%%%%%%%%%%%%
%%%%%%%%%%%%%%%%%%%%%%%%%%%%%%%%%%%%%%%%%%%%%%%%%%%%%%%%%%%%%%%%%%%%%%%%%%%%%%%%%%%%%%%%%%%%%%%%%%%%%%%%%%%
%%%%%%%%%%%%%%%%%%%%%%%%%%%%%%%%%%%%%%%%%%%%%%%%%%%%%%%%%%%%%%%%%%%%%%%%%%%%%%%%%%%%%%%%%%%%%%%%%%%%%%%%%%%
\begin{theorem}\label{igfp} 
Let $\sigma$ be bounded $2$-cell whose source is a boundary vertex.
Then the bitangent line segments of the sequence $\hourglassR{\sigma}$ (resp. $\hourglassL{\sigma}$) 
bound a pseudotriangle incident to the boundary of the convex hull 
along the bitangent line segment corresponding to the sink (resp. source) or 
the source (resp. sink) of $\sigma$ depending on whether 
the source of $\sigma$ is a left-left or a right-right  bitangent. \qed
\end{theorem}
%%%%%%%%%%%%%%%%%%%%%%%%%%%%%%%%%%%%%%%%%%%%%%%%%%%%%%%%%%%%%%%%%%%%%%%%%%%%%%%%%%%%%%%%%%%%%%%%%%%%%%%%%%%
%%%%%%%%%%%%%%%%%%%%%%%%%%%%%%%%%%%%%%%%%%%%%%%%%%%%%%%%%%%%%%%%%%%%%%%%%%%%%%%%%%%%%%%%%%%%%%%%%%%%%%%%%%%
%%%%%%%%%%%%%%%%%%%%%%%%%%%%%%%%%%%%%%%%%%%%%%%%%%%%%%%%%%%%%%%%%%%%%%%%%%%%%%%%%%%%%%%%%%%%%%%%%%%%%%%%%%%
%%%%%%%%%%%%%%%%%%%%%%%%%%%%%%%%%%%%%%%%%%%%%%%%%%%%%%%%%%%%%%%%%%%%%%%%%%%%%%%%%%%%%%%%%%%%%%%%%%%%%%%%%%%
%%%%%%%%%%%%%%%%%%%%%%%%%%%%%%%%%%%%%%%%%%%%%%%%%%%%%%%%%%%%%%%%%%%%%%%%%%%%%%%%%%%%%%%%%%%%%%%%%%%%%%%%%%%
%%\begin{proof}
%%In case the source of $\sigma$ is a boundary bitangent, then $p = p'=1$ : an example is depicted in Figure~\ref{hourglasshullone}.  
%%\end{proof}
Note that under the assumption that the source of $\sigma$ is a boundary bitangent 
then the pseudotriangle defined 
by $\hourglassR{\sigma}$ is denoted indifferently $\HRsink(\sigma)$
or $\HRsource(\sigma)$; similarly 
the pseudotriangle defined 
by $\hourglassL{\sigma}$ is denoted indifferently $\HLsink(\sigma)$
or $\HLsource(\sigma).$

%%%%%%%%%%%%%%%%%%%%%%%%%%%%%%%%%%%%%%%%%%%%%%%%%%%%%%%%%%%%%%%%%%%%%%%%%%%%%%%%%%%%%%%%%%%%%%%%%%%%%%%%%%%
%%%%%%%%%%%%%%%%%%%%%%%%%%%%%%%%%%%%%%%%%%%%%%%%%%%%%%%%%%%%%%%%%%%%%%%%%%%%%%%%%%%%%%%%%%%%%%%%%%%%%%%%%%%
%%%%%%%%%%%%%%%%%%%%%%%%%%%%%%%%%%%%%%%%%%%%%%%%%%%%%%%%%%%%%%%%%%%%%%%%%%%%%%%%%%%%%%%%%%%%%%%%%%%%%%%%%%%
%%%%%%%%%%%%%%%%%%%%%%%%%%%%%%%%%%%%%%%%%%%%%%%%%%%%%%%%%%%%%%%%%%%%%%%%%%%%%%%%%%%%%%%%%%%%%%%%%%%%%%%%%%%
Combining now Theorem~\ref{orbit-theorem} and the previous analysis we get the 
following key result for our purpose.
\begin{theorem}\label{truekey} Let $\sigma$ be bounded $2$-cell and let $e$ be a left $1$-cell 
of the left boundary chain  of $\sigma.$ 
Then the bitangent $\phiback(e)$ leaves the pseudotriangle $\HLsink(\sigma)$. 
(In case the source of $\sigma$ is a left-left boundary  bitangent and if $e$ is the first left $1$-cell of the left boundary of $\sigma$ 
then $\phiback(e)$ and the image under $\reverse$ of the source of $\sigma$ coincide.) \qed
\end{theorem} 

%%%%%%%%%%%%%%%%%%%%%%%%%%%%%%%%%%%%%%%%%%%%%%%%%%%%%%%%%%%%%%%%%%%%%%%%%%%%%%%%%%%%%%%%%%%%%%%%%%%%%%%%%%%
%%%%%%%%%%%%%%%%%%%%%%%%%%%%%%%%%%%%%%%%%%%%%%%%%%%%%%%%%%%%%%%%%%%%%%%%%%%%%%%%%%%%%%%%%%%%%%%%%%%%%%%%%%%
%%%%%%%%%%%%%%%%%%%%%%%%%%%%%%%%%%%%%%%%%%%%%%%%%%%%%%%%%%%%%%%%%%%%%%%%%%%%%%%%%%%%%%%%%%%%%%%%%%%%%%%%%%%
%%%%%%%%%%%%%%%%%%%%%%%%%%%%%%%%%%%%%%%%%%%%%%%%%%%%%%%%%%%%%%%%%%%%%%%%%%%%%%%%%%%%%%%%%%%%%%%%%%%%%%%%%%%
%%%%%%%%%%%%%%%%%%%%%%%%%%%%%%%%%%%%%%%%%%%%%%%%%%%%%%%%%%%%%%%%%%%%%%%%%%%%%%%%%%%%%%%%%%%%%%%%%%%%%%%%%%%
%%\subsection{The pseudotriangles $\LeaveHG{\sigma}$ and $\EnterHG{\sigma}$}

%\clearpage
\subsubsection{\ABzerpts} 
Let $\MAC$ be a maximal antichain of $\Vposet$ free of vertices.
We use the symbols
$\burop{}{J},\bulop{}{J},\bdrop{}{J},\bdlop{}{J}$ for the products $\urfop\circ \btoponecellop{}{J}$,
$\ulfop\circ\btoponecellop{}{J}$,
$\drfop\circ\bbotonecellop{}{J}$,
$\dlfop\circ\bbotonecellop{}{J}$ where  
$\urf{\sigma} = \sigmaforw$ ($\urfop$ for ``upper right face''),
$\ulf{\sigma} = \sigmaback$,
$\drf{\sigma} = \dlf{\sigma} = \sigmarigh$ for a left-left $0$-cell or a left $1$-cell $\sigma$, and where 
$\urf{\sigma} = \ulf{\sigma} = \sigmaleft$, $\drf{\sigma} = \sigmaback$, $\dlf{\sigma} = \sigmaforw$ 
for a right-right $0$-cell or a right $1$-cell $\sigma$.
(Note that $\urfop$ and $\dlfop$ on one hand and $\ulfop$ and $\drfop$ on the other hand are conjugate under $\shiftop$.)
The \ABzerpts\ of $\sigma$, $\sigma \in \facesG(\MAC)$, are denoted $\LeaveHG{\sigma}$ and $\EnterHG{\sigma}$ and are defined as follows   
\begin{equation}
\LeaveHG{\sigma} =  \begin{cases} 
                      \HLsink(\bdl{\sigma}{\MAC}) & \text{if $\sigma \in
\fdtG{\one}{\thrset}(\MAC)$} \\
                      \HRsource(\sigma) & \text{otherwise}, 
\end{cases}
\end{equation}
and 
%%%  $\EnterHG{\sigma} =  \HRsink(\ur{\sigma})$ if $\sigma \in \fdt{\thrset}{\one}(I)$; $\HLsource(\sigma)$ otherwise. 
\begin{equation}\label{definition}
\EnterHG{\sigma} = \begin{cases} 
                      \HRsink(\bur{\sigma}{\MAC}) & \text{if $\sigma \in
\fdtG{\thrset}{\one}(\MAC)$} \\
                     \HLsource(\sigma) & \text{otherwise} 
\end{cases}
\end{equation}
where, recall, $\fdt{i}{j}(\MAC)$, $i, j \in \{1,2,3\}$, denotes the set of $\sigma \in \faces(\MAC)$ such that
$\bbotonecell{\sigma}{\MAC} \in \rc_i(\sigma)$ and $\btoponecell{\sigma}{\MAC} \in \lc_j(\sigma)$; cf.  Theorem~\ref{face-description}.
(For $J,K$ subsets of $\{1,2,3\}$ we set  $\fdt{J}{K} (\MAC)= 
\bigcup_{i \in J, j\in K} \fdt{i}{j} (\MAC).$) Note that the operators $\opLeaveHG$ and $\opEnterHG$ are conjugate under $\reverse$.

%%%%%%%%%%%%%%%%%%%%%%%%%%%%%%%%%%%%%%%%%%%%%%%%%%%%%%%%%%%%%%%%%%%%%%%%%%%%%%%%%%%%%%%%%%%%%%%%%%%%%%%%%%%
%%%%%%%%%%%%%%%%%%%%%%%%%%%%%%%%%%%%%%%%%%%%%%%%%%%%%%%%%%%%%%%%%%%%%%%%%%%%%%%%%%%%%%%%%%%%%%%%%%%%%%%%%%%
%%%%%%%%%%%%%%%%%%%%%%%%%%%%%%%%%%%%%%%%%%%%%%%%%%%%%%%%%%%%%%%%%%%%%%%%%%%%%%%%%%%%%%%%%%%%%%%%%%%%%%%%%%%
%%%%%%%%%%%%%%%%%%%%%%%%%%%%%%%%%%%%%%%%%%%%%%%%%%%%%%%%%%%%%%%%%%%%%%%%%%%%%%%%%%%%%%%%%%%%%%%%%%%%%%%%%%%
\begin{theorem} \label{keyonezero}
Let $\MAC$ be a maximal antichain of $\Vposet$ free of vertices
and let $\sigma \in \facesG(\MAC)$ whose source 
is a  left-left (resp. right-right) boundary vertex. Then
its sink bitangent line segment is a right-right (resp. left-left) bitangent line segment and is one of the three sides of $\LeaveHG{\sigma}$
(resp.  $\EnterHG{\sigma}$).  \qed
\end{theorem}

%%%%%%%%%%%%%%%%%%%%%%%%%%%%%%%%%%%%%%%%%%%%%%%%%%%%%%%%%%%%%%%%%%%%%%%%%%%%%%%%%%%%%%%%%%%%%%%%%%%%%%%%%%%
%%%%%%%%%%%%%%%%%%%%%%%%%%%%%%%%%%%%%%%%%%%%%%%%%%%%%%%%%%%%%%%%%%%%%%%%%%%%%%%%%%%%%%%%%%%%%%%%%%%%%%%%%%%
%%%%%%%%%%%%%%%%%%%%%%%%%%%%%%%%%%%%%%%%%%%%%%%%%%%%%%%%%%%%%%%%%%%%%%%%%%%%%%%%%%%%%%%%%%%%%%%%%%%%%%%%%%%
%%%%%%%%%%%%%%%%%%%%%%%%%%%%%%%%%%%%%%%%%%%%%%%%%%%%%%%%%%%%%%%%%%%%%%%%%%%%%%%%%%%%%%%%%%%%%%%%%%%%%%%%%%%
%%%%%%%%%%%%%%%%%%%%%%%%%%%%%%%%%%%%%%%%%%%%%%%%%%%%%%%%%%%%%%%%%%%%%%%%%%%%%%%%%%%%%%%%%%%%%%%%%%%%%%%%%%%
\begin{theorem} \label{keyone}
Let $\MAC$ be a maximal antichain of $\Vposet$ free of vertices and let $\sigma \in \facesG(\MAC)$ whose source 
is an interior vertex. Then
the source bitangent line segment of $\sigma$  belongs to the boundary of $\LeaveHG{\sigma}$,
the pseudotriangle 
$\LeaveHG{\sigma}$ lies locally on the left side of the source bitangent
of $\sigma$,
and the sink bitangent line segment of $\sigma$ leaves $\LeaveHG{\sigma}$.  \qed
\end{theorem}
\begin{proof} 
Assume first that $\sigma\in \fdt{\reversetwoset}{\thrset}(\MAC).$ 
Then  $\LeaveHG{\sigma} = \HRsource(\sigma)$ and we conclude using the definition 
of $\HRsource(\sigma)$ and Theorem~\ref{igfp}.
Assume now that $\sigma \in \fdt{\one}{\thrset}(\MAC)$.   
Then one can easily check that 
\begin{enumerate}
\item $\bdl{\sigma}{\MAC} \notin \setbotface$; 
\item $\bbotonecell{\sigma}{\MAC}$ is a left $1$-cell of the left boundary of $\bdl{\sigma}{\MAC};$
%%\item $\dl{\sigma} \in \fdt{\thrset}{2}(I)$;
\item $\sour{\sigma} = \phibackstar(\bbotonecell{\sigma}{\MAC})$ and $\sink{\sigma}= \phiback(\bbotonecell{\sigma}{\MAC})$; 
\item $\touchop \circ \sour{\sigma}\in \hourglassLplus{\bdl{\sigma}{\MAC},\bbotonecell{\sigma}{\MAC}}$ (since $\sour{\sigma}$ is an interior vertex);
\item $\LeaveHG{\sigma} = \HLsink(\bdl{\sigma}{\MAC})$;
\end{enumerate} 
from which it follows that 
the source bitangent line segment of $\sigma$ appears in the boundary 
of $\LeaveHG{\sigma}.$
It remains to prove that $\sink{\sigma}$ leaves $\LeaveHG{\sigma}$; but this is exactly the statement 
of Theorem~\ref{truekey}.
\end{proof} 

%%%%%%%%%%%%%%%%%%%%%%%%%%%%%%%%%%%%%%%%%%%%%%%%%%%%%%%%%%%%%%%%%%%%%%%%%%%%%%%%%%%%%%%%%%%%%%%%%%%%%%%%%%%
%%%%%%%%%%%%%%%%%%%%%%%%%%%%%%%%%%%%%%%%%%%%%%%%%%%%%%%%%%%%%%%%%%%%%%%%%%%%%%%%%%%%%%%%%%%%%%%%%%%%%%%%%%%
%%%%%%%%%%%%%%%%%%%%%%%%%%%%%%%%%%%%%%%%%%%%%%%%%%%%%%%%%%%%%%%%%%%%%%%%%%%%%%%%%%%%%%%%%%%%%%%%%%%%%%%%%%%
%%%%%%%%%%%%%%%%%%%%%%%%%%%%%%%%%%%%%%%%%%%%%%%%%%%%%%%%%%%%%%%%%%%%%%%%%%%%%%%%%%%%%%%%%%%%%%%%%%%%%%%%%%%
%%%%%%%%%%%%%%%%%%%%%%%%%%%%%%%%%%%%%%%%%%%%%%%%%%%%%%%%%%%%%%%%%%%%%%%%%%%%%%%%%%%%%%%%%%%%%%%%%%%%%%%%%%%
Thanks to the conjugation relation $\opLeaveHG \circ \reverse  = \reverse \circ\opEnterHG$  
we see that $\LeaveHG{\sigma}$ and $\EnterHG{\sigma}$ are adjacent along 
the source bitangent line segment of $\sigma$ and that the sink bitangent line segment of $\sigma$
joins $\LeaveHG{\sigma}$ to $\EnterHG{\sigma}$. 
Walking in counterclockwise order around the boundary of $\LeaveHG{\sigma}$ 
starting from the tail of the source bitangent line segment of $\sigma$ 
we find successively the convex chains $\LeaHGone{\sigma}$, $\LeaHGtwo{\sigma}$,
$\LeaHGthr{\sigma}$ (which is reduced to an arc or a boundary bitangent line segment),
and $\LeaHGfou{\sigma}$. 
%%(which is NOT  reduced to the empty chain, or an arc or the concatenation of a bitangent and an arc).
Let 
$\LeaHGtt(\sigma)$ be the concatenation 
of the chains $\LeaHGtwo{\sigma}$ and $\LeaHGthr{\sigma}$. 
 We know that 
the sink bitangent line segment of $\sigma$ leaves the chain $\LeaHGtt(\sigma)$, 
that $\touch{\bbotonecell{\sigma}{\MAC}}$ is a subarc of an arc, say $\tau^-$,  of the chain
$\LeaHGtt(\sigma)$ : so we define $\LeaHGlea(\sigma)$
to be the suffix subchain of $\LeaHGtt(\sigma)$ 
starting at $\tau^-$. 
Similarly we define the chain $\EntHGent(\sigma).$
By construction the sink bitangent line segment of $\sigma$ leaves the chain $\LeaHGlea(\sigma)$ 
and reaches  the chain $\EntHGent(\sigma)$; however one can't use directly these chains 
to compute the sink bitangent of $\sigma$
because neither the source of the first arc of $\LeaHGlea(\sigma)$
nor the sink of its last arc are efficiently computable.

\subsubsection{\ABpts} 
We now assume that the convex bodies of $\disks$ are lifts of
convex bodies of a finite family $\groundbodies$ of pairwise disjoint convex bodies of the topological plane  $\pointset$. 
The central point of a convex body of $\disks$ is defined to be the branched 
point contained in that body if any; otherwise any point chosen arbitrarily in its interior. A {\it boundary tangent} is a tangent to the convex hull of convex bodies of $\disks$; all other tangents are said to be {\it interior tangents}.  The backward/forward view of a line is the first/last atom of its label.

\begin{figure}[!htb]
\begin{center}
\small
\psfrag{hull}{$v$}
\psfrag{gammapp}{$\gamma$}
\psfrag{a}{$t$}
\psfrag{half}{$\sigma$}
\psfrag{B}{$o'$}
\psfrag{U}{$o$}
\psfrag{BU}{$o=o'$}
\includegraphics[width= 0.857500065675\linewidth]{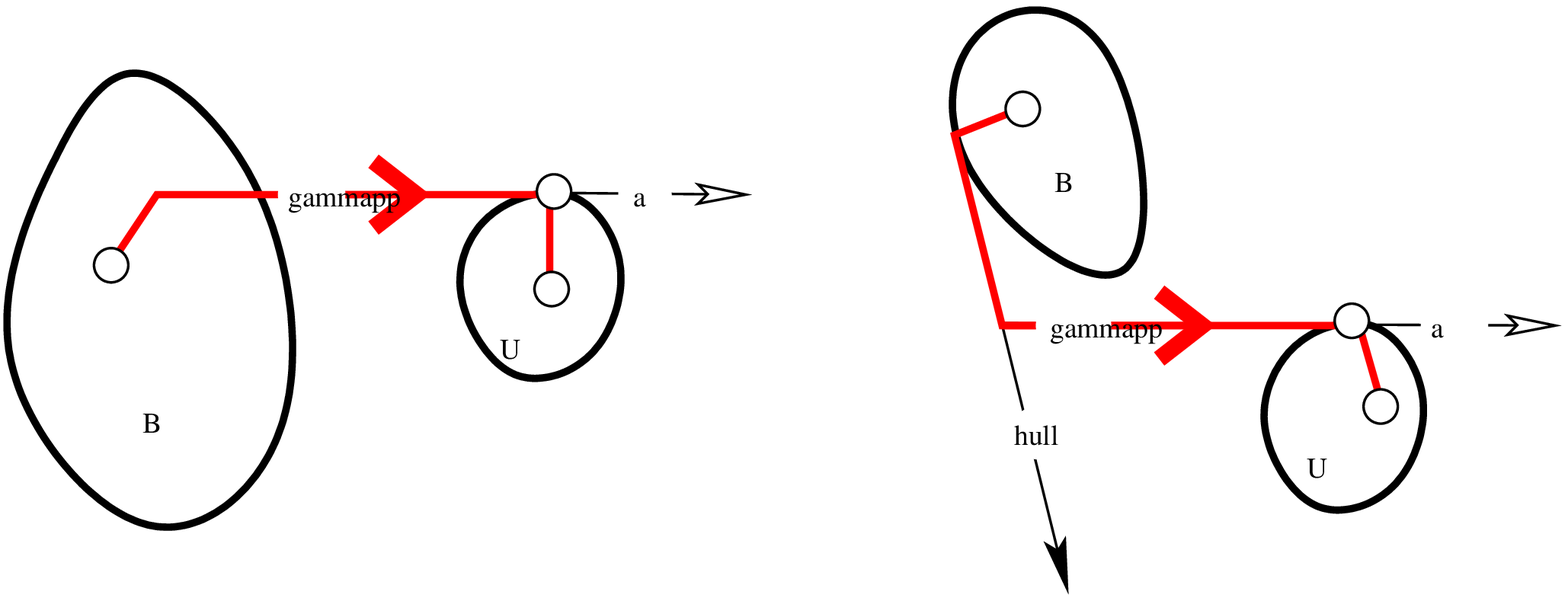}%%xfigfinal
\\
\includegraphics[width= 0.85750065675\linewidth]{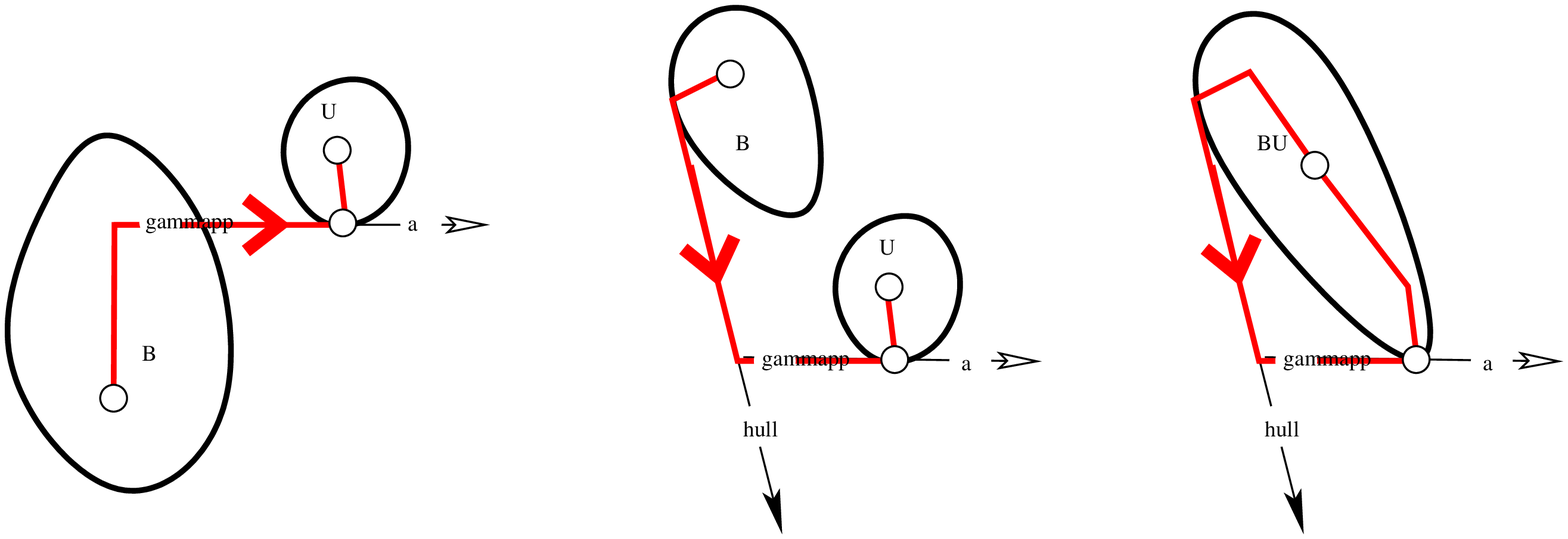}%%xfigfinal
\end{center}
\caption{\small \label{topsum}}
\end{figure}

Let $s$ be the initial segment of an interior tangent $t$ to a convex body $o$ of $\disks$ with backward view 
a convex body $o'.$ We set $\disks(s) = \{\overline{o},\overline{o}'\} \subseteq \groundbodies.$
Let $\gamma$ be a simple oriented curve in $o'\cup o\cup s$ joining the 
central point of $o'$ to the central point of $o$ with the property that its projection $\gamma'$ in $\pointset$ is
simple, as illustrated in the left top and left bottom  diagrams of Figure~\ref{topsum}. 
%%%%%%%%%%%%%%%%%%%%%%%%%%%%%%%%%%%%%%%%%%%%%%%%%%%%%%%%%%%%%%%%%%%%%%%%%%%%%%%%%%%%%%%%%%%%%%%%%%%%%%%%%%%
%%%%%%%%%%%%%%%%%%%%%%%%%%%%%%%%%%%%%%%%%%%%%%%%%%%%%%%%%%%%%%%%%%%%%%%%%%%%%%%%%%%%%%%%%%%%%%%%%%%%%%%%%%%
%%%%%%%%%%%%%%%%%%%%%%%%%%%%%%%%%%%%%%%%%%%%%%%%%%%%%%%%%%%%%%%%%%%%%%%%%%%%%%%%%%%%%%%%%%%%%%%%%%%%%%%%%%%
%%%%%%%%%%%%%%%%%%%%%%%%%%%%%%%%%%%%%%%%%%%%%%%%%%%%%%%%%%%%%%%%%%%%%%%%%%%%%%%%%%%%%%%%%%%%%%%%%%%%%%%%%%%
%%%%%%%%%%%%%%%%%%%%%%%%%%%%%%%%%%%%%%%%%%%%%%%%%%%%%%%%%%%%%%%%%%%%%%%%%%%%%%%%%%%%%%%%%%%%%%%%%%%%%%%%%%%
%%%%%%%%%%%%%%%%%%%%%%%%%%%%%%%%%%%%%%%%%%%%%%%%%%%%%%%%%%%%%%%%%%%%%%%%%%%%%%%%%%%%%%%%%%%%%%%%%%%%%%%%%%%
%%%%%%%%%%%%%%%%%%%%%%%%%%%%%%%%%%%%%%%%%%%%%%%%%%%%%%%%%%%%%%%%%%%%%%%%%%%%%%%%%%%%%%%%%%%%%%%%%%%%%%%%%%%
%%%%%%%%%%%%%%%%%%%%%%%%%%%%%%%%%%%%%%%%%%%%%%%%%%%%%%%%%%%%%%%%%%%%%%%%%%%%%%%%%%%%%%%%%%%%%%%%%%%%%%%%%%%
%%%%%%%%%%%%%%%%%%%%%%%%%%%%%%%%%%%%%%%%%%%%%%%%%%%%%%%%%%%%%%%%%%%%%%%%%%%%%%%%%%%%%%%%%%%%%%%%%%%%%%%%%%%
%%%%%%%%%%%%%%%%%%%%%%%%%%%%%%%%%%%%%%%%%%%%%%%%%%%%%%%%%%%%%%%%%%%%%%%%%%%%%%%%%%%%%%%%%%%%%%%%%%%%%%%%%%%
We construct a new branched covering 
$\Liftop{}{s}{(\CoSur{})}$ of $\pointset$ as follows.
We cut $\mathbb{C} = \CoSur{} \sqcup \pointset$ along $\gamma$ and $\gamma'$, call $\mathbb{C}_{\gamma}$ the resulting surface
and $\Map{q}{\mathbb{C}_{\gamma}}{\mathbb{C}}$ the induced projection;
we define $\Liftop{}{s}{(\CoSur{})}$ as the quotient space of $\mathbb{C}_{\gamma}$ 
%by identifying the left (resp. right) lift of $\gamma$ and the right (resp.  left) lift of $\gamma'$ under $q.$ 
by identifying, on one hand, the left lift (under $q$) of $\gamma$ and the right 
 lift of $\gamma'$ and, on the other hand, the right lift of $\gamma$ and the left 
 lift of $\gamma'$. 
By construction the set $\Liftop{}{s}(\disks)$ 
of connected components of 
$q^{-1}(\bigcup (\disks \cup \disks(s)))$ 
is a set of pairwise disjoint convex bodies 
of $\Liftop{}{s}(\CoSur{}).$  
A similar construction can be done with the terminal segment of the interior tangent in place of its initial segment.

\begin{theorem}\label{maintwofirst}
Let $s$ be the initial segment of an interior tangent to a convex body $o$ of $\disks$  with backward view a convex body $o'$,
let $\sigma = \dlf{s}$  and let $\sigma' = \dlf{s'}$ where $s'$ is the right
lift of $s$ in the visibility complex of $\Liftop{}{s}(\disks).$
Then
$\sink{\sigma'} = \sink{\sigma}$, $\hourglassR{\sigma'} = \hourglassR{\sigma}$, $\hourglassLplus{\sigma',s'}  = \hourglassLplus{\sigma,s}$,
and  $\hourglassLminus{\sigma', s'}$ is the right-left or right-right bitangent line segment joining $\overline{o}'$ to $\overline{o}$
depending on whether $s$ is a left or right tangent. %%%%({\sc attention $o$ et $o'$ peuvent coincider.})
A similar result holds for the terminal segment of an interior tangent with forward view a convex body. \qed
\end{theorem}
\begin{proof} By construction modulo the simple observation that the segment $s$ does not cross any of the bitangent line segments  of
$\hourglassR{\sigma}$ and $\hourglassL{\sigma}.$  
\end{proof} 

A similar construction is done in the case where the backward view of $t$ is the infinity symbol 
using the left-left boundary  bitangent line segment $v$ pierced by $s$ and the
convex body $o'$ that $v$ leaves, as illustrated in the right diagrams of Figure~\ref{topsum}.  
Note that in the case where $t$ is a left tangent one can have $o=o'$.

\begin{theorem}\label{maintwosecond} 
Let $s$ be the initial segment of an interior tangent whose backward view is the infinity symbol, let $\sigma = \dlf{s}$  and let $\sigma' = \dlf{s'}$ where $s'$ is the right
lift of $s$ in the visibility complex of $\Liftop{}{s}(\disks).$
Then
$\sink{\sigma'} = \sink{\sigma}$, $\hourglassR{\sigma'} = \hourglassR{\sigma}$, $\hourglassLplus{\sigma',s'}  = \hourglassLplus{\sigma,s}$,
and  $\hourglassLminus{\sigma', s'}$ is 
the left-right bitangent line segment joining $\overline{o}'$ to $\overline{o}$ if  $s$ is a  right tangent, 
the left-left bitangent line segment joining $\overline{o}'$ to $\overline{o}$ if $s$ is a left tangent and $o\neq o'$; 
the empty sequence otherwise.
A similar result holds for the terminal segment of an interior tangent whose forward view is the infinity symbol. \qed
\end{theorem}

We turn now to the definition of the \ABpts. 

Let  $\MAC$ be a maximal antichain free of vertices  of $\Vposet(\disks)$ with a distinguished line $\dline{e} \in e$ for each $e \in \edges(\MAC)$ and let $s$ be the initial segment of one 
of the $\dline{e}$, $e \in \edges(\MAC)$, with the property that the supporting line of $s$ is an interior tangent.
Let $\MACBIS$ be a maximal antichain free of vertices  of  $\Vposet(\disks(s))$ with a distinguished line $\dline{e} \in e$ for each $e \in \edges(\MACBIS)$ 
such that $\MAC$ and $\MACBIS$ agree along  the supporting line $t$ of $s$, that is, the projection of $t$ in $\pointset$
is a  $\dline{e'}$ for some $e'\in \edges(\MACBIS).$ 
In that case the lines $q^{-1}(\dline{e})$,
$ e \in \edges(\MAC) \sqcup \edges(\MACBIS)$, define a maximal antichain of the
visibility complex of $\Liftop{}{s}(\disks)$, denoted $\Liftop{}{s}(\MAC)$ thereafter.

Let $\sigma \in \fdt{x}{y}(\MAC)$ whose source is an interior vertex. 
We  define $\liftoperator^+(\CoSur{})$, $\liftoperator^+(\disks)$, $\liftoperator^+(\MAC)$, and $\liftoperator^+(\sigma)$ to be 
$\Liftop{}{s}(\CoSur{})$, $\Liftop{}{s}(\disks)$, $\Liftop{}{s}(\MAC)$, and $\dlf{s'}$,
 where $s$ is the initial segment of $\dline{\btoponecell{\sigma}{\MAC}}$ 
and $s'$ is the right lift in $\liftoperator^+(\CoSur{})$ of $s$, if $y \in \{2,3\}$;  $\CoSur{}$, $\disks$, $\MAC$, and $\sigma$ otherwise. 
We define $\liftoperator^-$ to be the conjugate of $\liftoperator^+$ under $\reverse$ and we set $\liftoperator = \liftoperator^- \circ \liftoperator^+$.
Applying twice Theorem~\ref{maintwofirst} we get the following theorem  where we write 
$\hourglassRplus{\sigma; \MAC}$, 
$\hourglassLplus{\sigma; \MAC}$,
$\hourglassRminus{\sigma; \MAC}$, 
$\hourglassLminus{\sigma; \MAC}$,
for $\hourglassRplus{\sigma, \bbotonecell{\sigma}{\MAC}}$, 
$\hourglassLplus{\sigma, \btoponecell{\sigma}{\MAC}}$, 
$\hourglassRminus{\sigma, \bbotonecell{\sigma}{\MAC}}$, and
$\hourglassLminus{\sigma, \btoponecell{\sigma}{\MAC}}$. 

%%%%%%%%%%%%%%%%%%%%%%%%%%%%%%%%%%%%%%%%%%%%%%%%%%%%%%%%%%%%%%%%%%%%%%%%%%%%%%%%%%%%%%%%%%%%%%%%%%%%%%%%%%%
%%%%%%%%%%%%%%%%%%%%%%%%%%%%%%%%%%%%%%%%%%%%%%%%%%%%%%%%%%%%%%%%%%%%%%%%%%%%%%%%%%%%%%%%%%%%%%%%%%%%%%%%%%%
%%%%%%%%%%%%%%%%%%%%%%%%%%%%%%%%%%%%%%%%%%%%%%%%%%%%%%%%%%%%%%%%%%%%%%%%%%%%%%%%%%%%%%%%%%%%%%%%%%%%%%%%%%%
%%%%%%%%%%%%%%%%%%%%%%%%%%%%%%%%%%%%%%%%%%%%%%%%%%%%%%%%%%%%%%%%%%%%%%%%%%%%%%%%%%%%%%%%%%%%%%%%%%%%%%%%%%%
%%%%%%%%%%%%%%%%%%%%%%%%%%%%%%%%%%%%%%%%%%%%%%%%%%%%%%%%%%%%%%%%%%%%%%%%%%%%%%%%%%%%%%%%%%%%%%%%%%%%%%%%%%%
\begin{theorem} \label{maintwo}
Let $\sigma \in \fdt{x}{y}(\MAC)$ whose source is an interior vertex, let $\MAC' = \liftoperator(\MAC)$, and let
$\sigma' = \liftoperator(\sigma)$. Then 
\begin{enumerate}
\item $\sigma'\in \fdt{x}{y}(\MAC')$;
\item $\sink{\sigma} = \sink{\sigma'}$;
\item $\hourglassRplus{\sigma'; \MAC'} = \hourglassRplus{\sigma;\MAC}$,  
$\hourglassLplus{\sigma';\MAC'} = \hourglassLplus{\sigma;\MAC}$; and 
\item 
$\hourglassRminus{\sigma';\MAC'}$, 
$\hourglassLminus{\sigma'; \MAC'}$, and $\sour{\sigma'}$ 
are computable in constant time.  \qed
\end{enumerate}
\end{theorem}
%%%%%%%%%%%%%%%%%%%%%%%%%%%%%%%%%%%%%%%%%%%%%%%%%%%%%%%%%%%%%%%%%%%%%%%%%%%%%%%%%%%%%%%%%%%%%%%%%%%%%%%%%%%
%%%%%%%%%%%%%%%%%%%%%%%%%%%%%%%%%%%%%%%%%%%%%%%%%%%%%%%%%%%%%%%%%%%%%%%%%%%%%%%%%%%%%%%%%%%%%%%%%%%%%%%%%%%
%%%%%%%%%%%%%%%%%%%%%%%%%%%%%%%%%%%%%%%%%%%%%%%%%%%%%%%%%%%%%%%%%%%%%%%%%%%%%%%%%%%%%%%%%%%%%%%%%%%%%%%%%%%
%%%%%%%%%%%%%%%%%%%%%%%%%%%%%%%%%%%%%%%%%%%%%%%%%%%%%%%%%%%%%%%%%%%%%%%%%%%%%%%%%%%%%%%%%%%%%%%%%%%%%%%%%%%
%%%%%%%%%%%%%%%%%%%%%%%%%%%%%%%%%%%%%%%%%%%%%%%%%%%%%%%%%%%%%%%%%%%%%%%%%%%%%%%%%%%%%%%%%%%%%%%%%%%%%%%%%%%
As announced in the introduction the \ABpts\  of $\sigma$  are the \ABzerpts\ of  $\liftoperator(\sigma)$; 
they are denoted $\Lea(\sigma)$ and $\Ent(\sigma)$ in the sequel.

%%%%%%%%%%%%%%%%%%%%%%%%%%%%%%%%%%%%%%%%%%%%%%%%%%%%%%%%%%%%%%%%%%%%%%%%%%%%%%%%%%%%%%%%%%%%%%%%%%%%%%%%%%%
%%%%%%%%%%%%%%%%%%%%%%%%%%%%%%%%%%%%%%%%%%%%%%%%%%%%%%%%%%%%%%%%%%%%%%%%%%%%%%%%%%%%%%%%%%%%%%%%%%%%%%%%%%%
%%%%%%%%%%%%%%%%%%%%%%%%%%%%%%%%%%%%%%%%%%%%%%%%%%%%%%%%%%%%%%%%%%%%%%%%%%%%%%%%%%%%%%%%%%%%%%%%%%%%%%%%%%%
%%%%%%%%%%%%%%%%%%%%%%%%%%%%%%%%%%%%%%%%%%%%%%%%%%%%%%%%%%%%%%%%%%%%%%%%%%%%%%%%%%%%%%%%%%%%%%%%%%%%%%%%%%%
%%%%%%%%%%%%%%%%%%%%%%%%%%%%%%%%%%%%%%%%%%%%%%%%%%%%%%%%%%%%%%%%%%%%%%%%%%%%%%%%%%%%%%%%%%%%%%%%%%%%%%%%%%%
\subsubsection{Summary} 
\begin{figure}[!htb]
\centering
\footnotesize
\psfrag{zero}{$\fdtthrthr$}
\psfrag{deux}{$\fdttwotwo$}
\psfrag{four}{$\fdtoneone$}
\psfrag{ureasy}{$\fdtthrtwo$}
\psfrag{urhard}{$\fdtthrone$}
\psfrag{ul}{$\fdtonetwo$}
\psfrag{dr}{$\fdttwoone$}
\psfrag{dleasy}{$\fdttwothr$}
\psfrag{dlhard}{$\fdtonethr$}
\includegraphics[width=0.8575\linewidth]{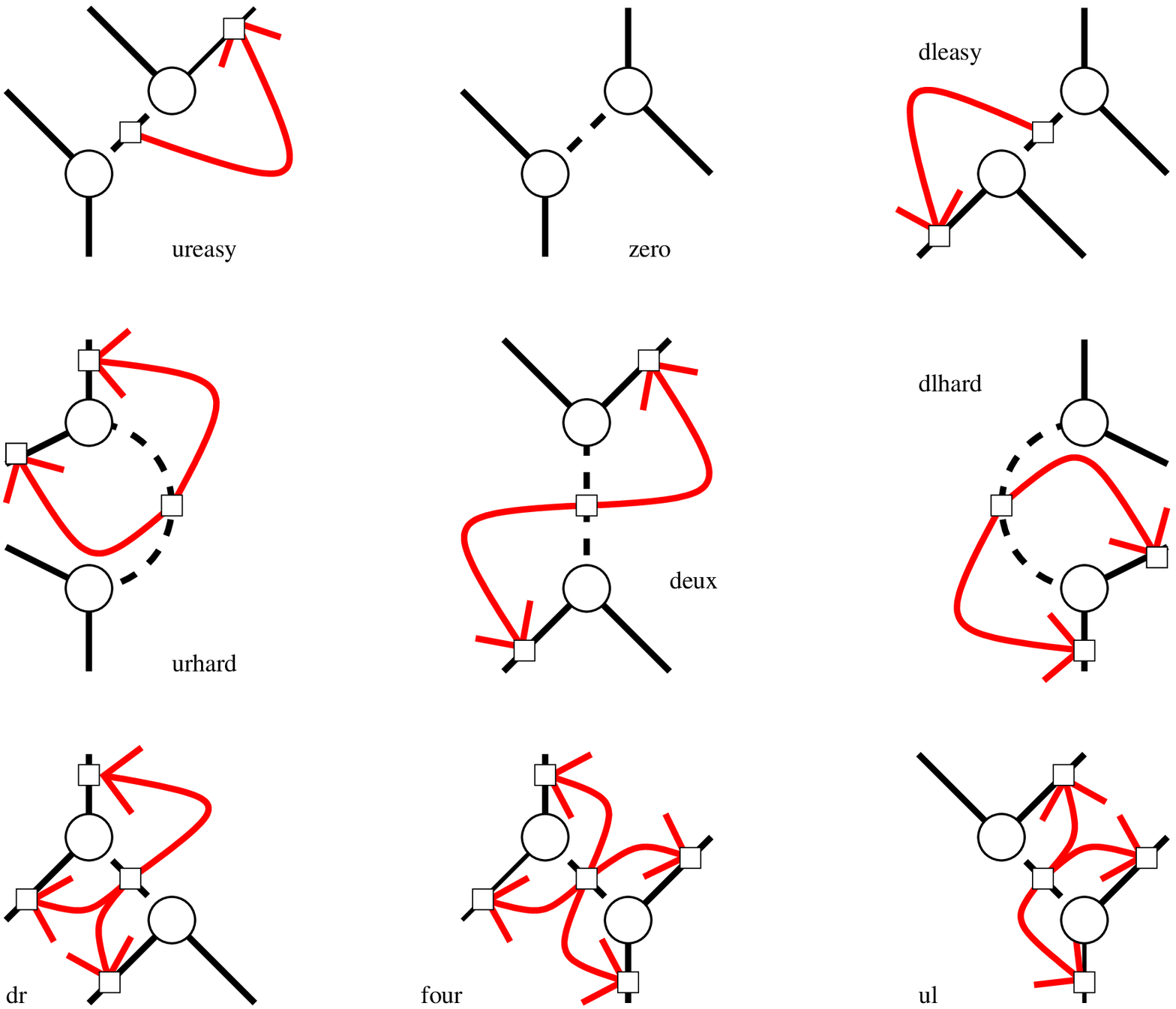}%%xfigfinal
\caption{\small \protect \label{keycells}}
\end{figure}

We
reformulate, in preparation to the description of our pseudotriangulation algorithm in the next section, Theorem~\ref{maintwo} using the following more adequate  
notations. 
For $\sigma \in \facesG(\MAC)$  we set 
\begin{equation}
\dgdl(\sigma) = \begin{cases}
\emptyset & \text{if $\sigma \in \fdt{\thr}{\thrset}(\MAC)$;}\\
\{\bdl{\sigma}{\MAC}\} & \text{if $\sigma \in \fdt{\two}{\thrset}(\MAC)$}; \\
\{\bdl{\sigma}{\MAC}, \burop{}{\MAC} \circ \bdl{\sigma}{\MAC}\} & \text{otherwise;}
\end{cases}
\end{equation}
similarly we define 
$\dgur(\sigma)$ to be the empty set 
if $\sigma \in \fdt{\thrset}{\thr}(\MAC)$; the singleton $\{\bur{\sigma}{\MAC}\}$ if 
$\sigma \in \fdt{\thrset}{\two}(\MAC)$; the pair $\{\bur{\sigma}{\MAC}, \bdlop{}{\MAC} \circ \bur{\sigma}{\MAC}\}$ otherwise;
note that $\dgur$ is the conjugate of $\dgdl$ under the shift operator $\shiftop$.
Thus Theorem~\ref{maintwo} can be read as follows.
%%%%%%%%%%%%%%%%%%%%%%%%%%%%%%%%%%%%%%%%%%%%%%%%%%%%%%%%%%%%%%%%%%%%%%%%%%%%%%%%%%%%%%%%%%%%%%%%%%%%%%%%%%%
%%%%%%%%%%%%%%%%%%%%%%%%%%%%%%%%%%%%%%%%%%%%%%%%%%%%%%%%%%%%%%%%%%%%%%%%%%%%%%%%%%%%%%%%%%%%%%%%%%%%%%%%%%%
%%%%%%%%%%%%%%%%%%%%%%%%%%%%%%%%%%%%%%%%%%%%%%%%%%%%%%%%%%%%%%%%%%%%%%%%%%%%%%%%%%%%%%%%%%%%%%%%%%%%%%%%%%%
%%%%%%%%%%%%%%%%%%%%%%%%%%%%%%%%%%%%%%%%%%%%%%%%%%%%%%%%%%%%%%%%%%%%%%%%%%%%%%%%%%%%%%%%%%%%%%%%%%%%%%%%%%%
%%%%%%%%%%%%%%%%%%%%%%%%%%%%%%%%%%%%%%%%%%%%%%%%%%%%%%%%%%%%%%%%%%%%%%%%%%%%%%%%%%%%%%%%%%%%%%%%%%%%%%%%%%%
\begin{theorem}\label{summary} Assume that the orbits of the sink bitangent line
segments of the 
$2$-cells of $\dgdl(\sigma) \cup \dgur(\sigma)$ under the operators $\pbackop$ and $\pforwop$ are known.
Then representations of the \ABpts\ $\Lea(\sigma)$ and $\Ent(\sigma)$ are computable in constant time.\qed
\end{theorem}
%%%%%%%%%%%%%%%%%%%%%%%%%%%%%%%%%%%%%%%%%%%%%%%%%%%%%%%%%%%%%%%%%%%%%%%%%%%%%%%%%%%%%%%%%%%%%%%%%%%%%%%%%%%
%%%%%%%%%%%%%%%%%%%%%%%%%%%%%%%%%%%%%%%%%%%%%%%%%%%%%%%%%%%%%%%%%%%%%%%%%%%%%%%%%%%%%%%%%%%%%%%%%%%%%%%%%%%
%%%%%%%%%%%%%%%%%%%%%%%%%%%%%%%%%%%%%%%%%%%%%%%%%%%%%%%%%%%%%%%%%%%%%%%%%%%%%%%%%%%%%%%%%%%%%%%%%%%%%%%%%%%
%%%%%%%%%%%%%%%%%%%%%%%%%%%%%%%%%%%%%%%%%%%%%%%%%%%%%%%%%%%%%%%%%%%%%%%%%%%%%%%%%%%%%%%%%%%%%%%%%%%%%%%%%%%
%%%%%%%%%%%%%%%%%%%%%%%%%%%%%%%%%%%%%%%%%%%%%%%%%%%%%%%%%%%%%%%%%%%%%%%%%%%%%%%%%%%%%%%%%%%%%%%%%%%%%%%%%%%
Let $G$ be the digraph whose set of nodes is the set of arcs of the cross-section and whose set of arcs is the set of pairs $(\sigma,\sigma')$, $\sigma'\in \dgdl(\sigma) \cup \dgur(\sigma)$, as
illustrated in Figure~\ref{keycells} where we have drawn $G$ on the canonical upward drawing of the cross-section. 
%%%%%%%%%%%%%%%%%%%%%%%%%%%%%%%%%%%%%%%%%%%%%%%%%%%%%%%%%%%%%%%%%%%%%%%%%%%%%%%%%%%%%%%%%%%%%%%%%%%%%%%%%%%
%%%%%%%%%%%%%%%%%%%%%%%%%%%%%%%%%%%%%%%%%%%%%%%%%%%%%%%%%%%%%%%%%%%%%%%%%%%%%%%%%%%%%%%%%%%%%%%%%%%%%%%%%%%
%%%%%%%%%%%%%%%%%%%%%%%%%%%%%%%%%%%%%%%%%%%%%%%%%%%%%%%%%%%%%%%%%%%%%%%%%%%%%%%%%%%%%%%%%%%%%%%%%%%%%%%%%%%
%%%%%%%%%%%%%%%%%%%%%%%%%%%%%%%%%%%%%%%%%%%%%%%%%%%%%%%%%%%%%%%%%%%%%%%%%%%%%%%%%%%%%%%%%%%%%%%%%%%%%%%%%%%
%%%%%%%%%%%%%%%%%%%%%%%%%%%%%%%%%%%%%%%%%%%%%%%%%%%%%%%%%%%%%%%%%%%%%%%%%%%%%%%%%%%%%%%%%%%%%%%%%%%%%%%%%%%
%%%%%%%%%%%%%%%%%%%%%%%%%%%%%%%%%%%%%%%%%%%%%%%%%%%%%%%%%%%%%%%%%%%%%%%%%%%%%%%%%%%%%%%%%%%%%%%%%%%%%%%%%%%
%%%%%%%%%%%%%%%%%%%%%%%%%%%%%%%%%%%%%%%%%%%%%%%%%%%%%%%%%%%%%%%%%%%%%%%%%%%%%%%%%%%%%%%%%%%%%%%%%%%%%%%%%%%
%%%%%%%%%%%%%%%%%%%%%%%%%%%%%%%%%%%%%%%%%%%%%%%%%%%%%%%%%%%%%%%%%%%%%%%%%%%%%%%%%%%%%%%%%%%%%%%%%%%%%%%%%%%
%%%%%%%%%%%%%%%%%%%%%%%%%%%%%%%%%%%%%%%%%%%%%%%%%%%%%%%%%%%%%%%%%%%%%%%%%%%%%%%%%%%%%%%%%%%%%%%%%%%%%%%%%%%
%%%%%%%%%%%%%%%%%%%%%%%%%%%%%%%%%%%%%%%%%%%%%%%%%%%%%%%%%%%%%%%%%%%%%%%%%%%%%%%%%%%%%%%%%%%%%%%%%%%%%%%%%%%
If $G$ is acyclic then, according to Theorem~\ref{summary}, any topological sort of $G$ provides a total order on the set of arcs
of the cross-section to compute their \ABpts\ in constant time per arc.  Unfortunately the digraph $G$ is not acyclic in general. 
The following (easy to check) theorem will be used in the next section to decide where to break the cycles of $G$ in order to be able to compute the \ABpts\ no longer 
in constant time but in constant amortized time per arc.

%%%%%%%%%%%%%%%%%%%%%%%%%%%%%%%%%%%%%%%%%%%%%%%%%%%%%%%%%%%%%%%%%%%%%%%%%%%%%%%%%%%%%%%%%%%%%%%%%%%%%%%%%%%
%%%%%%%%%%%%%%%%%%%%%%%%%%%%%%%%%%%%%%%%%%%%%%%%%%%%%%%%%%%%%%%%%%%%%%%%%%%%%%%%%%%%%%%%%%%%%%%%%%%%%%%%%%%
%%%%%%%%%%%%%%%%%%%%%%%%%%%%%%%%%%%%%%%%%%%%%%%%%%%%%%%%%%%%%%%%%%%%%%%%%%%%%%%%%%%%%%%%%%%%%%%%%%%%%%%%%%%
%%%%%%%%%%%%%%%%%%%%%%%%%%%%%%%%%%%%%%%%%%%%%%%%%%%%%%%%%%%%%%%%%%%%%%%%%%%%%%%%%%%%%%%%%%%%%%%%%%%%%%%%%%%
%%%%%%%%%%%%%%%%%%%%%%%%%%%%%%%%%%%%%%%%%%%%%%%%%%%%%%%%%%%%%%%%%%%%%%%%%%%%%%%%%%%%%%%%%%%%%%%%%%%%%%%%%%%
\begin{theorem}\label{table}
Let $\sigma$ be a left $1$-cell of $\MAC$. Then $\urf{\sigma} \in \fdt{\thr}{\thrset}(\MAC)$,
$\ulf{\sigma} \in \fdt{\one}{\thrset}(\MAC)$, and $\drf{\sigma} = \dlf{\sigma} 
\in \fdt{\thrset}{\two}(\MAC) \cup\setbotface$.
A similar result holds for right $1$-cell of $\MAC$  using
the conjugation relation $\fdt{i}{j} \circ \reverse = \reverse \circ \fdt{j}{i}$. \qed
\end{theorem}
%%%%%%%%%%%%%%%%%%%%%%%%%%%%%%%%%%%%%%%%%%%%%%%%%%%%%%%%%%%%%%%%%%%%%%%%%%%%%%%%%%%%%%%%%%%%%%%%%%%%%%%%%%%
%%%%%%%%%%%%%%%%%%%%%%%%%%%%%%%%%%%%%%%%%%%%%%%%%%%%%%%%%%%%%%%%%%%%%%%%%%%%%%%%%%%%%%%%%%%%%%%%%%%%%%%%%%%
%%%%%%%%%%%%%%%%%%%%%%%%%%%%%%%%%%%%%%%%%%%%%%%%%%%%%%%%%%%%%%%%%%%%%%%%%%%%%%%%%%%%%%%%%%%%%%%%%%%%%%%%%%%
%%%%%%%%%%%%%%%%%%%%%%%%%%%%%%%%%%%%%%%%%%%%%%%%%%%%%%%%%%%%%%%%%%%%%%%%%%%%%%%%%%%%%%%%%%%%%%%%%%%%%%%%%%%
%%%%%%%%%%%%%%%%%%%%%%%%%%%%%%%%%%%%%%%%%%%%%%%%%%%%%%%%%%%%%%%%%%%%%%%%%%%%%%%%%%%%%%%%%%%%%%%%%%%%%%%%%%%

\clearpage
%%%%%%%%%%%%%%%%%%%%%%%%%%%%%%%%%%%%%%%%%%%%%%%%%%%%%%%%%%%%%%%%%%%%%%%%%%%%%%%%%%%%%%%%%%%%%%%%%%%%%%%%%%%
%%%%%%%%%%%%%%%%%%%%%%%%%%%%%%%%%%%%%%%%%%%%%%%%%%%%%%%%%%%%%%%%%%%%%%%%%%%%%%%%%%%%%%%%%%%%%%%%%%%%%%%%%%%
%%%%%%%%%%%%%%%%%%%%%%%%%%%%%%%%%%%%%%%%%%%%%%%%%%%%%%%%%%%%%%%%%%%%%%%%%%%%%%%%%%%%%%%%%%%%%%%%%%%%%%%%%%%
%%%%%%%%%%%%%%%%%%%%%%%%%%%%%%%%%%%%%%%%%%%%%%%%%%%%%%%%%%%%%%%%%%%%%%%%%%%%%%%%%%%%%%%%%%%%%%%%%%%%%%%%%%%
%%%%%%%%%%%%%%%%%%%%%%%%%%%%%%%%%%%%%%%%%%%%%%%%%%%%%%%%%%%%%%%%%%%%%%%%%%%%%%%%%%%%%%%%%%%%%%%%%%%%%%%%%%%
\section{Our algorithm and its complexity analysis\label{sec4}}
We are now ready to describe our pseudotriangulation algorithm. 
Recall that our pseudotriangulation algorithm proceeds in three steps:
we first compute the  convex hull of the family of convex bodies,
then the cross-section of the visibility complex of the family of convex bodies with constraints
 assigned to a boundary bitangent line segment (not forgetting to add all the boundary bitangent line segments computed at the first step to the set of constraints), and 
finally the greedy pseudotriangulation associated with that cross-section.
The input of our pseudotriangulation algorithm is a finite planar family of pairwise
disjoint convex bodies together with a distinguished set of pairwise interior non-crossing 
 free bitangent line segments of the family: the constraints; the family is only given
by its chirotope, that is, for all triple of indices of the family the position vector of the
corresponding triple of convex bodies is computable in constant time;
equivalently, for any convex body of the family the relative counterclockwise circular order of any triple of bitangents 
tangent to that body is computable in constant time; cf. Appendix~\ref{TopPlanes}.
The family is denoted $\disks = \{o_0,o_1,\ldots,o_{n-1}\}$, the underlying
topological plane is denoted $\pointset$, and the left-left, left-right, right-left and right-right bitangent line segments joining the body $o_i$ to the body $o_{j}$
 are denoted $v_{ij}$, $v_{i\overline{j}}$, $v_{\overline{i}j}$ and $v_{\overline{ij}}$, respectively.

%%%%%%%%%%%%%%%%%%%%%%%%%%%%%%%%%%%%%%%%%%%%%%%%%%%%%%%%%%%%%%%%%%%%%%%%%%%%%%%%%%%%%%%%%%%%%%%%%%%%%%%%%%%
%%%%%%%%%%%%%%%%%%%%%%%%%%%%%%%%%%%%%%%%%%%%%%%%%%%%%%%%%%%%%%%%%%%%%%%%%%%%%%%%%%%%%%%%%%%%%%%%%%%%%%%%%%%
%%%%%%%%%%%%%%%%%%%%%%%%%%%%%%%%%%%%%%%%%%%%%%%%%%%%%%%%%%%%%%%%%%%%%%%%%%%%%%%%%%%%%%%%%%%%%%%%%%%%%%%%%%%
%%%%%%%%%%%%%%%%%%%%%%%%%%%%%%%%%%%%%%%%%%%%%%%%%%%%%%%%%%%%%%%%%%%%%%%%%%%%%%%%%%%%%%%%%%%%%%%%%%%%%%%%%%%
%%%%%%%%%%%%%%%%%%%%%%%%%%%%%%%%%%%%%%%%%%%%%%%%%%%%%%%%%%%%%%%%%%%%%%%%%%%%%%%%%%%%%%%%%%%%%%%%%%%%%%%%%%%
\subsection{Convex hull algorithm} Let $\Map{p}{\CoSur{}}{\pointset}$ be a (connected) 
4-sheeted branched covering of $\pointset$ ramified over the central point of
$o_0$, 
let $\tau$ be a generator of its automorphism group ($\approx \mathbb{Z}_4$),
and let $X_i$, $i\in \mathbb{Z}_4$, $\tau(X_i) = X_{i+1}$, be the four connected
components of the pre-image under 
$\CoSur{} \rightarrow \pointset$ of the complement in $\pointset$ of  
a curve $\gamma \subset o_0 \cup v_{01}$ joining the central
point of $o_0$ to the point at $+\infty$ on the bitangent $v_{01}.$
The sole lift $c$ of $o_0$ is called the central body, and 
the lift of $o_i$, $i \neq 0$, whose interior is entirely included in $X_k$, $k\in \mathbb{Z}_4$,
or intersects both $X_k$ and $X_{k+1}$ is denoted $\lift{o_i}{k}.$ 
%%%(or $o_i^k$ or $o_i^{\overbrace{'\cdots'}^{\text{$k$ times}}}$). 
We denote by $\straddle$ the family of $\lift{o_i}{k}$, $k \in \{0,1,2\}$, augmented with 
the central body. 
Our algorithm to compute the bitangent line segments of the convex hull of $\disks$ 
is based on the following three simple observations. 
\begin{enumerate}
\item The set of bitangent line segments of the family $\straddle$ and the set of pairs 
of crossing bitangent line segments of the family $\straddle$ 
depend only on the chirotope of the family $\disks$;  
\item The convex hull of the family $\disks$ can be extracted in linear time from the 
convex hull of the family $\straddle$; indeed, let $\CH$
be the counterclockwise linear sequence of bitangent line segments that appear in the boundary of 
the convex hull of the family $\straddle$
starting from the left-left bitangent line segment joining the central body to $\lift{o_1}{0}$, 
one can easily check (details are left to the reader) that   
the first bitangent line segment $v$ of $\CH$ entering an $\lift{o_i}{1}$ 
is well-defined, 
that $\tau(v)$ appears in the sequence $\CH$, and 
that the projection of the factor $v \ldots w$ of $\CH$ where $w$ is the bitangent 
line segment 
that precedes $\tau(v)$ in $\CH$ is the sequence of bitangent line segments
that appear in the boundary of the convex hull of $\disks.$
For example in the configuration depicted in  Figure~\ref{CCHplus} one has 
$\Sigma =\Sigma_1 v\Sigma_2 w\tau(v)\Sigma_3$
with $v = \ve{51'}$, $w\tau(v) = \ve{4'5'}\ve{5'1''}$,
$\Sigma_1 =\ve{01} \ve{12}\ve{23}\ve{34}\ve{45}$,
$\Sigma_2 =\ve{1'6}\ve{62'}\ve{2'3'}\ve{3'4'}$, and 
$\Sigma_3 =\ve{1''6'}\ve{6'2''}\ve{2''3''}\ve{3''4''}\ve{4''5''}\ve{5''6''}\ve{6''0}.$
The projection of the factor $v\Sigma_2w$ is the counterclockwise sequence of
boundary bitangent line segments;
%%%%%%%%%%%%%%%%%%%%%%%%%%%%%%%%%%%%%%%%%%%%%%%%%%%%%%%%%%%%%%%%%%%%%%%%%%%%%%%%%%%%%%%%%%%%%%%%%%%%%%%%%%%
%%%%%%%%%%%%%%%%%%%%%%%%%%%%%%%%%%%%%%%%%%%%%%%%%%%%%%%%%%%%%%%%%%%%%%%%%%%%%%%%%%%%%%%%%%%%%%%%%%%%%%%%%%%
%%%%%%%%%%%%%%%%%%%%%%%%%%%%%%%%%%%%%%%%%%%%%%%%%%%%%%%%%%%%%%%%%%%%%%%%%%%%%%%%%%%%%%%%%%%%%%%%%%%%%%%%%%%
%%%%%%%%%%%%%%%%%%%%%%%%%%%%%%%%%%%%%%%%%%%%%%%%%%%%%%%%%%%%%%%%%%%%%%%%%%%%%%%%%%%%%%%%%%%%%%%%%%%%%%%%%%%
%%%%%%%%%%%%%%%%%%%%%%%%%%%%%%%%%%%%%%%%%%%%%%%%%%%%%%%%%%%%%%%%%%%%%%%%%%%%%%%%%%%%%%%%%%%%%%%%%%%%%%%%%%%
\begin{figure}[!htb]
\footnotesize
\psfrag{v}{$v$}\psfrag{w}{$w$}
\psfrag{1}{$1$}\psfrag{2}{$2$}\psfrag{3}{$3$}\psfrag{4}{$4$}\psfrag{5}{$5$}\psfrag{6}{$6$}\psfrag{0}{$0$}
\psfrag{onep}{$1$}\psfrag{twop}{$2$}\psfrag{thrp}{$3$}\psfrag{foup}{$4$}\psfrag{fivp}{$5$}\psfrag{sixp}{}\psfrag{zerp}{$0$}
\psfrag{onepp}{$1'$}\psfrag{twopp}{$2'$}\psfrag{thrpp}{$3'$}\psfrag{foupp}{$4'$}\psfrag{fivpp}{$5'$}\psfrag{sixpp}{$6$}\psfrag{zerp}{$0$}
\psfrag{oneppp}{$1''$}\psfrag{twoppp}{$2''$}\psfrag{thrppp}{$3''$}\psfrag{fouppp}{$4''$}\psfrag{fivppp}{$5''$}\psfrag{sixppp}{$6'$}\psfrag{zerp}{$0$}
\psfrag{sixpppp}{$6''$}
\psfrag{gamma}{the curve $\gamma$}
\psfrag{plane}{the plane\ $\pointset$} \psfrag{sheet0}{$X_0$} \psfrag{sheet1}{$X_1$} \psfrag{sheet2}{$X_2$} \psfrag{sheet3}{$X_3$}
\begin{center}
\includegraphics[width=0.958575\linewidth]{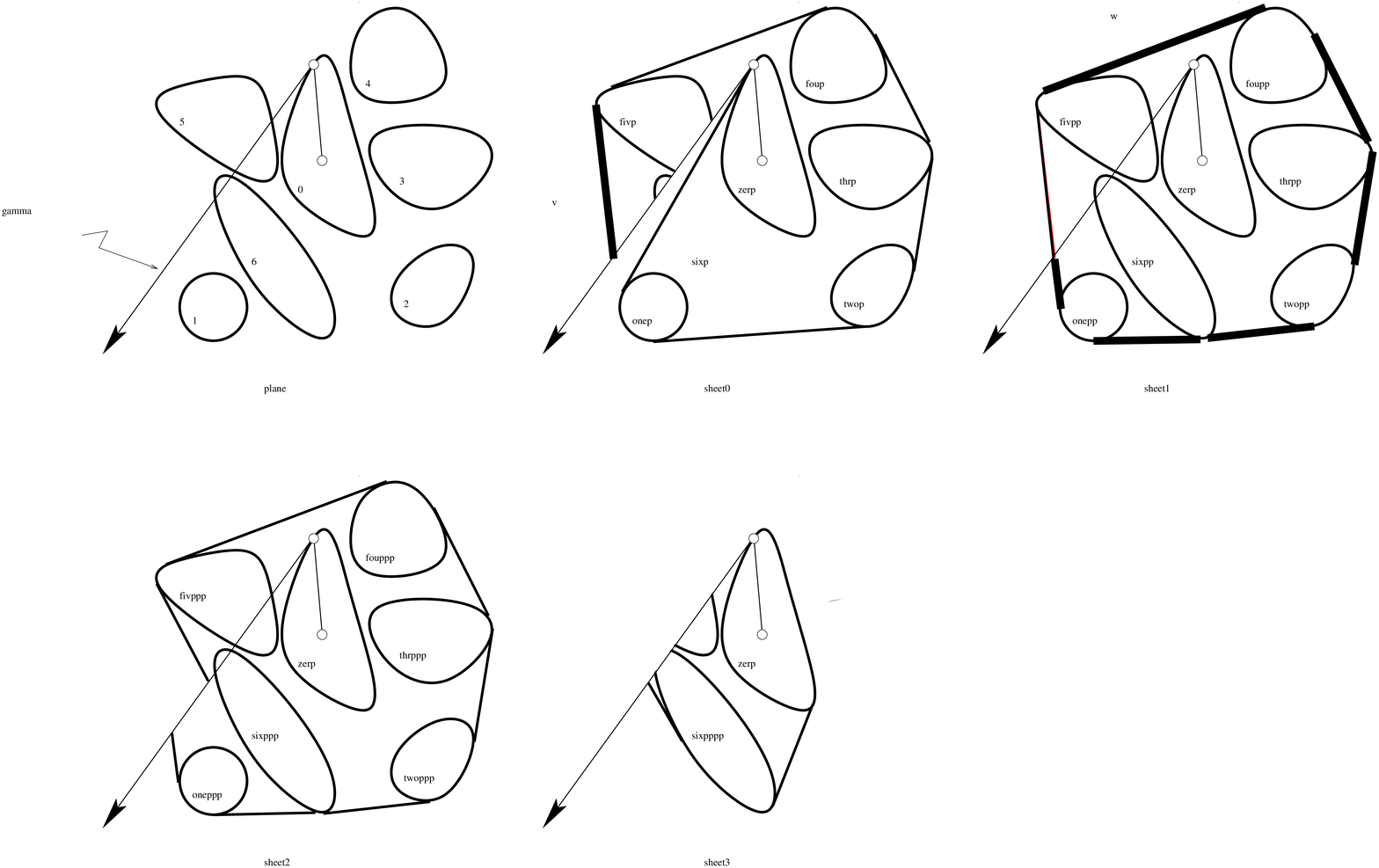}%%xfigfinal
\end{center}
\caption{\protect \small \label{CCHplus}}
\end{figure}
%%%%%%%%%%%%%%%%%%%%%%%%%%%%%%%%%%%%%%%%%%%%%%%%%%%%%%%%%%%%%%%%%%%%%%%%%%%%%%%%%%%%%%%%%%%%%%%%%%%%%%%%%%%
%%%%%%%%%%%%%%%%%%%%%%%%%%%%%%%%%%%%%%%%%%%%%%%%%%%%%%%%%%%%%%%%%%%%%%%%%%%%%%%%%%%%%%%%%%%%%%%%%%%%%%%%%%%
%%%%%%%%%%%%%%%%%%%%%%%%%%%%%%%%%%%%%%%%%%%%%%%%%%%%%%%%%%%%%%%%%%%%%%%%%%%%%%%%%%%%%%%%%%%%%%%%%%%%%%%%%%%
%%%%%%%%%%%%%%%%%%%%%%%%%%%%%%%%%%%%%%%%%%%%%%%%%%%%%%%%%%%%%%%%%%%%%%%%%%%%%%%%%%%%%%%%%%%%%%%%%%%%%%%%%%%
%%%%%%%%%%%%%%%%%%%%%%%%%%%%%%%%%%%%%%%%%%%%%%%%%%%%%%%%%%%%%%%%%%%%%%%%%%%%%%%%%%%%%%%%%%%%%%%%%%%%%%%%%%%

%%%%%%%%%%%%%%%%%%%%%%%%%%%%%%%%%%%%%%%%%%%%%%%%%%%%%%%%%%%%%%%%%%%%%%%%%%%%%%%%%%%%%%%%%%%%%%%%%%%%%%%%%%%
%%%%%%%%%%%%%%%%%%%%%%%%%%%%%%%%%%%%%%%%%%%%%%%%%%%%%%%%%%%%%%%%%%%%%%%%%%%%%%%%%%%%%%%%%%%%%%%%%%%%%%%%%%%
%%%%%%%%%%%%%%%%%%%%%%%%%%%%%%%%%%%%%%%%%%%%%%%%%%%%%%%%%%%%%%%%%%%%%%%%%%%%%%%%%%%%%%%%%%%%%%%%%%%%%%%%%%%
%%%%%%%%%%%%%%%%%%%%%%%%%%%%%%%%%%%%%%%%%%%%%%%%%%%%%%%%%%%%%%%%%%%%%%%%%%%%%%%%%%%%%%%%%%%%%%%%%%%%%%%%%%%
%%%%%%%%%%%%%%%%%%%%%%%%%%%%%%%%%%%%%%%%%%%%%%%%%%%%%%%%%%%%%%%%%%%%%%%%%%%%%%%%%%%%%%%%%%%%%%%%%%%%%%%%%%%
\item The Graham's scan to compute the convex hull of a family of points can be generalized  to families of
pairwise disjoint convex bodies provided that the bodies are sorted around a boundary body. 
Since the central body is, by construction, a boundary body of the family $\straddle$  we can apply the generalization of Graham's scan we have in mind 
to compute the convex hull of the family $\straddle$, and apply our second observation to derive the 
convex hull of the family $\disks$.
\end{enumerate}

%%%%%%%%%%%%%%%%%%%%%%%%%%%%%%%%%%%%%%%%%%%%%%%%%%%%%%%%%%%%%%%%%%%%%%%%%%%%%%%%%%%%%%%%%%%%%%%%%%%%%%%%%%%
%%%%%%%%%%%%%%%%%%%%%%%%%%%%%%%%%%%%%%%%%%%%%%%%%%%%%%%%%%%%%%%%%%%%%%%%%%%%%%%%%%%%%%%%%%%%%%%%%%%%%%%%%%%
%%%%%%%%%%%%%%%%%%%%%%%%%%%%%%%%%%%%%%%%%%%%%%%%%%%%%%%%%%%%%%%%%%%%%%%%%%%%%%%%%%%%%%%%%%%%%%%%%%%%%%%%%%%
%%%%%%%%%%%%%%%%%%%%%%%%%%%%%%%%%%%%%%%%%%%%%%%%%%%%%%%%%%%%%%%%%%%%%%%%%%%%%%%%%%%%%%%%%%%%%%%%%%%%%%%%%%%
%%%%%%%%%%%%%%%%%%%%%%%%%%%%%%%%%%%%%%%%%%%%%%%%%%%%%%%%%%%%%%%%%%%%%%%%%%%%%%%%%%%%%%%%%%%%%%%%%%%%%%%%%%%
We now explain our generalization of the Graham's scan.  

\begin{figure}[!htb]
\footnotesize
\psfrag{sweepline}{the sweep half-line $\ell$}
\psfrag{1}{$1$}\psfrag{2}{$2$}\psfrag{3}{$3$}\psfrag{4}{$4$}\psfrag{5}{$5$}\psfrag{6}{$6$}\psfrag{0}{$0$}
\psfrag{onep}{$1$}\psfrag{twop}{$2$}\psfrag{thrp}{$3$}\psfrag{foup}{$4$}\psfrag{fivp}{$5$}\psfrag{sixp}{}\psfrag{zerp}{$0$}
\psfrag{onepp}{$1'$}\psfrag{twopp}{$2'$}\psfrag{thrpp}{$3'$}\psfrag{foupp}{$4'$}\psfrag{fivpp}{$5'$}\psfrag{sixpp}{$6$}\psfrag{zerp}{$0$}
\psfrag{oneppp}{$1''$}\psfrag{twoppp}{$2''$}\psfrag{thrppp}{$3''$}\psfrag{fouppp}{$4''$}\psfrag{fivppp}{$5''$}\psfrag{sixppp}{$6'$}\psfrag{zerp}{$0$}
\psfrag{sixpppp}{$6''$}
\psfrag{gamma}{the curve $\gamma$}
\psfrag{plane}{the plane\ $\pointset$} \psfrag{sheet0}{$X_0$} \psfrag{sheet1}{$X_1$} \psfrag{sheet2}{$X_2$} \psfrag{sheet3}{$X_3$}
\begin{center}
\includegraphics[width=0.958575\linewidth]{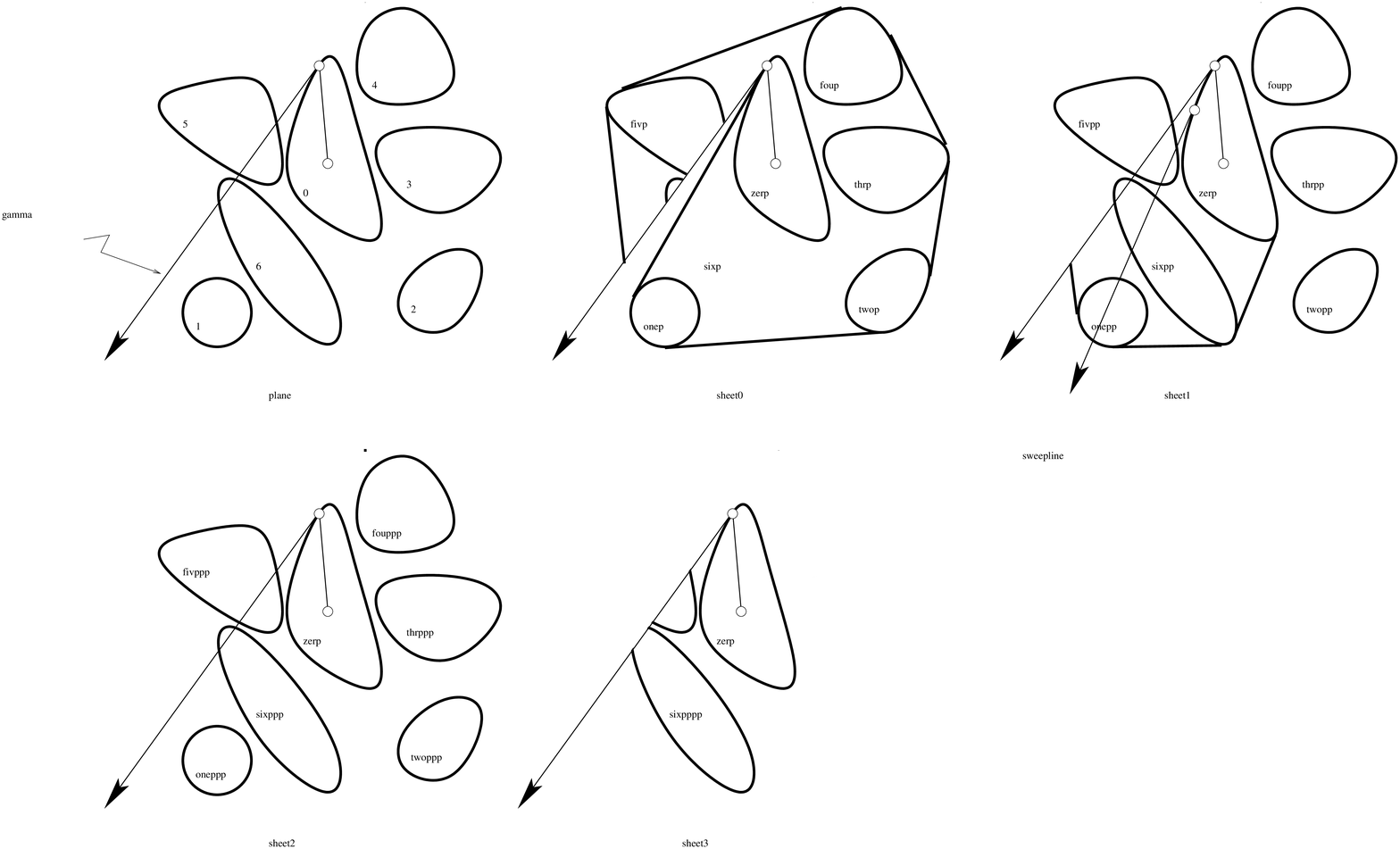}%%xfigfinal
\end{center}
\caption{\protect \small \label{GrahamScanCovering}}
\end{figure}

We perform a counterclockwise 
rotational sweep of the 4-sheeted branched covering space $\CoSur{}$ with a half-line whose supporting line 
is a left tangent to the central 
body $\central$ at its origin.
The lift in sheet $X_k$, $k \in \mathbb{Z}_4$,  of the half-line supporting the bitangent line segment
$v_{oi}$ with origin its tangency point upon $o_0$
is denoted $\lift{\ell_{i}}{k}.$  
The sweep starts at position $\lift{\ell_1}{0}$ and ends at position $\lift{\ell_1}{3}.$ 
During the sweep we maintain the convex hull 
of the subset of  bodies of $\straddle$ that have been entirely of partially swept by the sweeping 
half-line; to this end we keep track of a subset of the  bodies that intersect $\ell$;
therefore we update our data structures (to be defined in a second) when the sweeping half-line reaches the $\lift{\ell_{i}}{k}$:
an  enter event, and some of the $\lift{\ell_{\overline{i}}}{k}$: a leave event. 
For a given position $\ell$ of the sweep half-line, we define  
\begin{enumerate}
\item
$\CH(\ell)$ to be the linear sequence 
$v_1v_2\ldots v_k$ of bitangent line segments  encountered when walking counterclockwise 
along the boundary of the convex hull 
of the  bodies that has been reached so far by the sweep 
half-line,  starting from 
the left-left bitangent line segment joining the central body to $\lift{o_{1}}{0}$;
the body that the bitangent line segment $v_i$ reaches is denoted $o'_i$
(in particular  $o'_k$ is the central body);
\item $\vl{\ell}$ to be the bitangent line segment  $v_{j'}$ where $j'$ is the
minimal element of the subset of indexes~$j$ ($1 \leq j \leq k$) such that
an infinitesimal counterclockwise shift of $\ell$ pierces the  bodies $o'_i$ ($j\leq i \leq k$) in
the order $o'_k, o'_{k-1},\ldots, o'_j$; 
by construction $\ell$ pierces either $\vl{\ell}$, or its successor 
arc or its successor bitangent line segment; 
\item $\Queue(\ell)$ to be the list of bodies $o'_k o'_{k-1}\ldots o'_{j'}.$
%%(see Figure~\ref{GrahamScanCovering} for an illustration).
%% we denote by $\tau_k, \tau_{k-1}, \ldots, \tau_{j''}$  
%% the connected components of the complement of the union of the $o'_j$ 
%% in the intersection of the current convex hull with the 
%% ``left region'' defined by $\ell$, with the convention that  
%% $\tau_j$ is adjacent to $v_j$; note that $j'' = j'-1$ or $j'$ depending on
%% whether $\ell$ pierces transversely $\vl{\ell}$ or not; 
%% we note also that $\tau_j$ ($j\neq j'$) is included the convex hull 
%%of $o'_j$ and $o'_{j+1}$.  
%%(See Figure~\ref{pierce} for an illustration.) 
\end{enumerate}
For example in the configuration depicted in Figure~\ref{GrahamScanCovering} 
%%%%%%%%%%%%%%%%%%%%%%%%%%%%%%%%%%%%%%%%%%%%%%%%%%%%%%%%%%%%%%%%%%%%%%%%%%%%%%%%%%%%%%%%%%%%%%%%%%%%%%%%%%%
%%%%%%%%%%%%%%%%%%%%%%%%%%%%%%%%%%%%%%%%%%%%%%%%%%%%%%%%%%%%%%%%%%%%%%%%%%%%%%%%%%%%%%%%%%%%%%%%%%%%%%%%%%%
%%%%%%%%%%%%%%%%%%%%%%%%%%%%%%%%%%%%%%%%%%%%%%%%%%%%%%%%%%%%%%%%%%%%%%%%%%%%%%%%%%%%%%%%%%%%%%%%%%%%%%%%%%%
%%%%%%%%%%%%%%%%%%%%%%%%%%%%%%%%%%%%%%%%%%%%%%%%%%%%%%%%%%%%%%%%%%%%%%%%%%%%%%%%%%%%%%%%%%%%%%%%%%%%%%%%%%%
%%%%%%%%%%%%%%%%%%%%%%%%%%%%%%%%%%%%%%%%%%%%%%%%%%%%%%%%%%%%%%%%%%%%%%%%%%%%%%%%%%%%%%%%%%%%%%%%%%%%%%%%%%%
%%%%%%%%%%%%%%%%%%%%%%%%%%%%%%%%%%%%%%%%%%%%%%%%%%%%%%%%%%%%%%%%%%%%%%%%%%%%%%%%%%%%%%%%%%%%%%%%%%%%%%%%%%%
%%%%%%%%%%%%%%%%%%%%%%%%%%%%%%%%%%%%%%%%%%%%%%%%%%%%%%%%%%%%%%%%%%%%%%%%%%%%%%%%%%%%%%%%%%%%%%%%%%%%%%%%%%%
%%%%%%%%%%%%%%%%%%%%%%%%%%%%%%%%%%%%%%%%%%%%%%%%%%%%%%%%%%%%%%%%%%%%%%%%%%%%%%%%%%%%%%%%%%%%%%%%%%%%%%%%%%%
the set of convex bodies entirely or partially swept by the current sweeping  half-line $\ell$ is 
$\{0,1,2,3,4,5,6,1'\}$ and one has 
\begin{eqnarray*}
\CH(\ell) & = & \ve{01} \ve{12}\ve{23}\ve{34}\ve{45}\ve{51'}\ve{1'6}\ve{60},\\
\Queue(\ell) & = & 061',  \\
\vl{\ell} & = & \ve{51'}.
\end{eqnarray*}

%%%%%%%%%%%%%%%%%%%%%%%%%%%%%%%%%%%%%%%%%%%%%%%%%%%%%%%%%%%%%%%%%%%%%%%%%%%%%%%%%%%%%%%%%%%%%%%%%%%%%%%%%%%
%%%%%%%%%%%%%%%%%%%%%%%%%%%%%%%%%%%%%%%%%%%%%%%%%%%%%%%%%%%%%%%%%%%%%%%%%%%%%%%%%%%%%%%%%%%%%%%%%%%%%%%%%%%
%%%%%%%%%%%%%%%%%%%%%%%%%%%%%%%%%%%%%%%%%%%%%%%%%%%%%%%%%%%%%%%%%%%%%%%%%%%%%%%%%%%%%%%%%%%%%%%%%%%%%%%%%%%
%%%%%%%%%%%%%%%%%%%%%%%%%%%%%%%%%%%%%%%%%%%%%%%%%%%%%%%%%%%%%%%%%%%%%%%%%%%%%%%%%%%%%%%%%%%%%%%%%%%%%%%%%%%
%%%%%%%%%%%%%%%%%%%%%%%%%%%%%%%%%%%%%%%%%%%%%%%%%%%%%%%%%%%%%%%%%%%%%%%%%%%%%%%%%%%%%%%%%%%%%%%%%%%%%%%%%%%
During the sweep, we maintain the lists $\CH(\ell)$, $\Queue(\ell)$, and the bitangent line segment $\vl{\ell}$.
Initially 
$\ell = \lift{\ell_{1}}{0}$, 
$\vl{\ell}$ is the left-left bitangent line segment $v_{\central, \lift{o_{1}}{0}}$ joining the central body to $\lift{o_{1}}{0}$,
$\CH(\ell)
= [v_{\central, \lift{o_{1}}{0}}, v_{\lift{o_{1}}{0},\central}]$, 
 and
 $\Queue(\ell) = [\central, \lift{o_{1}}{0}]$.
We store $\Queue(\ell)$ in a binary search tree. 
Assume we are to process the enter-event $\next$ for the body $o$ successor
 of  the enter or leave event $\previous$.  Let $r$ be the
rightmost body of $\Queue(\previous)$, $r_*$ its predecessor along $\CH(\previous)$, and
$r^*$ its successor, if any, that is, if $r$ is  not the central body.
Now, we explain how to update $\CH$ and $\Queue$. First we locate $o$ in $\Queue$.
Assume first that $o$ is to the left of $r$. Let $\alpha$ and
$\alpha'$ be its left-hand and right-hand neighbors in $\Queue(\previous)$. If
$o$ does not intersect $v = v_{\alpha\alpha'}$ then $o$ is included in the convex
hull of $\alpha$ and $\alpha'$ and we just ignore $o$:
$\CH(\next)=\CH(\previous)$, $\Queue(\next)=\Queue(\previous)$ and  $\vl{\next}= \vl{\previous}$. 
Otherwise we insert $o$ into $\Queue$, and update $\CH$: we split $\CH$ at $v$, and
we regard the two resulting parts as stacks of arcs whose respective heads
are the arcs contributed by $\alpha$ and $\alpha'$.  Then we pop from the
left-hand stack until an arc $\beta$, say supported by body $o'$, is met
such that $o'$ is the central body or $v = v_{o,o'}$ reaches $\beta$.
 Similarly we pop
from the right-hand stack until an arc $\beta'$, say supported by body $o''$, is
met such that $v'' = v_{o''o}$ reaches $\beta'.$
Then, we shorten $\beta$ and 
$\beta'$: the source of $\beta$ and the sink of $\beta'$  are replaced with
$v_{oo'}$ and  $v_{o''o}$, respectively.  Then, to build $\CH(\next)$, we
concatenate what is left of the two stacks, with the arc, say $\delta$,
of $\partial o$ with source $v_{o''o}$ and sink $v_{oo'}$ in between. When
an arc that follows the arc $a(\previous)$ that $\vl{\ell}$ reaches (included) in $\CH(\previous)$ is popped, its
supporting body is removed from $\Queue$. 
If $r(\previous)$ is removed from $\Queue$, 
and  the body supporting the predecessor of $\delta$
along $\CH(\next)$ intersects
$\next$ at the right of $o$, we insert it into
$\Queue$, so that it becomes $r(\next)$ instead of $o$.
If $r(\previous)$ is removed from $\Queue$, 
and  
the body supporting the predecessor of $\delta$
along $\CH(\next)$ does not intersect
$\next$ at the right of $o$, $o$ becomes $r(\next)$.
Assume now that $o$ is to the right of $r$.
We discard $o$ when $o$ is included in the current convex hull, that is, if 
$o$ is included  in the convex hull of $r,r_*$  and $r^*$.
Otherwise, we proceed
as in the previous case, except that we split $\CH(\previous)$ through the arc $a$
that $\vl{\ell}$ reaches 
instead of bitangent $v_{\alpha'\alpha}$ (that is, there is one copy of $a$
at the head of both stacks), and a body is removed from $\Queue$ only if an
arc it supports is popped from the left-hand stack. The body $\liftoperator$
supporting the predecessor of $\delta$ along $\CH(\next)$ is inserted into $\Queue$
if $a(\previous)$ has been popped from the right-hand stack and $\liftoperator$ intersects
$\next$ at the right of $o.$
Finally concerning the leave events only leave events for $r$ need to be processed. The processing of
those events simply consists in removing $r$ from $\Queue.$ 

\begin{figure}[!htb]
\small
\psfrag{theplane}{the plane}
\psfrag{sheet1}{sheet 0}
\psfrag{sheet2}{sheet 1}
\psfrag{sheet3}{sheet 2}
\psfrag{sheet4}{sheet 4}
\begin{center}
\includegraphics[width=0.98575\linewidth]{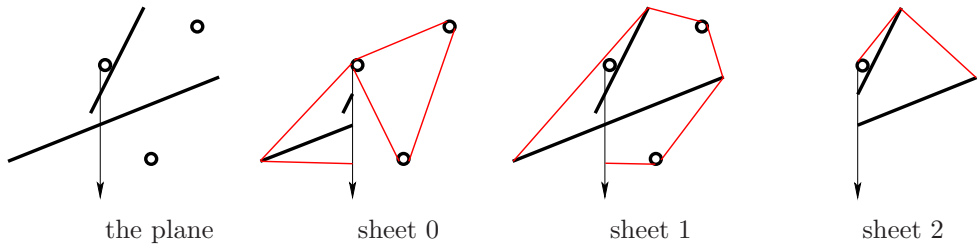}
\end{center}
\caption{\protect \small Three sheets are not enough.}
\label{notenough}
\end{figure}

%%%%%%%%%%%%%%%%%%%%%%%%%%%%%%%%%%%%%%%%%%%%%%%%%%%%%%%%%%%%%%%%%%%%%%%%%%%%%%%%%%%%%%%%%%%%%%%%%%%%%%%%%%%
%%%%%%%%%%%%%%%%%%%%%%%%%%%%%%%%%%%%%%%%%%%%%%%%%%%%%%%%%%%%%%%%%%%%%%%%%%%%%%%%%%%%%%%%%%%%%%%%%%%%%%%%%%%
%%%%%%%%%%%%%%%%%%%%%%%%%%%%%%%%%%%%%%%%%%%%%%%%%%%%%%%%%%%%%%%%%%%%%%%%%%%%%%%%%%%%%%%%%%%%%%%%%%%%%%%%%%%
%%%%%%%%%%%%%%%%%%%%%%%%%%%%%%%%%%%%%%%%%%%%%%%%%%%%%%%%%%%%%%%%%%%%%%%%%%%%%%%%%%%%%%%%%%%%%%%%%%%%%%%%%%%
%%%%%%%%%%%%%%%%%%%%%%%%%%%%%%%%%%%%%%%%%%%%%%%%%%%%%%%%%%%%%%%%%%%%%%%%%%%%%%%%%%%%%%%%%%%%%%%%%%%%%%%%%%%
Finally we mention that if, instead of working in a $4$-sheeted covering, we work in a $3$-sheeted
covering, and lift the bodies accordingly, then the convex hull of the
lifts still contains the convex hull of the bodies as a factor but our criteria to locate efficiently this factor breaks down; see
Figure~\ref{notenough} for an illustration.

%%%%%%%%%%%%%%%%%%%%%%%%%%%%%%%%%%%%%%%%%%%%%%%%%%%%%%%%%%%%%%%%%%%%%%%%%%%%%%%%%%%%%%%%%%%%%%%%%%%%%%%%%%%

%%%%%%%%%%%%%%%%%%%%%%%%%%%%%%%%%%%%%%%%%%%%%%%%%%%%%%%%%%%%%%%%%%%%%%%%%%%%%%%%%%%%%%%%%%%%%%%%%%%%%%%%%%%
%%%%%%%%%%%%%%%%%%%%%%%%%%%%%%%%%%%%%%%%%%%%%%%%%%%%%%%%%%%%%%%%%%%%%%%%%%%%%%%%%%%%%%%%%%%%%%%%%%%%%%%%%%%
%%%%%%%%%%%%%%%%%%%%%%%%%%%%%%%%%%%%%%%%%%%%%%%%%%%%%%%%%%%%%%%%%%%%%%%%%%%%%%%%%%%%%%%%%%%%%%%%%%%%%%%%%%%
%%%%%%%%%%%%%%%%%%%%%%%%%%%%%%%%%%%%%%%%%%%%%%%%%%%%%%%%%%%%%%%%%%%%%%%%%%%%%%%%%%%%%%%%%%%%%%%%%%%%%%%%%%%
%%%%%%%%%%%%%%%%%%%%%%%%%%%%%%%%%%%%%%%%%%%%%%%%%%%%%%%%%%%%%%%%%%%%%%%%%%%%%%%%%%%%%%%%%%%%%%%%%%%%%%%%%%%

\clearpage
%%%%%%%%%%%%%%%%%%%%%%%%%%%%%%%%%%%%%%%%%%%%%%%%%%%%%%%%%%%%%%%%%%%%%%%%%%%%%%%%%%%%%%%%%%%%%%%%%%%%%%%%%%%
%%%%%%%%%%%%%%%%%%%%%%%%%%%%%%%%%%%%%%%%%%%%%%%%%%%%%%%%%%%%%%%%%%%%%%%%%%%%%%%%%%%%%%%%%%%%%%%%%%%%%%%%%%%
%%%%%%%%%%%%%%%%%%%%%%%%%%%%%%%%%%%%%%%%%%%%%%%%%%%%%%%%%%%%%%%%%%%%%%%%%%%%%%%%%%%%%%%%%%%%%%%%%%%%%%%%%%%
%%%%%%%%%%%%%%%%%%%%%%%%%%%%%%%%%%%%%%%%%%%%%%%%%%%%%%%%%%%%%%%%%%%%%%%%%%%%%%%%%%%%%%%%%%%%%%%%%%%%%%%%%%%
\subsection{Cross-section algorithm}
As said in the introduction our cross-section algorithm is a sweep of the convex hull of the bodies by a half-line 
whose supporting line is a left tangent at the origin of the half-line to a distinguished boundary body. 
The sweep starts at a distinguished boundary bitangent line segment leaving the
distinguished boundary body (there might be several), and during the sweep we construct the cross-section of the visibility complex of the 
family of bodies with constraints assigned to the distinguished boundary bitangent line segment.
%% and---although this step is not strictly necessary as explain 
%%in the last paragraph (Final remarks) of this section---we identify the $2$-cells whose sink bitangent line segments are boundary bitangent line segments or constraints 
%%and compute their sink bitangent line segments, see Figure~\ref{CrossSection} for an illustration. Actually the main difficulty
%%is to construct a cross-section because one can easily see that identifying the $2$-cells whose sinks
%%are boundary bitangent line segments, and computing their sink bitangent line segments, is reducible to computing 
%%the cross-section of the visibility complex of the constrained family of convex bodies obtained by adding to the set of constraints the boundary bitangent line
%%segments. So let us concentrate on the problem of computing the cross-section. 
The construction of the cross-section boils down to maintaining during the sweep the ordered
sequence of forward and backward views{}\footnote{The views are the connected pieces of the boundary of free space cut at its cusp points--there is  one cusp point per endpoint of constraint---and at the contact points of the left-left and left-right bitangents joining the distinguished boundary body to the other bodies of the family.}
 of the lines of free space supported by
the sweeping half-line, that is,  whose second coordinate is the sweeping half-line. 
However it is asking to much of our chirotope because a view might start or (not exclusive) end at the endpoint of a constraint, and 
we have already observed that the relative positions of the endpoints of the constraints with respect to the bitangents are not
completely determined by the chirotope of the convex bodies. To overcome this
difficulty we are going to embed the views into larger boundary curves---called {\it paths} in the sequel---that start and
end only at the touching points of the sweep half-lines with the convex bodies, and not at the endpoints of the constraints. 
So we reduce the problem of computing the cross-section to the problem of maintaining during the sweep the ordered sequence of paths pierced 
by the sweeping half-line---each path being represented by a subpath of constant complexity, called its {\it window},  because contrary to the sum of the complexities of the views 
the sum of the complexities of the paths is not necessarily linear. We let the (standard) details of the maintenance of the pierced paths by the sweep half-line 
to the reader and we concentrate on the definition of the paths and on the definition of their associated windows.

\begin{figure}[!htb]
\begin{center}
\psfrag{oi}{$o_i$}\psfrag{taui}{$\tau_i$}
\psfrag{di}{$d_i$}\psfrag{gi}{$g_i$}
\psfrag{ai}{$a_i$}\psfrag{Ai}{$A_i$}
\psfrag{Mi}{$M_i$}\psfrag{mi}{$m_i$}
\psfrag{M}{$M$}\psfrag{m}{$m$}
\psfrag{A}{$A$}\psfrag{a}{$a$}
\includegraphics[width = 0.8575\linewidth]{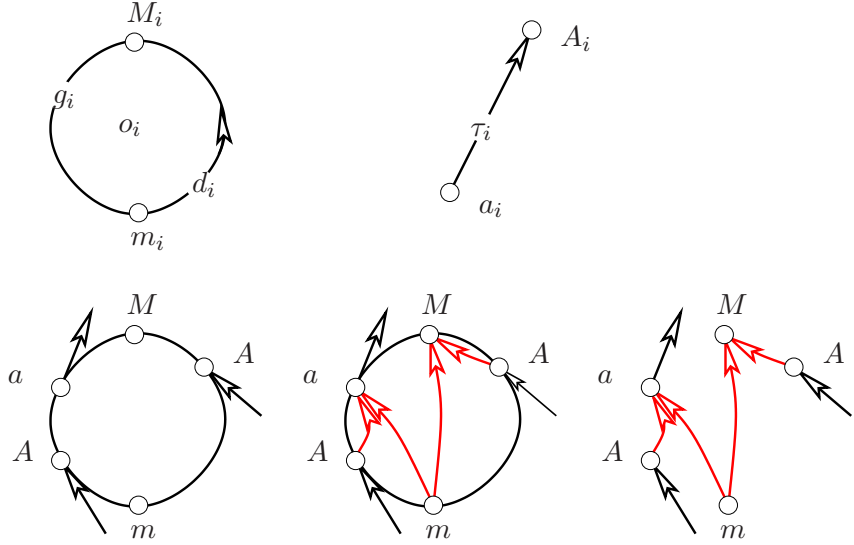}%xfigfinal
\caption{\protect \small Construction of the geometric graph $G'.$ 
\label{secondstep}}
\end{center}
\end{figure}

%%%%%%%%%%%%%%%%%%%%%%%%%%%%%%%%%%%%%%%%%%%%%%%%%%%%%%%%%%%%%%%%%%%%%%%%%%%%%%%%%%%%%%%%%%%%%%%%%%%%%%%%%%%
%%%%%%%%%%%%%%%%%%%%%%%%%%%%%%%%%%%%%%%%%%%%%%%%%%%%%%%%%%%%%%%%%%%%%%%%%%%%%%%%%%%%%%%%%%%%%%%%%%%%%%%%%%%
%%%%%%%%%%%%%%%%%%%%%%%%%%%%%%%%%%%%%%%%%%%%%%%%%%%%%%%%%%%%%%%%%%%%%%%%%%%%%%%%%%%%%%%%%%%%%%%%%%%%%%%%%%%
%%%%%%%%%%%%%%%%%%%%%%%%%%%%%%%%%%%%%%%%%%%%%%%%%%%%%%%%%%%%%%%%%%%%%%%%%%%%%%%%%%%%%%%%%%%%%%%%%%%%%%%%%%%
%%%%%%%%%%%%%%%%%%%%%%%%%%%%%%%%%%%%%%%%%%%%%%%%%%%%%%%%%%%%%%%%%%%%%%%%%%%%%%%%%%%%%%%%%%%%%%%%%%%%%%%%%%%
Assume without loss of generality that $o_0$ is a boundary body and that the sweep is done around $o_0$.
Let $G$ be the geometric graph union of the boundaries of the convex bodies $o_i$ (except $o_0$) 
and the constraints $\tau_i$ (we delete from the set of constraints the constraints
incident to the body $o_0$). 
Let $m_i$ and $M_i$ be the touching points with the body $o_i$ of the left-left and left-right
bitangents joining $o_0$ to $o_i$; similarly let $a_i$ and $A_i$ 
be the touching points with the constraint $\tau_i$ of the left-left and left-right bitangents joining 
$o_0$ to $\tau_i.$  The points $m_i$ and $M_i$ split the
boundary of $o_i$ into two arcs $g_i$ and $d_i$ where by convention $d_i$ joins $m_i$ to $M_i$ when walking counterclockwise around the boundary of $o_i$. 
The arcs $g_i$ and $d_i$ are
oriented from $m_i$ to $M_i$, and the  constraint $\tau_i$ is oriented from $a_i$
to $A_i.$ This turns the graph $G$ into a directed acyclic graph that is
monotone with respect to sweeping half-line coordinate. 
We construct a new geometric graph $G'$ in two steps: for every body $o_i$ we firstly add to $G$ the chords of $o_i$ joining  
(1) $m_i$ to $M_i$, (2) $m_i$ to every $a_j \in \partial o_i$, and 
(3) every $A_j \in \partial o_i$ to the first $a_k \in \partial o_i$ that follows, if any;
$A_j$ to $M_i$ otherwise---and secondly we delete the body boundary edges of $G$.
(See Figure~\ref{secondstep} for an  illustration.)
The unique maximal monotone path in $G'$ 
whose first constraint atom is $\tau_j$ is denoted $p(\tau_j)$;  
similarly the path reduced to the chord  joining $m_i$ to $M_i$ is denoted $p(o_i).$  
The set of paths $p(\tau_j)$ and $p(o_i)$ is denoted $\paths.$

%%%%%%%%%%%%%%%%%%%%%%%%%%%%%%%%%%%%%%%%%%%%%%%%%%%%%%%%%%%%%%%%%%%%%%%%%%%%%%%%%%%%%%%%%%%%%%%%%%%%%%%%%%%
%%%%%%%%%%%%%%%%%%%%%%%%%%%%%%%%%%%%%%%%%%%%%%%%%%%%%%%%%%%%%%%%%%%%%%%%%%%%%%%%%%%%%%%%%%%%%%%%%%%%%%%%%%%
%%%%%%%%%%%%%%%%%%%%%%%%%%%%%%%%%%%%%%%%%%%%%%%%%%%%%%%%%%%%%%%%%%%%%%%%%%%%%%%%%%%%%%%%%%%%%%%%%%%%%%%%%%%
%%%%%%%%%%%%%%%%%%%%%%%%%%%%%%%%%%%%%%%%%%%%%%%%%%%%%%%%%%%%%%%%%%%%%%%%%%%%%%%%%%%%%%%%%%%%%%%%%%%%%%%%%%%
%%%%%%%%%%%%%%%%%%%%%%%%%%%%%%%%%%%%%%%%%%%%%%%%%%%%%%%%%%%%%%%%%%%%%%%%%%%%%%%%%%%%%%%%%%%%%%%%%%%%%%%%%%%

%%%%%%%%%%%%%%%%%%%%%%%%%%%%%%%%%%%%%%%%%%%%%%%%%%%%%%%%%%%%%%%%%%%%%%%%%%%%%%%%%%%%%%%%%%%%%%%%%%%%%%%%%%%
%%%%%%%%%%%%%%%%%%%%%%%%%%%%%%%%%%%%%%%%%%%%%%%%%%%%%%%%%%%%%%%%%%%%%%%%%%%%%%%%%%%%%%%%%%%%%%%%%%%%%%%%%%%
%%%%%%%%%%%%%%%%%%%%%%%%%%%%%%%%%%%%%%%%%%%%%%%%%%%%%%%%%%%%%%%%%%%%%%%%%%%%%%%%%%%%%%%%%%%%%%%%%%%%%%%%%%%
%%%%%%%%%%%%%%%%%%%%%%%%%%%%%%%%%%%%%%%%%%%%%%%%%%%%%%%%%%%%%%%%%%%%%%%%%%%%%%%%%%%%%%%%%%%%%%%%%%%%%%%%%%%
%%%%%%%%%%%%%%%%%%%%%%%%%%%%%%%%%%%%%%%%%%%%%%%%%%%%%%%%%%%%%%%%%%%%%%%%%%%%%%%%%%%%%%%%%%%%%%%%%%%%%%%%%%%
We now introduce the window of a path assigned to a sweeping half-line. 
Let $p$ be a path of $\paths$ and let $\ell$ be a sweeping half-line piercing the path $p$, let $o'$ be the first body pierced by $\ell$ among the 
bodies---contributing for a chord to the path $p$---lying at the right of the path $p$ and let $c'$ be a chord of the path $p$ supported by $o'$; similarly let $o''$ be the last body pierced by $\ell$
among the bodies lying at the left of the path $p$ and let $c''$ be a chord of the path $p$ supported by $o''$.
We define a set of at most six  chords $\epsilon_j,\epsilon'_j$, $j=1,2,3$, as follows:
(1) $\epsilon_1$ is the highest chord of the path $p$ supported by a body below $\ell$,  
$\epsilon_2$ is the highest chord of  the path $p$ below $c'$ 
supported by a body pierced by $\ell$  and following (strictly) $o'$, and 
$\epsilon_3$ is the highest chord below $c''$ supported by a body
pierced by $\ell$ and preceding $o''$; and similarly 
(2) $\epsilon'_1$ is  the smallest chord of the path $p$ supported by a body above $\ell$,
$\epsilon'_2$ is the lowest chord above $c'$ supported by a body
pierced by $\ell$  and following (strictly) $o'$, and  
$\epsilon'_3$ is the lowest chord of the path $p$ above $c''$ supported by
a body pierced by $\ell$ and preceding $o''$.
The window of the path $p$ at $\ell$ is denoted $f(p,\ell)$ and is defined as the subpath of $p$ defined as the intersection of the paths $p$,
$p_+(\epsilon_j)$ and $p_{-}(\epsilon'_j)$, $j=1,2,3$, 
where for any chord $c$ of the path $p$, the subpath $p_+(c)$ 
is the suffix subpath of $p$ starting at the constraint following $c$
 and similarly the subpath $ p_{-}(c)$ is the prefix subpath of $p$  ending at the constraint
preceding $c$. A simple case analysis leads to the
following theorem.

\begin{theorem} Let $p$ be a path of $\paths$ pierced by the sweeping half-line
$\ell$. Then the subpath  $f(p,\ell)$ of $p$ is well-defined, contains at most two chords, and is pierced by the
half-line $\ell$.  \qed
\end{theorem}

%%%%%%%%%%%%%%%%%%%%%%%%%%%%%%%%%%%%%%%%%%%%%%%%%%%%%%%%%%%%%%%%%%%%%%%%%%%%%%%%%%%%%%%%%%%%%%%%%%%%%%%%%%%
%%%%%%%%%%%%%%%%%%%%%%%%%%%%%%%%%%%%%%%%%%%%%%%%%%%%%%%%%%%%%%%%%%%%%%%%%%%%%%%%%%%%%%%%%%%%%%%%%%%%%%%%%%%
%%%%%%%%%%%%%%%%%%%%%%%%%%%%%%%%%%%%%%%%%%%%%%%%%%%%%%%%%%%%%%%%%%%%%%%%%%%%%%%%%%%%%%%%%%%%%%%%%%%%%%%%%%%
%%%%%%%%%%%%%%%%%%%%%%%%%%%%%%%%%%%%%%%%%%%%%%%%%%%%%%%%%%%%%%%%%%%%%%%%%%%%%%%%%%%%%%%%%%%%%%%%%%%%%%%%%%%
%%%%%%%%%%%%%%%%%%%%%%%%%%%%%%%%%%%%%%%%%%%%%%%%%%%%%%%%%%%%%%%%%%%%%%%%%%%%%%%%%%%%%%%%%%%%%%%%%%%%%%%%%%%
It remains to recall the definition of the views and to embed the views into the paths. 
Let $S$ be the surface obtained by cutting the free part of the convex hull of the bodies along
the constraints---note that $S$ is not
necessarily connected. The cusp points of $\partial S$ (one cusp point $\hat{x}$ per endpoint $x$ of constraint) 
and the extreme points $m_i, M_i$ of the bodies induce a natural decomposition of $\partial S$ into 
monotone convex paths $v_i$, called the views. 
%%%%%%%%%%%%%%%%%%%%%%%%%%%%%%%%%%%%%%%%%%%%%%%%%%%%%%%%%%%%%%%%%%%%%%%%%%%%%%%%%%%%%%%%%%%%%%%%%%%%%%%%%%%
%%%%%%%%%%%%%%%%%%%%%%%%%%%%%%%%%%%%%%%%%%%%%%%%%%%%%%%%%%%%%%%%%%%%%%%%%%%%%%%%%%%%%%%%%%%%%%%%%%%%%%%%%%%
%%%%%%%%%%%%%%%%%%%%%%%%%%%%%%%%%%%%%%%%%%%%%%%%%%%%%%%%%%%%%%%%%%%%%%%%%%%%%%%%%%%%%%%%%%%%%%%%%%%%%%%%%%%
\begin{figure}[!htb]
\begin{center}
\psfrag{Fmi}{$F(m_i)$}
\psfrag{Bmi}{$B(m_i)$}
\psfrag{Aai}{$A(a_i)$}
\psfrag{Bai}{$B(a_i)$}
\psfrag{oi}{$o_i$}\psfrag{taui}{$\tau_i$}
\psfrag{di}{$d_i$}\psfrag{gi}{$g_i$}
\psfrag{ai}{$a_i$}\psfrag{Ai}{$A_i$}
\psfrag{aihat}{$\hat{a}_i$}\psfrag{Ai}{$A_i$}
\psfrag{Mi}{$M_i$}\psfrag{mi}{$m_i$}
\psfrag{M}{$M$}\psfrag{m}{$m$}
\psfrag{A}{$A$}\psfrag{a}{$a$}
\includegraphics[width = 0.8575\linewidth]{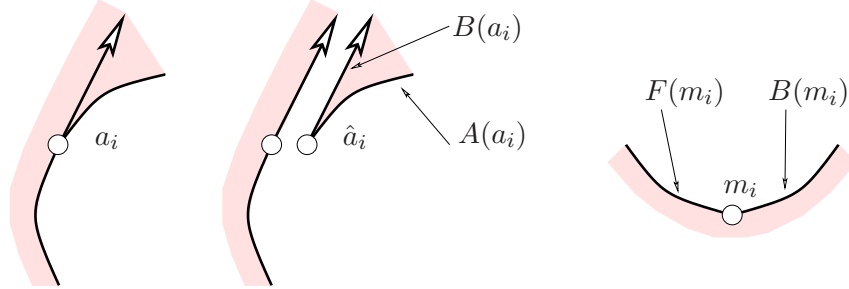}%xfig
\caption{\protect \small The views.  \label{views}}
\end{center}
\end{figure}
%%%%%%%%%%%%%%%%%%%%%%%%%%%%%%%%%%%%%%%%%%%%%%%%%%%%%%%%%%%%%%%%%%%%%%%%%%%%%%%%%%%%%%%%%%%%%%%%%%%%%%%%%%%
%%%%%%%%%%%%%%%%%%%%%%%%%%%%%%%%%%%%%%%%%%%%%%%%%%%%%%%%%%%%%%%%%%%%%%%%%%%%%%%%%%%%%%%%%%%%%%%%%%%%%%%%%%%
%%%%%%%%%%%%%%%%%%%%%%%%%%%%%%%%%%%%%%%%%%%%%%%%%%%%%%%%%%%%%%%%%%%%%%%%%%%%%%%%%%%%%%%%%%%%%%%%%%%%%%%%%%%
%%%%%%%%%%%%%%%%%%%%%%%%%%%%%%%%%%%%%%%%%%%%%%%%%%%%%%%%%%%%%%%%%%%%%%%%%%%%%%%%%%%%%%%%%%%%%%%%%%%%%%%%%%%
%%%%%%%%%%%%%%%%%%%%%%%%%%%%%%%%%%%%%%%%%%%%%%%%%%%%%%%%%%%%%%%%%%%%%%%%%%%%%%%%%%%%%%%%%%%%%%%%%%%%%%%%%%%
There are two views per vertex $a_i$---a view $B(a_i)$ whose first atom is the constraint leaving $\hat{a}_i$ 
and a view $A(a_i)$ whose first atom is an arc leaving $\hat{a}_i$---and also two views per vertex $m_i$---a view $F(m_i)$ whose first atom is a subarc of 
$g_i$ leaving $m_i$ and a view $B(m_i)$ whose first atom is an subarc of $d_i$ leaving $m_i$, see Figure~\ref{views} for an illustration.
%%% (Idem  with the $A_i$ and $M_i$.)
Let $\tau_i$ be a constraint leaving the convex body $o_j$.
The view $B(a_i)$ is assigned to the path $p(\tau_i)$ and the
view $A(a_i)$ is assigned to the path $p(\tau'_i)$ where $\tau'_i$ is the
constraint leaving  $o_i$ that follows $\tau_i$, if any; to the path $p(o_j)$
otherwise. The forward view $F(m_i)$ is assigned to the path $p(\tau_j)$ 
where $\tau_j$ is the first constraint leaving $g_i$ if any;  to the path
$p(o_i)$ otherwise. 
Similarly the view $B(m_i)$ is assigned to the path $p(\tau_j)$ 
where $\tau_j$ is the first constraint leaving $d_i$ if any;  to the path
$p(o_i)$ otherwise. Note that exactly two views are therefore assigned to a path.

%%%%%%%%%%%%%%%%%%%%%%%%%%%%%%%%%%%%%%%%%%%%%%%%%%%%%%%%%%%%%%%%%%%%%%%%%%%%%%%%%%%%%%%%%%%%%%%%%%%%%%%%%%%
%%%%%%%%%%%%%%%%%%%%%%%%%%%%%%%%%%%%%%%%%%%%%%%%%%%%%%%%%%%%%%%%%%%%%%%%%%%%%%%%%%%%%%%%%%%%%%%%%%%%%%%%%%%
%%%%%%%%%%%%%%%%%%%%%%%%%%%%%%%%%%%%%%%%%%%%%%%%%%%%%%%%%%%%%%%%%%%%%%%%%%%%%%%%%%%%%%%%%%%%%%%%%%%%%%%%%%%
%%%%%%%%%%%%%%%%%%%%%%%%%%%%%%%%%%%%%%%%%%%%%%%%%%%%%%%%%%%%%%%%%%%%%%%%%%%%%%%%%%%%%%%%%%%%%%%%%%%%%%%%%%%
%%%%%%%%%%%%%%%%%%%%%%%%%%%%%%%%%%%%%%%%%%%%%%%%%%%%%%%%%%%%%%%%%%%%%%%%%%%%%%%%%%%%%%%%%%%%%%%%%%%%%%%%%%%
\subsection{Greedy pseudotriangulation algorithm}
The input of the greedy pseudotriangulation algorithm is a cross-section $\Gamma(\MAC)$ of the visibility complex of the set of convex bodies 
$\disks$ and the restriction of the sink bitangent line segment operator 
to the subset of $2$-cells of that cross-section whose sink bitangent line segments are boundary bitangent line segments.

We denote by $\order$ the partial order  
on $\faces = \faces(\MAC)$  defined by 
$\sigma \order \sigma'$ if there is an edge-path in 
$\Gamma(\MAC)$ with source  $\sigma$ and sink $\sigma'$ (recall that the cross-section is acyclic). %%% (cf.  Theorem~\ref{acyclic})
Let now $<$ be a partial order on  $\faces$ 
compatible with $\order$ on the sets 
$\fdt{\twoset}{\thr}$ and $\fdt{\one}{\two}$, 
compatible with  the dual order $\orderstar$ on the sets
$\fdt{\thr}{\twoset}$ and $\fdt{\two}{\one}$,  
and such that
\begin{equation}
\fdt{\thr}{\thr} <{\fdt{\thr}{\twoset} \cup \fdt{\twoset}{\thr}} < 
{ \fdt{\two}{\two}} < {\fdt{\one}{\two} \cup \fdt{\two}{\one}} < {\fdt{\one}{\one}}.
\end{equation}
%%\begin{equation}
%%\fdt{\thr}{\thr}(\MAC) <{\fdt{\thr}{\twoset}(\MAC) \cup \fdt{\twoset}{\thr}}(\MAC) < 
%%{ \fdt{\two}{\two}}(\MAC) < {\fdt{\one}{\two}(\MAC) \cup \fdt{\two}{\one}}(\MAC) < {\fdt{\one}{\one}}(\MAC).
%%\end{equation}
Our algorithm maintains the directed graph $\Gamma(\MAC)$, the restriction 
$\sinkpar{\downset}$ of the $\sinkop$ operator to $\downset$ and 
the restrictions to $B(\downset)= \Bforw{\downset} \cup \Bback{\downset}$ 
 of the operators $\pforwop$ and $\pbackop$  
when $\downset$ describes a maximal chain of down-sets in the interval 
$[\emptyset,\faces]$ (here $\Bforw{\sigma}$ and $\Bback{\sigma}$ are the 
orbits of the sink bitangent line segment of $\sigma$ under $\pforwop$ and
$\pbackop$, respectively).  

Let $\downset$ be a  down-set of $(\faces,<)$. 
%%We set $\downs{i}{j} = \downset \cap \fdt{i}{j}$. 
Let $\sigma$ be a minimal element of $\faces\setminus \downset$  whose sink bitangent line segment $t$ is not a boundary bitangent 
line segment. 
We explain how to compute $B' = B(\downset \cup \{\sigma\})$ 
from $B =B(\downset)$. 
Thanks to the conjugation relations 
$\fdt{i}{j} \circ \reverse = \reverse \circ \fdt{j}{i}$ 
it is sufficient to examine the cases  
$\sigma \in \fdt{i}{j}$ with $i \leq j$.
Assume first that 
$\sigma\notin \fdt{\one}{\thr}.$  
In that case one can easily check that $B' = B \cup\{t\}$ 
and that linked representations of $\Lea(\sigma)$ and 
$\Ent(\sigma)$ are computable in constant time (cf. Theorem~\ref{summary} and Theorem~\ref{table}).
So it remains to explain how to compute $t$ efficiently. We postpone 
this point to the next paragraph. 
Assume now that
 $\sigma \in \fdt{\one}{\thr}$. 
In that case $B \cup \{t\}$ might be a proper subset of $B'$; so our goal is now not only to compute $t$ but also its orbit under $\pbackop$
(its orbit under $\pforwop$ reduces to $t$ since $t$ is right-right of left-right).
We proceed as follows.
Let $\sigma_1, \sigma_2, \ldots, \sigma_{k+1}$ with $k\geq 1$ be the sequence
 of $2$-cells defined by 
%%$\sigma_1 = \sigma \in \fdt{\one}{\thr}$,
$\sigma_1 = \sigma$,
$\sigma_{i+1}= \bdl{\sigma_i}{\MAC}$,
$\sigma_2,\ldots, \sigma_{k} \in \fdt{\one}{\two}$,
and $\sigma_{k+1} \in \fdt{\reversetwoset}{\two}$, or $ = \botface$. 
Assume to fix the ideas that $\sigma_{k+1} \in \fdt{\two}{\two}$, that its
forward view is a body, and that the
backward view of the $\sigma_i$ is a body (the other
cases can be treated similarly). In that case $\bdl{\sigma_{k+1}}{\MAC} \in
\fdt{{\{1,2,3\}}}{\thr}$ belongs to $\downset$, the bitangent line segments of the
sequence $\hourglassRplus{\sigma_{k+1};\MAC}$ are bitangent line segments of $B$, and a linked representation of $\Lea(\sigma_{k+1})$ is computable 
in constant time. 
%%Unfortunately that is not necessarily the case of $\Ent(\sigma_{k+1})$ which prevent us to compute the sink 
%%of $\sigma_{k+1}.$
Assume for a moment that the $2$-cells $\bur{\sigma_i}{\MAC}$ ($2\leq i \leq k+1$) 
are in $\fdt{\thr}{\thr}$. In that case the $\Ent(\sigma_i)$, $i =2,3,\ldots,k+1$, 
are computable in constant time. Therefore one can compute successively the sinks of  
the $\sigma_{i}$ as the bitangent line segments joining the $\Lea(\sigma_i)$ to the $\Ent(\sigma_i)$ 
for $i = k+1,k,\ldots,2,1.$ 
Now we drop the assumption that the $\sigma_i$ are elements of 
$\fdt{\thr}{\thr}$.  Let $s_i$ be the terminal segment of a line 
$t_i \in \btoponecell{\sigma_i}{\MAC}$, let $s'_i$ be the right lift of $s_i$ in 
the visibility complex of $\lambda_{s_2} \circ \cdots \circ
\lambda_{s_{k+1}}(\disks)$, and let $\widehat{\MAC}$ be the antichain $\lambda_{s_2} \circ \cdots \circ \lambda_{s_{k+1}}(\MAC).$ 
Let $\widehat{\sigma}_i = \drf{s'_i}$ ($i=2,3,\ldots,k+1$) and 
let $\widehat{\sigma}_1 = \ulf{s'_2}.$ 
Using similar arguments to the ones given in the proof of
Theorem~\ref{maintwofirst} one can prove that 
the orbit of the sink bitangent line segment of $\sigma_1$ under $\pbackop$ 
coincides 
with the orbit of the sink bitangent line segment of $\widehat{\sigma}_1$ under $\pbackop$; furthermore 
$\bur{\widehat{\sigma}_i}{\widehat{\MAC}} \in \fdt{3}{3}(\widehat{\MAC})$, $2\leq i\leq k$. 
Therefore one can proceed exactly as in the case $\bur{\sigma_i}{\MAC} \in \fdt{\thr}{\thr}$ but with 
the $\widehat{\sigma}_i$. The cost 
of this computation is a big-O of $k$ plus 
the size of the prefix part of $\hourglassRplus{\bdl{\sigma_{k+1}}{\MAC};\MAC}$ involved in the computation, which sum up to a $O(n)$ 
when $\sigma$ ranges over $\fdt{1}{3}$.

%%%%%%%%%%%%%%%%%%%%%%%%%%%%%%%%%%%%%%%%%%%%%%%%%%%%%%%%%%%%%%%%%%%%%%%%%%%%%%%
%%%%%%%%%%%%%%%%%%%%%%%%%%%%%%%%%%%%%%%%%%%%%%%%%%%%%%%%%%%%%%%%%%%%%%%%%%%%%%%
%%%%%%%%%%%%%%%%%%%%%%%%%%%%%%%%%%%%%%%%%%%%%%%%%%%%%%%%%%%%%%%%%%%%%%%%%%%%%%%
%%%%%%%%%%%%%%%%%%%%%%%%%%%%%%%%%%%%%%%%%%%%%%%%%%%%%%%%%%%%%%%%%%%%%%%%%%%%%%%
%%%%%%%%%%%%%%%%%%%%%%%%%%%%%%%%%%%%%%%%%%%%%%%%%%%%%%%%%%%%%%%%%%%%%%%%%%%%%%%
%%\paragraph{Computing the bitangent joining $\Lea(\sigma)$ to $\Ent(\sigma)$.} 
It remains to explain how to compute 
the bitangent that leaves $\Lea(\sigma)$ and enters $\Ent(\sigma)$ 
in $O(1)$ amortized time. We will charge the cost on the bitangent line segments 
of the greedy pseudotriangulation $\G(\MAC)$ associated with the cross-section. 
%% Recall (cf.~\cite{G-pv-tsvcp-96}) that 
%% a  bitangent $u\in \G(\MAC)$ is said to be right-minimal (resp. left-minimal) if $u$ is 
%% the $\prec$-minimal element of the set of bitangent line segments that appear 
%% in the boundary of the  pseudotriangle of $\G(\MAC)$ lying locally to the 
%% right (resp. left) of $u$; in that case this pseudotriangle is independent 
%% of $\MAC$ and is denoted $\R(u)$ (resp. $\L(u)$).
Recall (cf.~\cite{G-pv-tsvcp-96}) that 
a  bitangent $u\in \G(\MAC)$ is said to be right-minimal if $u$ is 
the $\prec$-minimal element of the set of bitangent line segments that appear 
in the boundary of the  pseudotriangle of $\G(\MAC)$ lying locally to the 
right of $u$; in that case this pseudotriangle is independent 
of $\MAC$ and is denoted $\R(u)$.
Similarly a  bitangent $u\in \G(\MAC)$ is said to be left-minimal if $u$ is 
the $\prec$-minimal element of the set of bitangent line segments that appear 
in the boundary of the  pseudotriangle of $\G(\MAC)$ lying locally to the 
left of $u$; in that case this pseudotriangle is independent 
of $\MAC$ and is denoted $\L(u)$. 
Clearly the sum of the 
complexities of the $\R(u)$ and $\L(u)$ for $u$ right or left-minimal in 
$\G(\MAC)$ is linear in the size of $\G(\MAC)$, that is, a $O(n)$.
%%Given two adjacent pseudotriangles $T$ and $T'$ 
%%we denote by  $\B\MACT{T}{T'}$ the bitangent that leaves $T$ and enters $T'$.
Let $e_1e_2 \ldots e_{k}$  and 
$e'_1e'_2\ldots e'_{k'}$ ($k, k'\geq 1$) 
be the sequences of (oriented) arcs of the chains $\Entent(\sigma)$ 
and $\Lealea(\sigma)$, 
and let $e_m$ and $e'_{m'}$ be the arcs that $t$ enters
and leaves. The source and the sink of $e_i$ are denoted $v_i$ and $v_{i+1}$.
Note that for $i= 2,3, \ldots, k$ the bitangent $v_i$ is an atom of the chain 
$\Enttwo(\sigma)$.
For $i = 2,\ldots,k$, let $T_i$ be one of the two pseudotriangles of size two  
adjacent to $\Ent(\sigma)$ 
along the bitangent $v_i$ and let $u_i$ be the bitangent joining ${T_i}$ to ${\Ent(\sigma)}$;
it is convenient to denote by $u_1$ the source of $\liftoperator(\sigma)$.
Clearly $u_i$ enters $\Entthr(\sigma)\Entfou(\sigma)$---which has constant
complexity{}\footnote{This property is no more valid in the presence of constraints if the family of bodies is not well-constrained.}---and $u_i$ is 
computable in constant time.
The bitangent line segments $u_i$ decompose the pseudopolygon 
$\Ent(\sigma) \cup T_2 \cup T_3 \cup \cdots \cup T_k$---which reduces to 
$\Ent(\sigma)$ 
in case $k=1$---into $k$ pseudotriangles 
$R_1,\ldots, R_{k}$ where $R_i$ lies locally to the left 
 of the bitangent $u_{i}$ ($1\leq i\leq k$).
Note that the number of bitangent line segments lying 
in the boundary of $R_i$ is a $O(1)$ (except for $R_1$ in the case 
$\sigma \in \fdt{\one}{\one}$). 
We define inductively the sequences $R^*_i, u^*_i$ 
 ($i=1,\ldots,k^*$) as follows: 
\begin{enumerate}
\item $R^*_1 = \Lea(\sigma)$ and we define $u_1^*$ as the bitangent joining $R^*_{1}$ to $R_1$, 
\item if $u_i^*$ enters the arc $e_i$ or the arc 
$e_k$ then $k^*=i$ and we are done;
 otherwise we define 
$R^*_{i+1}$ to be the right pseudotriangle of $u^*_i$ 
in the pseudoquadrangle $R^*_{i} \cup R_i$ 
and we define $u_{i+1}^* $ to be the bitangent joining $R^*_{i+1}$ to $R_{i+1}.$ 
\end{enumerate}
Clearly the sequences $R^*_i, u_i^*$ are well-defined;
$R^*_i$ and $R_i$ are adjacent along the bitangent $u_i$; 
$t$ is the last term of the sequence of $u^*_i$
(that is, $t= u^*_{k^*}$);
$u_i^*$ leaves one of the arcs, say $e'_{m^*_i}$, of 
the sequence $e'_1 \ldots e'_{k'}$;  
 $m^*_1 \geq m^*_2 \geq \cdots \geq m'.$  
Consequently $t$ is computable in time $O(m+m_1^*)$.  
Let now $t'$ be the sink of $\toponecell{\sigma}$. A simple case analysis 
shows that either $t = t'$, $m =1$ and $m'=m_1^*$;
or $t \neq t'$, $m\geq 2$, $t'$ is a bitangent of $\Ent(\sigma)$, 
$t'$ is left- or right-minimal in $\G(\MAC)$ 
depending on whether 
$\Ent(\sigma)$ lies locally
on the left or on the right side of $t'$, and 
$m$ and $m^*_1-m'$ are bounded by the size of $\L(t')$ or 
$\R(t')$ depending on whether 
$\Ent(\sigma)$ lies locally
on the left or on the right side of $t'$. 
Its follows that 
 $\sum_{\sigma \in \facesG(\MAC)} \{m(\sigma) + m_1^*(\sigma)\} = O(n)$.
This prove that $t$ is computable in constant amortized time. 

%%%%%%%%%%%%%%%%%%%%%%%%%%%%%%%%%%%%%%%%%%%%%%%%%%%%%%%%%%%%%%%%%%%%%%%%%%%%%%%
%%%%%%%%%%%%%%%%%%%%%%%%%%%%%%%%%%%%%%%%%%%%%%%%%%%%%%%%%%%%%%%%%%%%%%%%%%%%%%%
%%%%%%%%%%%%%%%%%%%%%%%%%%%%%%%%%%%%%%%%%%%%%%%%%%%%%%%%%%%%%%%%%%%%%%%%%%%%%%%
%%%%%%%%%%%%%%%%%%%%%%%%%%%%%%%%%%%%%%%%%%%%%%%%%%%%%%%%%%%%%%%%%%%%%%%%%%%%%%%
%%%%%%%%%%%%%%%%%%%%%%%%%%%%%%%%%%%%%%%%%%%%%%%%%%%%%%%%%%%%%%%%%%%%%%%%%%%%%%%

\clearpage
%%%%%%%%%%%%%%%%%%%%%%%%%%%%%%%%%%%%%%%%%%%%%%%%%%%%%%%%%%%%%%%%%%%%%%%%%%%%%%%
%%%%%%%%%%%%%%%%%%%%%%%%%%%%%%%%%%%%%%%%%%%%%%%%%%%%%%%%%%%%%%%%%%%%%%%%%%%%%%%
%%%%%%%%%%%%%%%%%%%%%%%%%%%%%%%%%%%%%%%%%%%%%%%%%%%%%%%%%%%%%%%%%%%%%%%%%%%%%%%
%%%%%%%%%%%%%%%%%%%%%%%%%%%%%%%%%%%%%%%%%%%%%%%%%%%%%%%%%%%%%%%%%%%%%%%%%%%%%%%
%%%%%%%%%%%%%%%%%%%%%%%%%%%%%%%%%%%%%%%%%%%%%%%%%%%%%%%%%%%%%%%%%%%%%%%%%%%%%%%
\section{Conclusion and open problems\label{sec5}}
We have  presented an efficient algorithm to compute a pseudotriangulation
of a planar (well-constrained) finite family  of pairwise disjoint convex  bodies with constraints presented by its chirotope. Its 
running time is a $O(n \log n)$ 
where $n$ is the number of convex bodies of the family.
We expect to prove its practical efficiency 
in a forthcoming  implementation. 
The design of our pseudotriangulation algorithm relies 
on an extension of the theory of pseudotriangulations and visibility complexes
to the setting of branched coverings of topological planes. 
%We are convinced that  branched coverings will find other applications in the design 
%of geometric algorithms. Preliminary investigations in this direction are 
%reported in~\cite{G-p-cvgls-07,G-p-ecrre-05}. 
We conclude  by a sequence of open problems raised by this work. 
Can we extend our pseudotriangulation algorithm to not necessarily  well-constrained family of convex bodies keeping the same time bound on its running time? 
Can we sweep efficiently the dual arrangement of a finite planar family of convex bodies presented by its chirotope using only linear space? 
According to the main result of the paper this problem boils down in $O(n \log n)$ time 
to sweeping the arrangement of the dual pseudolines of the pseudotriangles of a pseudotriangulation of the family of convex bodies;
this arrangement is an arrangement of pseudolines with contact points to which it is tempting to apply the topological sweep method of Edelsbrunner and Guibas~\cite{eg-tsa-89};
however the fact that the chirotope of the pseudolines is not computable in constant time (since the pseudolines do not have constant complexities) 
prevents a naive application of the topological sweep method.
Can we enumerate the pseudotriangulations of a planar family of convex bodies presented by its chirotope in sublinear time per pseudotriangulation using only linear space?
Does the complex of pseudotriangulations of a planar family of pairwise disjoint 
convex bodies is a (shellable) sphere? a matroid polytope?  a polytope? 
It is known to be a polytope for families of bodies lying in the classical real affine plane~\cite{rss-emppp-03}.
Finally we observe that these questions can be asked not only for
families of convex bodies of topological planes but also for families of convex bodies of branched 
coverings of topological planes.

%%%%%%%%%%%%%%%%%%%%%%%%%%%%%%%%%%%%%%%%%%%%%%%%%%%%%%%%%%%%%%%%%%%%%%%%%%%%%%%
%%%%%%%%%%%%%%%%%%%%%%%%%%%%%%%%%%%%%%%%%%%%%%%%%%%%%%%%%%%%%%%%%%%%%%%%%%%%%%%
%%%%%%%%%%%%%%%%%%%%%%%%%%%%%%%%%%%%%%%%%%%%%%%%%%%%%%%%%%%%%%%%%%%%%%%%%%%%%%%
%%%%%%%%%%%%%%%%%%%%%%%%%%%%%%%%%%%%%%%%%%%%%%%%%%%%%%%%%%%%%%%%%%%%%%%%%%%%%%%
%%%%%%%%%%%%%%%%%%%%%%%%%%%%%%%%%%%%%%%%%%%%%%%%%%%%%%%%%%%%%%%%%%%%%%%%%%%%%%%
\subsection*{Acknowledgments.}
The authors thank the referees for pointing several relevant references and for
their helpful comments and encouragements to improve the intuitive presentation.

\clearpage
%%%%%%%%%%%%%%%%%%%%%%%%%%%%%%%%%%%%%%%%%%%%%%%%%%%%%%%%%%%%%%%%%%%%%%%%%%%%%%%
%%%%%%%%%%%%%%%%%%%%%%%%%%%%%%%%%%%%%%%%%%%%%%%%%%%%%%%%%%%%%%%%%%%%%%%%%%%%%%%
%%%%%%%%%%%%%%%%%%%%%%%%%%%%%%%%%%%%%%%%%%%%%%%%%%%%%%%%%%%%%%%%%%%%%%%%%%%%%%%
%%%%%%%%%%%%%%%%%%%%%%%%%%%%%%%%%%%%%%%%%%%%%%%%%%%%%%%%%%%%%%%%%%%%%%%%%%%%%%%
%%%%%%%%%%%%%%%%%%%%%%%%%%%%%%%%%%%%%%%%%%%%%%%%%%%%%%%%%%%%%%%%%%%%%%%%%%%%%%%
{
%%\small 
%% \normalsize
\bibliographystyle{abbrv}
\bibliography{%%acm-biblio%
$HOME/BIB/total%
,$HOME/BIB/geom%
,$HOME/BIB/geomplus%
,$HOME/BIB/divers%
,$HOME/BIB/science%
,$HOME/BIB/algo%
}
}
%\end{document}

\clearpage
%\phantom{sautdepage}
\clearpage
\appendix

\section{Duality in topological planes}\label{TopPlanes}

%%%%%%%%%%%%%%%%%%%%%%%%%%%%%%%%%%%%%%%%%%%%%%%%%%%%%%%%%%%%%%%%%%%%%%%%%%%%%%%
%%%%%%%%%%%%%%%%%%%%%%%%%%%%%%%%%%%%%%%%%%%%%%%%%%%%%%%%%%%%%%%%%%%%%%%%%%%%%%%
%%%%%%%%%%%%%%%%%%%%%%%%%%%%%%%%%%%%%%%%%%%%%%%%%%%%%%%%%%%%%%%%%%%%%%%%%%%%%%%
%%%%%%%%%%%%%%%%%%%%%%%%%%%%%%%%%%%%%%%%%%%%%%%%%%%%%%%%%%%%%%%%%%%%%%%%%%%%%%%
\begin{figure}[!htb]
\centering
\small
\psfrag{p}{$p$}\psfrag{q}{$q$}\psfrag{pq}{$p\vee q$}
\psfrag{infty}{$\infty$}
\psfrag{points}{space of points $\approx \mathbb{R}^2$}
\psfrag{lines}{space of lines $\approx \mathbb{P}^2\setminus \{\infty\}$}
\includegraphics[width=0.75\linewidth]{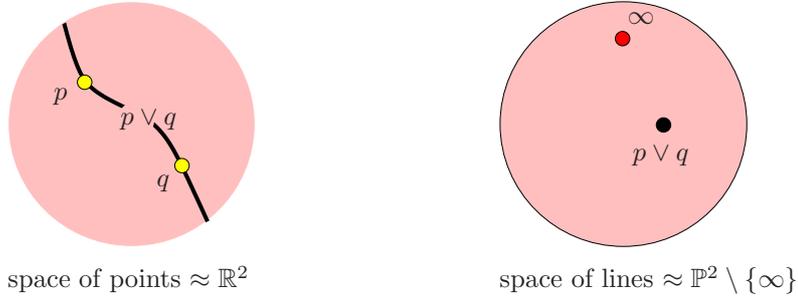}
 \caption{\protect A topological plane.  \label{topologicalplane}}
\end{figure}
%%%%%%%%%%%%%%%%%%%%%%%%%%%%%%%%%%%%%%%%%%%%%%%%%%%%%%%%%%%%%%%%%%%%%%%%%%%%%%%
%%%%%%%%%%%%%%%%%%%%%%%%%%%%%%%%%%%%%%%%%%%%%%%%%%%%%%%%%%%%%%%%%%%%%%%%%%%%%%%
%%%%%%%%%%%%%%%%%%%%%%%%%%%%%%%%%%%%%%%%%%%%%%%%%%%%%%%%%%%%%%%%%%%%%%%%%%%%%%%
%%%%%%%%%%%%%%%%%%%%%%%%%%%%%%%%%%%%%%%%%%%%%%%%%%%%%%%%%%%%%%%%%%%%%%%%%%%%%%%

%%%%%%%%%%%%%%%%%%%%%%%%%%%%%%%%%%%%%%%%%%%%%%%%%%%%%%%%%%%%%%%%%%%%%%%%%%%%%%%
%%%%%%%%%%%%%%%%%%%%%%%%%%%%%%%%%%%%%%%%%%%%%%%%%%%%%%%%%%%%%%%%%%%%%%%%%%%%%%%
%%%%%%%%%%%%%%%%%%%%%%%%%%%%%%%%%%%%%%%%%%%%%%%%%%%%%%%%%%%%%%%%%%%%%%%%%%%%%%%
%%%%%%%%%%%%%%%%%%%%%%%%%%%%%%%%%%%%%%%%%%%%%%%%%%%%%%%%%%%%%%%%%%%%%%%%%%%%%%%
\paragraph{\bf Topological planes.}
A topological plane on $\mathbb{R}^2$ is a surface homeomorphic to $\mathbb{R}^2$ endowed with a topological point-line incidence geometry 
whose  line space  is a subspace of the space of pseudolines of the surface satisfying the so-called linear space axiom: any two distinct points
$p$ and $q$ are contained in exactly one line $p\vee q$, their {\it joining line}.  
It is known that the  line space of topological plane is a open crosscap, that is, a surface homeomorphic to $\mathbb{RP}^2$ with one point deleted.
A topological plane is called {\it affine} if for 
every point-line pair $(p,l)$ there exists a unique line $k$ through the point $p$ that coincides with the line $l$ or has no point in common with it; the line $k$ is called
 the {\it parallel} to $l$ through $p$.
Classical examples of topological planes are 
\begin{enumerate}
\item the real affine plane   or classical topological plane 
defined as the affine topological plane whose set of points is $\mathbb{R}^2$ and  whose set of lines is the set of curves with equations 
$y=ax+b$, $a,b\in \mathbb{R}$, and $x =c$, $c\in \mathbb{R}$; the real affine plane  is the default plane in the field of discrete and  computational geometry;
\item the Moulton planes obtained by starting from the real affine plane and replacing the lines of the real affine plane with negative slope 
by the kinked lines $\{(x,y) \mid y=ax+b,  x\geq0\} \cup \{(x,y) \mid y=akx+b, x\leq 0\}$, $a\in \mathbb{R}^-, b \in \mathbb{R}$,
where $k >1$ is a fixed parameter;
\item the Klein model of the real hyperbolic plane defined as
the restriction of the real affine plane to the open unit disk; and 
\item the so-called {\it arc planes} which include examples of topological planes
with no embedding into any affine topological plane, e.g.,  
start with the real affine plane and replace the lines with positive slope by the shifted arcs $\{(x,y) \in \mathbb{R}^2 \mid y-n = e^{x-m}\}$, $m,n\in \mathbb{R}$. 
\end{enumerate}
The projective completion of  the real affine plane  
is the unique desarguesian
projective topological plane on $\mathbb{RP}^2$, a characterization due to
Hilbert, 1899.  The Moulton planes are examples of non  desarguesian affine
topological planes.  We refer to~\cite{b-gg-55,sbghl-cpp-95,ps-gs-01,gpwz-atp-94} for a more detailed background material on topological planes, and 
for the precise meaning of the terms used above (topological projective planes, projective completion, desarguesian, etc.) but not defined explicitly.

\begin{figure}[!htb]
\centering
\small
\psfrag{A}{$A$}
\psfrag{B}{$A'$}
\psfrag{C}{$B$}
\psfrag{D}{$B'$}
\psfrag{E}{$B''$}
\psfrag{join}{$1\vee 2$}
\psfrag{1}{$1$} \psfrag{un}{1} \psfrag{2}{$2$} \psfrag{3}{$3$} \psfrag{deux}{$2$} \psfrag{trois}{$3$} \psfrag{infty}{$\infty$}
\includegraphics[width = 0.75\linewidth]{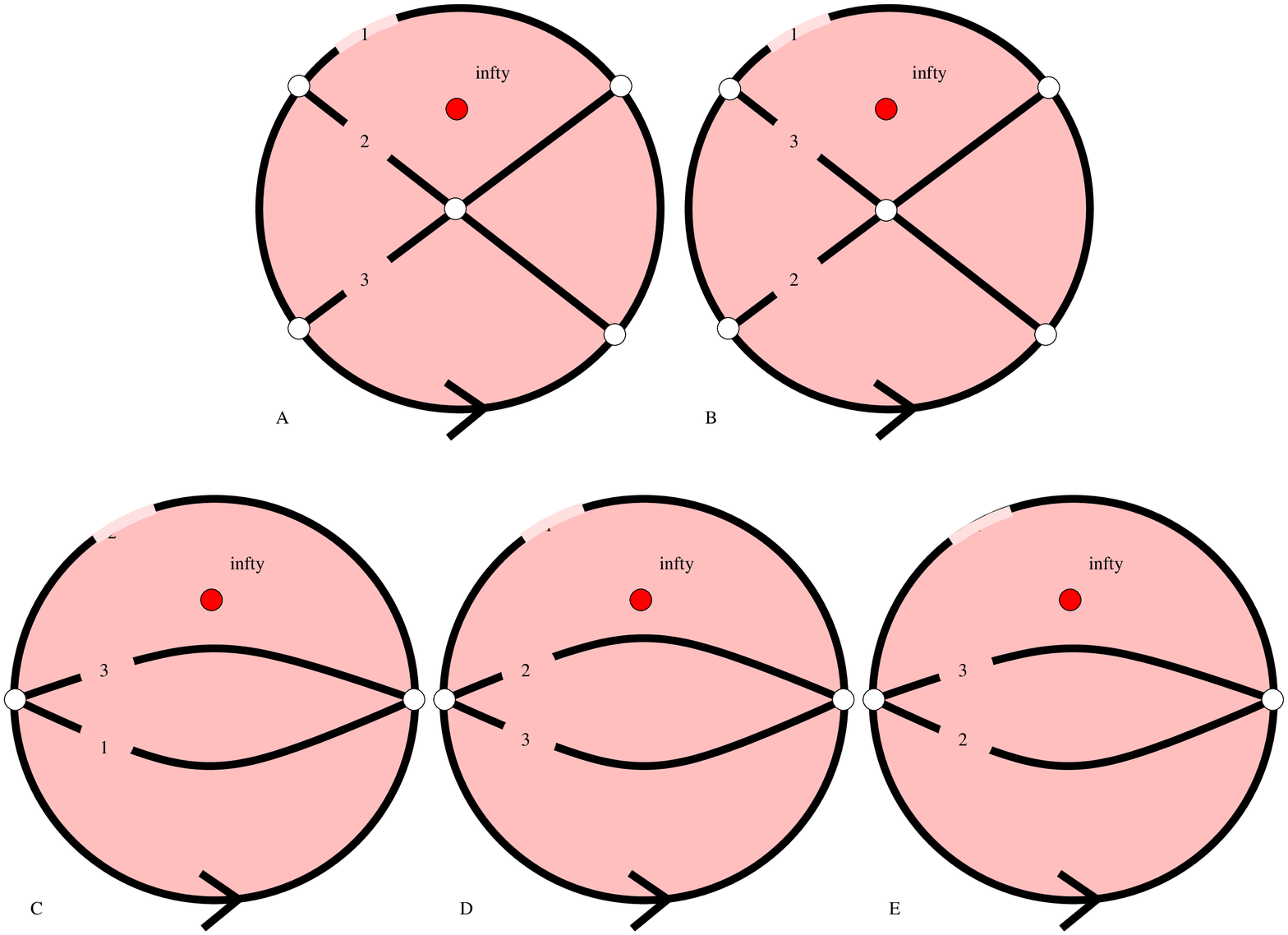}%%%xfigfinal
\caption{\protect \small 
%% Representatives of the five chirotopes of planar families of three points, and
%% representatives of the
%% corresponding five isotopy classes of
%% arrangements of three pseudolines defined on a given set of indices.
\label{chirotopesthrpoints}}
\end{figure}

\paragraph{\bf Duality.} We work in an oriented topological plane on $\mathbb{R}^2$.
The term duality refers to the map that assigns to a point of the plane the pencil of lines through that point,
embedded in the line space of the plane, and more generally to the map that assigns to a convex body of the plane its set of tangent lines. 
%%%% and oriented according to the orientation of the plane. 
The basic properties of the duality map for points relevant to our purpose are the following 
\begin{enumerate}
\item the dual of a point is a pseudoline, that is, a simple closed curve homotopic to a, hence any, generator of the fundamental group of the 
line space or, equivalently, a non-separating simple closed curve;  
\item the dual family of a family of points is an arrangement of pseudolines, that is, a finite family of pseudolines living in the same open crosscap 
with the property that any two cross in exactly one point; 
\item the isomorphism class 
of the dual family of a family of points depends only on the chirotope of the family of points and vice-versa; and 
\item  any 
arrangement of pseudolines is isomorphic to the dual family of a planar family of points. 
\end{enumerate}
Figure~\ref{chirotopesthrpoints} shows the dual arrangements of the five families of three points realizing the five possible chirotopes on a given indexing set of size three 
depicted in Figure~\ref{thrchipoints}.
%%%%%%%%%%%%%%%%%%%%%%%%%%%%%%%%%%%%%%%%%%%%%%%%%%%%%%%%%%%%%%%%%%%%%%%%%%%%%%
%%%%%%%%%%%%%%%%%%%%%%%%%%%%%%%%%%%%%%%%%%%%%%%%%%%%%%%%%%%%%%%%%%%%%%%%%%%%%%%
%%%%%%%%%%%%%%%%%%%%%%%%%%%%%%%%%%%%%%%%%%%%%%%%%%%%%%%%%%%%%%%%%%%%%%%%%%%%%%%
%%%%%%%%%%%%%%%%%%%%%%%%%%%%%%%%%%%%%%%%%%%%%%%%%%%%%%%%%%%%%%%%%%%%%%%%%%%%%%%
%%%%%%%%%%%%%%%%%%%%%%%%%%%%%%%%%%%%%%%%%%%%%%%%%%%%%%%%%%%%%%%%%%%%%%%%%%%%%%%
%%%%%%%%%%%%%%%%%%%%%%%%%%%%%%%%%%%%%%%%%%%%%%%%%%%%%%%%%%%%%%%%%%%%%%%%%%%%%%%
%%%%%%%%%%%%%%%%%%%%%%%%%%%%%%%%%%%%%%%%%%%%%%%%%%%%%%%%%%%%%%%%%%%%%%%%%%%%%%%
%%%%%%%%%%%%%%%%%%%%%%%%%%%%%%%%%%%%%%%%%%%%%%%%%%%%%%%%%%%%%%%%%%%%%%%%%%%%%%%
In this figure the line space is represented by a 
circular diagram with antipodal boundary points identified, one marked point ($\infty$) in the role of the deleted point,
 and one arrow on the boundary of the circular diagram to indicate which of the two generators of the fundamental group of the line space 
is the one that fits via duality the orientation of the topological plane. According to the third basic property of the duality map of points mentioned above 
the chirotope of a planar family of points can be coded by the isomorphism classes of the subarrangements of size three of its dual arrangement. In general 
the isomorphism class of an  arrangement of pseudolines  $\gamma_1,\gamma_2,\ldots, \gamma_n$
can be coded by the poset of its cells ordered by inclusion or, equivalently, by its {\it side cycles} : $C_1,C_2,\ldots, C_n$ 
where $C_i$ is the circular sequence of signed indices of the pseudolines crossed by the side wheel of a sidecar rolling on $\gamma_i$, according to its orientation, 
with the convention that an index is signed positively or negatively depending on whether the crossed pseudoline is  (locally)  
directed towards or away $\gamma_i$.  For example the side cycles of the arrangements of Figure~\ref{chirotopesthrpoints} labeled $A$ and  $B$ are  

\newcommand{\ione}{1}
\newcommand{\itwo}{2}
\newcommand{\ithr}{3}
\newcommand{\nione}{\overline{1}}
\newcommand{\nitwo}{\overline{2}}
\newcommand{\nithr}{\overline{3}}

$$  \begin{array}{ll}
1 : & \itwo\ithr\nitwo\nithr\\
2 : & \ithr\ione\nithr\nione\\
3 : & \ione\itwo\nione\nitwo
\end{array}
\qquad \text{and} \qquad
 \begin{array}{ll}
1 : & \nithr\ithr\nitwo\itwo\\
2 : & \ione\ithr\nithr\nione\\
3 : & \nione\ione\itwo\nitwo,
\end{array}
$$
respectively, and those of $A',B'$ and $B''$ are obtained from those of $A$ and $B$ by suitable permutations of the indices.

%%%%%%%%%%%%%%%%%%%%%%%%%%%%%%%%%%%%%%%%%%%%%%%%%%%%%%%%%%%%%%%%%%%%%%%%%%%%%%%
%%%%%%%%%%%%%%%%%%%%%%%%%%%%%%%%%%%%%%%%%%%%%%%%%%%%%%%%%%%%%%%%%%%%%%%%%%%%%%%
%%%%%%%%%%%%%%%%%%%%%%%%%%%%%%%%%%%%%%%%%%%%%%%%%%%%%%%%%%%%%%%%%%%%%%%%%%%%%%%
%%%%%%%%%%%%%%%%%%%%%%%%%%%%%%%%%%%%%%%%%%%%%%%%%%%%%%%%%%%%%%%%%%%%%%%%%%%%%%%

\begin{figure}[!htb]
\centering
\psfrag{ii}{$3$}
\psfrag{ji}{$2$}
\psfrag{ki}{$1$}
\psfrag{one}{1} \psfrag{two}{2} \psfrag{thr}{3}
\psfrag{fou}{4} \psfrag{fiv}{5} \psfrag{six}{6}
\psfrag{infty}{$\infty$}
\includegraphics[width=0.8575\linewidth]{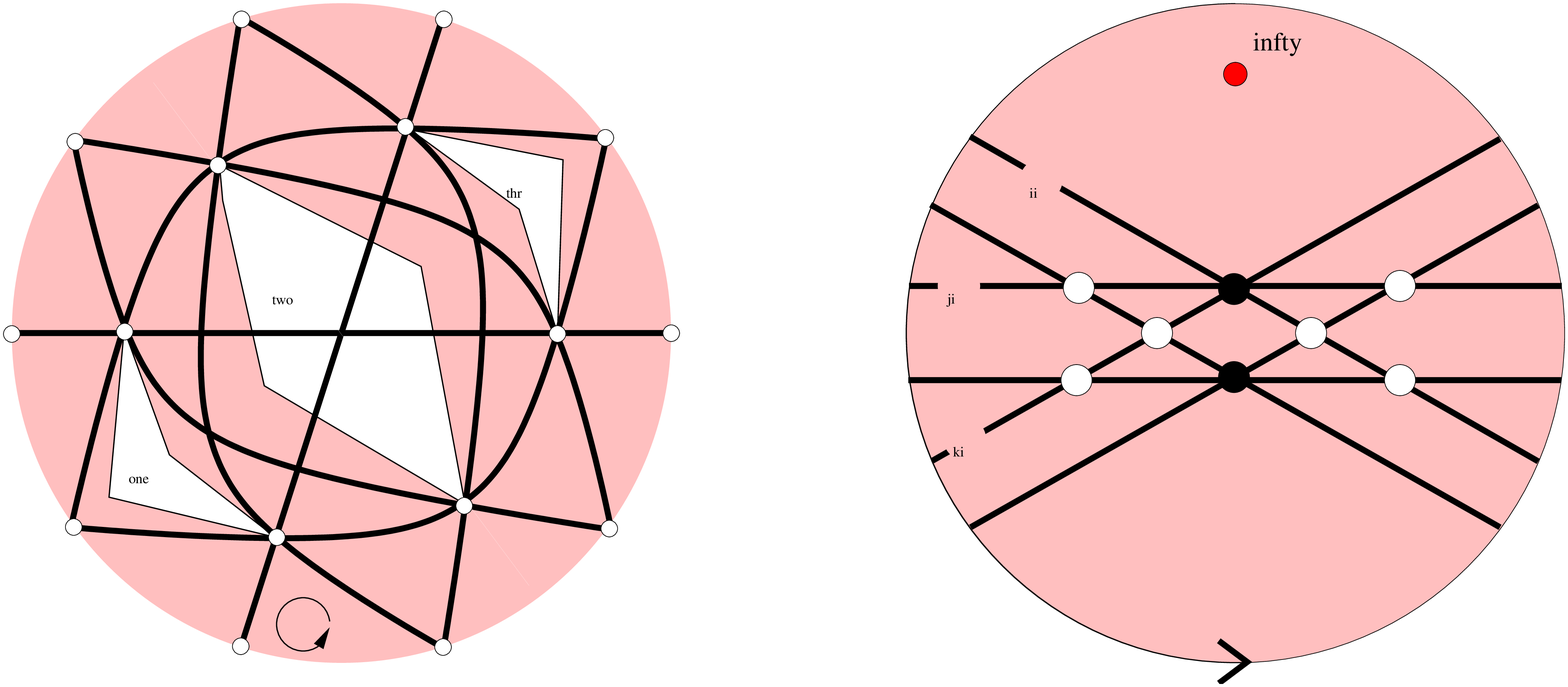}%%%%
\caption{\protect 
%%%The family of unit discs of the Euclidean plane centered at the points with coordinates $(2i,0)$, $i\in \mathbb{N}$. 
\label{infinitepencil}}
\end{figure}

In the companion paper~\cite{G-hp-adp-06,G-hp-adp-09} we show that the above properties of the duality map for points extend to the duality map for convex bodies 
 in the following terms 
\begin{enumerate}
\item the dual of a convex body is a double pseudoline, that is, a simple closed curve homotopic to the double of a, hence any, 
generator of the 
fundamental group of the line space;  
\item the dual family of a family of pairwise disjoint convex bodies is  
an arrangement of double pseudolines, that is,  a finite family  of double pseudolines living in the same open crosscap, with the property
that any two meet exactly four times, meet transversely exactly four times, and induce a cell structure on the one-point compactification of the crosscap; 
\item  the isomorphism class
of the dual family of a family of pairwise disjoint convex bodies depends only on the chirotope of the family of bodies and vice-versa; and 
\item any arrangement of double pseudolines is isomorphic 
to the dual family of a 
planar family of pairwise disjoint convex bodies.
\end{enumerate}
For example the arrangement of three double pseudolines depicted in the right diagram of Figure~\ref{infinitepencil} is  
a representative of the isomorphism class of the dual arrangement of 
the  family of three convex bodies depicted in the left diagram of Figure~\ref{infinitepencil}
%%%%%%%%%%%%%%%%%%%%%%%%%%%%%%%%%%%%%%%%%%%%%%%%%%%%%%%%%%%%%%%%%%%%%%%%%%%%%%%
%%%%%%%%%%%%%%%%%%%%%%%%%%%%%%%%%%%%%%%%%%%%%%%%%%%%%%%%%%%%%%%%%%%%%%%%%%%%%%%
%%%%%%%%%%%%%%%%%%%%%%%%%%%%%%%%%%%%%%%%%%%%%%%%%%%%%%%%%%%%%%%%%%%%%%%%%%%%%%%
%%%%%%%%%%%%%%%%%%%%%%%%%%%%%%%%%%%%%%%%%%%%%%%%%%%%%%%%%%%%%%%%%%%%%%%%%%%%%%%
%%%%%%%%%%%%%%%%%%%%%%%%%%%%%%%%%%%%%%%%%%%%%%%%%%%%%%%%%%%%%%%%%%%%%%%%%%%%%%%
%%%%%%%%%%%%%%%%%%%%%%%%%%%%%%%%%%%%%%%%%%%%%%%%%%%%%%%%%%%%%%%%%%%%%%%%%%%%%%%
%%%%%%%%%%%%%%%%%%%%%%%%%%%%%%%%%%%%%%%%%%%%%%%%%%%%%%%%%%%%%%%%%%%%%%%%%%%%%%%
%%%%%%%%%%%%%%%%%%%%%%%%%%%%%%%%%%%%%%%%%%%%%%%%%%%%%%%%%%%%%%%%%%%%%%%%%%%%%%%
(the two black vertices are the two tritangents of the family). As for arrangements of pseudolines  an arrangement of double pseudolines 
can be coded  by the poset of its cells ordered by inclusion or by  its side cycles which are defined similarly, except that there are now two side cycles per index since a double pseudoline has, contrary to a pseudoline,
 two sides : a crosscap side and a disk side; thus the side cycle of crosscap type  of the arrangement of Figure~\ref{infinitepencil} is
$$  \begin{array}{ll}
1 : & \nitwo\nithr\nithr\nitwo\itwo\ithr\ithr\itwo\\
2 : & \nithr\ione\nithr\ione\nione\ithr\nione\ithr\\
3 : & \itwo\ione\ione\itwo\nitwo\nione\nione\nitwo
\end{array}
$$
and the one of disk type is 
$$  \begin{array}{ll}
1 : & \itwo\ithr\itwo\ithr\nithr\nitwo\nithr\nitwo\\
2 : & \ithr\ithr\nione\nione\ione\ione\nithr\nithr\\
3 : & \nione\nitwo\nione\nitwo\itwo\ione\itwo\ione.
\end{array}
$$
Observe that for a simple arrangement the side cycle of disk type and the side cycle of crosscap type are the negatives of each other.   
 
%Dual representatives of the $118$ simple chirotopes are implicitly  depicted in 
Figure~\ref{mobiuslist} depicts representatives of the $22$ isomorphism classes of non indexed arrangements of three double pseudolines: 
%%%%%%%%%%%%%%%%%%%%%%%%%%%%%%%%%%%%%%%%%%%%%%%%%%%%%%%%%%%%%%%%%%%%%%%%%%%%%%%
%%%%%%%%%%%%%%%%%%%%%%%%%%%%%%%%%%%%%%%%%%%%%%%%%%%%%%%%%%%%%%%%%%%%%%%%%%%%%%%
%%%%%%%%%%%%%%%%%%%%%%%%%%%%%%%%%%%%%%%%%%%%%%%%%%%%%%%%%%%%%%%%%%%%%%%%%%%%%%%
%%%%%%%%%%%%%%%%%%%%%%%%%%%%%%%%%%%%%%%%%%%%%%%%%%%%%%%%%%%%%%%%%%%%%%%%%%%%%%%
\newcommand{\miror}{\star}
\begin{figure}[!htbp]
\footnotesize
\def\factor{0.131500023}
\def\factor{0.15315015000023}
\def\factor{0.18315015000023}
\def\factor{0.21315015000023}
\def\factor{0.24315015000023}
\def\chtw{2}
\def\chth{3}
\def\chfo{4}
\centering
\psfrag{8}{\small 8} \psfrag{7}{\small 7} \psfrag{6}{\small 6} \psfrag{5}{\small 5} \psfrag{4}{\small 4} \psfrag{3}{\small 3} \psfrag{2}{\small 2}
\psfrag{8}{} \psfrag{7}{} \psfrag{6}{} \psfrag{5}{} \psfrag{4}{} \psfrag{3}{} \psfrag{2}{}
\psfrag{AA}{}
\psfrag{A}{$049000$}
\psfrag{A}{$\nameM_{04}$}
\psfrag{B}{$073300$}
\psfrag{B}{$\nameM_{07}$}
\psfrag{C}{$181030$}
\psfrag{C}{$\nameM_{18}$}
\psfrag{D}{$252310$} %%% comme K
\psfrag{F}{$073300$} %%% comme B
\psfrag{G}{$370003$}
\psfrag{G}{$\nameM_{37}$}
\psfrag{H2}{$154300$}
\psfrag{H2}{$\nameM_{15(\chtw)}$}
\psfrag{H}{$154300$}
\psfrag{H}{$\nameM_{15(\chth)}$}
\psfrag{Hm}{$\nameM_{15(\chth)}^\miror$}
\psfrag{I}{\textcolor{red}{$252310$}}
\psfrag{J}{$433021$}
\psfrag{J}{$\nameM_{43}$}
\psfrag{K}{$252310$}
\psfrag{Km}{$252310\miror$}
\psfrag{L}{$333310$}
\psfrag{Lm}{$333310\miror$}
\psfrag{L}{$\nameM_{33}$}
\psfrag{Lm}{$\nameM_{33}^\miror$}
\psfrag{M}{$326020$}
\psfrag{N}{$252310$}
\psfrag{N2}{$\nameM_{25(\chtw)}$}
\psfrag{N3}{$\nameM_{25(\chth)}$}
\psfrag{Nm}{$252310\miror$}
\psfrag{N2m}{$\nameM_{25(\chtw)}^{\miror}$}
\psfrag{Nstar}{$252310^*$}
\psfrag{Nstar3}{$\nameM_{25_1(\chth)}$}
\psfrag{Nstar2}{$\nameM_{25_1(\chtw)}$}
\psfrag{Nstarm}{$252310^*\miror$}
\psfrag{Nstar3m}{$\nameM_{25_1(\chth)}^\miror$}
\psfrag{Nstar2m}{$\nameM_{25_1(\chtw)}^\miror$}
\psfrag{O}{$326020$}
\psfrag{Om}{$326020\miror$}
\psfrag{O}{$\nameM_{32}$}
\psfrag{Om}{$\nameM_{32}^{\miror}$}
\psfrag{P}{$228010$}
\psfrag{P4}{$\nameM_{22(\chfo)}$}
\psfrag{P2}{$\nameM_{22(\chtw)}$}
\psfrag{Q}{$252310$}
\psfrag{R}{$360040$}
\psfrag{R}{$\nameM_{36}$}
\psfrag{Z}{$6400003$}
\psfrag{i1}{$A$}
\psfrag{i2}{$B$}
\psfrag{i3}{$C$}
\psfrag{tr}{$3$}
\psfrag{un}{$1$}
\psfrag{de}{$2$}
\includegraphics[width = \factor\linewidth]{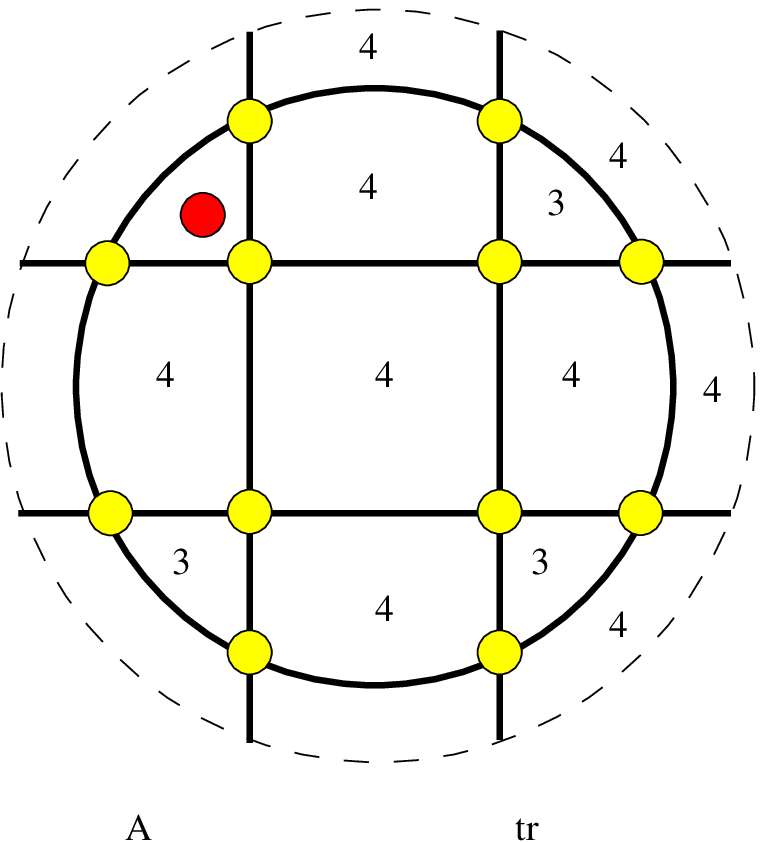}
\includegraphics[width = \factor\linewidth]{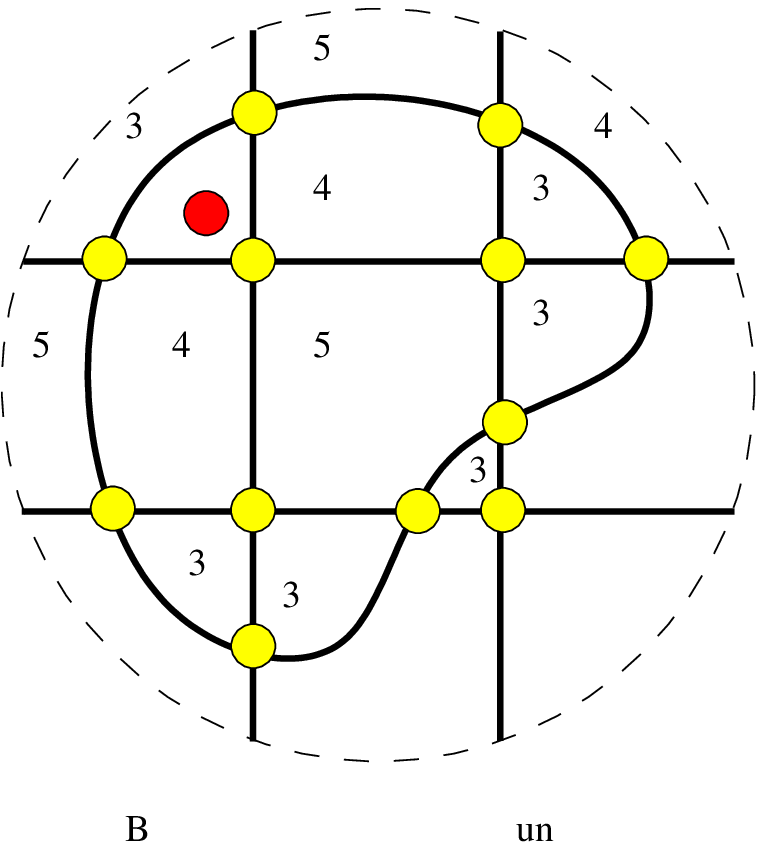}
\includegraphics[width = \factor\linewidth]{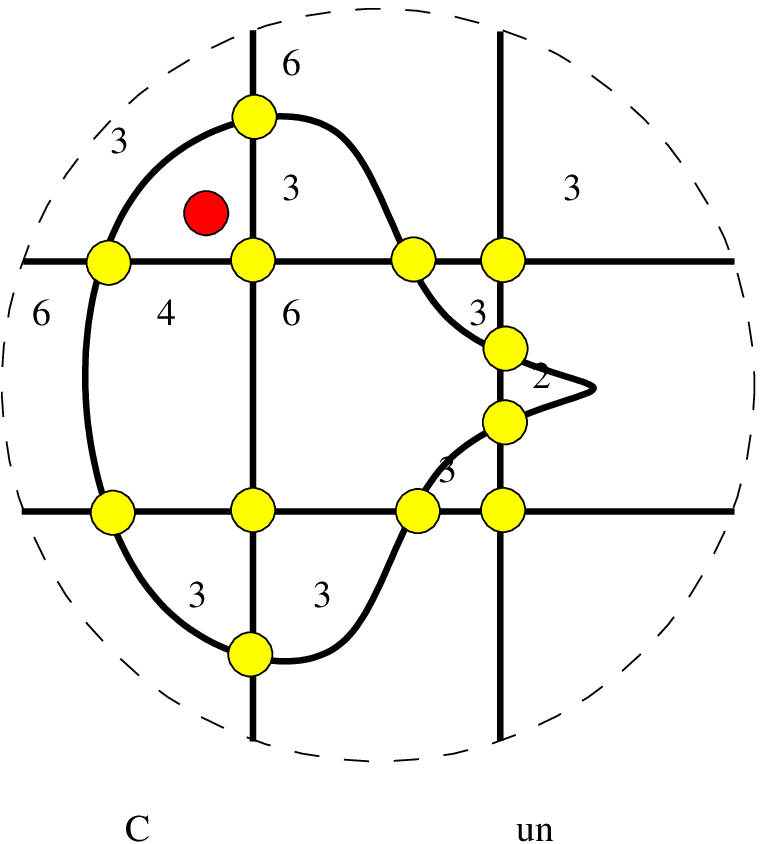}
\includegraphics[width = \factor\linewidth]{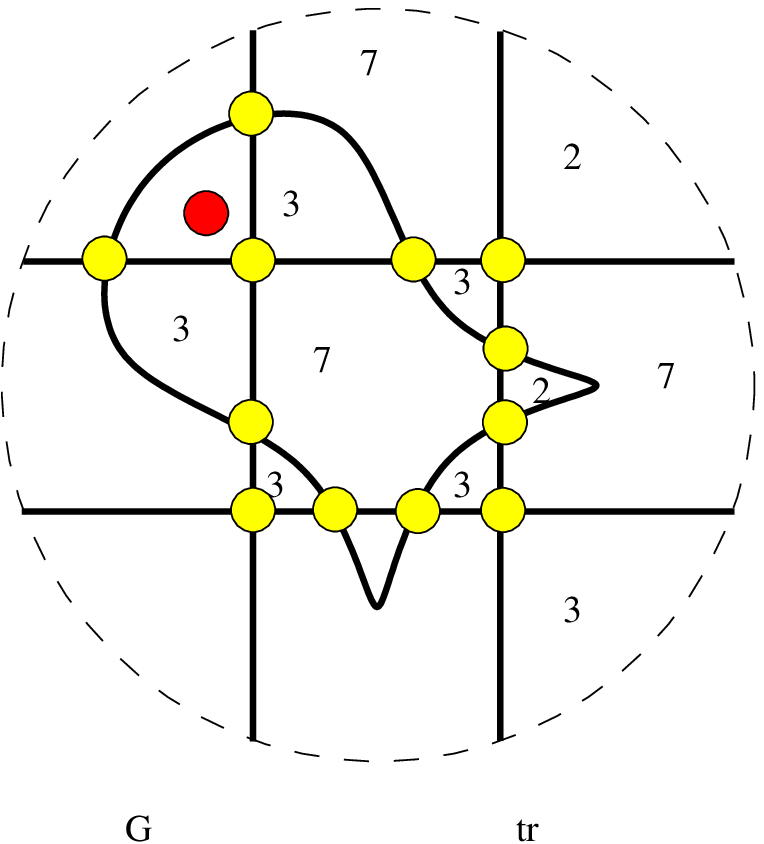}
\includegraphics[width = \factor\linewidth]{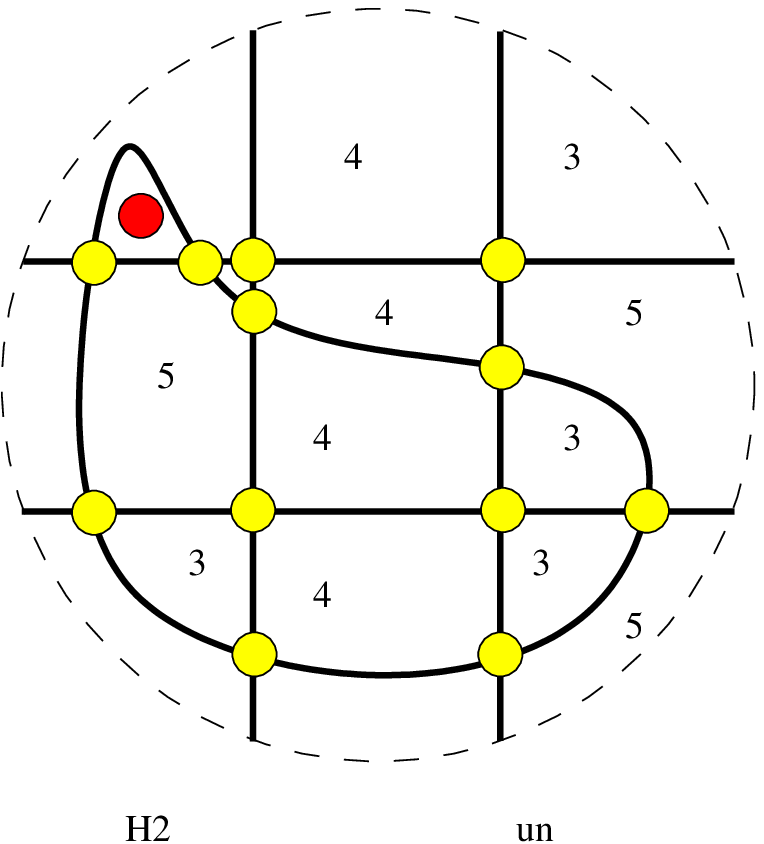}
\includegraphics[width = \factor\linewidth]{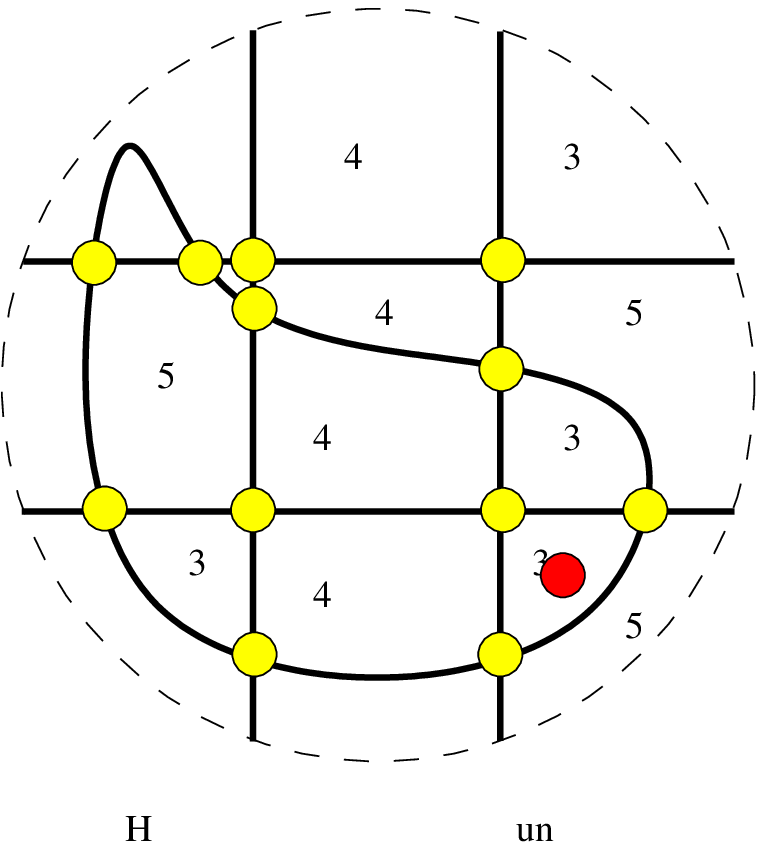}
\includegraphics[width = \factor\linewidth]{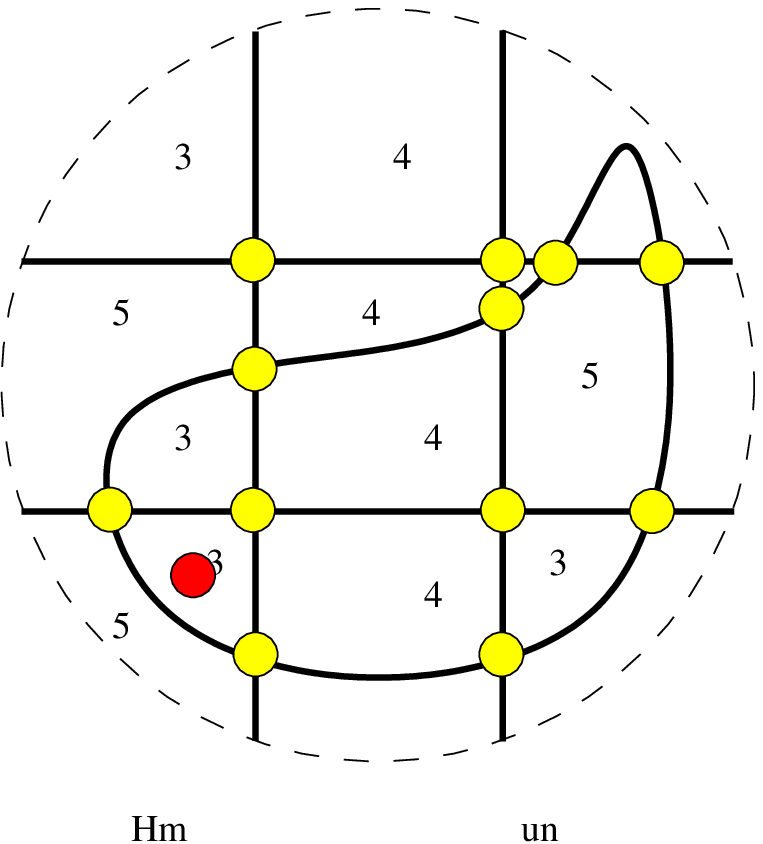}
\includegraphics[width = \factor\linewidth]{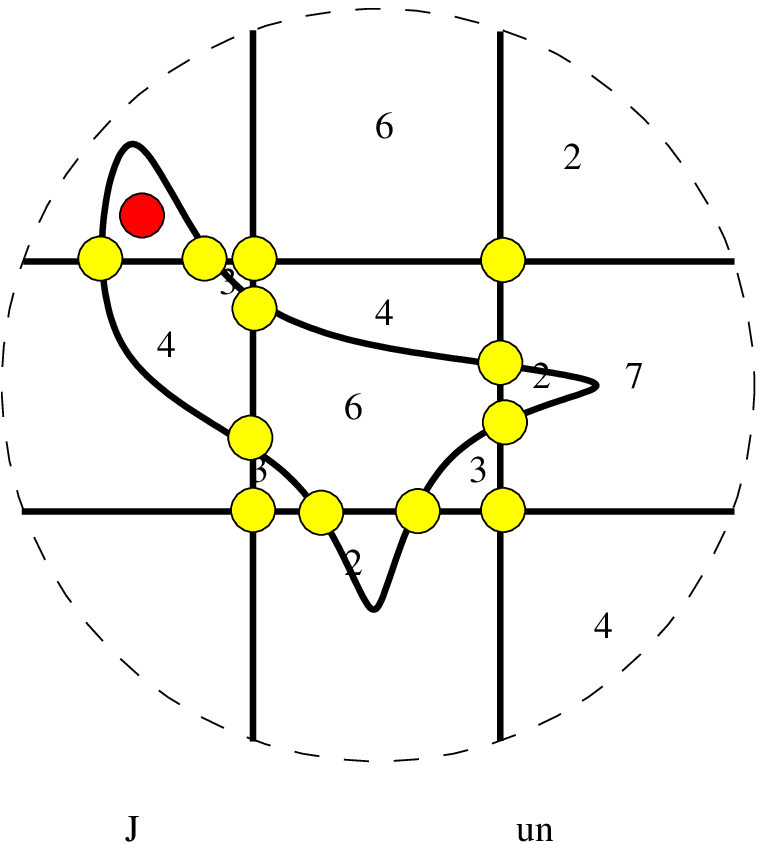}
\includegraphics[width = \factor\linewidth]{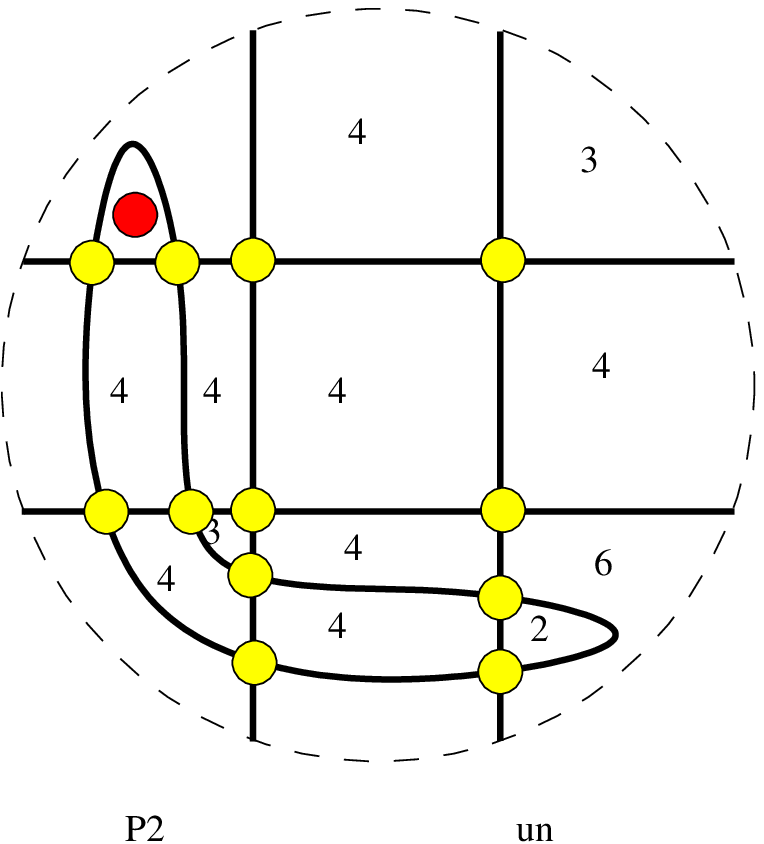}
\includegraphics[width = \factor\linewidth]{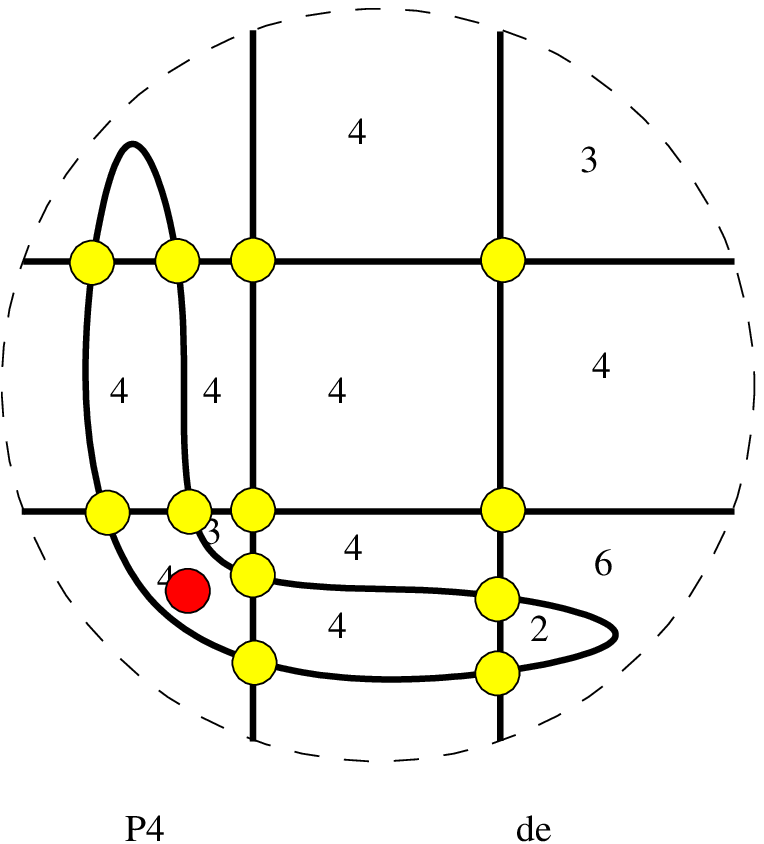}
\includegraphics[width = \factor\linewidth]{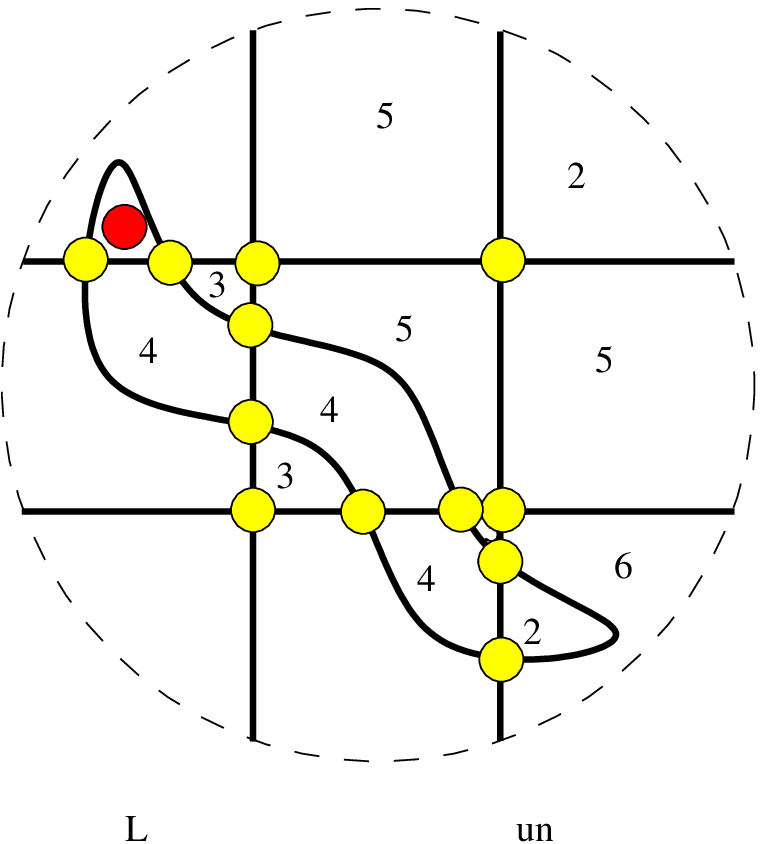}
\includegraphics[width = \factor\linewidth]{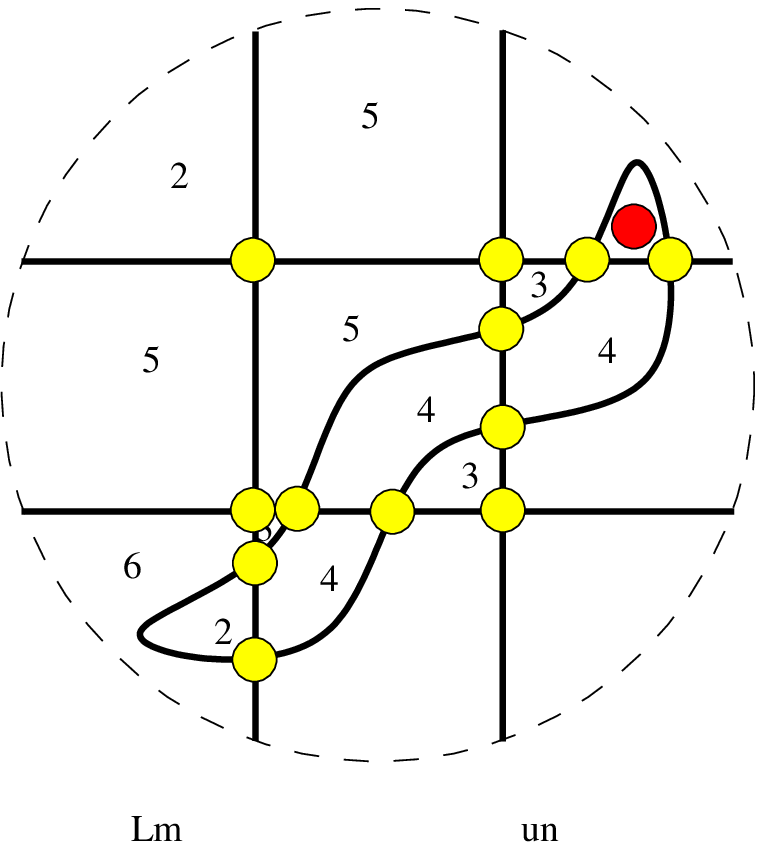}
\includegraphics[width = \factor\linewidth]{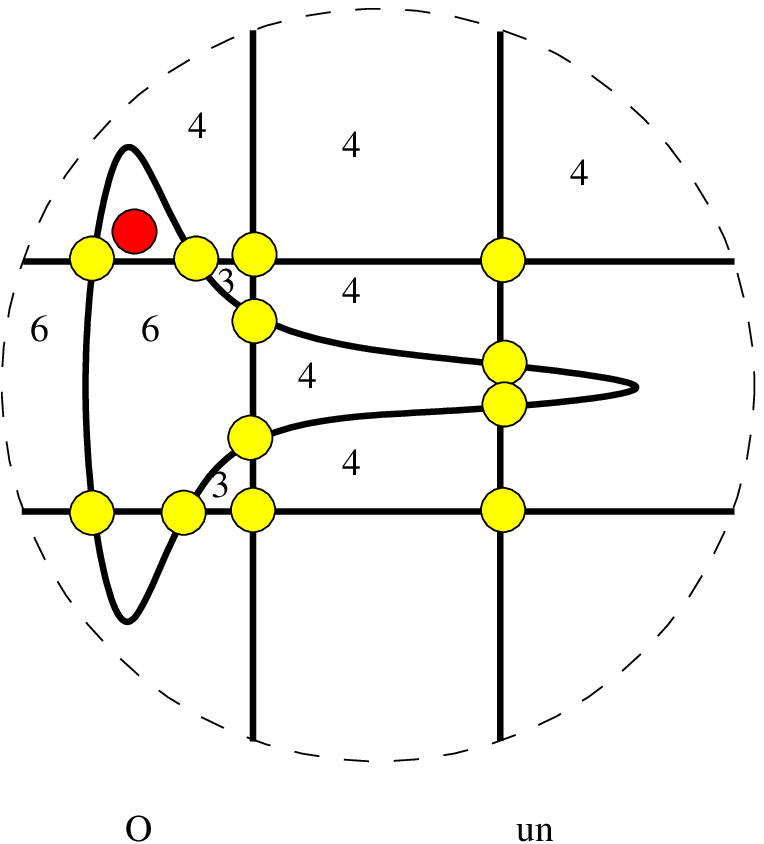}
\includegraphics[width = \factor\linewidth]{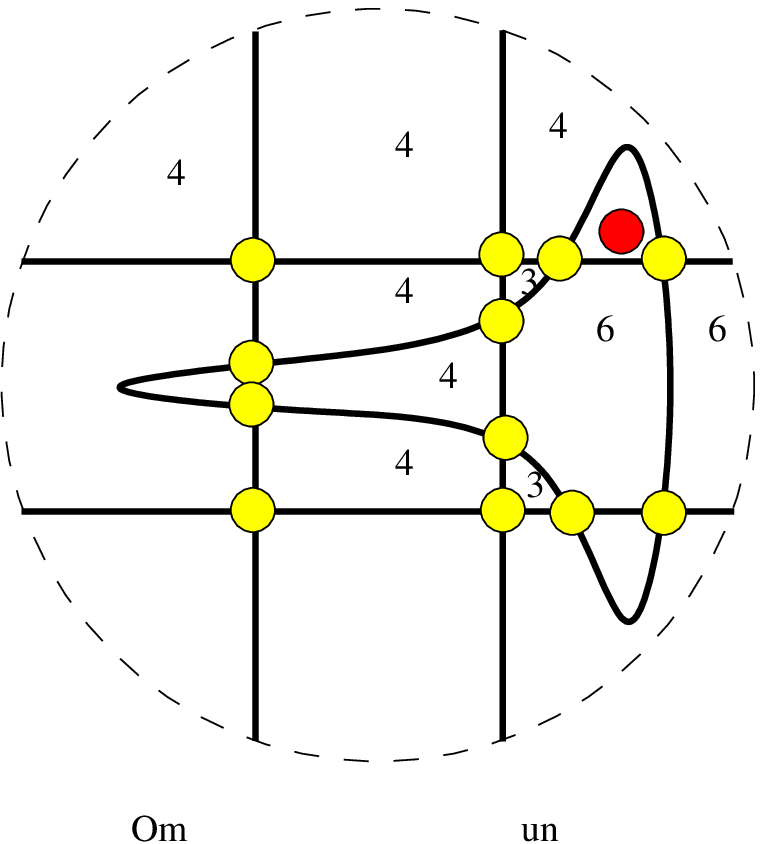}
\includegraphics[width = \factor\linewidth]{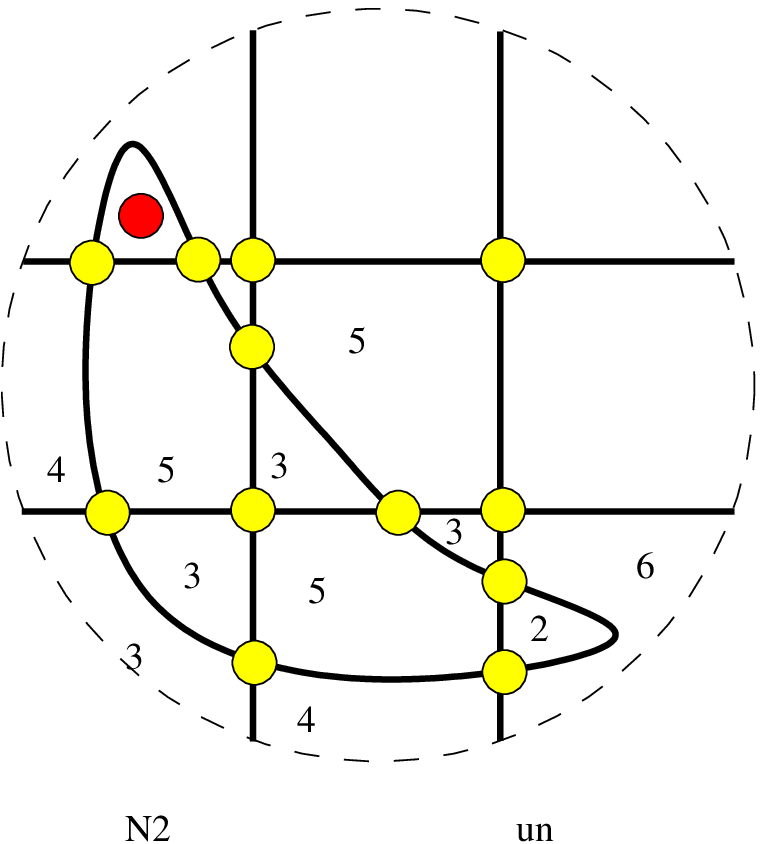}
\includegraphics[width = \factor\linewidth]{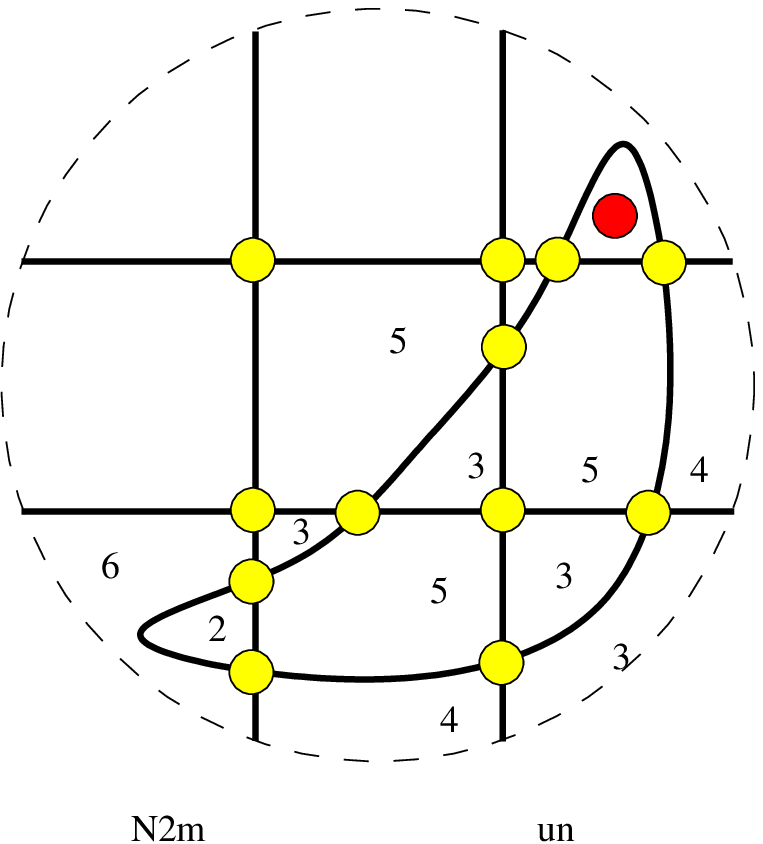}
\includegraphics[width = \factor\linewidth]{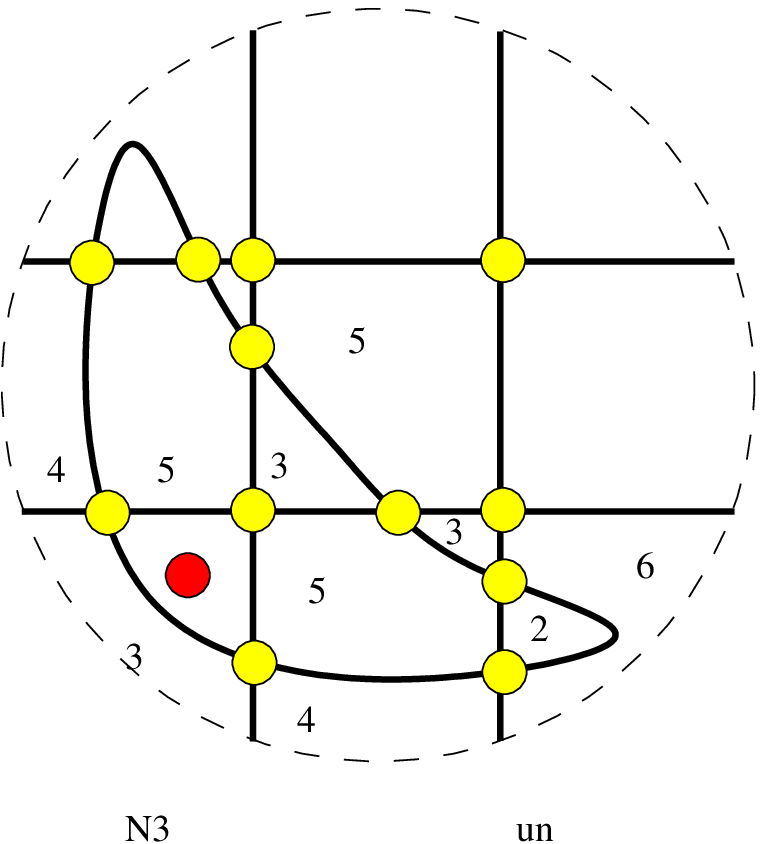}
\includegraphics[width = \factor\linewidth]{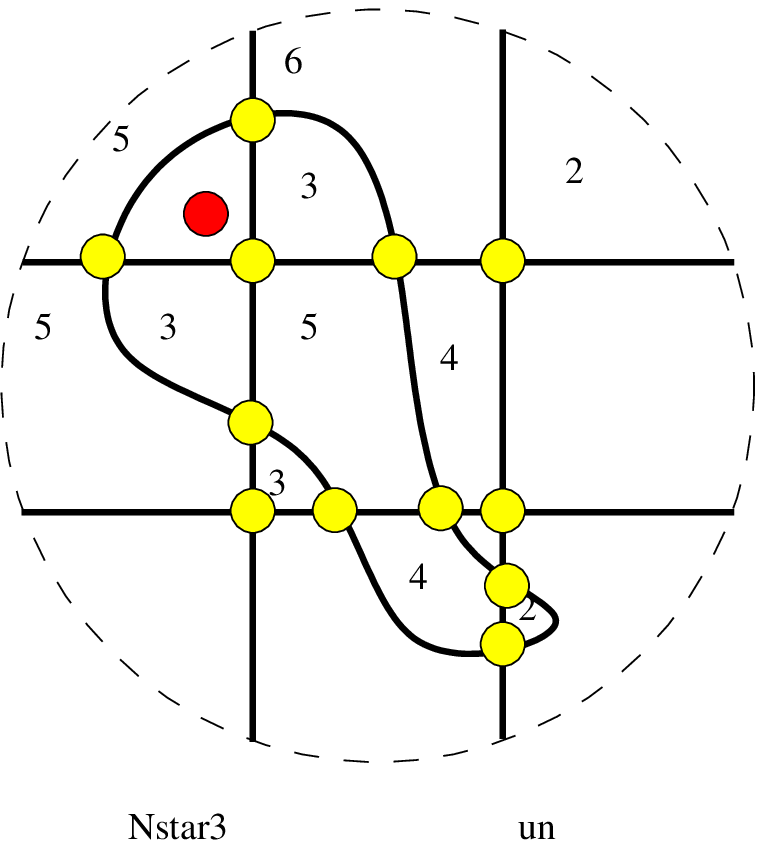}
\includegraphics[width = \factor\linewidth]{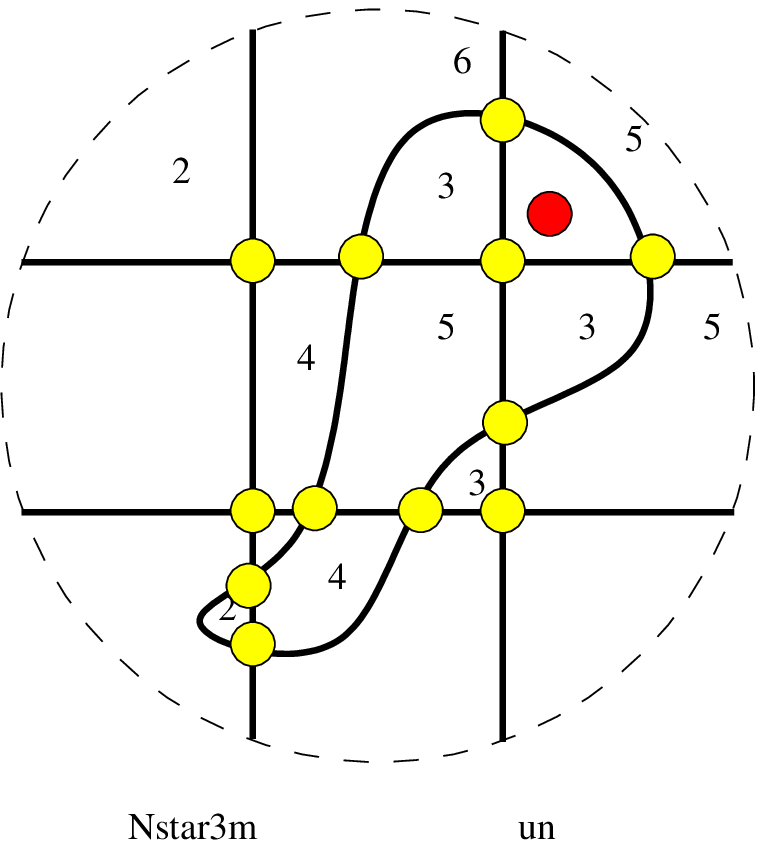}
\includegraphics[width = \factor\linewidth]{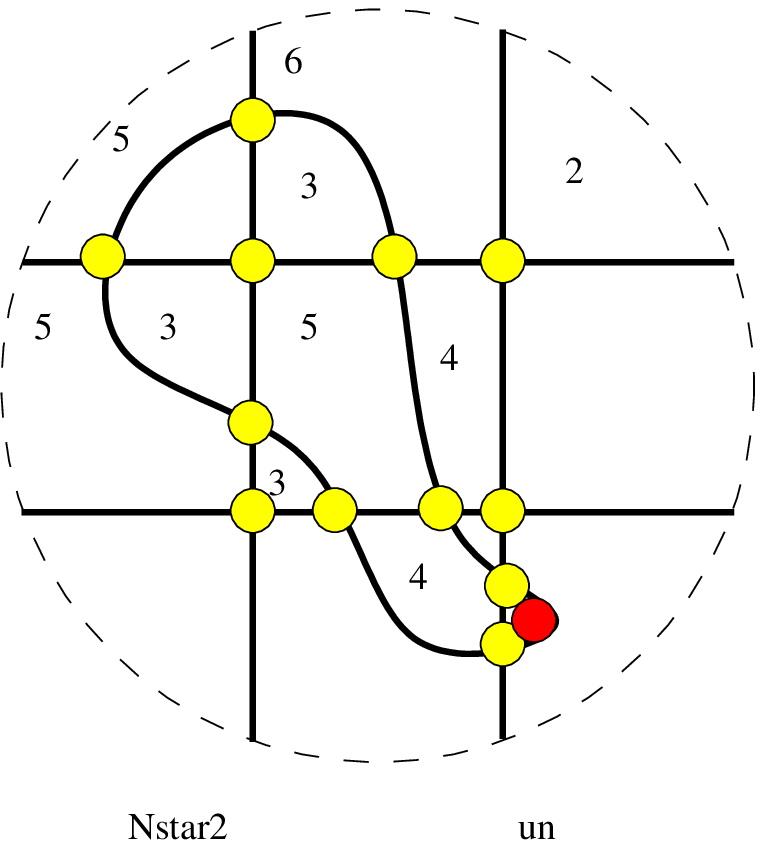}
\includegraphics[width = \factor\linewidth]{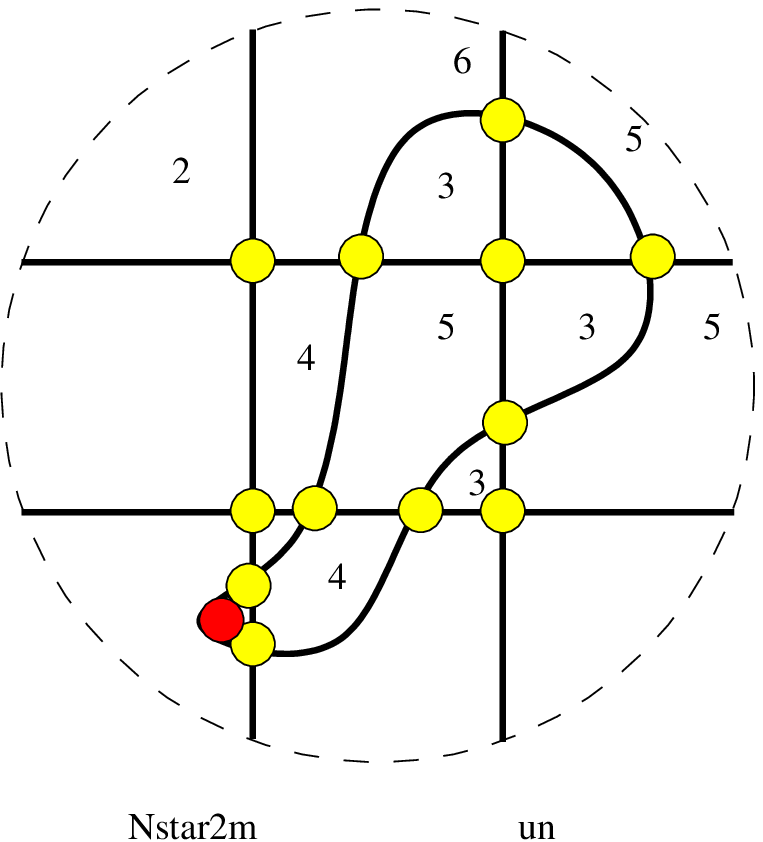}
\includegraphics[width = \factor\linewidth]{000000}
\includegraphics[width = \factor\linewidth]{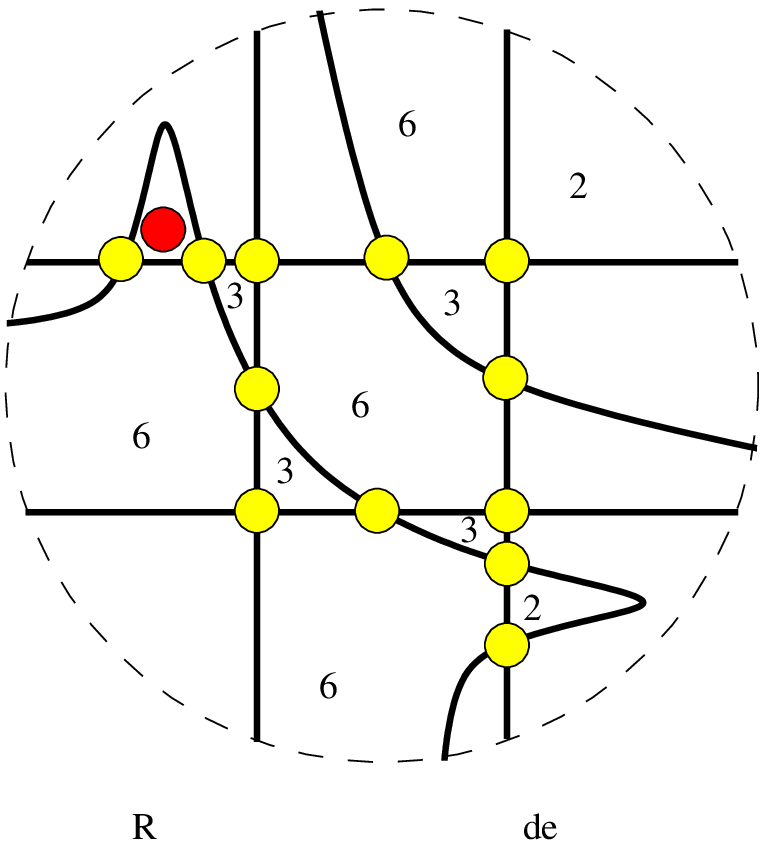}
\includegraphics[width = \factor\linewidth]{000000}
\caption{ 
\small 
%%Representatives of the 22 isomorphism classes of simple non indexed arrangements on three double pseudolines. 
%% the vector that appears below each arrangement is termed the characteristic vector of the arrangement and is defined as the sequence 
%% of its numbers of two-cells of size 2,3,4, etc.  
\label{mobiuslist}}
\end{figure}
%%%%%%%%%%%%%%%%%%%%%%%%%%%%%%%%%%%%%%%%%%%%%%%%%%%%%%%%%%%%%%%%%%%%%%%%%%%%%%%
%%%%%%%%%%%%%%%%%%%%%%%%%%%%%%%%%%%%%%%%%%%%%%%%%%%%%%%%%%%%%%%%%%%%%%%%%%%%%%%
%%%%%%%%%%%%%%%%%%%%%%%%%%%%%%%%%%%%%%%%%%%%%%%%%%%%%%%%%%%%%%%%%%%%%%%%%%%%%%%
%%%%%%%%%%%%%%%%%%%%%%%%%%%%%%%%%%%%%%%%%%%%%%%%%%%%%%%%%%%%%%%%%%%%%%%%%%%%%%%
Each diagram is labeled at its left bottom corner with a symbol to name it 
(of type $\nameM_\alpha$ where $\alpha$ is the $2$-sequence of its numbers of $2$-cells of size $2$ and $3$ possibly followed, in brackets, 
with the size of the unbounded $2$-cell of the arrangement in the case where there are several arrangements with the same $2$-sequence; $\nameM_{\alpha}$ and $\nameM_{\alpha}^\miror$ 
are mirror images of one another) and is labeled at its 
right bottom corner with the size of its automorphism group; thus the number ($118$) of simple chirotopes of families of three pairwise disjoint convex bodies 
on a given indexing set of size $3$  can be computed as the sum 
$$\sum_{k\geq 1} \frac{3!}{k} g_k = 
\frac{6}{1} \times 18
+
\frac{6}{2} \times 2 
+ 
 \frac{6}{3}\times 2 
$$
where $g_k$ is the number of arrangements of Figure~\ref{mobiuslist}  with group of automorphisms  of order~$k$.  

\clearpage
%%\section{Perturbations guided by the visibility}\label{Perturbations}
%%\section{Perturbations guided by the visibility}\label{Perturbations}
\section{Perturbation scheme}\label{Perturbations}
%%%%%%%%%%%%%%%%%%%%%%%%%%%%%%%%%%%%%%%%%%%%%%%%%%%%%%%%%%%%%%%%%%%%%%%%%%%%%%%
%%%%%%%%%%%%%%%%%%%%%%%%%%%%%%%%%%%%%%%%%%%%%%%%%%%%%%%%%%%%%%%%%%%%%%%%%%%%%%%
%%%%%%%%%%%%%%%%%%%%%%%%%%%%%%%%%%%%%%%%%%%%%%%%%%%%%%%%%%%%%%%%%%%%%%%%%%%%%%%
%%%%%%%%%%%%%%%%%%%%%%%%%%%%%%%%%%%%%%%%%%%%%%%%%%%%%%%%%%%%%%%%%%%%%%%%%%%%%%%
%%%%%%%%%%%%%%%%%%%%%%%%%%%%%%%%%%%%%%%%%%%%%%%%%%%%%%%%%%%%%%%%%%%%%%%%%%%%%%%
Our  purpose of this section is to give a detailed  proof of the following (intuitively clear) theorem.
%%%%%%%%%%%%%%%%%%%%%%%%%%%%%%%%%%%%%%%%%%%%%%%%%%%%%%%%%%%%%%%%%%%%%%%%%%%%%%%
%%%%%%%%%%%%%%%%%%%%%%%%%%%%%%%%%%%%%%%%%%%%%%%%%%%%%%%%%%%%%%%%%%%%%%%%%%%%%%%
%%%%%%%%%%%%%%%%%%%%%%%%%%%%%%%%%%%%%%%%%%%%%%%%%%%%%%%%%%%%%%%%%%%%%%%%%%%%%%%
\begin{theorem} \label{theo:perturbation}
Let $\chi$  be a non-simple chirotope of planar families of pairwise disjoint convex bodies. 
Then there exists a simple chirotope $\chi^*$, 
computable in constant time,
 such that $\chi$ and $\chi^*$ have the same set of pseudotriangulations.  \qed
\end{theorem}
%%%%%%%%%%%%%%%%%%%%%%%%%%%%%%%%%%%%%%%%%%%%%%%%%%%%%%%%%%%%%%%%%%%%%%%%%%%%%%%
%%%%%%%%%%%%%%%%%%%%%%%%%%%%%%%%%%%%%%%%%%%%%%%%%%%%%%%%%%%%%%%%%%%%%%%%%%%%%%%
%%%%%%%%%%%%%%%%%%%%%%%%%%%%%%%%%%%%%%%%%%%%%%%%%%%%%%%%%%%%%%%%%%%%%%%%%%%%%%%
Our proof is based on 
(1) the one-to-one and onto correspondence, induced by the  duality map,
between the class of chirotopes of finite planar families of disjoint convex bodies and the class of chirotopes of 
arrangements of double pseudolines~\cite{G-hp-adp-06}; cf.  Appendix~\ref{TopPlanes}; 
(2) the dual characterization of the class of pseudotriangulations of
a finite planar family of disjoint convex bodies of Pilaud and Pocchiola~\cite{pp-mppsn-12};
%the extension to any finite family of disjoint convex bodies of the dual characterization of the class of pseudotriangulations of
%a finite planar family of disjoint convex bodies in general position of~\cite{pp-mppsn-10},
%observation (of independant interest) that the dual characterization of pseudotriangulations of 
%a finite planar family of disjoint convex bodies in general position of~\cite{pp-mppsn-10} works as well for not necessarily in general position 
%finite families of disjoint convex bodies, 
and on (3)
the classical interpretation of simple pseudoline arrangements as non redundant primitive sorting 
networks~\cite[section 5.3.4]{k-acpss-73},~\cite[page 29]{k-ah-92},~\cite{b-sms-74}. 
%%% (which incidently proves that the pseudotriangulations depends only of the chirotope).
%%, first observed for  planar family of disjoint convex bodies in general position in~\cite{pp-mppsn-10}. %%% (which incidently proves that the pseudotriangulations depends only of the chirotope).
%Let $\Gamma$ be a (m{\"o}bius) arrangement of $n$ double pseudolines (cf.~\cite{G-hp-adp-09}).  

\begin{figure}[!htb]
\begin{center}
\small
\psfrag{zero}{$0$}
\psfrag{twopi}{$\exp{2\pi}$}
\psfrag{1}{$1$}
\psfrag{2}{$2$}
\psfrag{3}{$3$}
\psfrag{cp1}{$z_{+1}$}
\psfrag{cp2}{$z_{+2}$}
\psfrag{cp3}{$z_{+3}$}
\psfrag{cm1}{$z_{-1}$}
\psfrag{cm2}{$z_{-2}$}
\psfrag{cm3}{$z_{-3}$}
\psfrag{cp1}{$1$}
\psfrag{cp2}{$2$}
\psfrag{cp3}{$3$}
\psfrag{cm1}{$1$}
\psfrag{cm2}{$2$}
\psfrag{cm3}{$3$}
\psfrag{a}{$a$}
\psfrag{b}{$b$}
\psfrag{c}{$c$}
\psfrag{d}{$d$}
\psfrag{za}{$a$}
\psfrag{zb}{$b$}
\psfrag{zc}{$c$}
\psfrag{zd}{$d$}
\psfrag{pi}{$\pi$}
\includegraphics[width = 0.9875\linewidth]{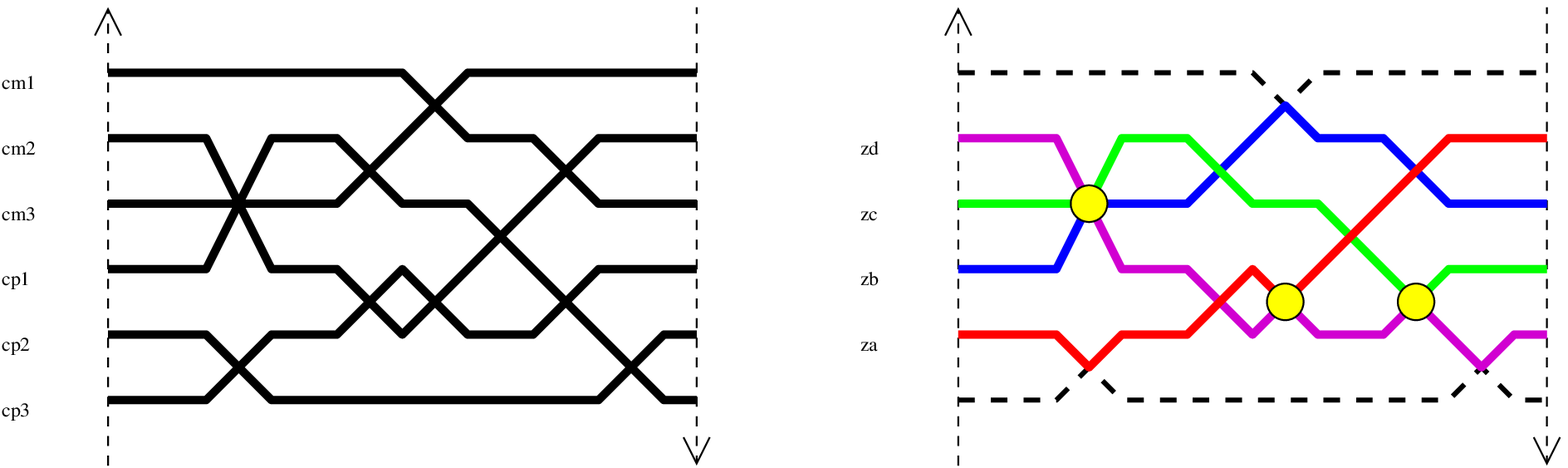}%xfigfinal
\end{center}
\caption{\label{fig:dualcharacpt}}
\end{figure}

By a {\it pseudotriangulation} of a  double pseudoline arrangement $\nabla$ living in an open crosscap
we mean a pseudoline arrangement with contact points $\cal L$, that is, a finite family of pseudolines that intersect pairwise in a finite number of points 
of which exactly one is a transversal intersection point, with the property that $\bigcup {\cal L}$ covers exactly the topological closure of $\bigcup \nabla$ minus the 
 first level of the arrangement $\nabla$ with respect to its noncompact face.
% It has been observed in~\cite{TR-94-4} that 
%the dual arrangement of the pseudotriangles of a pseudotriangulation of a finite planar family of disjoint convex bodies is a pseudotriangulation of its dual arrangement.
%In~\cite{pp-mppsn-10} it is proved that the converse is true for simple arrangements. Actually the converse is true even for non-simple arrangements. 
%% say dual arrangement of a family of convex bodies $\Delta$.
Figure~\ref{fig:dualcharacpt} depicts a non simple arrangement of three double pseudolines $1,2$ and $3$ 
and one of its pseudotriangulation $a,b,c,d$ (red, blue, green and purple in pdf color): the vertices that support a contact point are marked with a disk (yellow in pdf color); 
 the non simple vertex supports only one contact point; this contact point is a contact point between the pseudolines $b$ and $c$.
%%  a crossing point between $d$ and $c$ and also a crossing point between $d$ and $b$.

%%%%%%%%%%%%%%%%%%%%%%%%%%%%%%%%%%%%%%%%%%%%%%%%%%%%%%%%%%%%%%%%%%%%%%%%%%%%%%%
%%%%%%%%%%%%%%%%%%%%%%%%%%%%%%%%%%%%%%%%%%%%%%%%%%%%%%%%%%%%%%%%%%%%%%%%%%%%%%%
%%%%%%%%%%%%%%%%%%%%%%%%%%%%%%%%%%%%%%%%%%%%%%%%%%%%%%%%%%%%%%%%%%%%%%%%%%%%%%%
%%%%%%%%%%%%%%%%%%%%%%%%%%%%%%%%%%%%%%%%%%%%%%%%%%%%%%%%%%%%%%%%%%%%%%%%%%%%%%%
%%%%%%%%%%%%%%%%%%%%%%%%%%%%%%%%%%%%%%%%%%%%%%%%%%%%%%%%%%%%%%%%%%%%%%%%%%%%%%%
%%%%%%%%%%%%%%%%%%%%%%%%%%%%%%%%%%%%%%%%%%%%%%%%%%%%%%%%%%%%%%%%%%%%%%%%%%%%%%%
%%%%%%%%%%%%%%%%%%%%%%%%%%%%%%%%%%%%%%%%%%%%%%%%%%%%%%%%%%%%%%%%%%%%%%%%%%%%%%%
%%%%%%%%%%%%%%%%%%%%%%%%%%%%%%%%%%%%%%%%%%%%%%%%%%%%%%%%%%%%%%%%%%%%%%%%%%%%%%%

%%\begin{theorem}[\cite{TR-94-4},\cite{pp-mppsn-10}] Let $\Delta$ be a finite planar family of disjoint convex bodies, let $\Gamma$ be its dual arrangement, and let $\Phi$ be the map that assigns to a pseudotriangulation of $\Delta$ 
\begin{theorem}[Pilaud and Pocchiola~\cite{pp-mppsn-12}]
 Let $\Delta$ be a finite planar family of pairwise disjoint convex bodies, let $\nabla$ be its dual arrangement, 
and let $\Phi$ be the map that assigns to a pseudotriangulation of $\Delta$ 
the arrangement of the dual pseudolines of its pseudotriangles. Then $\Phi$ realizes a one-to-one and onto  correspondence between the set of pseudotriangulations of $\Delta$ 
and the set of pseudotriangulations of $\nabla$. \qed
\end{theorem}
\begin{proof} For completeness and clarity we repeat the proof of~\cite{pp-mppsn-12} (that consider only convex bodies in general position).
That $\Phi$ is well-defined and one-to-one and that its range is included in the set of pseudotriangulations of $\nabla$ was explicitly observed in~\cite{TR-94-4}. 
%%%~\cite{G-pv-ptta-96,G-pv-vc-96,TR-94-4}. 
%\begin{enumerate}
%\item the dual of a pseudotriangle is a pseudoline;
%\item the intersection point of two tangents to a pseudotriangle is included in that pseudotriangle;
%\item a tangent to a convex body of $\Delta$ not in the set of supporting lines of the bitangent line segments of $\EPTG$ is tangent to the convex hull of $\Delta$ or (exclusive) to a pseudotriangle of $\EPTG$.   
%\end{enumerate}
To prove that the range of $\Phi$ is exactly the set of pseudotriangulations of $\nabla$ we use a simple counting argument involving the tangency and winding numbers. 
Let $\Delta_1,\Delta_2,\ldots,\Delta_n$ be the convex bodies of $\Delta$ and let 
 $\EPTG_1,\EPTG_2\ldots,\EPTG_{2n-2}$ be the envelopes of the  pseudolines 
%%$\PTG_1,\ldots,\PTG_{2n-2}$ of a pseudotriangulation of $\Gamma$.  
of a pseudotriangulation of $\nabla$.  
%Thus if we write $e_1v_1e_2v_2\ldots e_mv_m$ for the decomposition of the pseudoline $\PTG_i$ induced by its  breakpoints $v_i$ and 
%$\Gamma_{\sigma(i)}$  for the  supporting double pseudoline of $e_i$ then 
%$\EPTG_i$ is the (closed) curve  $\env{e}_1\env{v}_1\env{e}_2\env{v}_2\ldots \env{e}_m\env{v}_m$ where $\env{e}_i$ is the set of contact points of the lines of $e_i$ with the  
%convex body $\Delta_{\sigma(i)}$ and where $\env{v}_i$ is the bitangent line segment joining $\nabla_{i}$ to $\nabla_{i+1}$ supported by the line $v_i$. 
%The contact points of $\env{v}_i$ with the convex bodies split $\env{v}_i$ into a sequence of bitangent line segments $b_{i1},b_{i2},\ldots,b_{i{k_i}}$ free of contact 
%points except at their endpoints.
Our goal is to prove that the $\EPTG_i$ %%%, $1\leq i\leq 2n-2$, 
are the boundaries of the pseudotriangles of a pseudotriangulation of~$\Delta$. 
%Our goal is to prove that the set of $b_{ik_j}$ is a  pseudotriangulation of $\Delta$. 
%Write $w(x, \EPTG_i)$ for the winding number of a point $x$ with respect to $\EPTG_i$.
This boils down to proving that for any point $x$ avoiding the boundaries of the  $\Delta_i$ and $\EPTG_i$ 
\begin{equation}\label{count}
\sum_{i=1}^{2n-2} w(x, \EPTG_i) = \begin{cases} 
                    1 & \text{if $x \in \Delta_0\setminus \bigcup_{i=1}^n \Delta_i$} \\
                    0 & \text{otherwise}
\end{cases}
\end{equation} 
%%where $w(x, \EPTG)$ is the sum of the winding numbers $w(x, \EPTG_i)$ of $x$ with respect to the $\EPTG_i$ and where $\Delta_0$ is the convex hull of the $\Delta_i$.
where $w(x, \EPTG_i)$ is the winding number of $x$ with respect to $\EPTG_i$ and where $\Delta_0$ is the convex hull of the $\Delta_i$.
Introduce in the picture the tangency number $t(x,\EPTG_i)$ of a point $x$ with respect to $\EPTG_i$ as the number of tangents to $\EPTG_i$ through $x$ or, to put it differently,
as the number of intersection points between the dual pseudolines of $x$ and $\EPTG_i$. 
Similarly introduce in the picture the tangency number $t(x,\Delta_i)$ of $x$ with respect to $\Delta_i.$ 
%%as the number of tangents to $\Delta_i$ through $x$.
The reader will easily check that 
\begin{equation}\label{precount}
\sum_{i =1}^n t(x,\Delta_i) = t(x,\Delta_0) + \sum_{i = 1}^{2n-2} t(x,\EPTG_i);
\end{equation} 
that $t(x,\Delta_i) =0$ or $2$ depending on whether $x$ belongs to the interior or to the exterior of $\Delta_i$;
and finally that $t(x,\EPTG_i) = 1+ 2 w(x,\EPTG_i)$.
Equation~(\ref{count})  follows easily.
\end{proof}

\begin{proof}[Proof of Theorem~\ref{theo:perturbation}] 
Let $\Gamma$ be a realization  of $\chi$ as a double pseudoline arrangement. 
We define $\chi^*$ as the chirotope of an arrangement of double pseudolines $\Gamma^*$ obtained $\Gamma$ by a sequence of mutations. 
Let $v$ be a non-simple vertex of $\Gamma$ with degree $2k$, $k \geq 3$. 
In the vicinity of $v$ the arrangement $\Gamma$ is isomorphic 
to a pencil of pseudolines $\ell_1,\ell_2,\ldots,\ell_k$  of the unit disk as indicated in the left diagram of Figure~\ref{fig:perturbationfou} where each pseudoline 
is oriented from left to right 
and where each pseudoline 
(except $\ell_1$ and $\ell_k$ for which the information is not relevant) 
bears a distinguished side---indicated by a rectangle in the figure---which corresponds to the  M{\"o}bius side of the corresponding double pseudoline.  
%%%%%%%%%%%%%%%%%%%%%%%%%%%%%%%%%%%%%%%%%%%%%%%%%%%%%%%%%%%%%%%%%%%%%%%%%%%%%%%
%%%%%%%%%%%%%%%%%%%%%%%%%%%%%%%%%%%%%%%%%%%%%%%%%%%%%%%%%%%%%%%%%%%%%%%%%%%%%%%
%%%%%%%%%%%%%%%%%%%%%%%%%%%%%%%%%%%%%%%%%%%%%%%%%%%%%%%%%%%%%%%%%%%%%%%%%%%%%%%
%%%%%%%%%%%%%%%%%%%%%%%%%%%%%%%%%%%%%%%%%%%%%%%%%%%%%%%%%%%%%%%%%%%%%%%%%%%%%%%
\begin{figure}[!htb]
\begin{center}
\psfrag{un}{$1$}
\psfrag{de}{$2^+$} \psfrag{tr}{$3^+$} \psfrag{qu}{$4^-$} \psfrag{ci}{$5^+$} \psfrag{si}{$6^-$} \psfrag{se}{$7$}
\psfrag{de}{$2$} \psfrag{tr}{$3$} \psfrag{qu}{$4$} \psfrag{ci}{$5$} \psfrag{si}{$6$} \psfrag{se}{$7$}
\psfrag{A}{$A_i$}
\psfrag{C}{$C_i$}
\psfrag{AA}{$\overline{A}_i$}
\psfrag{lun}{$l_1$}
\psfrag{lk}{$l_k$}
\psfrag{li}{$l_i$}
\psfrag{north}{$N$}
\psfrag{south}{$S$}
\includegraphics[width = 0.8575\linewidth]{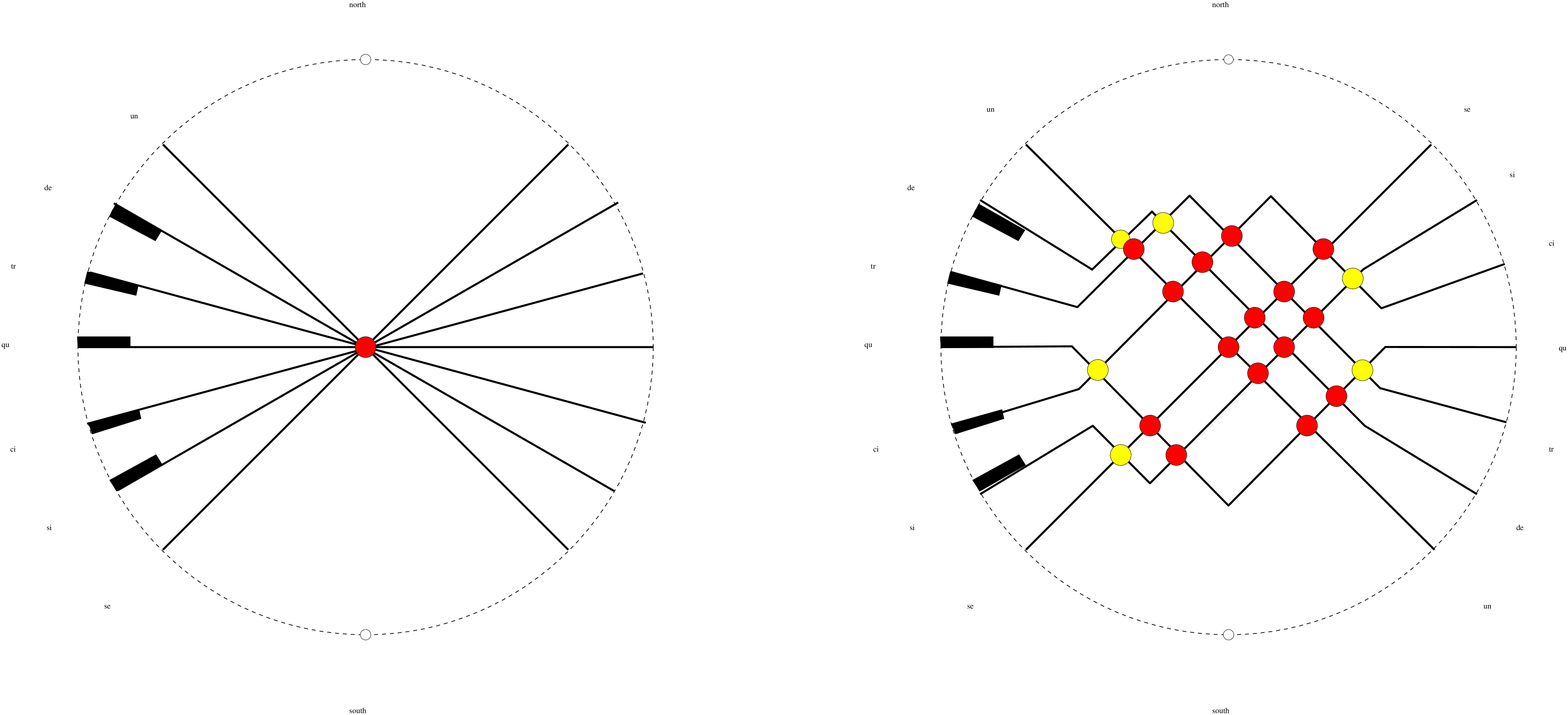}%xfigfinal
\caption{\protect \small \label{fig:perturbationfou}}
\end{center}
\end{figure}
%%%%%%%%%%%%%%%%%%%%%%%%%%%%%%%%%%%%%%%%%%%%%%%%%%%%%%%%%%%%%%%%%%%%%%%%%%%%%%%
%%%%%%%%%%%%%%%%%%%%%%%%%%%%%%%%%%%%%%%%%%%%%%%%%%%%%%%%%%%%%%%%%%%%%%%%%%%%%%%
%%%%%%%%%%%%%%%%%%%%%%%%%%%%%%%%%%%%%%%%%%%%%%%%%%%%%%%%%%%%%%%%%%%%%%%%%%%%%%%
%%%%%%%%%%%%%%%%%%%%%%%%%%%%%%%%%%%%%%%%%%%%%%%%%%%%%%%%%%%%%%%%%%%%%%%%%%%%%%%
%%%%%%%%%%%%%%%%%%%%%%%%%%%%%%%%%%%%%%%%%%%%%%%%%%%%%%%%%%%%%%%%%%%%%%%%%%%%%%%
We perturb the pencil $\ell_i$ in the vicinity of its vertex into the unique cyclic arrangement $\ell'_i$ 
with the property that the pseudoline $\ell'_i$ contributes to the upper or lower envelope 
of the arrangement depending on whether its  M{\"o}bius side contains the  south  or north pole of the unit disk, as illustrated in the right diagram of Figure~\ref{fig:perturbationfou}. 
Carrying back this perturbation on $\Gamma$ we get an arrangement $\Gamma'$  whose number of non-simple vertices is one less than the number of non-simple vertices 
of $\Gamma$.  
This perturbation is clearly compatible with the isomorphism relation and stable under taking sub-arrangement, that is, for any $J \subset I$ the restriction to $J$ of the 
arrangement $\Gamma'$ is isomorphic to the perturbed version of the restriction of $\Gamma$ to $J$: more formally $(\Gamma')_J =  (\Gamma_J)'$.
In particular the chirotope of the perturbation of $\Gamma$ is the perturbation of the chirotope of $\Gamma$; this proves that $\chi'$ is computable in constant time.  
Repeating this perturbation at each non-simple vertex of the arrangement $\Gamma$ we get a simple arrangement $\Gamma^*$ whose chirotope $\chi^*$ is computable in constant time. 
It remains to show that $\chi$ and $\chi'$ (hence $\chi^*$) have the same set of pseudotriangulations.  
%%%%%%%%%%%%%%%%%%%%%%%%%%%%%%%%%%%%%%%%%%%%%%%%%%%%%%%%%%%%%%%%%%%%%%%%%%%%%%%
%%We first examine the case where the vertex $v$ is not a vertex of the one-level of the arrangement $\Gamma$.
Let $\cal L$ be a pseudotriangulation of $\Gamma$.  
We construct a pseudotriangulation ${\cal L}'$ of $\Gamma'$ which coincides with ${\cal L}$ except in the vicinity of $v.$
Let $\sigma$ be the permutation of $\{1,2,\ldots,k\}$ that maps the index $i$ on the index $j$ defined by the condition that the pseudoline of $\cal L$ (or the first level of $\Gamma$)
 entering  $v$ along $\ell_i$ leaves $v$ along $\ell_{j}$.
%%%; note that the bitangent line segments $A_uA_{u+1},A_{u+1}A_{u+2},\ldots, A_{v-1,v}$, where $u = \min\{i,i'\}$ and $v = \max\{i,i'\}$ are bitangent line segments of the pseudotriangulation. 
Now we interpret the arrangement of lines $\ell'_i$ as a primitive sorting network to carry the keys $\sigma(i)$ from the left endpoints of the $\ell_i$ to the right endpoints
of the $\ell_{\sigma(i)}$ (\cite[page 29]{k-ah-92}); this yields to a touching or crossing status to each vertex of the arrangement $\ell'_i$ depending on whether 
the corresponding comparator is feeded with entries in sorted order or not and, therefore, to a pseudotriangulation 
${\cal L}'$ of $\Gamma'$ which coincides with ${\cal L}$ except in the vicinity of $v.$    
For example if, in the example of Figure~\ref{fig:perturbationfou},  we take  
$$
\sigma = 
\begin{pmatrix}
1 & 2  & 3 & 4 & 5 & 6 & 7 \\
4 & 1 & 2 & 3 & 6 & 5 & 7  
\end{pmatrix}   
$$
then we sort (in decreasing order)  the array $[4,1,2,3,6,5,7]$ with the sorting network
$$
\includegraphics[width = 0.750\linewidth]{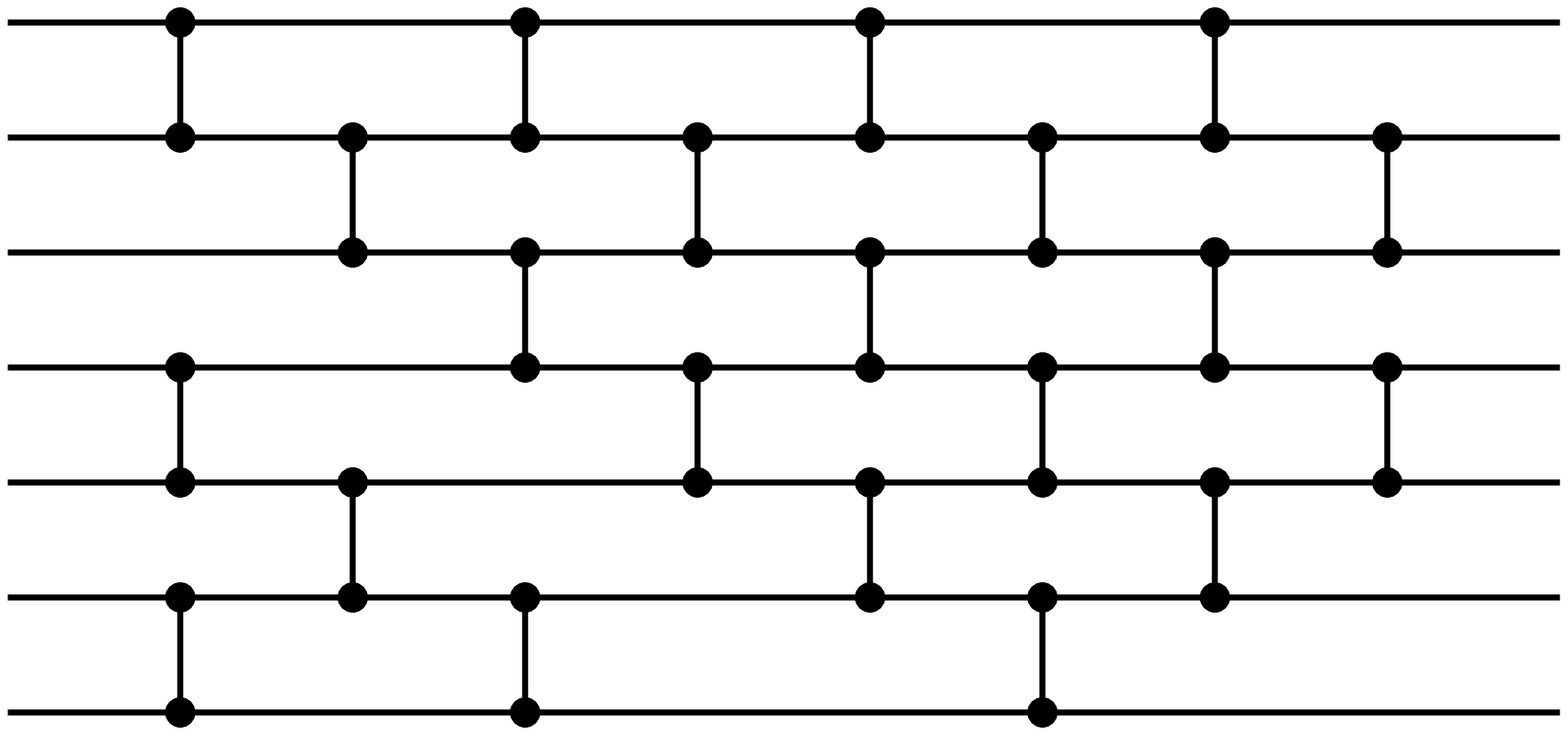}
$$
corresponding to the arrangement $\ell'_i$. 
The sequence of comparisons and arrays obtained during the sorting process is then 
$$
\psfrag{un}{$1$}
\psfrag{de}{$2$}
\psfrag{tr}{$3$}
\psfrag{qu}{$4$}
\psfrag{ci}{$5$}
\psfrag{si}{$6$}
\psfrag{se}{$7$}
\includegraphics[width = 0.750\linewidth]{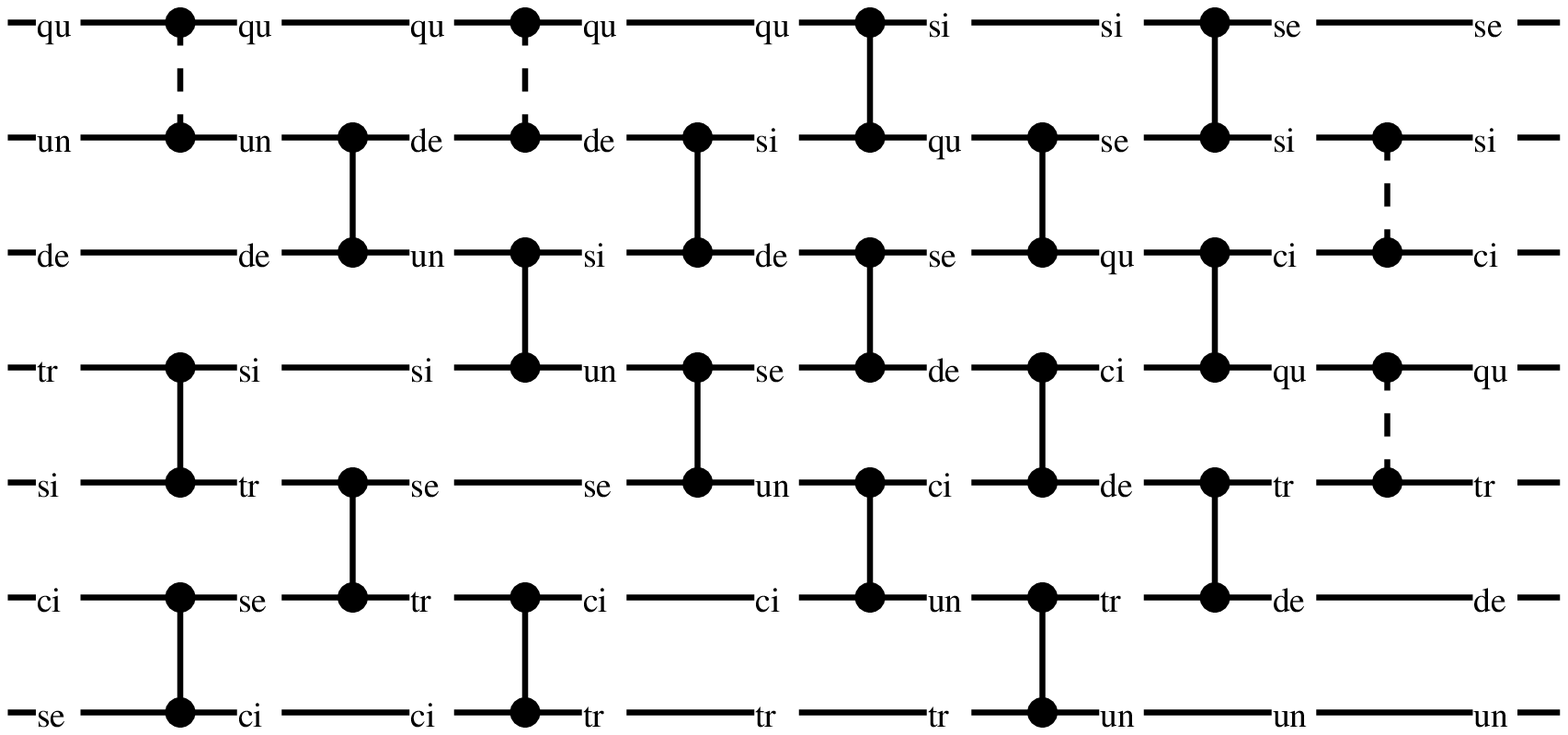}
$$
(the comparators feeded with a pair of indices in sorted order are drawn dashed) 
which yields to a touching status for the vertices $v_{12},v_{23},v_{34}$ and $v_{56}$ of the arrangement $\ell'_i$ and to a crossing status for the other vertices, as 
indicated in the following figure where the touching vertices are marked with a little (yellow in pdf color) disk.
  %%%%%%%%%%%%%%%%%%%%%%%%%%%%%%%%%%%%%%%%%%%%%%%%%%%%%%%%%%%%%%%%%%%%%%%%%%%%%%%
%%%%%%%%%%%%%%%%%%%%%%%%%%%%%%%%%%%%%%%%%%%%%%%%%%%%%%%%%%%%%%%%%%%%%%%%%%%%%%%
%%%%%%%%%%%%%%%%%%%%%%%%%%%%%%%%%%%%%%%%%%%%%%%%%%%%%%%%%%%%%%%%%%%%%%%%%%%%%%%
%%%%%%%%%%%%%%%%%%%%%%%%%%%%%%%%%%%%%%%%%%%%%%%%%%%%%%%%%%%%%%%%%%%%%%%%%%%%%%%
\begin{figure}[!htb]
\begin{center}
\psfrag{un}{$1$}
\psfrag{de}{$2$}
\psfrag{tr}{$3$}
\psfrag{qu}{$4$}
\psfrag{ci}{$5$}
\psfrag{si}{$6$}
\psfrag{se}{$7$}
\psfrag{A}{$A_i$}
\psfrag{C}{$C_i$}
\psfrag{AA}{$\overline{A}_i$}
\psfrag{lun}{$l_1$}
\psfrag{lk}{$l_k$}
\psfrag{li}{$l_i$}
\psfrag{north}{$N$}
\psfrag{south}{$S$}
\includegraphics[width = 0.4258575\linewidth]{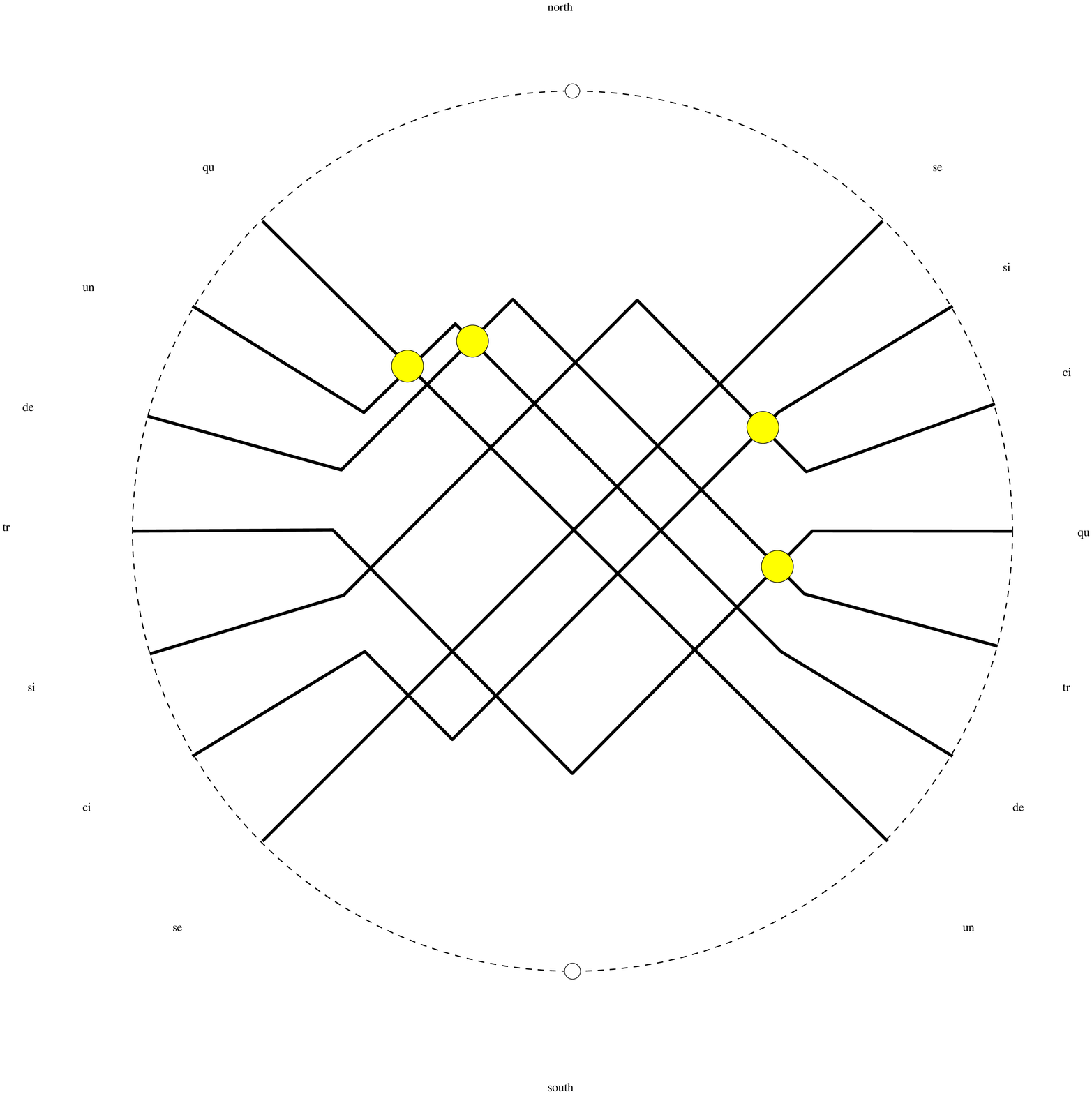}%xfigfinal
\caption{\protect \small 
%% Cutting  the complement in the plane of the interiors of the convex bodies along a given set of constraints we get a set of  surfaces 
%%whose cuffs contain exactly one cusp point per endpoint of constraint. 
%%These cusp points split the cuffs of these surfaces into a collection  of convex curves, called the views of the family of bodies and constraints.
\label{fig:perturbationfou}}
\end{center}
\end{figure}
%%%%%%%%%%%%%%%%%%%%%%%%%%%%%%%%%%%%%%%%%%%%%%%%%%%%%%%%%%%%%%%%%%%%%%%%%%%%%%%
%%%%%%%%%%%%%%%%%%%%%%%%%%%%%%%%%%%%%%%%%%%%%%%%%%%%%%%%%%%%%%%%%%%%%%%%%%%%%%%
%%%%%%%%%%%%%%%%%%%%%%%%%%%%%%%%%%%%%%%%%%%%%%%%%%%%%%%%%%%%%%%%%%%%%%%%%%%%%%%
%%%%%%%%%%%%%%%%%%%%%%%%%%%%%%%%%%%%%%%%%%%%%%%%%%%%%%%%%%%%%%%%%%%%%%%%%%%%%%%
%%%%%%%%%%%%%%%%%%%%%%%%%%%%%%%%%%%%%%%%%%%%%%%%%%%%%%%%%%%%%%%%%%%%%%%%%%%%%%%
It remains to show that if ${\cal L}''$ is a pseudotriangulation of $\Gamma'$ which coincides with ${\cal L}$ except in the vicinity of $v$, 
then ${\cal L}'={\cal L}''.$    
Let $\Delta'$ be a realization of $\Gamma'$ as a planar family of convex bodies and let $\Delta'_i$ be the convex body corresponding to the double pseudoline supported by $\ell'_i$. 
By construction the bitangent line segment $b_{ij}$ 
supported by the line $v_{ij}$ intersection of the curves $\ell'_i$ and $\ell'_j$, $i<j$, is free with respect to the $\Delta'_i$ 
if and only if $j=i+1$ and the $b_{ii+1}$ are pairwise interior noncrossing. This prove that ${\cal L}'={\cal L}''$ since the complex 
of pseudotriangulations is strongly flag connected. 
%Let ${\cal L}_i$ be the pseudoline of $\cal L$ that enters the  vertex $v$ supported by the line $\ell_i$ and let 
%${\cal L}_{i'}$ be the pseudoline of $\cal L$ that leaves the  vertex $v$ supported by the line $\ell_{k-i+1}.$  Clearly the 
%ap $i \mapsto i'$ is a permutation $\sigma$  of $\{1,2,\ldots,k\}$. 
%Let ${\cal L}_i$ be the pseudoline of $\cal L$ that enters the  vertex $v$ supported by the line $\ell_i$ and let 
%$\ell_{i'}$ be the pseudoline of that leaves the  vertex $v$ supported by the pseudoline  ${\cal L}_i.$  
%Clearly the map $i \mapsto i'$ is a permutation $\tau$  of $\{1,2,\ldots,k\}$. 
%As we shall see in the next paragraph this is a simple consequence 
%of the following dual definition of a pseudotriangulation that was first given in the setting of families of convex bodies in general position in~\cite{pp-mppsn-10}. 
%Without loss of generality we can assume that arrangement $\Gamma$ lies in the standard M{\"o}bius strip, quotient of $\mathbb{R}^2$ undre the applcation that assigns to the pair $(x,y)$ the 
%pair $(x+1,-y)$,  and that its double pseudolines are monotone with respect to the core circle $x=0$. 
%We perturb locally around $v$ the arrangement $\Gamma$ into an arrangement $\Gamma^*$ such that, locally around $v$,
% $\Gamma^*_i$ contributes for one edge to the upper or lower envelope of $\Gamma^*$ depending on whether
%the M{\"o}bius side of $\Gamma_i$ lies below or above $\Gamma_i$.
%%Introduire les cocycles; lors d'une mutation seuls les cocycles des sommets concernés par la mutation peuvent changer de valeur.
\end{proof}

%%%%%%%%%%%%%%%%%%%%%%%%%%%%%%%%%%%%%%%%%%%%%%%%%%%%%%%%%%%%%%%%%%%%%%%%%%%%%%%
%%%%%%%%%%%%%%%%%%%%%%%%%%%%%%%%%%%%%%%%%%%%%%%%%%%%%%%%%%%%%%%%%%%%%%%%%%%%%%%
%%%%%%%%%%%%%%%%%%%%%%%%%%%%%%%%%%%%%%%%%%%%%%%%%%%%%%%%%%%%%%%%%%%%%%%%%%%%%%%
%%%%%%%%%%%%%%%%%%%%%%%%%%%%%%%%%%%%%%%%%%%%%%%%%%%%%%%%%%%%%%%%%%%%%%%%%%%%%%%
%%%%%%%%%%%%%%%%%%%%%%%%%%%%%%%%%%%%%%%%%%%%%%%%%%%%%%%%%%%%%%%%%%%%%%%%%%%%%%%
\clearpage
\section{Computing the visibility graph of a set of line segments}\label{CompVisGraphs}
\begin{figure}[!htb]
%%\begin{figure}[p]
\begin{center}
\psfrag{P}{$P = \zeta(O)$}
\psfrag{O}{$O = \zeta^{-1}(P)$}
\psfrag{H}{$\etasection(K)$}
\psfrag{K}{$K$}
\psfrag{shelling}{a permutation $q$ of $P$}
%\psfrag{EK}{$E_K$}
\psfrag{EK}{visibility graph of $K$}
%%\psfrag{BH}{$B_{\etasection(K)}$}
\psfrag{BH}{visibility graph of $\etasection(K)$}
\includegraphics[width = 0.8575\linewidth]{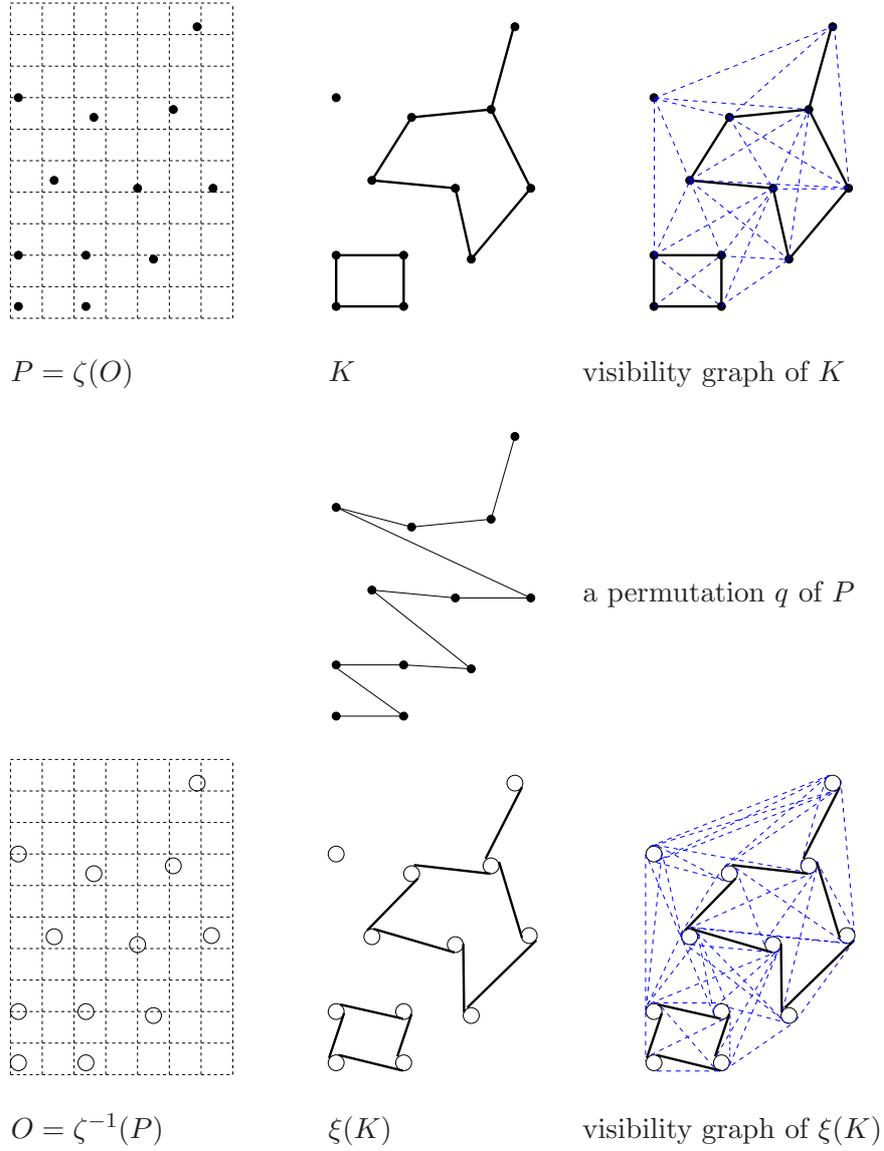}%%%xfigfinal
\end{center}
\caption{\protect \small This figure suggests how visibility graphs of finite planar families of interior disjoint line segments fit into 
the theory of visibility complexes of well-constrained families of convex bodies.
\label{visibilitygraphs}}
\end{figure}

In this section we show how visibility graphs of finite planar families of pairwise interior disjoint line segments (the so-called ``polygons with holes" in the computational geometry literature~\cite{g-vap-07,m-spn-04,or-v-04,m-gspno-00,ags-vp-00})
fit into the theory of visibility complexes of families of pairwise disjoint convex bodies with constraints. 
%To explain this point we now clarify the embedding of the class of chirotopes of families of points 
%into the class of chirotopes of families of convex bodies
%mentioned at the beginning of this section. 
To this end we embed the class of chirotopes of families of points into the class of chirotopes of families of convex bodies as follows.
Define a {\it thin} chirotope as a chirotope of a family of convex bodies with the property that its restrictions to 
subfamilies of convex bodies pierced by a line are chirotopes of  pencils of convex bodies (that is, families of disks  
with same radius and aligned centers of the classical topological plane), 
and let $\zeta$ be the map that assigns to a thin chirotope  
the conjunction  of its left-left and right-right components---we call this way  the maps that assign to a 
triple of indices the subvectors of its associated position vector whose entries are those  defined with respect to the left-left and right-right bitangents, respectively.
It is a simple exercise  to check that the map $\zeta$ realizes a constant time computable one-to-one and onto correspondence between 
the class of thin chirotopes and the class of chirotopes of families of points.
Furthermore for any thin chirotope, say realized by the family of convex bodies $O= \{o_i\}$, the four-to-one map $\eta$ 
that assigns to a bitangent line segment joining the bodies $o_i$ to $o_j$ 
the line segment joining the points $\zeta(o_i)$ to $\zeta(o_j)$ 
has an efficiently computable section $\etasection$ mapping a line segment on an interior bitangent line segment such that  the visibility
graph of the family of points $\zeta(O)$ with respect to a family of constraints $K$ is the image under
$\eta$ of the visibility graph of the family of convex bodies $O$ with respect to the family of constraints $\etasection(K).$
To see this we introduce a permutation  $q = q_1q_2\ldots q_n$ of the $p_i$ with the property 
that there exists a line through $q_{i}$ that separates the convex hull of 
the $q_j$, $1\leq j\leq i$, from the convex hull of the $q_{j'}$, $i<j'\leq
n$---such a permutation is computable in $O(n \log n)$ time as we shall see in the next section---and we define  $\etasection$ as the map that assigns to the line segment joining $\zeta(o_i)$ to
$\zeta(o_j)$ the bitangent line segment joining $o_i$ to $o_j$ whose oriented
version  with respect to the permutation $q$ is right-left.
The reader will easily check that the section $\etasection$ thus defined satisfies the
property mentioned above.
See Figure~\ref{visibilitygraphs}  for an illustration.
Therefore we get the following theorem. 
%%%%%%%%%%%%%%%%%%%%%%%%%%%%%%%%%%%%%%%%%%%%%%%%%%%%%%%%%%%%%%%%%%%%%%%%%%%%%%%
%%%%%%%%%%%%%%%%%%%%%%%%%%%%%%%%%%%%%%%%%%%%%%%%%%%%%%%%%%%%%%%%%%%%%%%%%%%%%%%
%%%%%%%%%%%%%%%%%%%%%%%%%%%%%%%%%%%%%%%%%%%%%%%%%%%%%%%%%%%%%%%%%%%%%%%%%%%%%%%
%%%%%%%%%%%%%%%%%%%%%%%%%%%%%%%%%%%%%%%%%%%%%%%%%%%%%%%%%%%%%%%%%%%%%%%%%%%%%%%
%%%%%%%%%%%%%%%%%%%%%%%%%%%%%%%%%%%%%%%%%%%%%%%%%%%%%%%%%%%%%%%%%%%%%%%%%%%%%%%
%%%%%%%%%%%%%%%%%%%%%%%%%%%%%%%%%%%%%%%%%%%%%%%%%%%%%%%%%%%%%%%%%%%%%%%%%%%%%%%
%%%%%%%%%%%%%%%%%%%%%%%%%%%%%%%%%%%%%%%%%%%%%%%%%%%%%%%%%%%%%%%%%%%%%%%%%%%%%%%
%%%%%%%%%%%%%%%%%%%%%%%%%%%%%%%%%%%%%%%%%%%%%%%%%%%%%%%%%%%%%%%%%%%%%%%%%%%%%%%
%%%%%%%%%%%%%%%%%%%%%%%%%%%%%%%%%%%%%%%%%%%%%%%%%%%%%%%%%%%%%%%%%%%%%%%%%%%%%%%
%%%%%%%%%%%%%%%%%%%%%%%%%%%%%%%%%%%%%%%%%%%%%%%%%%%%%%%%%%%%%%%%%%%%%%%%%%%%%%%

%%%%%%%%%%%%%%%%%%%%%%%%%%%%%%%%%%%%%%%%%%%%%%%%%%%%%%%%%%%%%%%%%%%%%%%%%%%%%%%
%%%%%%%%%%%%%%%%%%%%%%%%%%%%%%%%%%%%%%%%%%%%%%%%%%%%%%%%%%%%%%%%%%%%%%%%%%%%%%%
%%%%%%%%%%%%%%%%%%%%%%%%%%%%%%%%%%%%%%%%%%%%%%%%%%%%%%%%%%%%%%%%%%%%%%%%%%%%%%%
%%%%%%%%%%%%%%%%%%%%%%%%%%%%%%%%%%%%%%%%%%%%%%%%%%%%%%%%%%%%%%%%%%%%%%%%%%%%%%%
%%%%%%%%%%%%%%%%%%%%%%%%%%%%%%%%%%%%%%%%%%%%%%%%%%%%%%%%%%%%%%%%%%%%%%%%%%%%%%%
\begin{theorem} 
The visibility graph of a planar family of $n$ pairwise
interior disjoint line segments presented by the chirotope of the endpoints of the line segments is computable in $O(\text{size of the visibility graph + $n \log n$})$ time and 
linear working space. \qed
\end{theorem}
%%%%%%%%%%%%%%%%%%%%%%%%%%%%%%%%%%%%%%%%%%%%%%%%%%%%%%%%%%%%%%%%%%%%%%%%%%%%%%%
%%%%%%%%%%%%%%%%%%%%%%%%%%%%%%%%%%%%%%%%%%%%%%%%%%%%%%%%%%%%%%%%%%%%%%%%%%%%%%%
%%%%%%%%%%%%%%%%%%%%%%%%%%%%%%%%%%%%%%%%%%%%%%%%%%%%%%%%%%%%%%%%%%%%%%%%%%%%%%%
%%%%%%%%%%%%%%%%%%%%%%%%%%%%%%%%%%%%%%%%%%%%%%%%%%%%%%%%%%%%%%%%%%%%%%%%%%%%%%%
%%%%%%%%%%%%%%%%%%%%%%%%%%%%%%%%%%%%%%%%%%%%%%%%%%%%%%%%%%%%%%%%%%%%%%%%%%%%%%%

We should mention that this mechanism of interpretation of visibility graphs of families of line segments by visibility complexes of families of convex bodies with constraints
 (that is, the embedding $\zeta$, the map $\eta$ and one of its sections $\etasection$) is implemented in the visibility complex package of the CGAL library~\cite{G-ecg:ap-civc-03} 
(see also~\cite[Remark 5]{ap-sstvc-03}) and explicitly used to design an enumeration algorithm for pointed pseudotriangulations of families of points in general position~\cite{G-bkps-ceppg-06}. 
%%%%%%%%%%%%%%%%%%%%%%%%%%%%%%%%%%%%%%%%%%%%%%%%%%%%%%%%%%%%%%%%%%%%%%%%%%%%%%%
%%%%%%%%%%%%%%%%%%%%%%%%%%%%%%%%%%%%%%%%%%%%%%%%%%%%%%%%%%%%%%%%%%%%%%%%%%%%%%%
%%%%%%%%%%%%%%%%%%%%%%%%%%%%%%%%%%%%%%%%%%%%%%%%%%%%%%%%%%%%%%%%%%%%%%%%%%%%%%%
%%%%%%%%%%%%%%%%%%%%%%%%%%%%%%%%%%%%%%%%%%%%%%%%%%%%%%%%%%%%%%%%%%%%%%%%%%%%%%%
%%%%%%%%%%%%%%%%%%%%%%%%%%%%%%%%%%%%%%%%%%%%%%%%%%%%%%%%%%%%%%%%%%%%%%%%%%%%%%%

%%%%%%%%%%%%%%%%%%%%%%%%%%%%%%%%%%%%%%%%%%%%%%%%%%%%%%%%%%%%%%%%%%%%%%%%%%%%%%%
%%%%%%%%%%%%%%%%%%%%%%%%%%%%%%%%%%%%%%%%%%%%%%%%%%%%%%%%%%%%%%%%%%%%%%%%%%%%%%%
%%%%%%%%%%%%%%%%%%%%%%%%%%%%%%%%%%%%%%%%%%%%%%%%%%%%%%%%%%%%%%%%%%%%%%%%%%%%%%%
%%%%%%%%%%%%%%%%%%%%%%%%%%%%%%%%%%%%%%%%%%%%%%%%%%%%%%%%%%%%%%%%%%%%%%%%%%%%%%%
%%%%%%%%%%%%%%%%%%%%%%%%%%%%%%%%%%%%%%%%%%%%%%%%%%%%%%%%%%%%%%%%%%%%%%%%%%%%%%%
\clearpage
\section{Computing a pseudotriangulation of a set of points}\label{ptpoints}
\begin{figure}[!htb]
%%\begin{figure}[p]
\centering
\psfrag{pi}{$p_i$}
\psfrag{qi}{$\qq_i$}
\psfrag{Uqccw}{$\CCWT{\qq}{\epsilon}$}
\psfrag{Uqcw}{$\CCWT{\qq}{-\epsilon}$}
\psfrag{Umqccw}{$\CCWT{-\qq}{\epsilon}$}
\psfrag{PTone}{$\CCWT{\qq}{\pm \epsilon}$}
\psfrag{PQone}{$\CCWT{\pm \qq}{\epsilon}$}
\psfrag{GPT}{$\CCWT{\pm \qq}{\epsilon} \cup \MISS{\qq}{\epsilon}$}
\psfrag{MIS}{$\MISS{\qq}{\epsilon}$}
\psfrag{un}{$1$} \psfrag{de}{$2$} \psfrag{tr}{$3$} \psfrag{qu}{$4$}
\psfrag{ci}{$5$} \psfrag{si}{$6$} \psfrag{se}{$7$} \psfrag{hu}{$8$}
\psfrag{ne}{$9$} \psfrag{di}{$10$} \psfrag{on}{$11$} \psfrag{do}{$12$}
\psfrag{tre}{$13$} \psfrag{qua}{$14$}

\includegraphics[width = 0.8575 \linewidth]{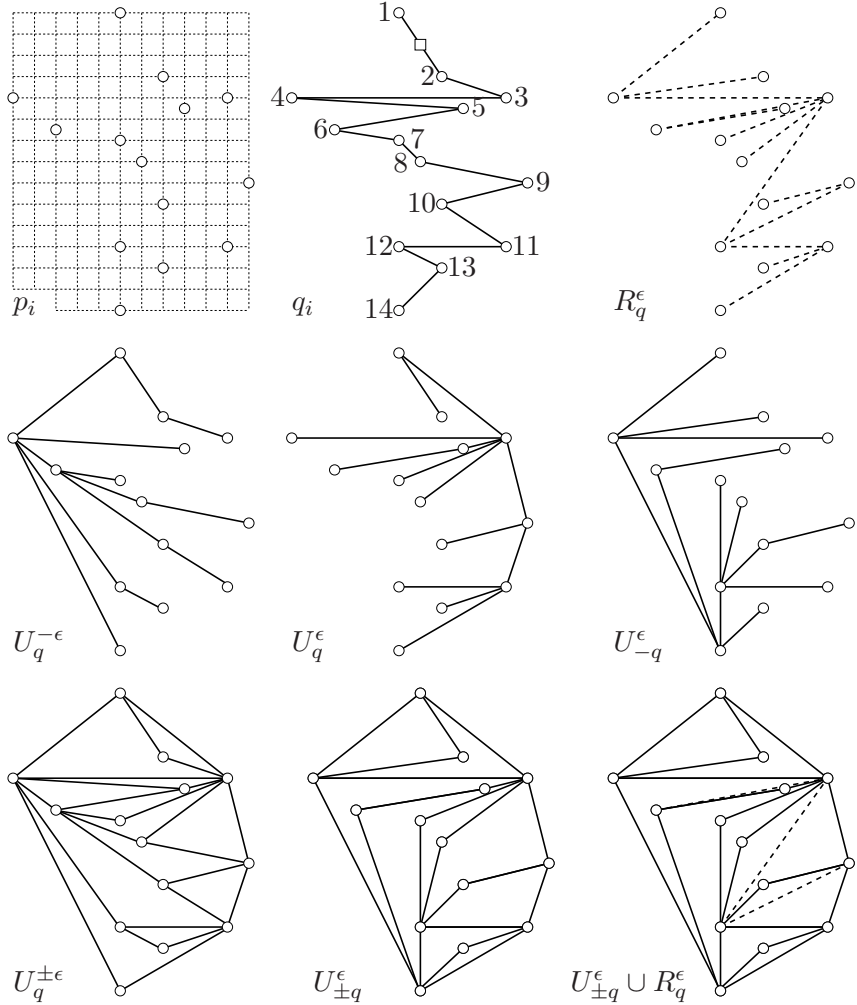}
\caption{\label{ComputingPTP}\protect \small Computing a greedy pseudotriangulation of a set of points.}
\end{figure}

%%%%%%%%%%%%%%%%%%%%%%%%%%%%%%%%%%%%%%%%%%%%%%%%%%%%%%%%%%%%%%%%%%%%%%%%%%%%%%%
%%%%%%%%%%%%%%%%%%%%%%%%%%%%%%%%%%%%%%%%%%%%%%%%%%%%%%%%%%%%%%%%%%%%%%%%%%%%%%%
%%%%%%%%%%%%%%%%%%%%%%%%%%%%%%%%%%%%%%%%%%%%%%%%%%%%%%%%%%%%%%%%%%%%%%%%%%%%%%%
In this section we explain how to compute a pseudotriangulation of a planar family of points presented by its chirotope. 
Let $p_1,p_2,\ldots,p_n$ be a finite planar  family of $n$ points,
let  $\qq = \qq_1\qq_2\ldots\qq_n$  be a permutation of the $p_i$ such that for any index $i$ there is line through $\qq_i$ that separates the convex hull of the $\qq_j$, 
$1\leq j\leq i$, from the convex hull of the $q_{j'}$, $i<j'\leq n$, and let~$\epsilon$ be an orientation of the underlying plane.
Note that such a permutation is computable in $O(n \log n)$ time~: pick an extreme point in $O(n \log n)$ time using your favorite convex hull algorithm and sort the remaining points angularly around it. 
The Graham's scan applied to the sequence $\qq$ boils down to computing explicitly the map 
$\Map{\ccwop{\qq} = \ccwop{\qq}_{\qq,\epsilon}}{\{\qq_2,\qq_3,\ldots,\qq_n\}}{\{\qq_1,\qq_2,\ldots,\qq_{n-1}\}}$ defined inductively by the relations  
\begin{equation}
\begin{cases}
\ccw{}{\qq_2}{\qq} = \qq_1; &  \\
\ccw{}{\qq_i}{\qq} = \ccw{r_i}{\qq_{i-1}}{\qq} 
\end{cases}
\end{equation}
where $r_i$ is the first natural number $r\geq 0$ such that the triangle spanned by ordered triple of points  $\qq_{i}$,$ \ccw{r}{\qq_{i-1}}{\qq}$, 
and $\ccw{r+1}{\qq_{i-1}}{\qq}$ is counterclockwise
or is not defined. 
%%%%%%%%%%%%%%%%%%%%%%%%%%%%%%%%%%%%%%%%%%%%%%%%%%%%%%%%%%%%%%%%%%%%%%%%%%%%%%
%%%%%%%%%%%%%%%%%%%%%%%%%%%%%%%%%%%%%%%%%%%%%%%%%%%%%%%%%%%%%%%%%%%%%%%%%%%%%%%
%%%%%%%%%%%%%%%%%%%%%%%%%%%%%%%%%%%%%%%%%%%%%%%%%%%%%%%%%%%%%%%%%%%%%%%%%%%%%%%
%%%%%%%%%%%%%%%%%%%%%%%%%%%%%%%%%%%%%%%%%%%%%%%%%%%%%%%%%%%%%%%%%%%%%%%%%%%%%%%
For example the following table  (where we write $i$ for  $q_i$) 
\begin{table}[!htb]
$$
\begin{array}{|c||r|r|r|r|r|r|r|r|r|r|r|r|r|}
\hline
i    & 2 & 3 & 4 & 5 & 6 & 7 & 8 & 9 & 10& 11 & 12 & 13 & 14\\
\hline
r_i &0&1&0&1&0&2&1&1&0&1&0&1&1\\
u(i) & 1 & 1 & 3 & 3 & 5 & 3 & 3 & 3 & 9 & 9  & 11 & 11 & 11\\
\hline
s_i &0& 0& 2& 0& 1& 0& 1& 0& 2& 0& 3& 0& 2\\
v(i) & 1 & 2 & 1 & 4 & 4 & 6 & 6 & 8 & 6 & 10 &  4 & 12 & 4\\
\hline 
\end{array}
$$
%\caption{}
\end{table}
depicts  the map $u$ and its relative $v$, obtained by reversing the orientation of the plane, 
associated with the family of 14 points $p_i$ of the real affine plane with  coordinates
$$(0,10),(2,8.5),(5,0),(5,3),(5,8),(5,14),(6,7),(7,2)$$
$$(7,5),(7,11),(8,9.5),(10,3),(10,10),(11,6)$$
and the permutation of the $p_i$ obtained by sorting them according to the decreasing values of their vertical coordinates, 
as depicted at the left top corner of Figure~\ref{ComputingPTP}.
%%%%%%%%%%%%%%%%%%%%%%%%%%%%%%%%%%%%%%%%%%%%%%%%%%%%%%%%%%%%%%%%%%%%%%%%%%%%%%
%%%%%%%%%%%%%%%%%%%%%%%%%%%%%%%%%%%%%%%%%%%%%%%%%%%%%%%%%%%%%%%%%%%%%%%%%%%%%%%
%%%%%%%%%%%%%%%%%%%%%%%%%%%%%%%%%%%%%%%%%%%%%%%%%%%%%%%%%%%%%%%%%%%%%%%%%%%%%%%
%%%%%%%%%%%%%%%%%%%%%%%%%%%%%%%%%%%%%%%%%%%%%%%%%%%%%%%%%%%%%%%%%%%%%%%%%%%%%%%
%%%%%%%%%%%%%%%%%%%%%%%%%%%%%%%%%%%%%%%%%%%%%%%%%%%%%%%%%%%%%%%%%%%%%%%%%%%%%%%
%%%%%%%%%%%%%%%%%%%%%%%%%%%%%%%%%%%%%%%%%%%%%%%%%%%%%%%%%%%%%%%%%%%%%%%%%%%%%%%
%%%%%%%%%%%%%%%%%%%%%%%%%%%%%%%%%%%%%%%%%%%%%%%%%%%%%%%%%%%%%%%%%%%%%%%%%%%%%%%
%%%%%%%%%%%%%%%%%%%%%%%%%%%%%%%%%%%%%%%%%%%%%%%%%%%%%%%%%%%%%%%%%%%%%%%%%%%%%%%
%%%%%%%%%%%%%%%%%%%%%%%%%%%%%%%%%%%%%%%%%%%%%%%%%%%%%%%%%%%%%%%%%%%%%%%%%%%%%%%
%%%%%%%%%%%%%%%%%%%%%%%%%%%%%%%%%%%%%%%%%%%%%%%%%%%%%%%%%%%%%%%%%%%%%%%%%%%%%%%
%%%%%%%%%%%%%%%%%%%%%%%%%%%%%%%%%%%%%%%%%%%%%%%%%%%%%%%%%%%%%%%%%%%%%%%%%%%%%%%
From the map $\ccwop{\qq}_{\qq,\epsilon}$  and its relatives 
$\ccwop{\qq}_{\qq,-\epsilon}$,
$\ccwop{\qq}_{-\qq,\epsilon}$,
$\ccwop{\qq}_{-\qq,-\epsilon}$,
 obtained by reversing the
orientation of the plane or the permutation of the $p_i$ or both, one can easily deduce,
thanks to~\cite[Theorem 12, Claim 3]{G-pv-tsvcp-96}, not
only the convex hull of the $p_i$---under the form of the sequence $\qq_{n}, \ccw{}{\qq_{n}}{\qq}, \ccw{2}{\qq_{n}}, \ldots, \qq_1$ and its relative obtained by reversing the orientation of the plane---but 
two pseudotriangulations (both  pseudotriangulations are pointed in the case where the family of points is in general position).  Indeed 
let $\CCWT{\qq}{\epsilon}$ be the set of line segments $\qq_i\ccw{}{\qq_{i}}{\qq}$, 
%%%%  or equivalently the set of $\qq_i\ccw{r_i}{\qq_{i-1}}{\qq}$, 
$2\leq i \leq n$, 
%%%let $\TRIA{\qq}{\epsilon}$ be the set of line segments $\qq_i\ccw{r}{\qq_{i-1}}{\qq}$, $0\leq r\leq r_i,2\leq i \leq n$. 
let $v_i$ be the undirected version of the right-left bitangent line segment joining
the convex hull of the $q_j$, $i+1\leq j\leq n$, to the convex hull of the $q_j$, $1\leq j \leq i$ and 
let $\MISS{\qq}{\epsilon}$ be the set of $v_i$, $1\leq i \leq  n-1$.
Then, as illustrated in  Figure~\ref{ComputingPTP},
\begin{enumerate}
\item $\CCWT{\qq}{\epsilon}$ is a  spanning tree of the family of $p_i$
%%, $\TRIA{\qq}{\epsilon}$ is a triangulation of the left side of the 
%%path $\qq_1\qq_2\ldots \qq_n$ in the convex hull of the family of $p_i$, $\TRIA{\qq}{\pm \epsilon}$ is a triangulation of the convex hull of the family of $p_i$, 
and  $\CCWT{\qq}{\pm \epsilon}$ is a pseudotriangulation of the family of $p_i$; 
in the case where the points are in general position the tree 
$\CCWT{\qq}{\epsilon}$ is the primal version of the upper horizon tree of Edelsbrunner and Guibas~\cite{eg-tsa-89}; 

\item $\CCWT{\pm\qq}{\epsilon}$ induces a decomposition of the convex hull of
the family of $p_i$ into pseudotriangles and pseudoquadrangles, and its completion by $\MISS{\qq}{\epsilon}$ is a greedy  
pseudotriangulation of the family of $p_i$  (this was first observed in \cite{G-p-htvpt-97}; see also~\cite[Lemma 11]{G-bkps-ceppg-06}).
\end{enumerate}
It is also interesting to mention that if $\TRIA{\qq}{\epsilon}$ denotes  the set of line segments $\qq_i\ccw{r}{\qq_{i-1}}{\qq}$, $0\leq r\leq r_i,2\leq i \leq n$, 
then $\TRIA{\qq}{\pm \epsilon}$ is a triangulation of the convex hull of the family of $p_i$. 

\end{document}